**Athena (Advanced Telescope for High ENergy Astrophysics) Assessment Study Report for ESA Cosmic Vision 2015-2025**

X. Barcons, D. Barret, A. Decourchelle, J.-W. den Herder, T. Dotani, A.C. Fabian, R. Fraga-Encinas, H. Kunieda, D. Lumb, G. Matt, K. Nandra, L. Piro, N. Rando, S. Sciortino, R.K. Smith, L. Strüder, M.G. Watson, N.E. White, R. Willingale, with the help from Athena's Science Working Group, Instrument Working Group Telescope Working Group, Ground Segment Working Group and ESA Study Team.

*Athena* is an X-ray observatory-class mission concept, developed from April to December 2011 as a result of the reformulation exercise for L-class mission proposals, as requested by ESA in the framework of Cosmic Vision 2015.

*Athena*'s science case is that of the Universe of extremes, from Black Holes to Large-scale structure. The specific science goals are structured around three main pillars: "Black Holes and accretion physics", "Cosmic feedback" and "Large-scale structure of the Universe". Underpinning these pillars, the study of hot astrophysical plasmas offered by *Athena* broadens its scope to virtually all corners of Astronomy. *Athena* is conceived to address some of the most important questions envisaged for Astrophysics in the 2020's.

The *Athena* concept consists of two co-aligned X-ray telescopes, with focal length 12 m, angular resolution of 10" or better, and totalling an effective area of 1 $m^2$ at 1 keV (0.5 $m^2$ at 6 keV). At the focus of one of the telescopes there is a Wide Field Imager (WFI) providing a field of view of 24'×24', 150 eV spectral resolution at 6 keV, and capability to handle very high count rates. At the focus of the other telescope there is the revolutionary X-ray Microcalorimeter Spectrometer (XMS), a cryogenic instrument based on superconducting sensors, offering a spectral resolution of 3 eV over a field of view of 2.3' × 2.3'.

At its meeting on May 2, 2012, ESA's Science Programme Committee decided not to select Athena as the L1 class mission for Cosmic Vision 2015-2025. It however resolved to continue with technological developments that will further enhance the scientific performance of the *Athena* payload. Both the science goals and the concept of *Athena*, developed as a true X-ray observatory-class mission, remain very valid and conform the basis of our next proposal to ESA as a flagship mission for the benefit of Astrophysics at large.

# Athena

## The extremes of the Universe: from black holes to large-scale structure

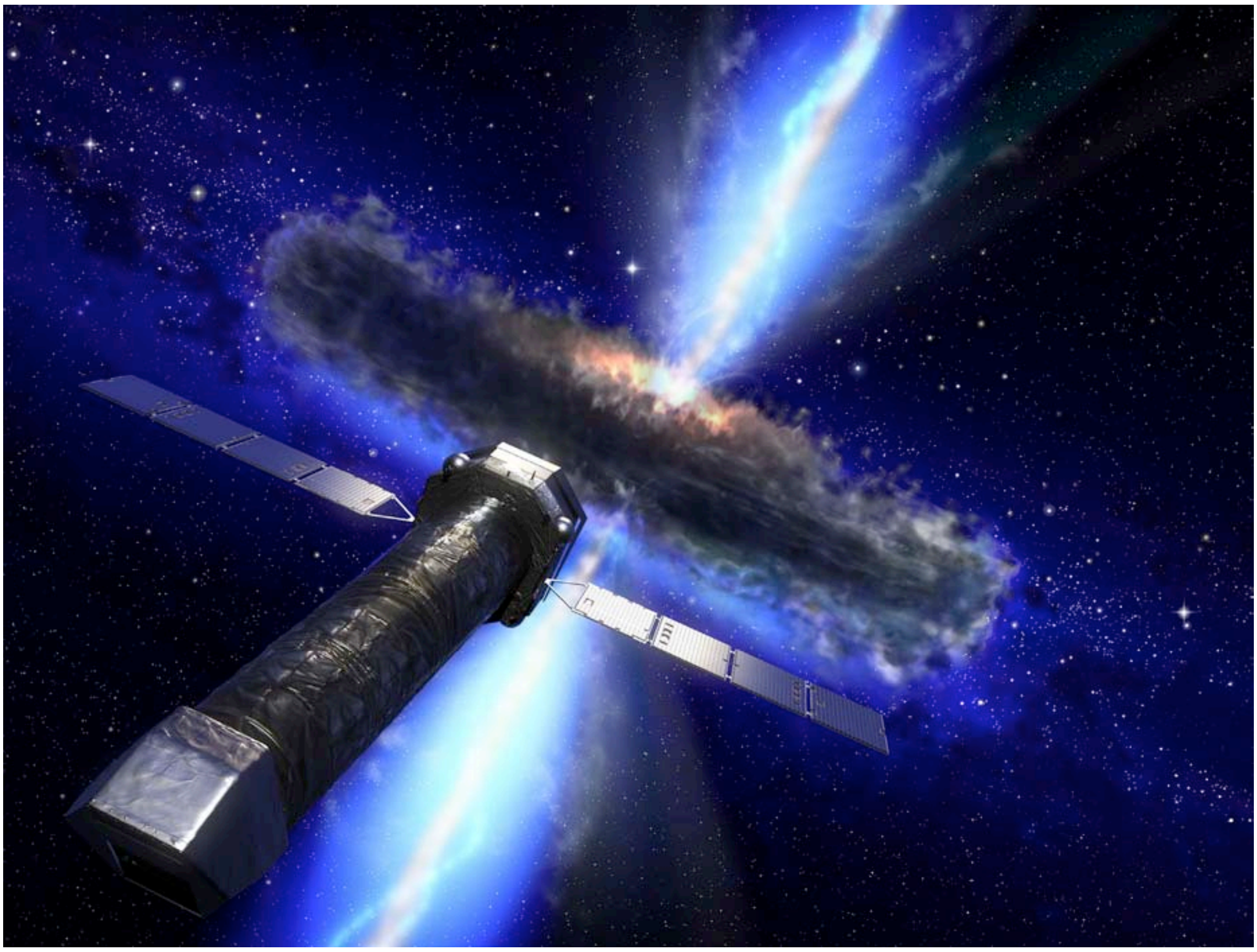

Assessment Study Report

# Authorship

## Athena Study Team

*Chair:* D.H. Lumb (ESA), *Deputy Chair:* K. Nandra (D).
X. Barcons (E), D. Barret (F), A. Decourchelle (F), J-W. den Herder (NL), T. Dotani (JP), H. Kunieda (JP), G. Matt (I), L. Piro (I), N. Rando (ESA), S. Sciortino (I), L. Strüder (D), M.G. Watson (UK), N.E. White (USA), R. Willingale (UK).

## Science Working Group

*Chair:* X. Barcons (E).
D. Barret (F), H. Böhringer (D), J. Bregman (USA), M. Cappi (I), A. Comastri (I), A. Decourchelle (F), J. de Plaa (NL), S. Ettori (I), A.C. Fabian (UK), R. Fraga-Encinas (E), J. Kaastra (NL), K. Nandra (D), T. Ohashi (JP), L. Piro (I), E. Pointecouteau (F), S. Sciortino (I), R.K. Smith (USA), J. Vink (NL), J. Wilms (D).

## Instrument Working Group

*Co-chairs:* J.-W. den Herder (NL), L. Strüder (D).
D. Barret (F), E. Costa (I), A. Stefanescu (D), R. den Hartog (NL), L. Duband (F), G. Fraser (UK), S. Herrmann (D), A. Holland (UK), R. Kelley (USA), P. Lechner (D), O. Limousin (F), D. Martin (ESA), K. Mitsuda (JP), C. Pigot (F), L. Piro (I), S. Triqueneaux (F), J. Wilms (D).

## Telescope Working Group

*Chair:* R. Willingale (UK).
M. Bavdaz (ESA), F. Christensen (DK), M. Collon (NL), P. Friedrich (D), R. Hudec (CZ), H. Kunieda (JP), G. Pareschi (I).

## Ground Segment Working Group

*Chair*: M.G. Watson (UK)
M. Ceballos (E), P. Giommi (I), F. Haberl (D), L. Metcalfe (ESA), C. Motch (F), J. Osborne (UK), S. Paltani (CH).

## ESA Study Team

M. Bavdaz (ESA), D.H. Lumb (ESA), D. Martin (ESA), T. Oosterbroek (ESA), N. Rando (ESA), P. Verhoeve (ESA).

## Additional Science Contributors

J. Aird (USA), C. Argiroffi (I), M. Audard (F), C. Badenes (IS), D. Ballantyne (USA), L. Ballo (E), T. Belloni (I), S. Bianchi (I), F. Bocchino (I), T. Boller (D), V. Braito (UK), E. Branchini (I), N. Brandt (USA), G. Branduardi-Raymont (UK), L. Brenneman (USA), F. Brighenti (I), M. Brusa (D), E. Cackett (UK), N. Cappelluti (I), F.J. Carrera (E), G. Chartas (USA), G. Chon (D), E. Constantini (NL), S. Corbel (F), J. Croston (UK), M. Dadina (I), M. De Becker (BE), N. Degenaar (NL), B. De Marco (I), M. Díaz-Trigo (ESO), C. Done (UK), E.D. Feigelson (USA), W. Forman (USA), F. Fürst (D), L. Gallo (CA), M. García (USA), R. Gilli (I), M. Giustini (I), A. Goldwurm (F), E. Gosset (BE), N. Grosso (F), F. Haberl (D), S. Heinz (USA), R. Hudec (CZ), P. Humphrey (USA), I. Kreykenbohm (D), S. Mateos (E), B. McNamara (Ca), M. Mendez (NL), S. Mereghetti (I), M. Middleton (UK), J.M. Miller (USA), G. Miniutti (E), S. Mohanti (UK), C. Motch (F), Y. Naze (Be), J. Nevalainen (FI), F. Nicastro (I), P. Nulsen (USA), P. O'Brien (UK), L.M. Oskinova (D), A. Paizis (IT), W.N. Pietsch (D), J. Pittard (UK), P. Plucinsky (USA), T. Ponman (UK), G. Ponti (I), D. Porquet (F), D. Proga (USA), J. Pye (UK), P. Ranalli (I), D. Rapetti (DK), G. Rauw (BE), J.N. Reeves (UK), T. Reiprich (D), C.S. Reynolds (USA), G. Risaliti (I), T. Roberts (UK), J. Rodriguez (F), P. Romano (I), R. Salvaterra (I), J. Sanders (UK), M. Sasaki (D), C. Schmid (D), R. Schmidt (D), A. Schwope (D), F. Shankar (D), D. Steeghs (UK), B. Stelzer (I), I. Stevens (UK), Y. Takei (JP), V. Tatischeff (F), R. Terrier (F), F. Tombesi (USA), J.M Torrejón (E), I. Traulsen (D), E. Ursino (I), P. Uttley (UK), L. Valencic (US), C. Vignali (I), D.J. Walton (UK), N. Werner (USA), P. Wheatley (UK), R. Wijnands (NL), D. Wilkins (UK), D. Worrall (UK), A. Young (UK), A. Zoghbi (UK).

# TABLE OF CONTENTS

# Athena Mission Summary

## CORE SCIENCE OBJECTIVES

| Black holes & accretion physics | Cosmic feedback | Large-scale structure of the Universe |
|---|---|---|
| Map the innermost region around accreting black holes and other compact objects; probe matter under strong gravity and high density conditions. | Reveal the physics of feedback from AGN and starbursts on all scales; quantify supermassive black hole growth and its relationship to galaxy evolution. | Trace the formation and evolution of large-scale structure via hot baryons in galaxy clusters, groups and the intergalactic medium comprising the cosmic web. |

Diagnose hot cosmic plasmas in all astrophysical environments.

## X-RAY TELESCOPE

- Two co-aligned X-ray telescopes
- 12 m Focal length
- 1 $m^2$ effective area at 1 keV
- 10” baseline resolution (with a 5” goal)
- 0.5 $m^2$ effective area at 6 keV
- Mirror assembly 650 kg
- Hierarchical assembly of ~ 500 modules
- Silicon pore optics already demonstrated in flight configuration

Two X-ray telescopes created from co-aligned silicon pore optics modules

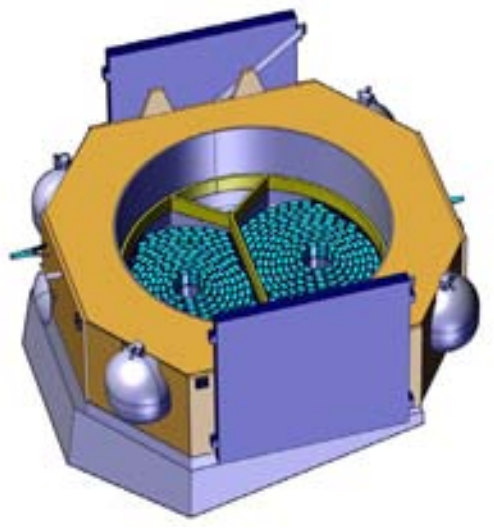

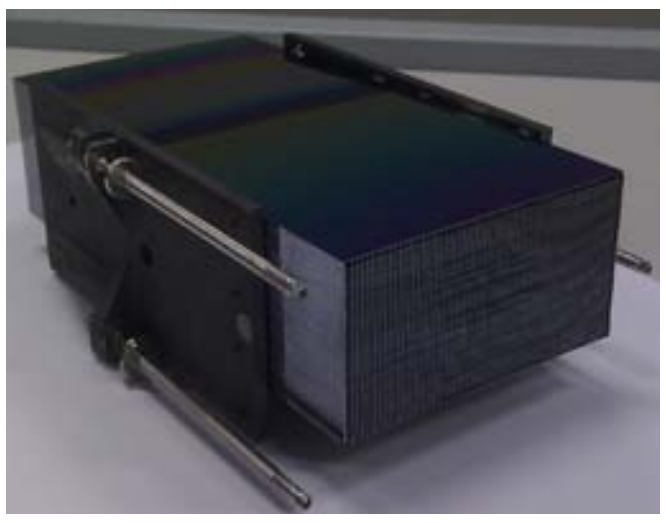

Silicon pore optics module

## INSTRUMENT PAYLOAD

| Instrument | Detector Type | Field of View (arcmin) | Energy Resolution (eV FWHM @ 6 keV) | Bandpass (keV) |
|---|---|---|---|---|
| X-ray Microcalorimeter Spectrometer (**XMS**) | Transition edge sensor/bolometer array | 2.3 x 2.3 ( goal of 3 x 3) | 3 ( goal of 2.5 ) | 0.3-12 |
| Wide Field Imager (**WFI**) | DePFET Active Pixel Sensor | 24 x 24 (goal of 28 x 28) | 150 (goal of 125) | 0.1-15 |

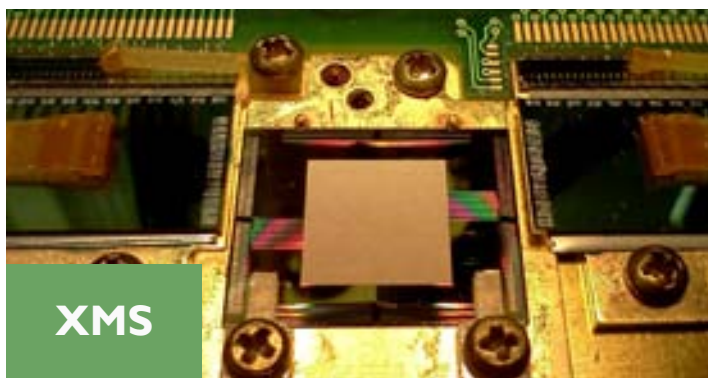

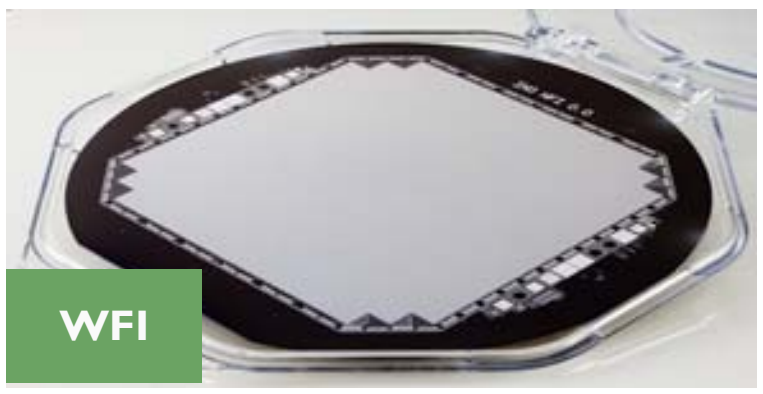

## SYSTEMS

| *System* | *Mass (Kg)* | *System* | *Power (W)* |
|---|---|---|---|
| Optics/Mirror assembly | 760 | Optics - thermal | 1450 |
| Instruments | 490 | Instruments (max) | 1190 |
| Structure/Thermal | 1150 | TT & C (max) | 100 |
| Avionics & Power | 390 | Data Handling | 180 |
| Propulsion | 90 | Propulsion | 60 |
| Spacecraft dry mass | 2880 | AOCS | 120 |
| Propellant | 405 | Structures, thermal, mechanism | 895 |
| Launch Adapter | 200 | Power Subsystem | 100 |
| Total system and maturity margin | 865 | Total system and maturity margin | 910 |
| **Total** | **4150** | **Total** | **5005** |

## SPACECRAFT

**Launcher:** Ariane V **Launch Date:** 2022

**Orbit:** L2 halo, 750,000 km amplitude

**Lifetime:** Design 5 yrs, consumables 10 yrs

**Optical Bench:** XMM-like metering structure

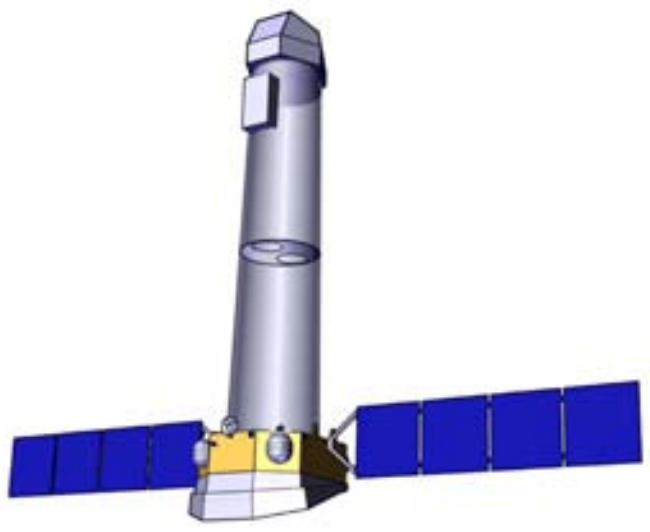

**Pointing:** 3-axes stabilised 1.5” measurement

**High Gain Antenna:** Steerable X-band

**Data Rate:** ~100 Gbit/day

**Observations:** ~500/year

**Ground Station:** New Norcia

**Launch capability to L2:** 6.6 tonnes

# 1 Executive Summary

*Athena* is an observatory designed to address a suite of key questions in modern astrophysics. It will determine how the process of accretion works by mapping the innermost flows around black holes, measure their spins, and determine the equation of state of ultradense matter in the cores of neutron stars; it will measure the energy flows giving rise to cosmic feedback, quantify the growth of supermassive black holes and the evolution of nuclear obscuration over cosmic time, and determine velocity and metallicity flows due to starburst superwinds; it will measure the evolution of the intracluster medium through temperature, metallicity and turbulent velocity changes with redshift, constrain dark energy as a function of redshift using clusters of galaxies and reveal the missing baryons at low redshift locked in the warm and hot intergalactic medium; in addition it will determine the physical conditions in hot plasmas covering a wide range of objects and phenomena, with profound impacts on astrophysics, from stars and planets, through supernovae and the Galactic Centre, out to other galaxies and the distant Universe.

Achieving these ambitious goals requires an X-ray observatory-class mission delivering a major leap forward in high-energy observational capabilities. Thanks to its revolutionary optics technology and the most advanced X-ray instrumentation, *Athena* will deliver superior high resolution X-ray imaging, timing and spectroscopy capabilities, far beyond those of any existing or approved future facilities. Like *XMM-Newton* today, *Athena* will play a central role in astrophysical investigations in the next decade. *Athena* will provide a unique perspective on the Universe, only accessible from space, but will be complemented by the suite of major facilities, both on the ground and in space, which will be operating at other wavelengths in the same timeframe. No other observatory-class X-ray facilities are programmed for that timeframe, and therefore *Athena* will provide our only view of the hot and extreme Universe, leaving a major legacy for the future.

The heart of the mission is the X-ray optical system utilising the innovative silicon pore optics (SPO) technology pioneered in Europe. The *Athena* science goals require a 1 $m^2$ effective collecting area with 10 arcsec angular resolution, achieved using an assembly of two fixed 12 m focal length telescopes. With respect to the current science goals, the science capabilities of *Athena* would be significantly boosted by a better angular resolution, and therefore a goal of 5 arcsec is kept. The considerable investment in the SPO technology by ESA has already demonstrated sub-10 arcsec angular resolution in a flight-like assembly, consistent with the mission scientific requirements and with a clear development plan towards the goal of 5 arcsec.

The baseline instrument complement consists of the X-ray microcalorimeter spectrometer (XMS) for high-resolution spectroscopic imaging, and a Wide Field Imager (WFI) consisting of an active pixel sensor camera with high count-rate capability and moderate resolution spectroscopic capability.

*Athena* will be placed in orbit at L2, which provides uninterrupted viewing and an ideal thermal environment. The design assumes a 5-year mission lifetime, but has consumables for at least 10 years. Both ESA and industry studies have concluded that the *Athena* spacecraft can be built with mature technologies. All spacecraft subsystems use established hardware with substantial flight heritage, with most components being "off-the-shelf." The spacecraft concept is robust, with all *Athena* resource margins meeting or exceeding requirements.

The assessment studies have confirmed that the technology readiness levels of both the instruments and the spacecraft are appropriate to proceed to a definition phase in 2012 and will reach TRL ≥ 5 by 2013 for all mission elements. If the *Athena* Phase A starts in 2012, then a mission implementation schedule allows for a launch in 2022. The schedule and management approach are also described in this report.

The costing has been performed on a system design that has evolved significantly since the IXO report. The simpler design and the technological developments substantially reduces estimated risk, and removes the need for a major strategic investment from partner agencies. The instruments will be developed by ESA member states. Whilst a clear path to a European-only solution has been defined, contributions from both NASA and JAXA to the XMS can bring an improvement in TRL and heritage to some elements of the instrument. A Technology Development Plan is in place to improve the performance of the telescope optics towards the goal of 5 arcsec, and ensure that TRL $\geq$ 5, will be achieved by the end of 2013.

*Athena* will address several of the most prominent themes posed by ESA's Cosmic Vision 2015-2025, especially *Matter under Extreme Conditions, The Evolving Violent Universe* and *The Universe taking Shape*. Following the outstanding contributions of current facilities like ESA's *XMM-Newton*, NASA's *Chandra* and JAXA's *Suzaku*, *Athena* is the mission that contemporary astrophysics needs to secure progress in the science areas uniquely addressed in the X-ray band, but beyond the reach of current or planned X-ray observatories.

# 2 Athena Science Objectives

## The Extremes of the Universe: from Black Holes to Large-scale structure

*Athena* is an observatory class X-ray astrophysics mission which will provide a spectacular leap forward in scientific capabilities compared with its predecessors (e.g. *XMM-Newton, Chandra, Suzaku, ASTRO-H**). This advance is required to address fundamental questions about the physical processes which shape our Universe. *Athena* will impact astrophysics in the broadest sense, from the solar system to stars, black holes and compact objects, supernovae and their remnants, galaxies, clusters of galaxies, and cosmology. While retaining the extremely broad scope of an observatory, *Athena* has been conceived to address the following specific goals:

- Determine how matter accretes onto black holes and other compact objects, in environments where strong gravity effects described by General Relativity apply.
- Reveal the physics underpinning cosmic feedback, and show how it relates the growth of supermassive black holes to the evolution of galaxies.
- Trace the formation and evolution of large-scale structure via the properties of hot baryons in clusters of galaxies and the cosmic web.
- Determine the physical conditions in hot cosmic plasmas in all astrophysical environments, with fundamental impacts ranging from the solar system, to stars, galaxies and beyond.

These are amongst the most important topics in the modern astrophysical landscape, and have inspired a number of ESA's Cosmic Vision (CV) science objectives (*Cosmic Vision: Space Science for Europe* 2015-2025, ESA BR-247): *(Q3.3) Matter Under Extreme Conditions, (Q4.3) The Evolving Violent Universe, (Q4.2) The Universe taking shape*. *Athena* will also have a major impact on several other CV themes, including *(Q4.1) The Early Universe* (which includes the investigation of the Dark Energy that dominates the content of today's Universe), by enabling measurements of dark matter and dark energy using galaxy clusters. It will also shed light onto *(Q2.1) From the Sun to the edge of the Solar System*, by probing the solar wind across the solar system through its effects in cometary tails, and *(Q2.2) The giant planets and their environments,* via spectroscopic measurements of the atmospheres and exospheres of Jupiter, Saturn and other planets. A large X-ray observatory like *Athena* also features prominently in the Science Vision for European Astronomy put forward by Astronet (www.astronet-eu.org).

* *Astro-H* is a Japanese X-ray mission scheduled for launch in 2014 and which includes important European contributions

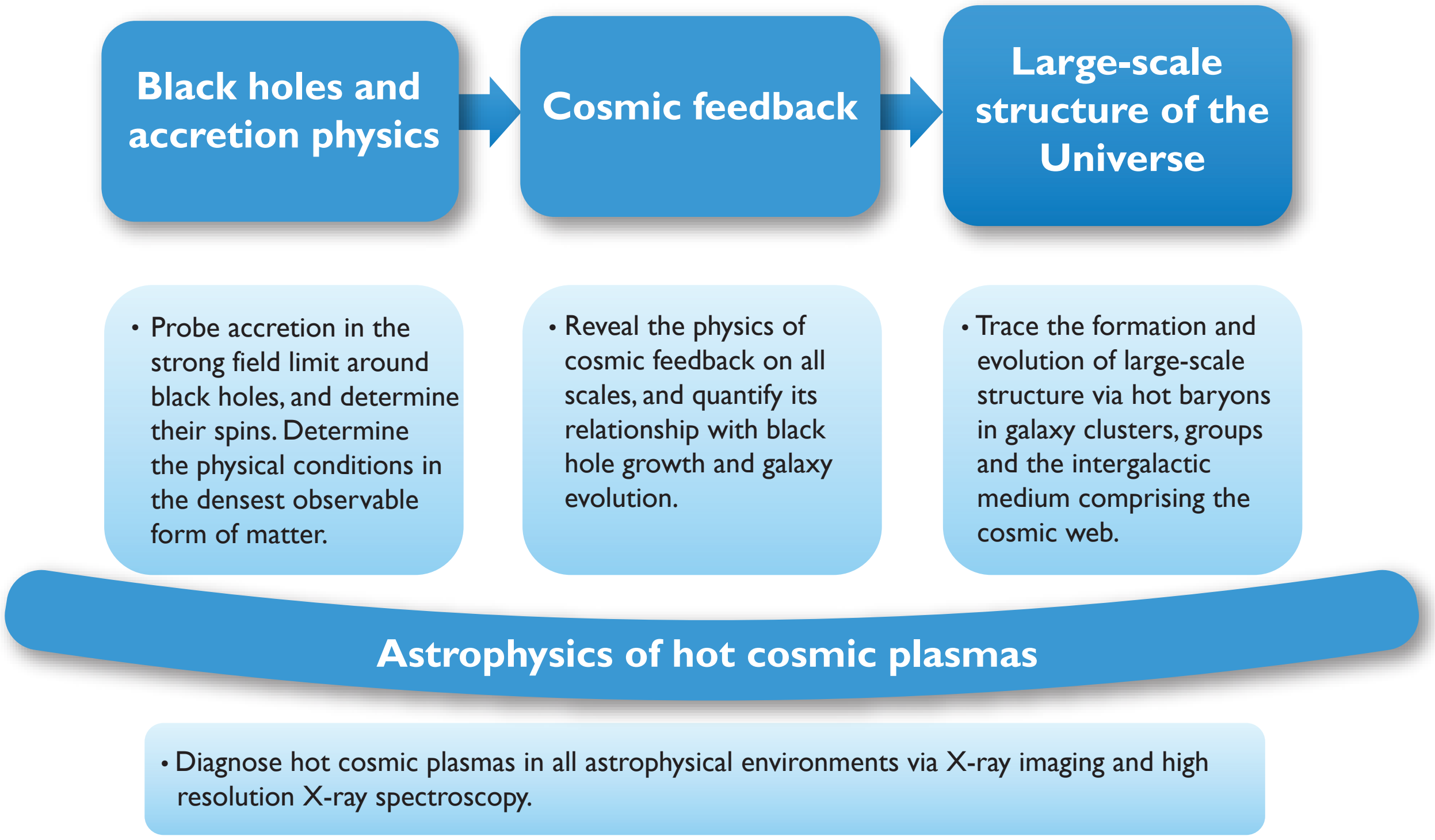

**Figure 2.1.** *This figure shows how the Athena science goals are presented in this section, based on the three main pillars - from small scales on the left to the largest scales on the right, with cosmic feedback in the middle providing the link. Underpinning the above goals, the study of hot plasmas on all scales supports our understanding of those and other fundamental scientific questions in contemporary astrophysics.*

The science case of *Athena* is structured around three main scientific pillars, which drive the mission requirements. The first is the physics of accretion around black holes, neutron stars and other compact objects. The second is cosmic feedback, the process by which accretion and star formation are linked to the formation and evolution of galaxies. The third topic is the physical nature and evolution of the large-scale structure itself, as mapped through the hot baryonic component. In addition to these core science pillars, *Athena* will provide a major leap forward in our understanding of the nature and influences of hot cosmic plasmas in all classes of astrophysical objects, through spatially resolved high-resolution X-ray spectroscopy and sensitive wide field X-ray imaging. The scheme of the science goals is shown in Figure 2.1.

X-ray observations conducted by *Athena*'s instruments will deliver the sky position, energy and arrival time for each single photon detected. These data will enable astronomers to obtain not only X-ray images of the sky, but spectra to derive the physical conditions at work in the sources under investigation as well as to study their variability, offering enormous potential for new astrophysical investigations (see, e.g. Figure 2.4).

**Figure 2.2.** *Artist's illustration of Athena's science goals, linking phenomena occurring around black holes to the large-scale structure of the Universe through cosmic feedback. The cartoon also illustrates some local and distant environments where Athena's observations will reveal the state of astrophysical hot plasmas.*

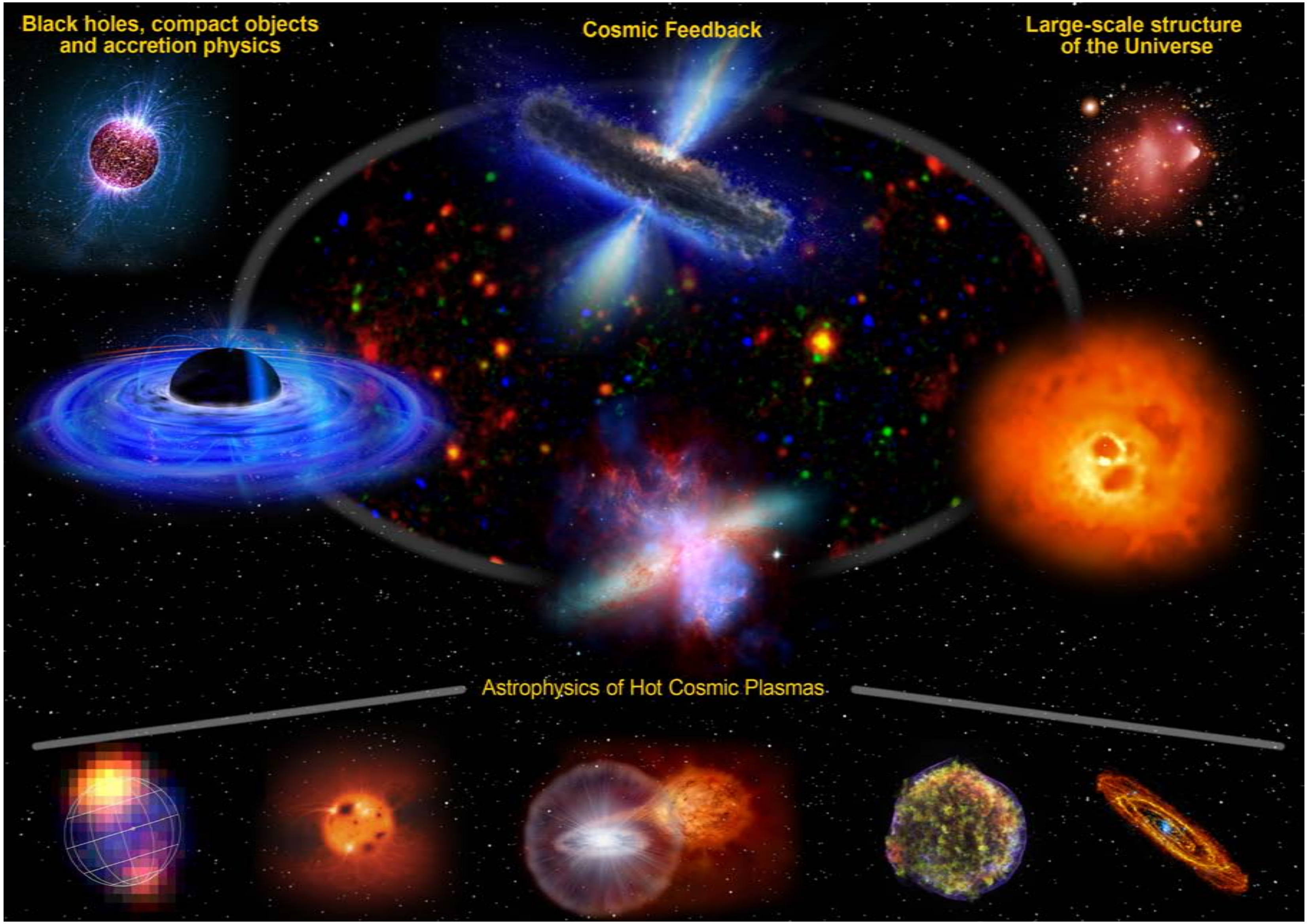

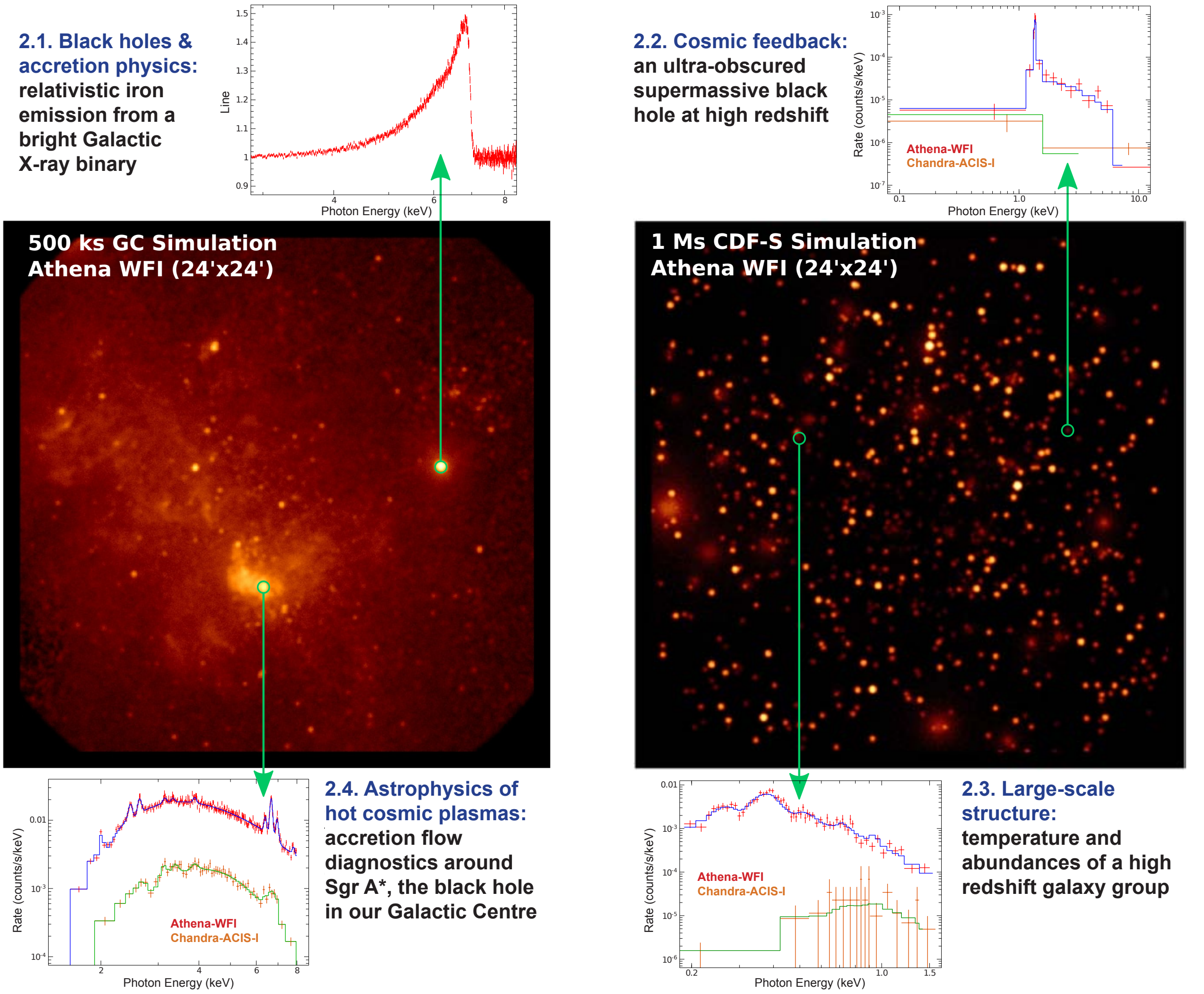

**Figure 2.3.** *Athena will provide a new perspective on both the Galactic and extragalactic sky at high energies. Simulated WFI images are shown of a field centred on Sgr A*, the black hole in Galactic Centre (GC, left, 500 ks exposure) and an extragalactic survey field, the Chandra Deep Field South (CDFS, right, 1 Ms). Compared to current facilities, Athena will move beyond simple source detection into the spectral domain, where the real astrophysics of the sources is revealed. Four examples of simulated Athena spectra are shown in the outer panels (red points, with blue model), compared where appropriate to Chandra-ACIS spectra (orange points, green model). Note that XMM-Newton often cannot access the objects of interest because of source confusion either due to crowding (GC) or extreme faintness (CDFS). Moving clockwise, it is shown:* ***Top Left:*** *a broad iron Kα line, whose shape reflects that it has been generated in the general relativistic environment of a black hole (Section 2.1). No Chandra simulation is shown, as the source is too bright to be observed by ACIS without pileup. This requires the high countrate capability of Athena-WFI;* ***Top Right:*** *an ultra-obscured supermassive black hole in the centre of a distant galaxy (Section 2.2). The telltale signature of obscuration, an intense iron Kα line, is revealed by Athena but invisible with Chandra.* ***Bottom Right:*** *a high redshift galaxy group, barely detected by Chandra, but with clear emission lines in the WFI spectrum which trace the thermal and chemical evolution of hot baryons throughout cosmic time (Section 2.3).* ***Bottom Left:*** *the Galactic Centre itself. Athena will reveal the astrophysics of hot cosmic plasmas (Section 2.4), including those associated with the black hole in our own back yard.*

Beyond the science goals described in this document, *Athena* will therefore have enormous potential for breakthrough discoveries that we cannot foresee given our current knowledge of the Universe. *Athena*'s instruments will have a survey capability far better than *XMM-Newton and Chandra*, thanks to its unique combination of angular resolution, effective area and field of view (see Figure 2.4). The limiting flux will be a factor of ~5 fainter than that of *XMM-Newton* for the same exposure time. The deep field sensitivity is around 4 x $10^{-17}$ erg $cm^{-2}$ $s^{-1}$ in 1 Ms, one order of magnitude deeper than deepest confusion-limited *XMM-Newton* exposures and similar to *Chandra*, but over a much larger field of view. *Athena*-WFI will detect about 3 times more sources than *Chandra* in a single pointing for the same exposure time and, with many more photons at a given flux, will provide for the first time meaningful spectral information for faint, deep Universe objects (Figure 2.3). *Athena*-WFI will have time resolution a factor >10 better than *XMM-Newton* (and indeed *Chandra* and *ASTRO-H*), and a count rate capability for sources 50 times brighter, therefore enabling spectral-timing studies of the bright sources in the X-ray sky for the first time (Figure 2.3)

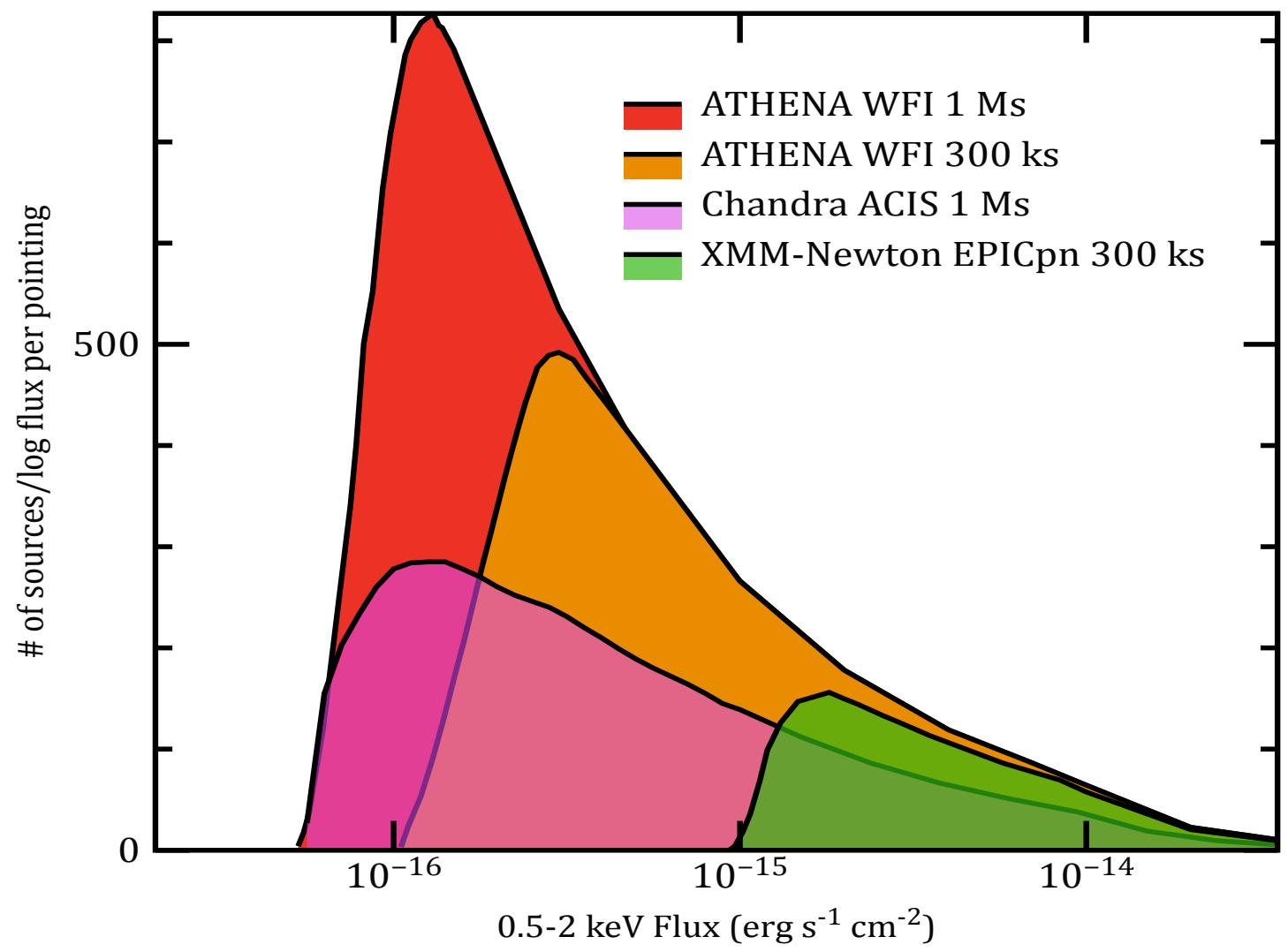

**Figure 2.4.** *Number of sources per logarithmic flux interval expected in single Athena-WFI pointings at high galactic latitudes, compared to Chandra and XMM-Newton. Flux sensitivity has been taken into account across the field of view, and source counts have been adopted from Georgakakis et al. (2008).*

In terms of high-resolution spectroscopy, *Athena* will be capable of detecting spectral lines more than ten times weaker than its existing or planned predecessors (see Figure 2.5). The XMS provides a huge advance over the *XMM-Newton* and *Chandra* gratings, and while the *ASTRO-H*-SXS calorimeter is a revolutionary instrument, it will also be severely photon-limited. True spectroscopic studies at high resolution prior to *Athena* will thus be limited largely to the local Universe.With its far greater collecting area, *Athena*-XMS will reach at least an order of magnitude fainter, opening up X-ray spectroscopy to sources at meaningful cosmological distances.

**Figure 2.5.** *Figure of Merit for weak spectral line detection of X-ray high-spectral resolution spectrometers, derived from the number of counts per independent spectral bin. The gratings line represents the best of the current XMM-Newton or Chandra gratings at each energy.*

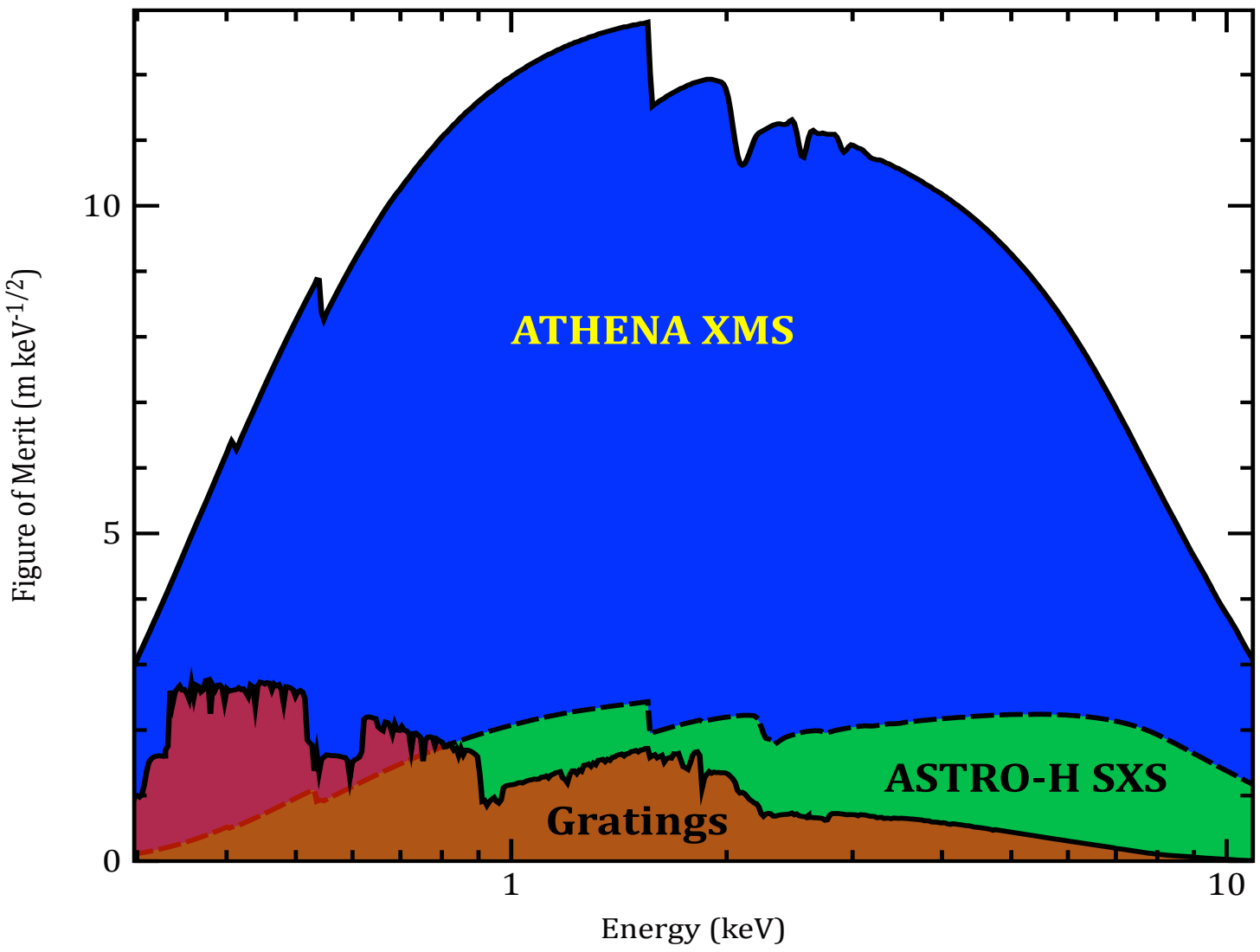

Unlike current *XMM-Newton* and *Chandra* gratings, *Athena* will be able to perform high energy resolution spatially resolved spectroscopy on extended sources. In comparison with the spatially resolved high-resolution spectroscopy that *ASTRO-H* calorimeters will enable for the first time, *Athena*-XMS will not only feature significantly higher spectral resolution as shown in Figure 2.6, but also much higher spatial resolution. This enhancement in high spatial resolution, high-resolution spectroscopy, will bring new science only attainable with *Athena* (e.g. feedback in cluster cores, the thermal history and velocity structure of clusters, chemical enrichment, supernova remnants, etc.)

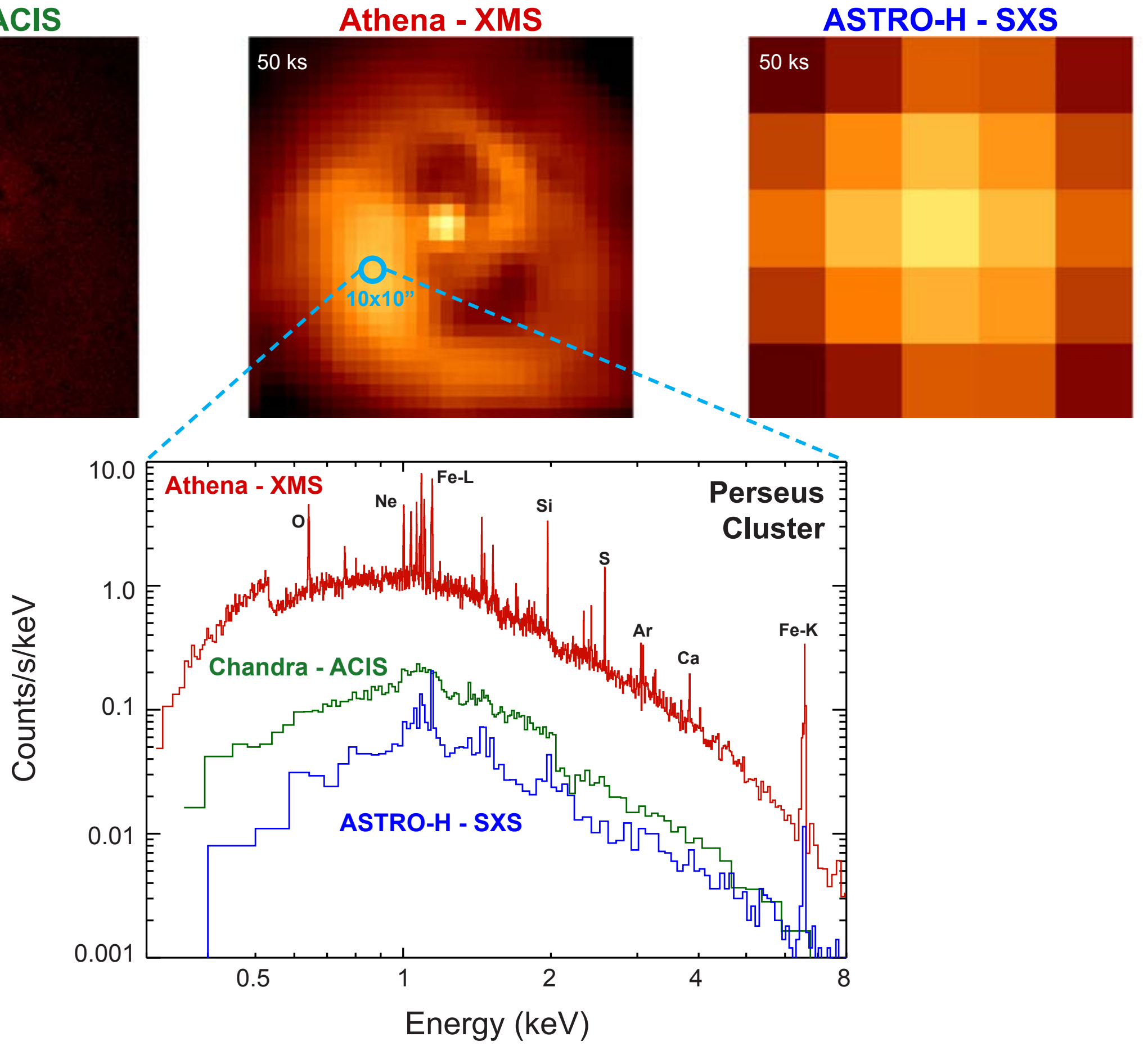

**Figure 2.6.** *Athena will bring unique new capabilities for high resolution imaging spectroscopy. The Perseus cluster core, in which the effects of cosmic feedback are directly observed, is shown for 50 ks observations by: Chandra ACIS-I (**top left**, real data), Athena XMS (**top centre**, simulation) and ASTRO-H SXS (**top right,** simulation). The **bottom panel** shows the spectrum from a 10x10 arcsec region scaled with the spectral resolution and effective area for each instrument (note that in reality ASTRO-H will not be able to resolve a region this small). While Chandra's angular resolution is sufficient to reveal the cavities sculpted in the interstellar medium by the black hole, the spectral resolution is too coarse to reveal the emission lines. Although ASTRO-H has excellent spectral resolution, it cannot resolve the cavities spatially, and has insufficient effective area to detect the bulk of the emission lines. Only Athena will deliver the unique combination of imaging, spectroscopy and effective area necessary to spatially and spectrally resolve the cluster core, revealing the gas motions and turbulence caused by the central AGN.*

While *Athena* has been designed to answer the questions posed in this science case, previous experience has shown that order-of-magnitude leaps forward in capabilities will unquestionably make new breakthroughs in astrophysical understanding.

## 2.1 Black holes, compact objects and accretion physics

Some of the most extreme physical conditions in the Universe are found in and around compact objects. Black holes (BH) create the strongest gravitational fields, while the densest observable form of matter is found in the centres of neutron stars (NS). Matter trapped in the extreme gravitational potential of these objects is spun and heated, causing it to radiate copiously in X-rays. X-ray observations thus provide a unique probe of the innermost regions around BHs and NSs, revealing how the laws of General Relativity (GR) apply in the strong-field limit and how matter behaves under such intense gravity and density. The physical process of accretion is the second most powerful source of energy since the Big Bang, after nuclear fusion in stars, and is 10-30 times more efficient at releasing energy than nuclear fusion. Through feedback (see Section 2.2), it can have a profound effect on galactic and even cosmological scales. In spite of its enormous influence and power, however, we have only rudimentary knowledge of how the accretion process operates.

With a well-balanced set of capabilities, *Athena* will bring about a corresponding revolution in our understanding of the innermost regions around BHs and NSs, and the physical processes which govern their properties and behaviour. A key design goal of the telescope system is the delivery of a large effective area in the critical 6 keV energy regime, where distinctive features are imprinted on the X-ray spectrum and sculpted in the extreme space-time of black holes and neutron stars. The *Athena* focal plane instruments then work in tandem: the XMS instrument provides high spectral resolution to measure sharp spectral features, while the WFI provides sufficient spectral resolution to measure broad features and a broad bandpass to determine the continuum with maximum sensitivity. The WFI also has the necessary high count rate capability and time resolution needed to measure strong gravitational effects for the brightest black hole and neutron star binaries. *Athena*'s specific goals are to:

- **Determine how black holes work, and measure their spins**

  X-ray emission from material just before its final plunge into the event horizon of a black hole shows distinctive spectral and timing properties which depend on the strong gravity environment, the nature of the accretion flow, and the black hole spin. The high spectral resolution of the XMS and the unique spectral and timing capabilities of the WFI will allow us to map the flow in the strong gravity regime in both stellar mass objects and supermassive black holes (SMBHs) in galactic nuclei. The space-time around the black hole depends on its spin, and this alters the spectral and timing signatures. By revealing these effects, *Athena* will measure the spins of hundreds of black holes, again over the full range of black hole masses. Pure spectroscopy (XMS prime), pure timing (WFI prime), and spectral-timing (WFI/XMS combined) will yield multiple redundant spin measurements.

- **Reveal the physics of accretion onto all types of compact object**

  Accretion is a process of fundamental importance in astrophysics and *Athena*'s unique range of capabilities will allow us to probe the physics of that process over the full range of masses, accretion rates and compact object classes. *Athena's* angular resolution is sufficient to pinpoint X-ray from Sgr A*, the least luminous known SMBH, situated in our own galaxy. The spatial resolution and wide field of view will reveal ultraluminous X-ray sources in external galaxies, which apparently push past the Eddington limit at the very opposite end of the accretion rate scale. The XMS spectral resolution will tell us how black holes are fed, by measuring velocity fields of gas at parsec scales in active galactic nuclei (AGN), and winds in compact X-ray binaries.

- **Determine the Equation of State of ultradense matter in neutron stars**

  The same techniques used to map black hole accretion flows can be applied to accreting neutron stars, yielding the radius of the NS both in physical units, and in terms of the gravitational radius, $R_{g.}$ This yields constraints on the mass-radius relationship, and hence the Equation of State of the supra-nuclear material in NS cores, with major implications for Quantum Chromodynamics theory. With many NS binaries being extremely bright (especially in outburst), high count-rate capability provided by the WFI will be crucial for these measurements. Novel techniques such as the measurement of gravitational redshifts during X-ray bursts will also be widely applicable for the first time, thanks to the spectral resolution of the XMS.

The ability of *Athena* to perform these investigations into compact objects and accretion physics is described in more detail in the following sections.

### 2.1.1 How do black holes work? Mapping the innermost accretion flows

Most of the power from quasars originates from matter spiralling into the black hole through a geometrically thin, dense accretion disc. The 1/r potential means that most of the power emerges from the smallest radii closest to the black hole. Quasi-blackbody radiation from the disc emerges in the UV. A hot coronal region, probably powered by disc magnetic fields, lies above the inner parts of the disc, scattering some of the blackbody photons up in energy to form a power-law continuum in the X-ray band. The corona is seen both directly and indirectly, by irradiating and being back-scattered by the disc (Figure 2.7). The "reflection" spectrum has a wealth of spectral and timing features, particularly around the iron-K line at 6.4 keV.

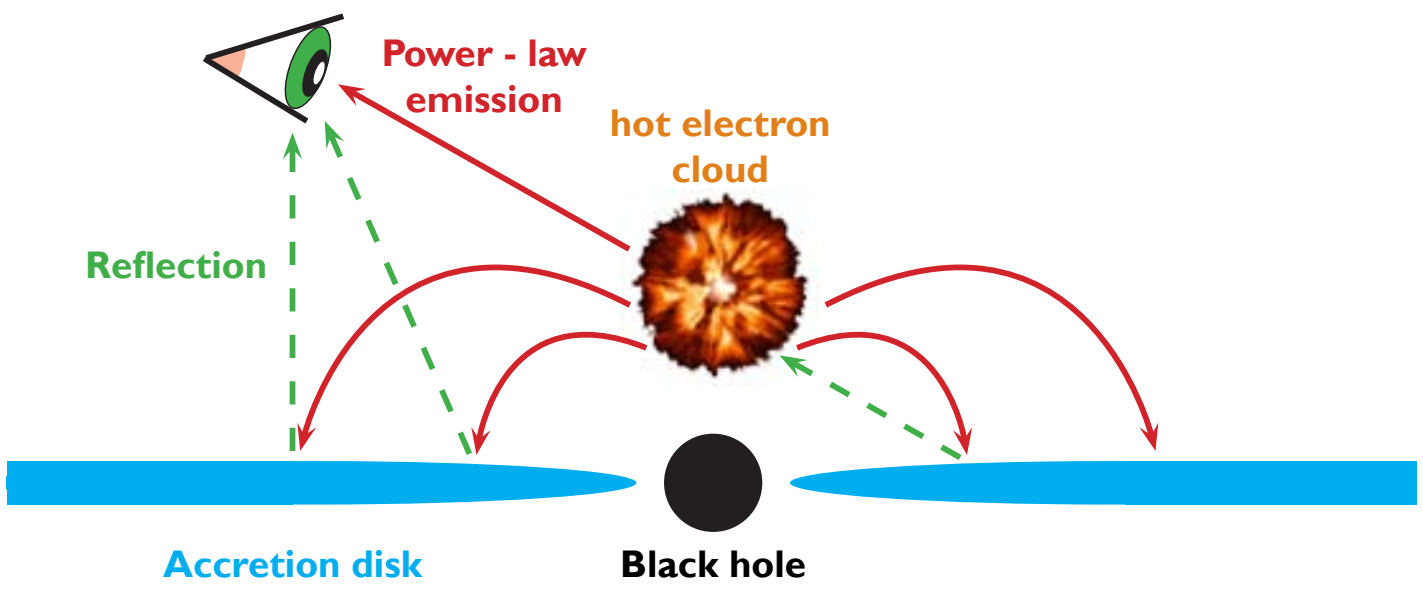

**Figure 2.7.** *Cartoon of a black hole and accretion disk (blue) above which the power-law continuum is generated by Comptonisation of thermal disk photons in a hot electron cloud (orange). Power-law emission (red) irradiates the disk producing the reflection spectrum (green). Intrinsic variations in the power-law source create delayed, blurred, changes (reverberation) in reflection. Thanks to Athena it will be possible to translate these signals into a physical picture of the inner region around the black hole. The reflection features and thermal disk emission both provide an independent estimate of the inner accretion disk radius via the maximum gravitational redshift of emission features, and the disk temperature.*

These features contain unique signatures of the strong gravitational field of the BH and the accretion flow in the inner regions yielding on a time average broad, skewed emission lines (Fabian et al. 1989; Laor 1991), the shape of which is controlled by extreme Doppler effects caused by the swirling gas, large gravitational redshifts due to the enormous depth of the potential well, and strong gravitational light bending. The extent of the skewness depends on the location of the innermost stable circular orbit (ISCO) around the BH, which in turn depends on the black hole spin (see Figure 2.8).

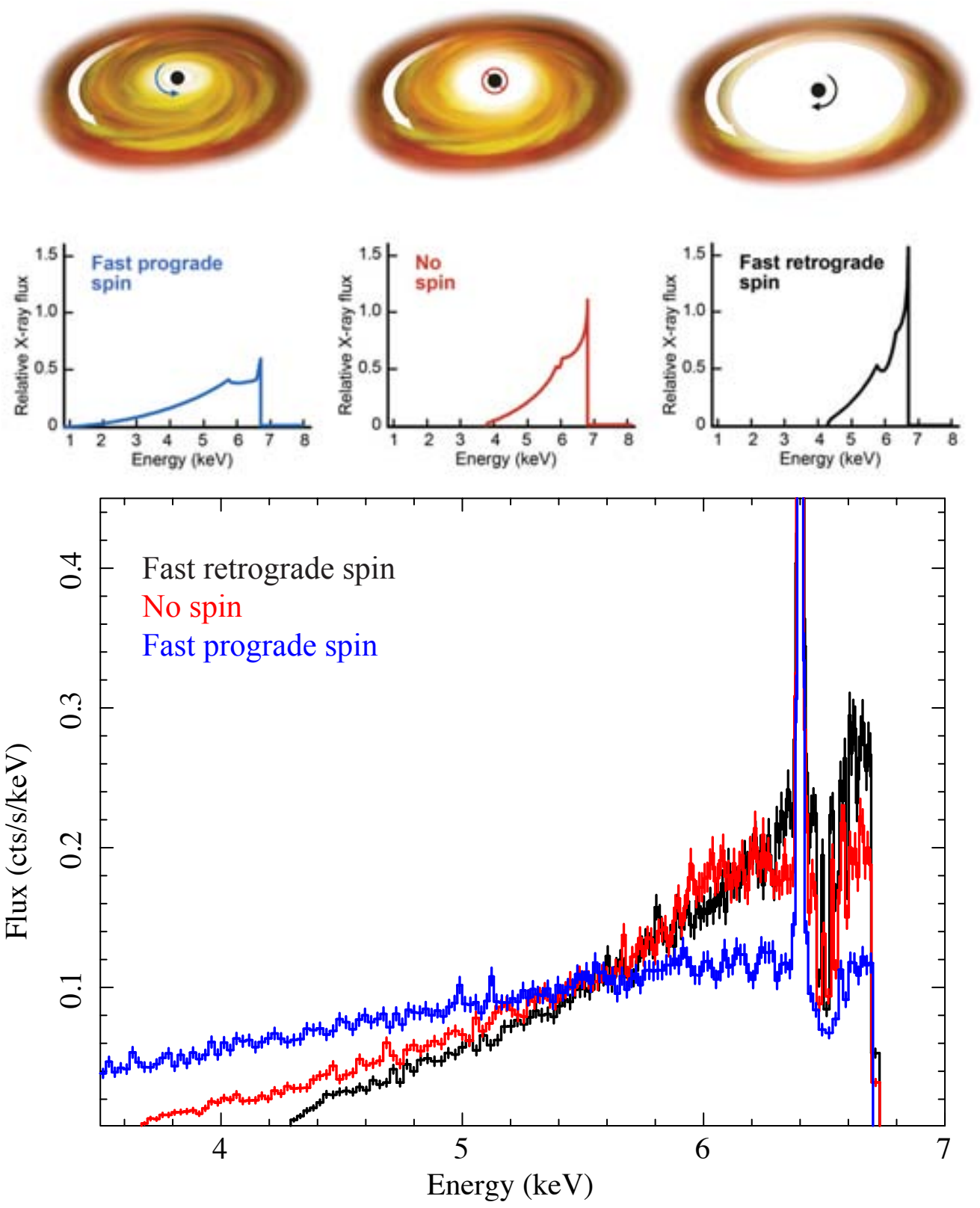

**Figure 2.8.** ***Top:*** *Artist's impression of accretion disks rotating as shown by the white arrows, around black holes with different spins, from maximally rotating (left), non rotating (centre) to maximally counter-rotating (right). The inner disk radius varies as a function of spin and can therefore be measured through the emission line profiles as affected by strong gravity. Image courtesy of Sky and Telescope.* ***Bottom****: Simulated Athena-XMS spectra corresponding to these situations, superposed on the narrow emission and absorption features emitted from a distant reflector (such as a molecular torus, not shown in the top illustration) plus ionised absorption from a wind. The XMS will separate any narrow features from those produced by strong gravity. Note that Athena will be able to distinguish between non-rotating and counter-rotating black holes despite the rather large innermost accretion disk radii in both cases and thence more modest strong gravity effects.*

The structure of the inner accretion flow, especially the nature of the hard coronal X-ray source, is poorly known, but is expected to be inhomogeneous and turbulent, giving rise to modulations of both the continuum and reflection components on the orbital timescale. Hot spots or bright rings of emission can produce relatively narrow, variable and energy-shifted features in the iron-K line profile, distinct from the main (disk-integrated) broadened emission (Reynolds et al. 1999, Dovciak et al. 2004). A major goal of *Athena* will be to probe these strong-gravity phenomena by either tracking the line modulations on the orbital timescale or by measuring the "reverberation" of the reflected emission (Figure 2.9), characterised by a time lag between variations in the driving power-law continuum and the response of the reflection spectrum.

## Hot Spot tracking

*Athena* will be able to map the inner regions of BH disks in the energy-time plane. Any non-axisymmetry in the iron line profile (e.g. associated with the expected turbulence in the disk and/or the formation of hot spots) will lead to a characteristic variability of the iron line, with "arcs" being traced out on the time-energy plane (Figure 2.9). Hints of such features are seen in *XMM-Newton* data (De Marco et al. 2010), and they are of immense diagnostic power because they can be used to trace out the inner turbulent flow of the disk in the strong gravity environment.

GR makes specific predictions for the form of these arcs, and the ensemble of arcs can be fitted for the mass and spin of the BH, and the inclination at which the accretion disk is being viewed. This type of iron line variability gives us our best view yet of the strong gravitational field close to the BH. *Athena*'s large effective area at 6 keV is crucial for enabling the detection and tracking (in energy and flux) of these faint features on the time scale of the innermost disk, yielding new information on the geometry and physics of this region.

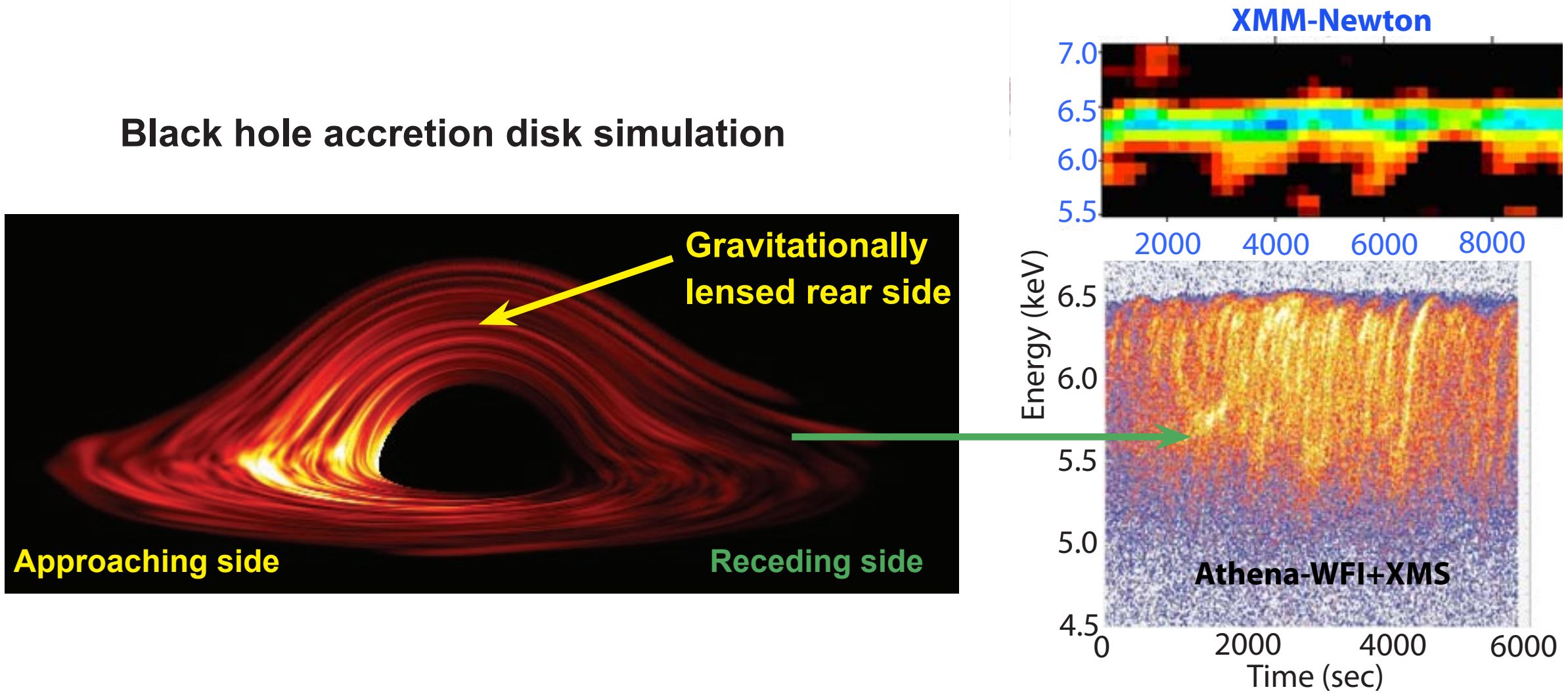

**Figure 2.9.** ***Left:*** *Snapshot from a time-dependent Magneto-Hydrodynamic (MHD) simulation of an accretion disk around a BH (Armitage & Reynolds 2003). Rings and hotspots of emission are seen due to turbulence, the emission from which should be modulated on the orbital timescale.* ***Right:*** *Because the features are variable in both flux and energy, the disk can be mapped out in the time-energy plane. The first hints of this behaviour have been seen in XMM-Newton data (Iwasawa et al 2004; upper right), but the factor ~4 improvement in throughput and ~40 in energy resolution offered by Athena are needed to sample weaker and (possibly) narrower features on suborbital timescales, allowing us to map out the inner accretion flow. Only unusually strong and slow flares can be detected by XMM-Newton.*

## Reverberation mapping

Measuring the light travel time between flux variations in the hard X-ray continuum and the lines that it excites in the accretion disk (light echo or "reverberation") provides a model-independent way to map the inner accretion flow, as the time delay simply translates into distance for a given geometry. Such reverberation lags have now been seen with *XMM-Newton* in a few rather special sources. Thanks to a week-long observation, a reverberation lag of 30s has been seen in the Narrow Line Seyfert 1 (NLS1) galaxy 1H0707-495 (Fabian et al. 2009). Similar lags appear in other bright NLS1 (Emmanoulopoulos et al. 2011). Measuring these lags is possible in these cases due to strong Fe-L emission accompanying the relativistic reflection, which is delayed by light echo compared to the direct X-ray continuum. While this strong Fe-L emission is relatively rare, relativistic Fe K$\alpha$ lines are much more common in both AGN (Nandra et al. 2007) and Galactic binaries (Miller et al. 2007) providing a more robust path towards reverberation measurements. *Athena* will, for the first time, provide the combination of photon statistics and spectral resolution to perform these reverberation experiments at iron K, for several 10s of AGN and black hole binaries. These measurements offer unique and fundamental information about the strong gravity environment.

For example, if the BH has a low spin parameter, iron emission lines in the 6.4-6.97 keV range should have a characteristic lag of approximately 6 $GM/c^3$; if the BH is rapidly spinning, lags can be as short as 1 $GM/c^3$. Furthermore, close to spinning BHs, the path light takes will be strongly impacted by space-time curvature. When very close to the BH, an otherwise isotropic source of hard X-ray emission will have its flux bent downward onto the disk. An observable consequence of these light-bending effects is a particular non-linear relationship between hard X-ray emission and iron emission lines (Miniutti & Fabian 2004). At present, there is tantalising evidence for this effect in some Seyfert AGN (e.g. Ponti et al. 2006) and Galactic black holes (e.g. Rossi et al. 2005). The combination of *Athena*-XMS and WFI have the time resolution, broad energy range, energy resolution, and flux tolerance needed to make careful studies of lags in galactic black holes (GBHs), NSs and AGN. These can lead to spin measurements and clear detections of gravitational light bending. Again, the unprecedented total effective area of *Athena* around the iron K$\alpha$ line at 6 keV is the key capability required to enable these measurements, as that feature is extremely common in accreting compact sources. The 30-s time lag between the continuum and Fe L reflection feature observed in 1H0707-495 is nonetheless also exciting because it offers the possibility of measuring spin and reverberation together in the soft X-ray band, where the effective area of *Athena* is even higher (Figure 2.10).

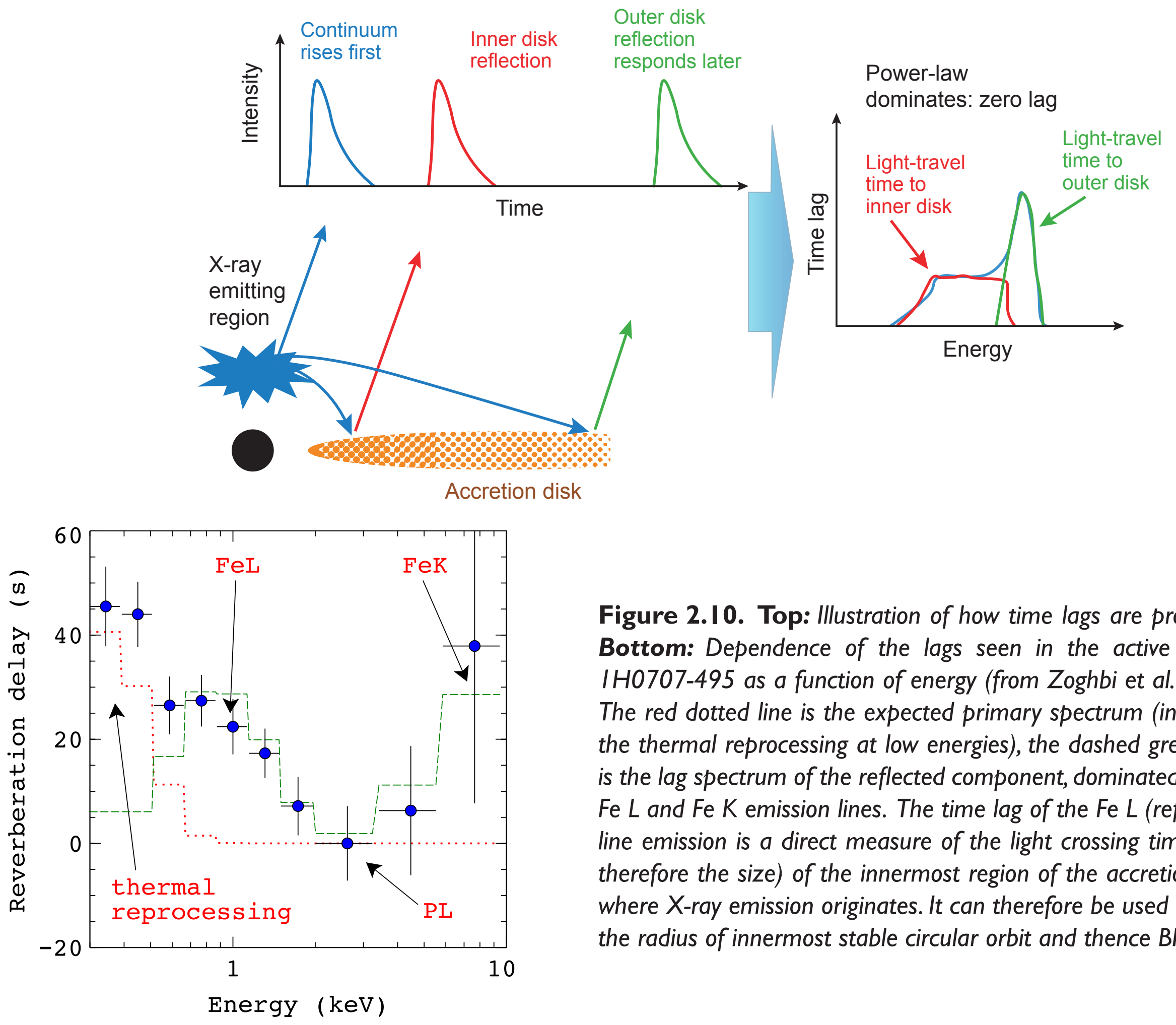

**Figure 2.10. Top:** *Illustration of how time lags are produced.* ***Bottom:*** *Dependence of the lags seen in the active galaxy 1H0707-495 as a function of energy (from Zoghbi et al. 2011). The red dotted line is the expected primary spectrum (including the thermal reprocessing at low energies), the dashed green line is the lag spectrum of the reflected component, dominated by the Fe L and Fe K emission lines. The time lag of the Fe L (reflected) line emission is a direct measure of the light crossing time (and therefore the size) of the innermost region of the accretion disk, where X-ray emission originates. It can therefore be used to infer the radius of innermost stable circular orbit and thence BH spin.*

ATHENA

*Athena* will observe matter in the strong gravity regime within a few gravitational radii of the event horizon of black holes. Large gravitational redshifts, extreme light bending and time delay effects introduce distinctive spectral and timing signatures in the X-ray emission. These general relativistic effects will be readily measured by *Athena*.

## X-ray "tomography"

Time-resolved spectroscopy of a few bright AGN has revealed eclipses of the X-ray source lasting a few hours (Risaliti et al 2010). They are thought to be due to obscuring clouds with column densities of $10^{23}$-$10^{24}$ cm$^{-2}$ crossing the line of sight with velocities in excess of $10^3$ km/s. This opens up the possibility for *Athena* to perform a novel experiment of "tomography" of the X-ray source: while passing in front of the central regions, the cloud covers and uncovers different parts of the accretion disk and corona, enabling various physical properties of each to be constrained (e.g., coronal geometry, disk emissivity pattern). If an occultation by a Compton-thick cloud is observed in an AGN with a broad relativistic line, the line profile is expected to change during the eclipse as the approaching and receding parts of the accretion disk are covered and uncovered, respectively (Figure 2.11). Conservative estimates predict that these measurements would be relatively straightforward with *Athena* for several sources with great accuracy, providing another opportunity to probe the relativistic effects due to strong gravity and fast orbital motion in the innermost regions of AGN.

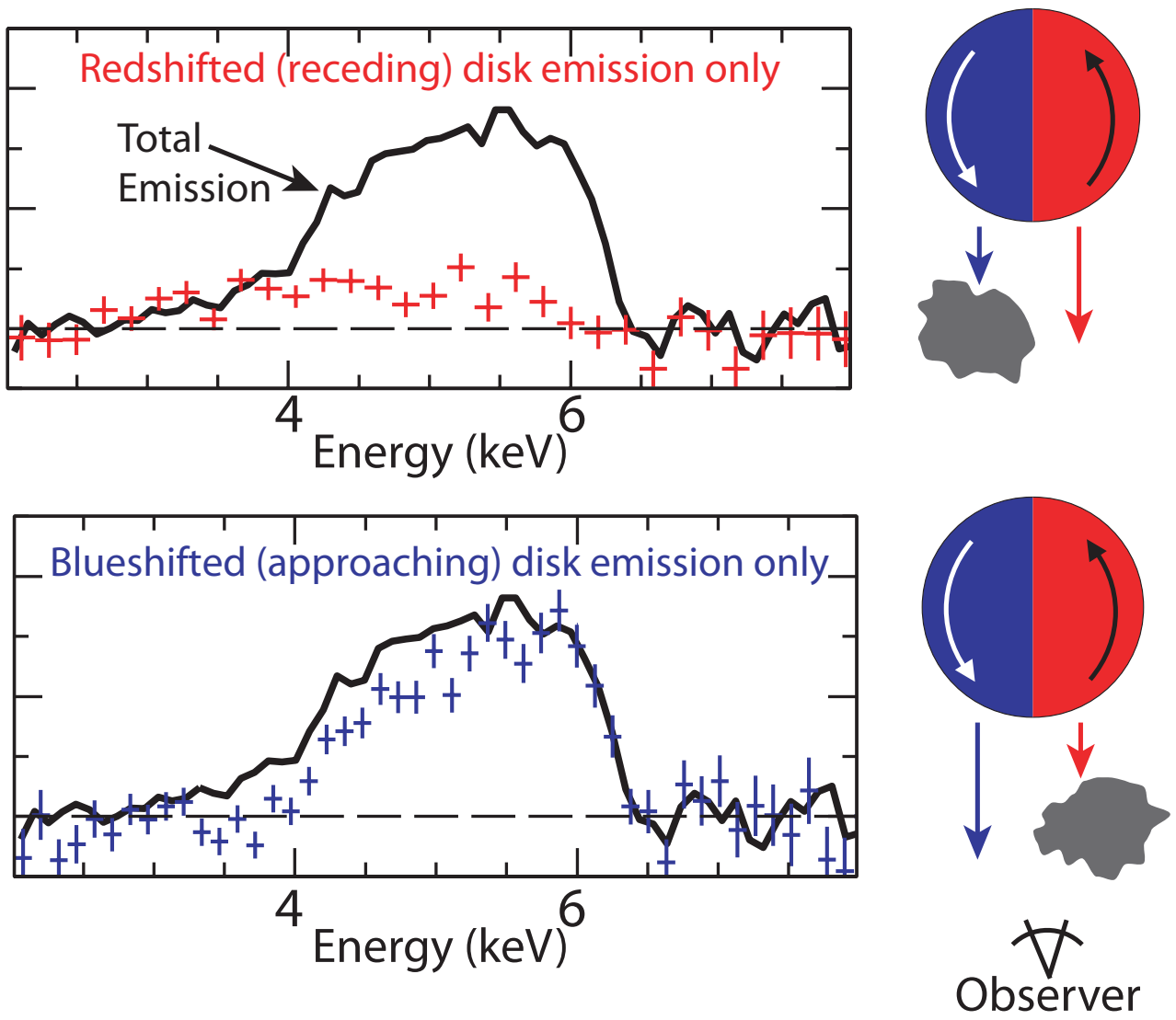

**Figure 2.11.** *X-ray tomography of the accretion disk around the black hole in NGC1365. The red and blue half circles indicate the approaching and receding disk, with the grey cloud indicating times when the disk is occulted by an orbiting cloud that eclipses the whole disk for 60 ksec, a model based on XMM-Newton and Suzaku data. From top to bottom, the black curve is the disk line profile obtained from the non-eclipsed part of the observation, while the red and blue profiles correspond to the "receding" and "approaching" halves of the X-ray disk. Athena will time-resolve these eclipses, providing a direct measure of the General Relativity effects in the vicinity of the central black hole.*

## Measuring black hole spin

According to the "no hair" theorem, astrophysical black holes are characterised by just two parameters: mass and spin. There has been significant progress in black hole mass measurement in recent years, both in actively accreting BHs and dormant BHs in galaxy centres. Determining the spin is far more challenging, as the effects of spin are only evident in the very innermost regions, as probed by the X-ray emission.

Ever since the seminal works of Penrose (1969) and Blandford & Znajek (1977), it has been realized that BH spin may be an important energy source in astrophysics, potentially allowing the extraction of energy and angular momentum from BHs themselves. BH spin may also play a role in the apparent radio-loud/radio-quiet 'dichotomy' in AGN, the powering of jets and/or winds in AGN and GBHs (Rees et al. 1982), and the production of gravitational waves from merging BHs. Spin may also encode information about the final stage of stellar evolution (supernovae) in GBHs and about galaxy merger and accretion histories in SMBHs. Despite its importance, however, we are only now gaining our first tantalising glimpses of BH spin in a few objects. Unlike measurements of BH mass, spin measurements require observables that originate within a few gravitational radii of the BH. The observational signatures needed to determine it are found in the X-ray band.

*Athena* will measure BH spin in several tens of bright GBHs and AGN using several independent techniques. These independent methods will be then used to "calibrate" BH spin measurements with (time-averaged) iron-K lines for application to larger samples of faint AGN, to constrain their merger and accretion histories, providing an important link with the co-evolution of SMBHs and their host galaxies (Section 2.2.2). In addition, the outflows and radiation thought to drive AGN feedback in evolving galaxies ultimately have their origin in physical processes close to the central BH. It is only by probing this region directly with *Athena* that we can hope to understand the remarkable link between these processes, which exists in spite of a factor more than $10^{10}$ difference in size scale between the SMBH and its host galaxy (Section 2.2.1). In contrast to SMBH, the spin of a stellar mass BH in a binary system is determined at birth, as it cannot accrete enough mass or angular momentum for the spin to change significantly in the accretion lifetime. It is therefore an indicator of how the core of the progenitor star collapsed.

Measurement of BH spin via hot-spot mapping and reverberation has already been discussed above. Here we highlight three *further* independent methods of spin measurement possible with *Athena*.

## Time-averaged fitting of broad Fe lines

The simplest and most widely applicable way of measuring BH spin is via time-averaged fitting of relativistic disk lines (Figure 2.8). The ubiquity of these lines demonstrates that they are an intrinsic property of accretion onto compact objects, and estimates of BH spin have already been made in a few AGN and Galactic BH using this method. The effects of spin on the disk reflection spectrum are not subtle, but the disk spectrum must be decomposed from other complexity such as continuum curvature, Comptonisation broadening and the effects of photoionised absorbers associated with absorbing, emitting and scattering gas around the accretion disc. To make this method robust we therefore have to understand whether and/or how much any complex/ionized absorption along the line of sight may affect the spin measurements (e.g. Miller et al. 2010). This necessitates a combination of large throughput combined with high energy resolution, as offered by *Athena*-XMS. Observations with *Athena* will eliminate the remaining uncertainties and allow us to measure the spin in perhaps hundreds of AGN, GBHs, NSs, opening the way to the first statistical studies of BH spin (see Section 2.2.2).

> ATHENA
> Astrophysical black holes are parameterised by just two quantities: mass and spin. Spin can only be measured through observations within the immediate environment of the black hole, where the X-rays are produced. *Athena* will measure the spins of both stellar mass and supermassive black holes using multiple independent techniques.

## Accretion disk thermal continuum fitting

Thermal continuum emission from the accretion disk may be used to measure the spin of GBHs. An accretion disk around a spinning BH is expected to be hotter and more luminous than a disk around a BH with low spin, because the ISCO is deeper within the gravitational well. New spectral models have recently been developed that exploit the corresponding changes in the shape of the continuum to measure spin. If the mass, distance and disk inclination to a BH are known, these models may be applied to spectra in order to measure the spin of a GBH or a neutron star (see, e.g., McClintock et al. 2006). By virtue of its flux tolerance and energy coverage down to 0.1 keV, the *Athena*-WFI is ideally suited to the measurement of BH spins using the disk continuum. As shown in Figure 2.12, the simulated statistical uncertainty on the BH spin measured through continuum fitting is about an order of magnitude smaller than that derived by fitting the relativistic iron line profile, but both measurements will be available simultaneously with *Athena*, allowing a cross-calibration of the techniques.

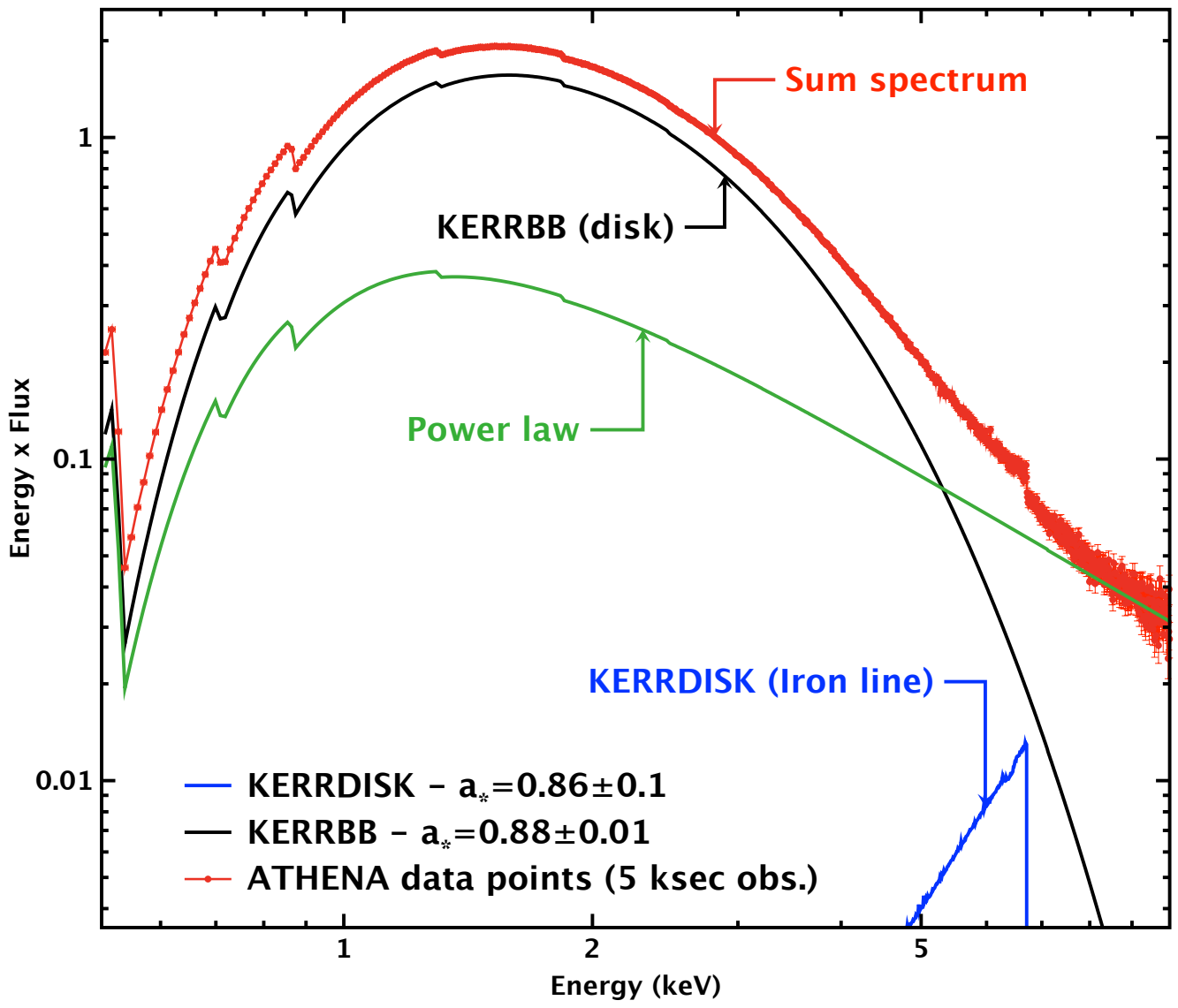

**Figure 2.12.** *A typical high state spectrum of a GBH (e.g. GX339-4) as observed by the Athena-WFI. Both a disk component and a weak iron line are present in the spectrum, together with a power law (all spectra are absorbed by photoelectric absorption from the interstellar medium along the line of sight). The statistical accuracy on the spin measurement from the disk component is about ten times smaller than for the relativistic iron line, as shown in the ststistical errors quoted for the spin parameter in the legend of the plot. Thanks to the Athena-WFI count rate capabilities, two independent measures of the BH spin will be obtained from a large sample of bright X-ray binaries.*

## Quasi-periodic oscillations

The X-ray flux from accreting BHs and NSs is sometimes modulated at frequencies commensurate with Keplerian frequencies close to the compact object. The oscillations are quasi-periodic, due to small variations in frequency and phase as expected for gas orbiting in a real fluid disk with internal viscosity. These quasi-periodic oscillations (QPOs) probe the strongly curved space-time around compact objects and constrain their mass, spin and radius. Most current models associate the QPOs with the general relativistic regime, so they hold further potential to probe strong field GR, including spin measurements (Figure 2.13).

*Athena* will make a large leap forward in sensitivity, opening the way to the detection of strong QPOs on timescales closer to the coherence time of the underlying oscillator, and to detect the weakest features predicted in models, allowing us to unambiguously identify the origin of the peaks in the power spectrum. Moreover *Athena* will reveal QPOs in a much wider variety of objects, such as ultra-luminous X-ray sources, which may possibly harbour intermediate mass BHs (Section 2.1.2), and AGN, for which a QPO detection was recently reported (Gierlinski et al. 2008). QPOs are now beginning to appear in some time-dependent simulations of magneto-rotational instability turbulence, and further model developments will provide the necessary framework for exploiting the potential of QPOs for probing GR, compact object parameters and accretion disk physics.

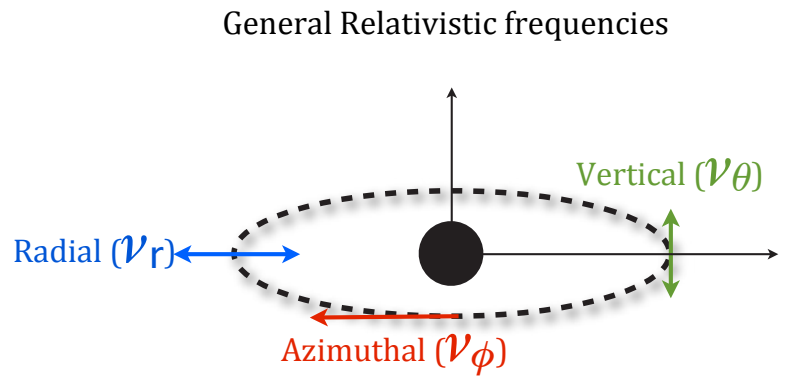

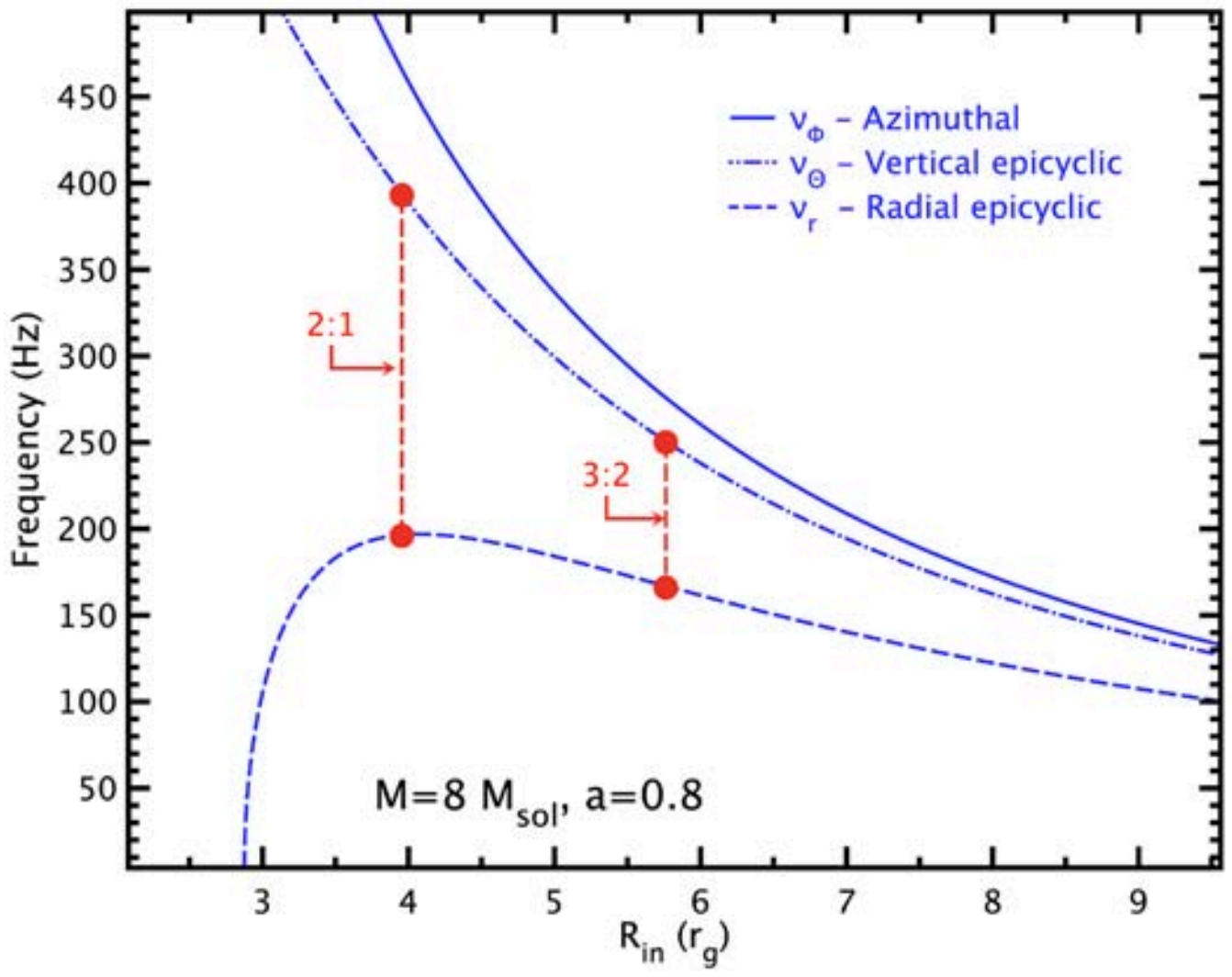

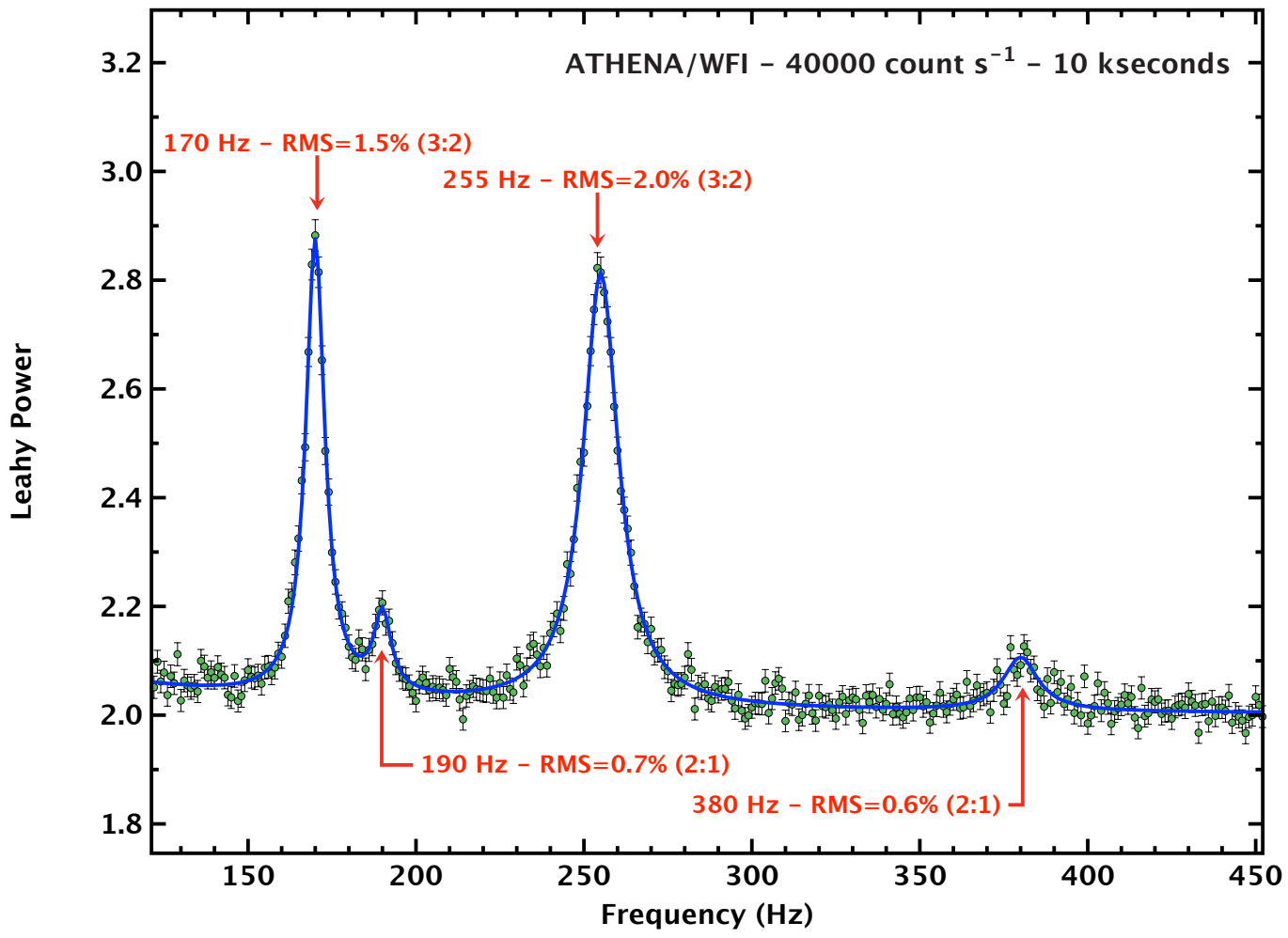

**Figure 2.13.** ***Top:*** *Azimuthal, vertical and radial motion does not occur at the same frequency in General Relativity, leading to epicyclic orbits.* ***Upper left:*** *The three GR frequencies around an 8* $M_\odot$ *black hole with a dimensionless spin parameter of 0.8. There exist two radii in the disk where the ratio of the vertical and radial epicyclic frequencies are equal to 3:2 and 2:1. In a particular class of models, resonance should be excited at these particular radii.* ***Lower left:*** *The power density spectrum simulated for an Athena-WFI observation of 10 ks (0.8 Crab source), in which the 4 QPOs are simultaneously detected at a significance level larger than* $6\sigma$*. As of today, only the QPOs corresponding to the 3:2 ratio have been reported from RXTE. If the mass of the black hole is known (from the dynamical mass function estimate of the binary system), solving the GR frequency equations for the black hole spin should yield an estimate with an error of about 10%. For those objects, measurement of the black hole spin via X-ray spectroscopy will also be possible with Athena (either through fitting the disk continuum emission or the iron line). A consistent estimate of the black hole spin from X-ray timing and spectroscopy would give strong confidence in the interpretation of broad iron lines in AGN, in the context of measuring the black hole spin evolution across cosmic time (Section 2.2.2).*

### 2.1.2 The physics of accretion

Due to its unique new capabilities, *Athena* will provide us with a holistic view of strong gravity and the accretion phenomenon spanning almost the full range of size scales, cosmological distances, masses and accretion rates, and central object types. For example, many lightly-magnetised accreting neutron stars show relativistically broadened reflection features with broad skewed iron lines similar to those seen in accreting black holes. The redshifts are less extreme, however, indicating innermost orbits compatible with the radii of neutron stars. *Athena* observations of accreting neutron stars therefore provide another perspective on the accretion process and on strong gravity. In principle, differences between GBH and NS binaries can provide evidence for a key prediction of GR: the existence of the event horizon (e.g. Garcia et al. 2001). Here we highlight further examples of "extreme" accreting systems, for which *Athena* will provide unique information:

#### Sgr A* : the black hole in the Galactic Centre

Sgr A*, the 4 x $10^6$ $M_\odot$ black hole at the nucleus of the Galaxy, is the closest SMBH to us, and the one with the lowest known accretion rate. It can therefore provide a unique link between actively accreting SMBHs powering AGN and the largely dormant SMBHs at the centres of most massive galaxies. The extremely low luminosity of this radio, infrared and X-ray source in its quiescent state ($<10^{-8}$ its Eddington luminosity) would render it undetectable in most external galaxies, so Sgr A* represents the best case study for black holes in quiescence. Indeed, all major models of radiatively inefficient accretion flows (RIAF) around BHs have been tested on, and even conceived for, Sgr A*.

A striking and unique feature of Sgr A* is its erratic activity in X-rays, characterised by flares that occur about once a day, lasting between 30 min to 3 hours and with luminosities that can increase by up to a factor of 200 over the quiescent level of 2 x $10^{33}$ erg/s. These events allow us to probe the accretion processes occurring within a few tens of gravitational radii from the horizon of our nearest SMBH (Baganoff et al. 2001, Goldwurm et al. 2003, Porquet et al. 2008). The X-ray flares are always accompanied by near infrared (NIR) flares, which are now interpreted as optically thin synchrotron emission from a non-thermal population of electrons. The origin of the associated X-ray emission however is still not understood. In particular the standard paradigm that X-rays are due to inverse Compton scattering of the same electrons producing the IR photons on the sub-millimetre radiation of the accretion flow does not seem compatible with recent *XMM*-VLT observations which instead favour a synchrotron nature for the X-rays. These fundamental issues remain open because of the uncertainties in the X-ray spectral slope, and the non-detection either of spectral changes or of a spectral break in the X-ray emission. In addition to the emission mechanism (Compton vs. synchrotron; Dodds-Eden et al. 2009), these uncertainties hamper the identification of the heating and cooling mechanisms, and the study of the presumed plasma adiabatic expansion (Trap et al. 2011).

*Athena* will enormously improve the determination of the spectral parameters of the Sgr A* X-ray flares. For a flare with peak luminosity 50 times quiescence, the spectral slope can be determined with an accuracy of 5% or better, an energy break up to about 8 keV can be detected (and therefore electron maximum energy determined) as well as spectral changes, giving crucial information on the heating/cooling mechanisms. Such measurements, along with simultaneous multi-wavelength observations, will constrain several physical parameters (electron energies, magnetic field, emitting region size) and will tell us whether the acceleration takes place in the inner accretion flow or in a compact jet.

Thanks to *Athena's* sensitivity, the spectral properties of Sgr A* 's flares will be well constrained even for relatively weak flares (~10 times the quiescent level) and for the brightest ones time-resolved spectroscopy will be possible for the first time. Such strong spectral constraints will allow us to determine whether all X-ray flares are produced by the same physical process and provide very stringent constraints to discriminate between models proposed for the formation of the flares. *Athena* will thus enable us to study with unprecedented capabilities matter close to the event horizon around the closest SMBH.

An exciting new development is the possibility of high resolution X-ray spectroscopy of Sgr A*. Spectroscopy could prove crucial in discriminating between various models for astrophysical processes in the central region around our closest SMBH. Radiatively inefficient accretion flow (RIAF) models predict emission lines in the spectrum from hot plasma forming a so-called "advective" flow (Shcherbakov & Baganoff 2010). Evidence for these high ionisation lines is indeed seen in the *Chandra* ACIS spectrum. *Athena* will resolve these lines (which probably represent a blend) into their constituent components and measure any velocity widths and shifts. A critical issue is the presence (or otherwise) of a 6.4 keV component which would indicate a significant population of spun-up, late-type stars within the Bondi radius of Sgr A*, which can be confused with the accreting source itself (Sazonov et al. 2011). On larger scales tentative evidence for a bipolar outflow has also been seen with *Chandra*. *Athena* will reveal the 2-D velocity structure of the diffuse emission in the Galactic centre in exquisite detail yielding new information on both the nature and consequences of RIAFs.

Finally, the study of the Galactic Centre molecular cloud X-ray emission will tell us about past activity in Sgr A* and therefore about the accretion history of our nearest SMBH (Koyama et al. 1996; see Section 2.4.5).

### Ultraluminous X-ray sources

At the other extreme in terms of accretion rate, Ultraluminous X-ray sources (ULXs) remain as one of the compelling mysteries of X-ray astronomy. Their observed luminosities exceed the Eddington limit for the luminosity of the stellar-mass black holes we know of in our own Galaxy; yet their extra-nuclear locations rule out a central SMBH. Hence we are left with a choice of exotic scenarios – either ULXs contain a hitherto unseen class of accreting intermediate-mass black holes (IMBHs), or they embody a new regime of accretion physics in which the flow of accreting material somehow appears to exceed the Eddington limit. While a combination of environment and counterparts currently supports the presence of super-Eddington accretion in most ULXs (e.g. Roberts 2007, Gladstone et al. 2009), a number of the brightest sources show properties consistent with harbouring IMBHs (e.g. Farrell et al. 2009). Either case has cosmological implications: IMBHs may be the relics of the population of black holes that seeded the first supermassive objects, whilst physical processes allowing the accretion of material at super-Eddington rates appear necessary for the rapid appearance of the first generation of QSOs at very high redshift.

*Athena* will provide the first data with sufficient quality to examine the detailed physics of the putative super-Eddington accretion states in ULXs, and to separate out the familiar sub-Eddington processes expected if IMBHs are present. In particular, deep observations of the brightest ULXs with the *Athena*-XMS offer the opportunity to search for blue-shifted narrow absorption features from highly ionised species that are undetectable in current data, but that would provide confirmation of the presence of an outflowing wind, a key prediction of super-Eddington models (e.g. Poutanen et al. 2007, Ohsuga & Mineshige 2011). The unprecedented data quality will also permit the implementation of techniques hitherto only possible for Galactic objects and AGN, such as time-lag reverberation mapping (e.g., Fabian et al. 2009), to provide new insight into ULX accretion physics. A critical issue here is the *Athena* spatial resolution, which allows the separation of the ULX emission from other sources in the crowded regions of external galaxies. Targeted observations of ULXs with the *Athena*-XMS permit high resolution spectroscopy, while the WFI provides the wide field necessary to discover new ULXs and indeed study several ULXs simultaneously in a single observation (Figure 2.14).

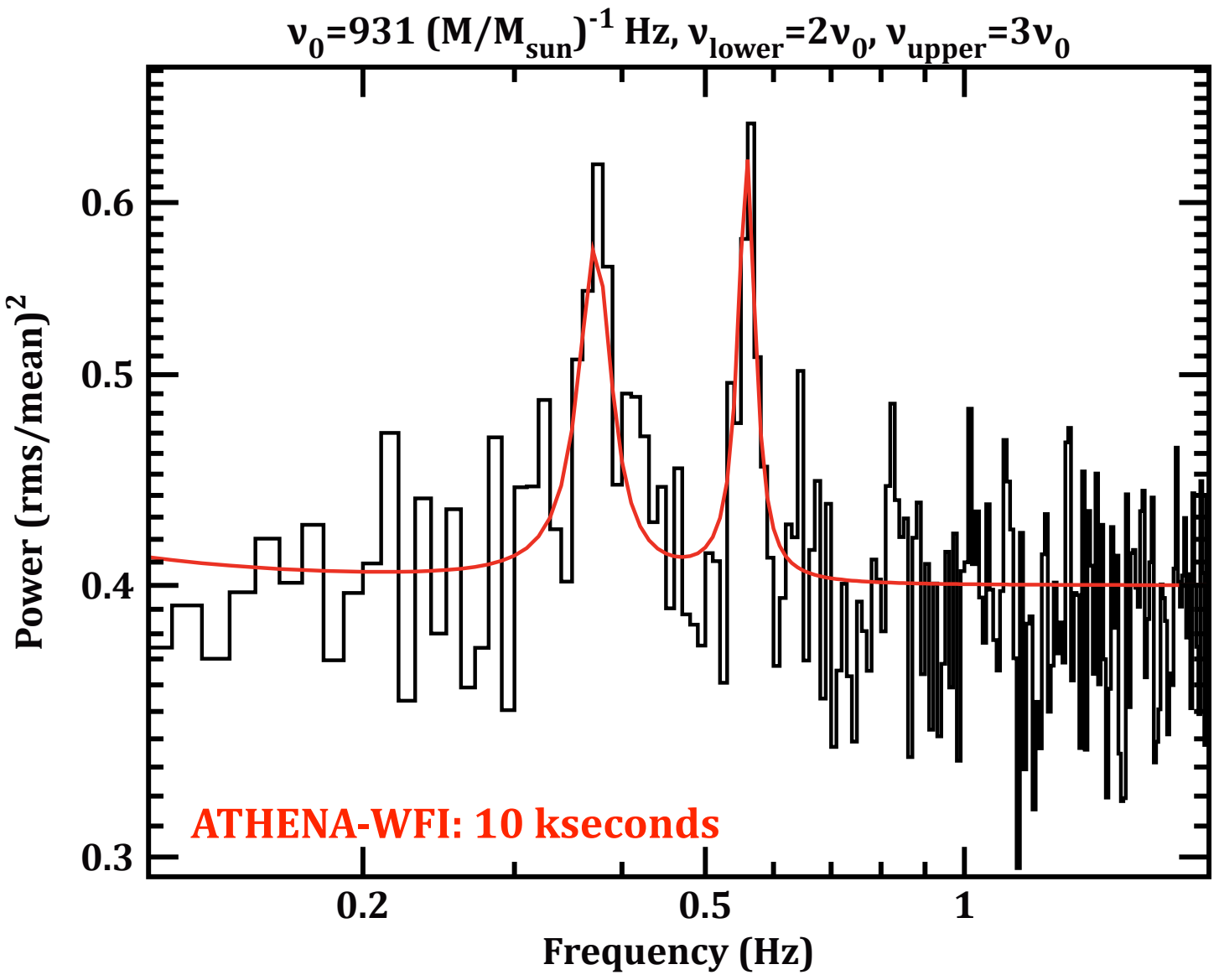

**Figure 2.14.** *Athena will detect those QPOs in a significant sample of ULXs (tens of systems), thanks to its improved timing sensitivity over XMM-Newton and its better angular resolution. Detecting such frequencies in the brightest ULXs, likely harbouring intermediate mass black holes, would provide insights on the mass of the black hole, its spin, and enable us to study accretion in strong gravity across the whole black hole mass range, by coupling these observations with those of stellar mass black hole binaries and AGNs. black binaries and AGN.*

The key evidence for ULXs as super-Eddington accretors currently comes from their unusual X-ray characteristics derived from the energy, fluctuation power density and RMS-energy spectra in the brightest ~10 objects (Gladstone et al. 2009; Heil et al. 2009), and how these change with source flux (Kajava & Poutanen 2009; Feng & Kaaret 2009). With multi-epoch *Athena* data we will be able to roll out these diagnostic tools to a far wider (100+) range of ULXs, providing more insight into their physical mechanisms. Crucially, such a source sample would include a number of the extremely rare, most luminous members of this class that are the best candidates for hosting IMBHs. *Athena* will have the power to identify whether these objects show super-Eddington accretion modes, consistent with other ULXs; or whether their accretion resembles the familiar sub-Eddington states, so marking them as excellent IMBH candidates. One particular issue in IMBH identification is the presence of QPOs in the power spectra of ULXs. Although only detected in a handful of ULXs, the relatively low frequencies of these QPOs argue for IMBHs on the simple basis of light-time-crossing arguments (Strohmayer & Mushotzky 2009). However, this interpretation may be flawed if the objects are super-Eddington accretors where different physical processes are present (Middleton et al. 2011). *Athena* will readily be able to detect QPOs in ULX (see Section 2.1.1) and allow us to distinguish between sub- and super-Eddington states, and so determine whether the use of ULX QPO frequency to determine mass is valid.

Ultraluminous X-ray sources are either stellar-mass black holes radiating well above their Eddington limit, or a new population of intermediate-mass black holes. *Athena*'s unique combination of sensitivity, spatial resolution, spectral and timing capabilities will allow us to discern between these novel scenarios.

## Feeding the monster: gas flows on large scales

In addition to probing the innermost regions around BHs, where the velocities are large, the high effective area and spectral resolution of *Athena* will provide detailed measurements of the gas flows at larger radii. This is crucial, as it will help us understand the fuelling mechanism for accreting systems.

In the case of AGN, using emission and absorption features produced by the circumnuclear medium, combined with techniques routinely used in optical astronomy, we can measure the sizes and velocity structures of the emitting regions. A narrow component of the iron Kα line is almost invariably present in the X-ray spectra of AGN (e.g. Yaqoob & Padmanabhan 2004; Nandra et al. 2007). This line originates from reflection of the primary component by the circumnuclear gas; hence it is a powerful tool to establish the location of this gas (either the Broad Line Region, a parsec-scale torus, or the Narrow Line Region), depending on the line width. In current observations the line is typically unresolved, with FWHM upper limits of several $10^3$ km/s, compatible with any of the possible locations. The unprecedented spectral resolution of the *Athena*-XMS, together with the large area at 6 keV, will allow us to easily measure the FWHM and any bulk velocity flows using the neutral iron Kα line in tens of sources (Figure 2.15). Reverberation mapping of these lines (see Section 2.1.1) will reveal their physical location and, combined with line profiles and shifts, can be used to determine the large-scale flows responsible for fuelling of SMBHs in the centres of galaxies.

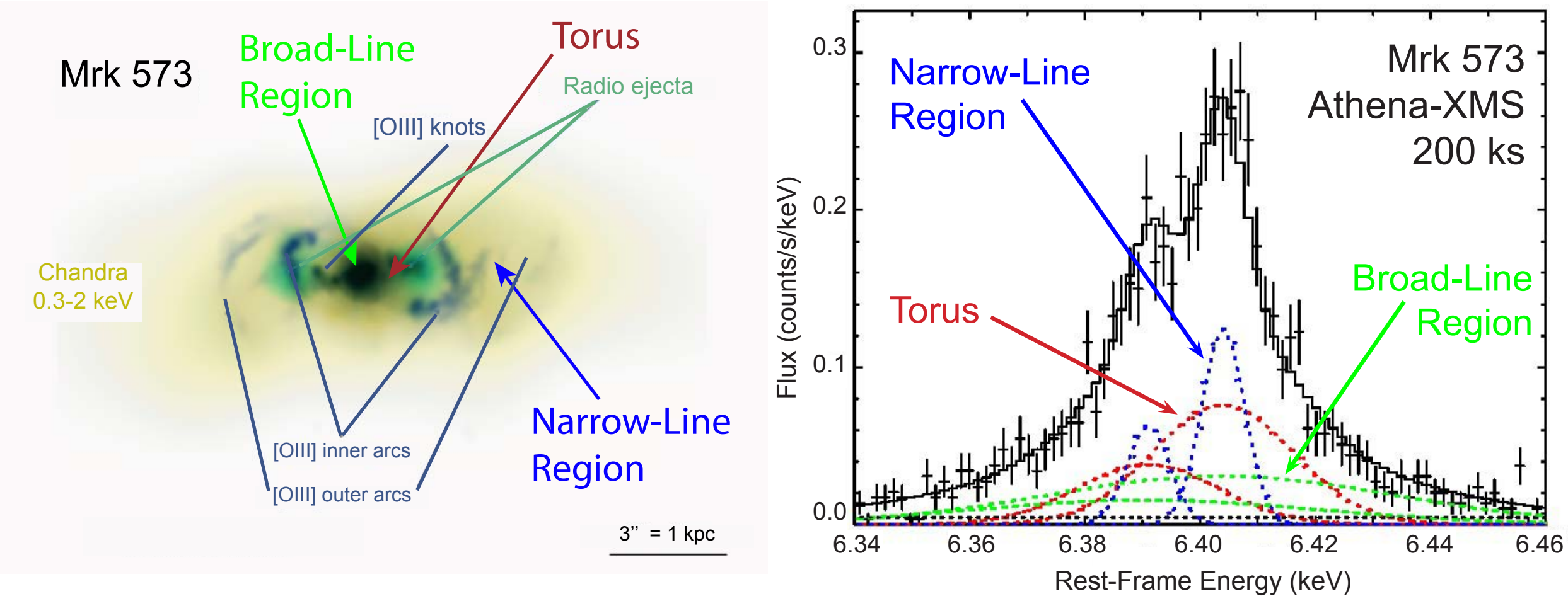

**Figure 2.15.** *The Seyfert 2 Mrk573.* ***Left:*** *multi-wavelength image combining soft X-ray (Chandra: yellow), radio 6 cm (VLA: green), and [OIII] emission (HST: dark blue). The approximate locations of the main line-emitting regions are shown (some of them are not actually resolved in this image).* ***Right:*** *Simulated 200 ks Athena-XMS observation of the neutral iron Kα line (actually a doublet) arising from material at different distances from the black hole. The three components can be easily deconvolved if present in a total profile.*

## Accretion across the mass scale

*Athena* will provide unique information on the fundamentally important physics of accretion across all mass scales and object types. Here we highlight four separate cases for which *Athena* will provide breakthrough observations in accretion physics: High Mass X-ray Binaries (HXMBs), Low-mass X-ray Binaries (LMXBs), white dwarf binaries, and Young Stellar Objects (YSOs).

HMXBs accrete mainly from the strong stellar winds of their companions. As the X-ray continuum from the BH or NS passes through the stellar wind, absorption lines containing a wealth of information on the physical state of the wind are imprinted. X-ray studies of these lines allows analysis of the wind structure, such as the clumping predicted by models for early-type stellar winds, as well as the overall wind structure via the P Cyg profiles of lines produced in the stellar wind (see Figure 2.16).

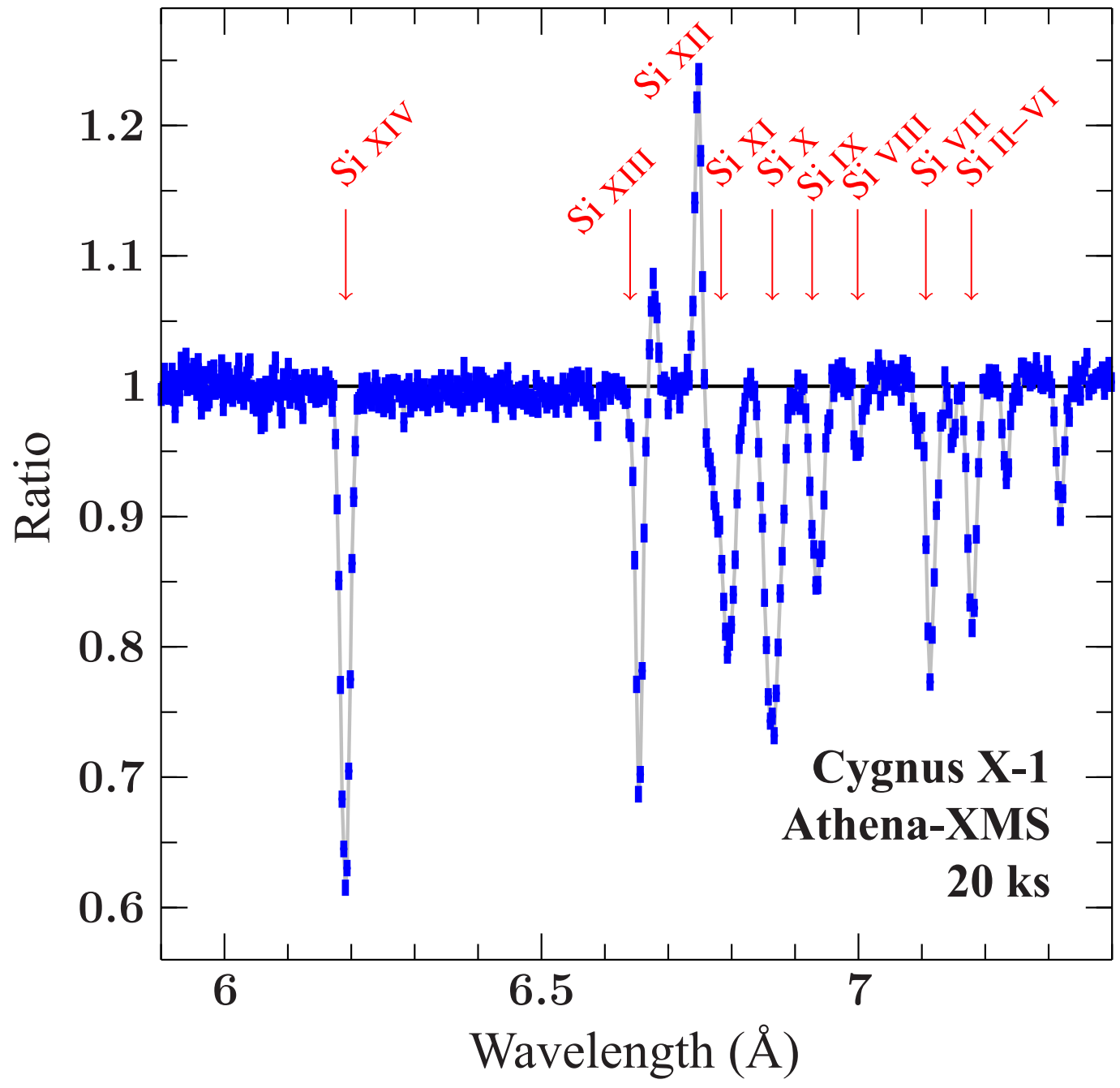

**Figure 2.16.** *Simulated 20 ks long Atena-XMS observation of the silicon region during a deep dip in the black hole binary Cygnus X-1 (data have not been binned).The figure shows the complex resonant absorption and emission lines from Si VIII to Si XIV caused by dense material in the stellar wind passing through the line of sight (model parameters are based on Chandra-HETGS observations of such events). The very high signal-to-noise ratio of these observations will allow us to perform, for the firt time, time resolved high-resolution spectroscopy of such events at a resolution of a few ks. Through such observations, Athena will reveal the structure of clumps in the photoionised winds of the early type donor stars of high mass X-ray binaries. Although an important ingredient in all current wind models for early type stars, the structure of the clumps is not known as such studies are not possible with current X-ray instrumentation, which lacks the large effective area required for measuring such spectra on the dipping timescale.*

LMXBs on the other hand accrete matter through Roche lobe overflow from a late-type companion. LMXBs seen (almost) edge-on show absorption dips caused by the edge of the accretion disk, and a multitude of absorption lines produced by an accretion disk corona and/or accretion disk wind. Through phase-dependent spectroscopy, *Athena* will reveal the ionisation structure of the disk, and time-resolved spectroscopy during absorption dips will distinguish (partial) absorption of the central emission source from changes in the ionisation state of the absorber (Díaz Trigo et al. 2006).

Interacting white dwarf (WD) binaries, because they are nearby and relatively bright systems, provide a key natural laboratory in which to study a range accretion processes which have much wider importance within astrophysics. X-ray observations of such systems provide unique plasma diagnostics of the disk-WD boundary layer in normal cataclysmic variables (CVs), and the magnetically-channelled accretion flows in polars and intermediate polars, systems in which the accreting WD has a very strong intrinsic magnetic field. In magnetic CVs most of the X-ray emission is produced in shocked regions – "accretion columns" – close to the magnetic poles of the WD. For such systems in particular, *Athena*-XMS will bring, for the first time, sufficient spectral and temporal resolution to make it possible to determine dynamical measurements of the accretion flows through emission line shifts, as illustrated in Figure 2.17. Coupled with the geometrical constraints on the accretion flow already available, for example from the highly-structured rotational light curves, this dynamical information will permit the 3-dimensional reconstruction of the flows, allowing an unprecedented detailed view of the complex interplay between the streaming accreted matter and the strong magnetic field of the accreting WD.

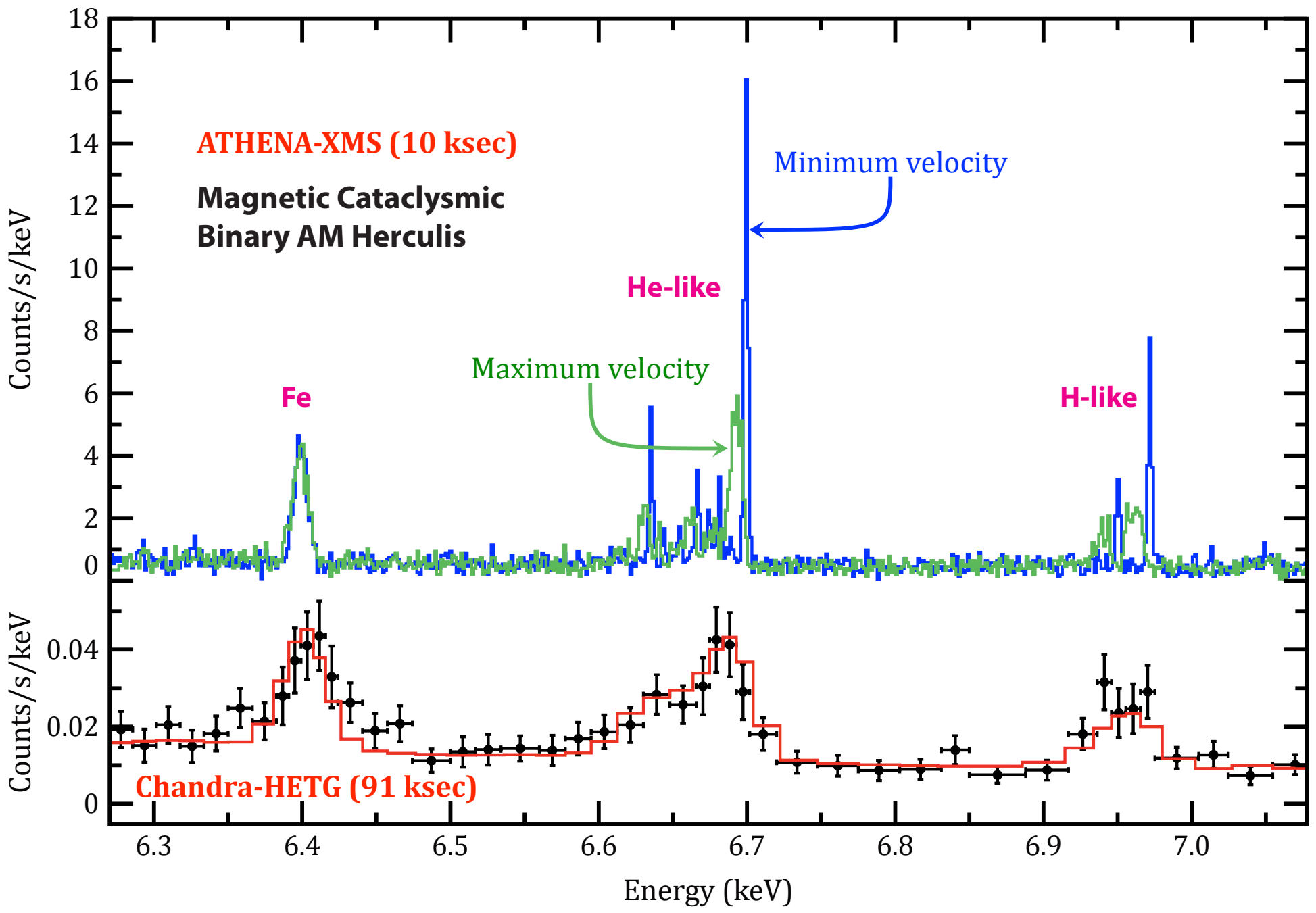

**Figure 2.17.** *Chandra-HETG and Athena-XMS high-resolution spectroscopy of the iron line complex in the prototypical cataclysmic binary star AM Herculis. The emission line system consists of highly ionised species originating from a stratified post-shock accretion plasma and a neutral line from reflection at the white dwarf surface. The lines are probing the relativistic plasma in a strongly magnetic environment. The red model curve applied to the observed Chandra data (90 ks spectrum averaged over the orbit) was used as input for Athena simulation (time-resolved spectroscopy with 10 ks time bins) which involves detailed accretion physics, gravitational redshift, the binary orbital motion and binary accretion geometry.*

Studies of YSOs are key to understanding the early phases of stellar evolution and illuminate key issues in planetary formation from protoplanetary disks, and accretion processes in YSOs are of critical importance in these studies. X-ray studies of YSOs provide one of the best ways of probing the complex interactions occurring in these objects and in particular of providing constraints on the accretion process taking place in these systems (see Figure 2.18). *Athena*'s high throughput will be key to further progress, making it possible for the first time to measure short-timescale variability in diagnostic lines covering the entire range of temperatures found in the accretion stream, revealing how inner disk instabilities feed the accretion stream and shedding light on its interaction with stellar atmospheres and coronae. Particularly crucial is obtaining iron Kα spectral diagnostics on a timescale significantly shorter than the rotational timescale (in a manner analogous to black hole accretion flow mapping). High quality time-resolved X-ray spectroscopic studies will uncover the connection between the central continuum X-ray source and the iron Kα line. If the emission is directly related to the intense flares, the time delays, expected to be of 50-100 sec (for a inner disk radius of 0.1-0.2 AU), measured with the reverberation mapping approach, will give a firm handle on the physical processes at work as well as on the geometry of the system. Given the extent of *Athena*-WFI field of view, high quality time-resolved spectral studies of large samples of YSOs can be obtained simultaneously in all nearby SFRs, allowing the detailed characterisation of the X-ray irradiation of YSO circumstellar disks.

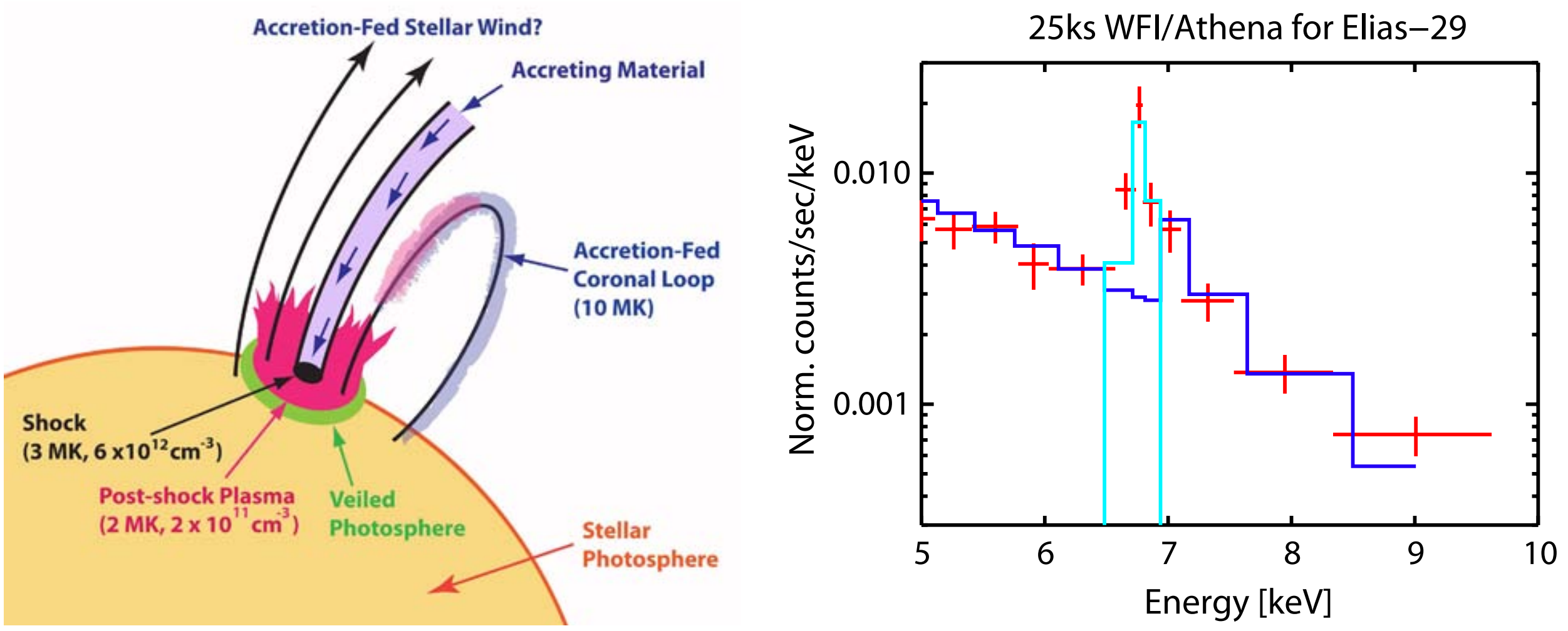

**Figure 2.18.** ***Left:*** *The accreting material in young stellar objects heats the stellar atmosphere, feeds coronal loops, and may drive the stellar wind (from Brickhouse et al. 2010). Improving our understanding of these processes requires spectral timing studies, only possible with Athena, that can detect variations in the accretion column due to inherent instabilities in the accretion disk.* ***Right:*** *Simulated Athena-WFI spectrum (red) based on an observed XMM-Newton spectrum of the young stellar object Elias 29. The iron Kα line (turquoise) is clearly detected superposed on the thermal spectrum (blue). With Athena-WFI this line will be detected in only 25 ksec with higher significance than in the more than ten times longer XMM-Newton observation.*

### 2.1.3 The Equation of State of ultradense matter

Among the most exotic physical conditions in the Universe are found in the cores of neutron stars. Initial attempts to understand these extreme conditions were based on the assumption that the matter could be adequately described as a degenerate gas of free neutrons, but it has become progressively clear that the cores of neutron stars must in fact be regions where intricate and complex collective behaviour of the constituent particles takes place.

Over most of the density/temperature phase plane, Quantum Chromodynamics (QCD) is believed to correctly describe the fundamental behaviour of matter, from the subnuclear scale up. While QCD has been tested in various ways on the earth, the limit of high densities and low (near zero, compared to the neutron Fermi energy) temperature QCD can only be tested in the extreme astrophysical environment of NS cores. Here QCD predicts rich behaviour. Exotic excitations such as hyperons, or Bose condensates of pions or kaons may appear, and it has also been suggested that at very high densities there may be a phase transition to strange quark matter. The key observable to distinguish between the various models is the Equation of State, which in astrophysical terms can be translated into the mass-radius relationship for neutron stars. *Athena*, with the high count rate capability of the WFI and the high resolution spectroscopic capability of the XMS will provide multiple, complementary and/or redundant measurements of the neutron star mass/radius relation. Crucially, this relation will be explored over the wide range of masses required to distinguish between the various competing models, and in a wide range of environments and conditions (e.g. isolated stars, accreting binaries, quiescent cooling systems, Eddington-limited X-ray bursters).

Thanks to its combination of spectral, timing and high count-rate capabilities, *Athena* will enable multiple independent constraints to be obtained on the mass and radius of neutron stars, giving insights into the densest form of observable matter in the Universe and testing Quantum Chromodynamics.

## The Mass-Radius relation of neutron stars

The relation between pressure and density, the equation of state, is the simplest way to parameterise the bulk behaviour of matter. The Equation of State (EoS) governs the mechanical equilibrium structure of bound stars and, conversely, measurements of quantities such as the mass and the radius, or the mass and the moment of inertia, of a star probe the EoS. Figure 2.19 shows the mass-radius plane for NSs, with a number of illustrative mass-radius relations based on various assumptions concerning the EoS (Lattimer & Prakash 2007). NSs have been the subject of intensive radio observations for forty years and precise radio pulse arrival time measurements on double neutron star binaries have produced a series of exquisite mass determinations, with a weighted average mass of $M_{NS} = 1.41 \pm 0.03\ M_{\odot}$. Such a mass can be accommodated by most EoS: definitive constraints can only be derived from simultaneous measurement of masses and radii of individual neutron stars. Accurate measurement of these parameters is also required. As an example, effective discrimination between different families of hadronic equations of state will require a relative precision of order 10% in mass and radius, and similar requirements apply to the strange matter EoS.

Looking at Figure 2.19, it is also clear that one needs to probe higher neutron star masses where the differences in the EoS predictions are more striking. Neutron stars in mass-transferring binaries will give us access to a wider range of neutron star masses (of order a solar mass of material can be transferred over the lifetime of a low-mass X-ray binary), to address this fundamental issue. Recently a neutron star as massive as $1.97 \pm 0.04\ M_{\odot}$ has been found from radio timing observations of the binary millisecond pulsar J1614-2230, through a strong Shapiro delay signature (Demorest et al. 2010), thus confirming the idea that massive neutron stars do exist.

The X-rays generated around NSs will be used to constrain their masses and radii, using multiple and complementary diagnostics for a wide range of objects. Those diagnostics, which require both high and medium spectral resolution and extreme count rate capabilities, are now fully implemented in the WFI and XMS designs. Hereafter we describe each measurement technique of M and R, bearing in mind that several techniques will be possible for the same object.

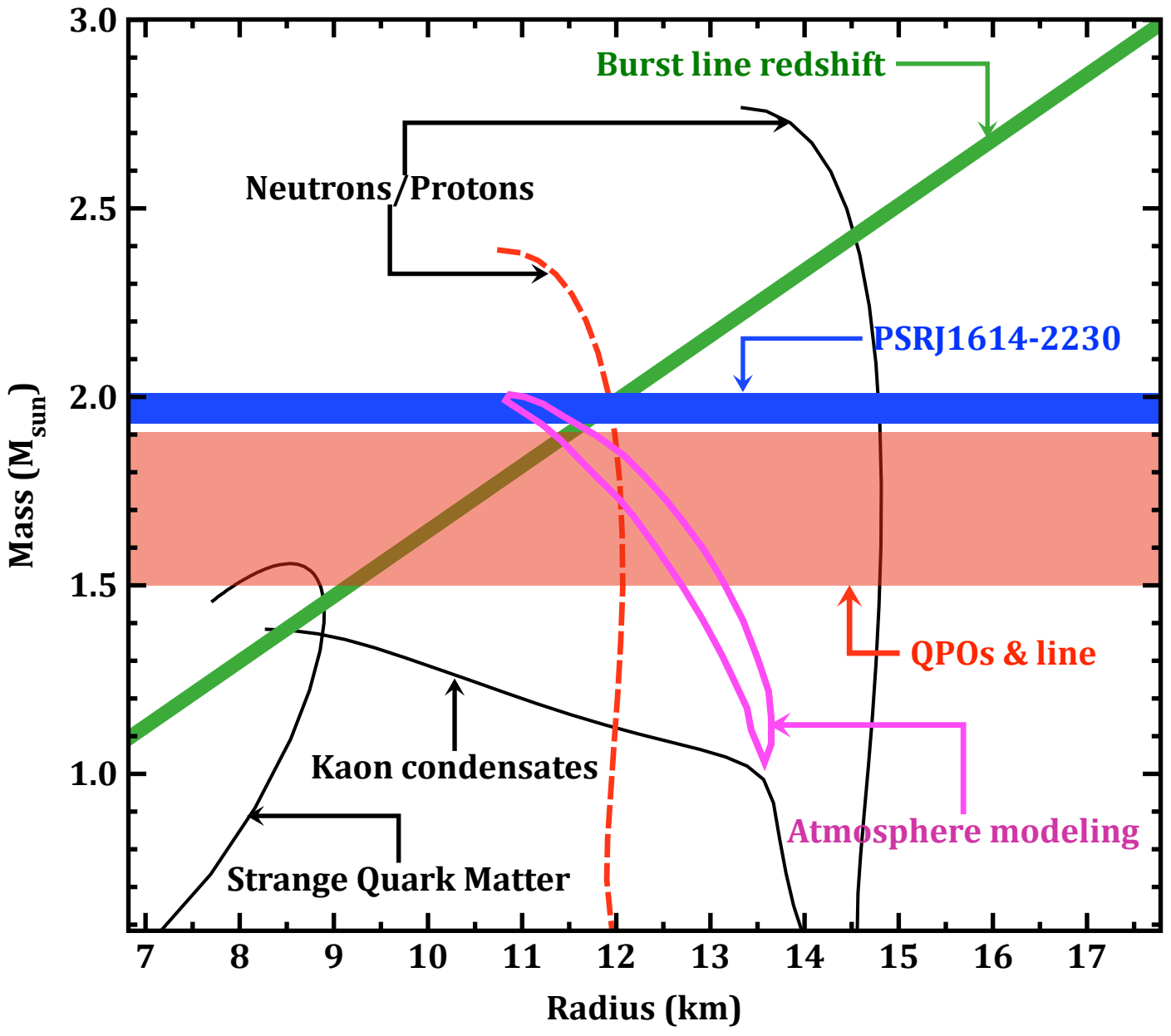

**Figure 2.19.** *The mass-radius relationship for neutron stars reflects the equation of state for cold superdense matter. Trajectories for typical EoSs are shown as black curves, for standard nucleonic matter (AP3, MS0), self-bound quark stars (SQM3) and Kaon condensates (GS1). PSRJ1614-2230 is one of the most massive neutron stars known today. Atmosphere modelling of the X-ray spectra of quiescent neutron star low mass X-ray binaries in Globular Cluster would yield the mass/radius constraint shown by the region within the purple boundary at the 90% confidence level (500 ksec observation). Constraints from a redshifted feature at the neutron star surface are shown as the green line. Finally, a mass determination for the neutron star could be obtained by a simultaneous measurement of a QPO frequency and a broad iron line (orange area).*

## Neutron star atmosphere modeling

A promising way to constrain the mass-radius relation for neutron stars is through observations of thermal emission from transiently accreting NSs when they are in their quiescent state, especially for those NSs for which a reliable distance can be estimated, e.g. for systems in globular clusters. Whether the energy res-ervoir is the heat deposited deep in the NS crust during the accretion phase of the transients or sustained by a low-level of residual accretion, the thermal X-rays originate from the atmosphere of the NSs in those systems. The exact spectrum will depend on the actual composition of the neutron-star atmosphere and the strength and structure of the surface magnetic field. The most current versions of the non-magnetic models, assuming a hydrogen atmosphere, have been shown to provide adequate fits to the quiescent X-ray spectra for all systems studied in detail, all providing plausible values for the NS radius (typically around 10-15 km) with uncertainties in the values of several tens of percents if one assumes a certain mass of the star. When simultaneously trying to constrain M and R, one typically obtains a large allowed region in the M-R diagram. The gain in sensitivity offered by *Athena* will allow for simultaneous constraints on both the mass and the radius with uncertainties which are small enough to be able to exclude a large region of the M-R plane (see Figure 2.19). In addition, if the thermal X-rays are indeed due to residual accretion onto the surface, absorption lines from high-Z elements are expected in the X-ray spectra of those sources. Those lines should be gravitational red-shifted and, if detected by *Athena*, will immediately constrain the mass-radius relation of those systems.

## Emission line modeling and fast time variability

Neutron star binaries exhibit broadened emission lines (e.g. Fe K$\alpha$) due to Doppler and GR effects, just as in AGN and stellar mass black holes (e.g. Cackett et al. 2010). The *Athena*-WFI will have the capability to observe bright X-ray binaries without suffering from any pile-up up to the intensity of the Crab, yet providing an energy resolution comparable to *XMM-Newton* and *Chandra*. As in the case of BHs, modeling the line profile yields the inner disk radius, which in turn sets an upper limit on the neutron star radius (hence a constraint on the NS EoS). Simultaneously with the line emission measurements kHz QPOs will be detected at a better sensitivity than that achieved by the RXTE/PCA (thanks to the increase of count rate by a factor of 2), yet with spectral resolution about an order of magnitude better (see Figure 2.20). Most models attribute the higher frequency QPO to an orbital frequency at the inner disk radius. Combining the kHz QPO frequency with the simultaneous radius given by a line measurement yields a mass determination (Cackett et al. 2010). If the link between the two quantities can be confirmed, then for a few strong line emitters and luminous QPO sources, the mass may be derived with an accuracy of 10-20%.

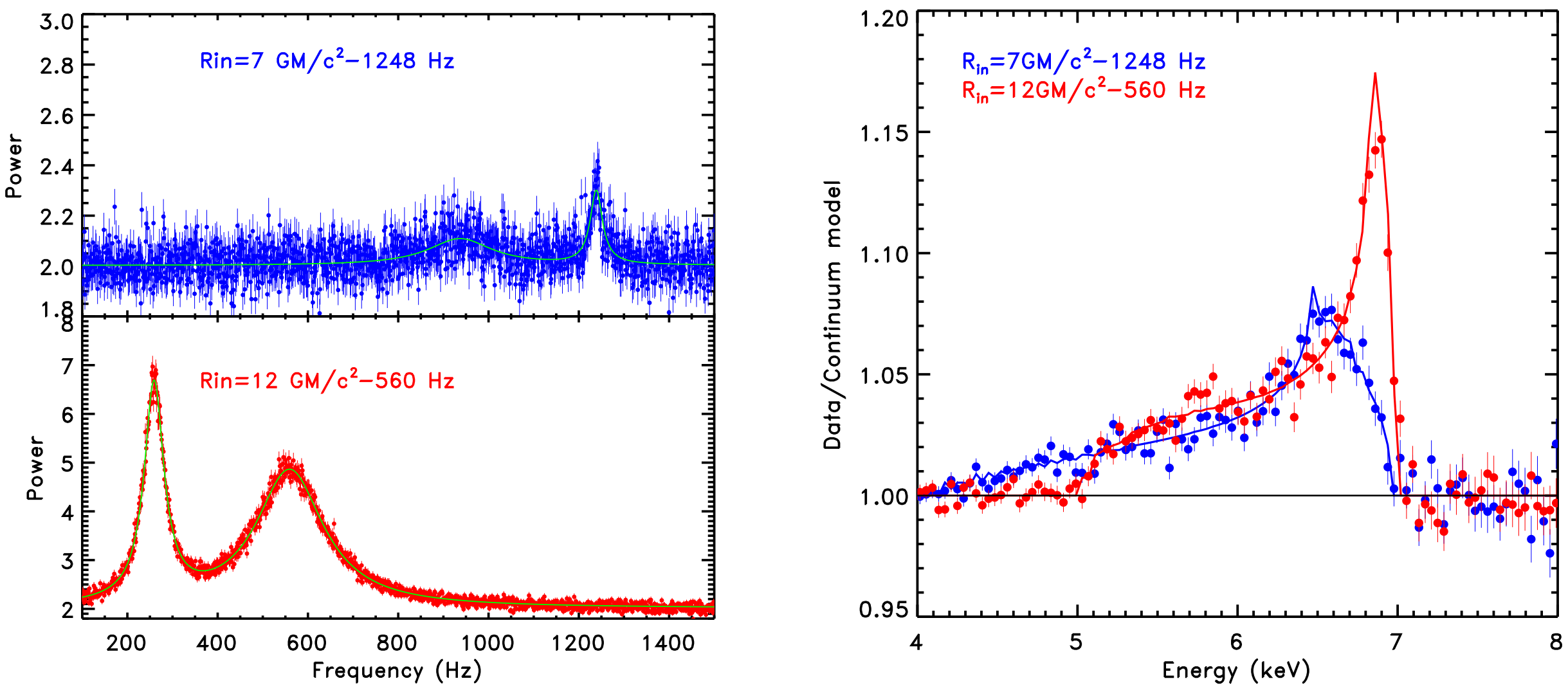

**Figure 2.20.** ***Left:*** *Simulated Athena-WFI power density spectra of a fiducial neutron star X-ray binary showing twin kHz QPOs, with the upper one associated with a Keplerian orbital frequency at two different radii, in units of GM/c$^2$ (2ks observation).* ***Right:*** *Ratio of data to continuum model for two relativistically smeared iron lines, associated with the same inner disk radii as the upper kHz QPOs. Thanks to the high count rate capability with large throughput and excellent spectral resolution of Athena-WFI, two probes of the inner disk radius can be obtained simultaneously. If the link between the QPO and iron line can be confirmed, then for a significant sample of strong line emitters and luminous QPO sources, the mass of the neutron star may be derived with an accuracy of 10-20%.*

### Gravitational redshifts during X-ray bursts

Weak (10 eV equivalent width) absorption lines (e.g. Fe Lyα) have been predicted in type I X-ray bursts (Chang et al. 2005). The profile of the line is distorted by contributions from magnetic (Zeeman or Paschen-Back) splitting by the star's magnetic field, longitudinal and transverse Doppler shifts, special relativistic beaming, gravitational redshifts, light bending, and frame dragging (Chang et al. 2005). Thanks to its improved sensitivity, spectroscopic observations of X-ray bursts with the high spectral resolution of the XMS of slowly rotating neutron stars and with the moderate spectral resolution of the WFI for rapidly rotating neutron stars will enable us to detect weak absorption features, if present (see Figure 2.21). Despite the surface lines being broad and asymmetric, an estimate of the redshift ($z_{grav}$~GM/($Rc^{2}$)) can still be derived with a few per cent accuracy (see Figure 2.19). In addition to providing the redshift, if the line broadening is mostly rotational, then we will be able to constrain the stellar radius through the measured surface velocity (proportional to ΩR). On the other hand, if the neutron star is slowly rotating, then Stark pressure broadening, proportional to $M/R^2$ is likely to dominate. This implies that from a single detection of a gravitational redshift, in either the rotational or pressure broadening limits, the two unknowns M and R can be determined uniquely.

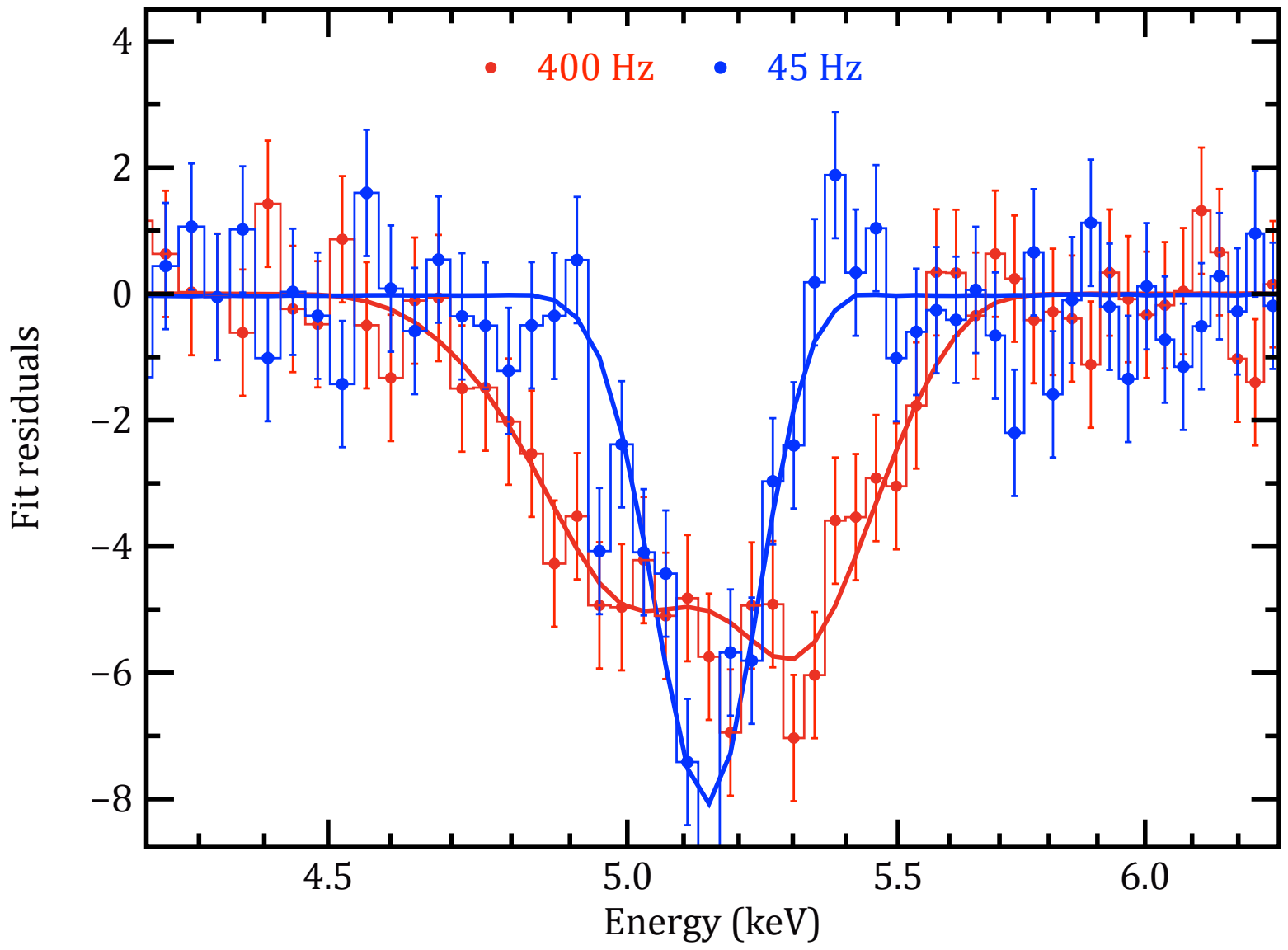

**Figure 2.21.** *Absorption-line spectrum (Fe XXVI Lya) of a 1.4 solar mass and 11.5 km neutron star observed with the WFI during a moderately bright (1 Crab) X-ray burst. The line profile includes the effects of light bending, rotational Doppler splitting, Doppler boosting and gravitational redshift. The figure shows the fit residuals (in units of sigma) with respect to the underlying continuum spectrum (excluding the line). The blue curve is for a 70 s exposure and neutron star spin of 45 Hz, while the red curve is for a 140 s exposure and a NS spin of 400 Hz. Thanks to the Athena-WFI count rate capabilities, from these measurements an estimate of the redshift (M/R) can still be derived with a few % accuracy, yielding insights on the equation of state of dense matter. Such absorption lines will be searched with Athena in a few tens of type I X-ray bursters.*

As has been emphasised above, actively accreting systems, whether SMBH at the centres of galaxies or compact objects in binary systems, offer an enormous opportunities to study strong GR environments, the physical process of accretion, and the properties of ultradense matter. *Athena* has been designed to exploit these opportunities, but will also produce enormous advances in our understanding of accretion at the lowest end of the mass scale, in accreting white dwarfs and young stellar systems such as T-Tauri stars. With *Athena*, we aim for a comprehensive understanding of the physical process of accretion in all classes of object. The importance of the accretion process for our wider understanding of the Universe is also within *Athena*'s grasp, and is described in the next section.

## 2.2 Cosmic feedback

Energy injection from growing SMBHs - AGN feedback - is a key ingredient in models of the formation and evolution of galaxies. Heating and/or expulsion of cold gas from galaxies by the AGN may drive the process of "downsizing" (i.e. massive galaxies forming first, smaller ones forming later), by quenching star formation in the largest galaxies. This process also explains the rapid move of galaxies from blue star-forming systems to "red and dead", passively evolving remnants, and the suppression of "cooling flows" in cluster cores. Star formation itself also acts as a feedback mechanism, through supernovae; this is especially important in starburst galaxies where spectacular superwinds expel large amounts of material out of the galaxy. Despite its apparently great importance, and its near-universal invocation in galaxy evolution modelling, the underlying physical processes driving AGN feedback are not well understood. *Athena* will address key outstanding issues in our understanding of feedback. The combination of high spectral and spatial resolution provided by the XMS will reveal the influence of both mechanical and radiatively-driven feedback in cluster cores, and also determine the detailed physical conditions in powerful AGN winds launched at small scales. High throughput, wide-field surveys with the *Athena* WFI will reveal which galaxies are undergoing their feedback phase, by providing a complete census of black hole growth throughout the main period of galaxy formation, including the most obscured systems. These WFI surveys will also yield the first glimpses of the importance of AGN growth and feedback in the early Universe, where the first galaxies and black holes formed. The XMS will provide unique insight into supernova-driven feedback and heavy element enrichment in starforming galaxies, again via spatially resolved spectroscopy. Specifically, *Athena* will:

- **Reveal the physics of feedback by measuring the energy flows due to SMBHs**

  *Athena*'s XMS will measure bulk velocities and line broadening measurements in the cool core region of clusters, where feedback is observed directly, revealing how the AGN couples to the intracluster gas, the energy content and timescales. The XMS will also measure the detailed physical conditions in AGN outflows launched as a consequence of the accretion processes in the innermost regions, where the feedback power ultimately originates.

- **Determine the growth of SMBH and the evolution of nuclear obscuration**

  Deep, wide area surveys with the *Athena* WFI will perform a census of galaxies growing their SMBH across cosmic time, giving clues to the triggering mechanism. Crucially, *Athena* will move beyond mere detection of these galaxies in X-rays and deliver high quality spectra, required to reveal the most obscured AGN and make the census complete.

- **Measure velocity and metallicity flows due to starburst superwinds**

  Spatially-resolved spectroscopy of nearby star-forming galaxies with the *Athena*-XMS will give an in-situ picture of the metal-enrichment of the intergalactic medium via the velocity flow and elemental abundances huge, galaxy-scale supernova-driven outflows.

### 2.2.1 The physics of feedback

Simple galaxy formation models based on the current standard cosmological paradigm over-predict both the number of local low luminosity/mass galaxies and the number of high luminosity galaxies. Feedback from supernovae and from AGN are believed to be the main ingredients missing in this simplistic picture at small and high masses respectively. But while the effects of feedback are very obvious and can be phenomenologically parameterised, their very origin, how much energy these processes generate, and how this energy is deposited in the surrounding medium are not well understood. Determining the physical causes of feedback, and the importance of AGN output in shaping the evolution of galaxies, are areas where *Athena* will provide an enormous leap forward.

There is no question that a growing black hole could drastically affect its host galaxy (Silk & Rees 1998). Whether and how it does so, however, is an open question that depends on how much of the energy released actually interacts with the matter in the galaxy. If the energy is in electromagnetic radiation and the matter largely stars, then very little interaction is expected. If the matter consists of gas, particularly with embedded dust, then the radiative output of the black hole can expel it via radiation pressure. Alternatively, if significant AGN power emerges in powerful winds or jets (see Figure 2.22), mechanical heating and momentum provide the link. Either form of interaction can be sufficiently strong that gas can be driven out of the galaxy entirely.

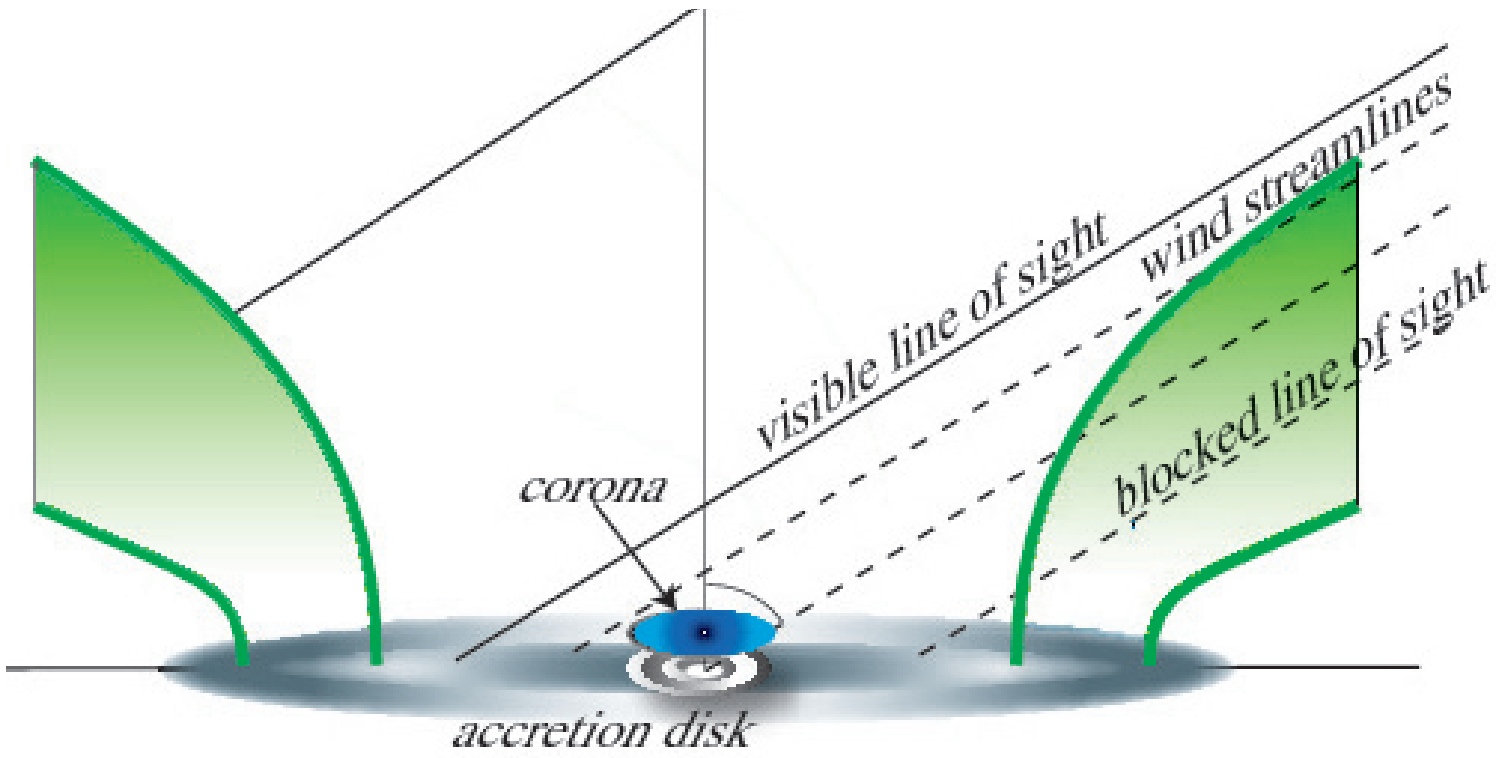

**Figure 2.22.** *Schematic view of the Compton-thick outflowing material around the Cloverleaf quasar, which blocks the direct view to the supermassive black hole and accretion disk, although the optically active broad line region can be seen (Chartas et al. 2007). The Compton-thick wind blocks the view of the near side of the accretion disk, but scattered and fluorescent emission from the far side and the outflow can reach the observer.*

The radiative form of feedback is most effective when the black hole is accreting close to its Eddington limit. The mechanical form associated with jets, on the other hand, continues to operate at accretion rates well below Eddington. X-ray observations are essential for studying all forms of feedback. In the cores of clusters, feedback is observed directly, via cavities, bubbles and ripples sculpted in the hot intracluster medium by the AGN. Ultimately, the energy source is close to the event horizon, and X-ray observations have shown clear evidence, via absorption line spectroscopy, of powerful winds being launched as part of the accretion process in nearby AGN. The next major step in understanding AGN feedback of both varieties will be provided by *Athena*. These measurements will probe over 9 decades in radial scale, from the inner accretion flow where the outflows are generated (~$10^{-4}$ pc, see Figure 2.23) to the halos of galaxies and clusters (~100 kpc or more), where the outflows deposit their energy. The study of feedback processes enabled by *Athena* will reveal the remarkable physical link between accretion around compact objects and the formation and evolution of galaxies.

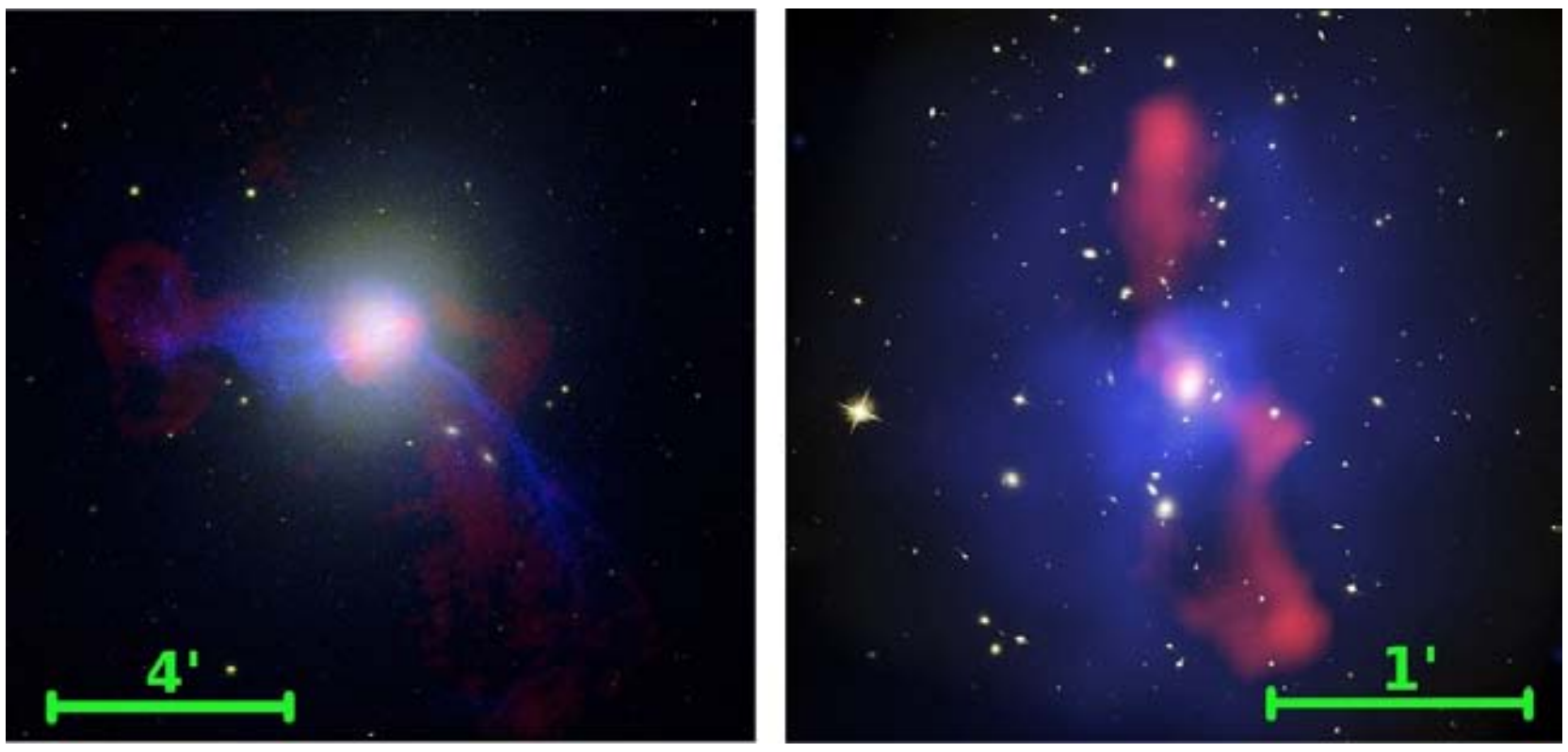

**Figure 2.23.** ***Left***: *X-ray emission (blue), radio emission (red) superposed on an optical image of M87. The X-ray structure was induced by ~$10^{58}$ erg outburst that began 10 Myr ago (Forman et al. 2005). The persistence of the delicate, straight-edge X-ray feature (part of the South West filament) indicates a lack of strong turbulence.* ***Right***: *X-ray emission (blue), 320 MHz radio emission (red) superposed on an HST optical image of the z=0.21 cluster MS0735.6+7421 (McNamara et al. 2005). Giant cavities each 200 kpc (1 arcmin) in diameter were excavated by the AGN. With a mechanical energy of $10^{62}$ erg, MS0735 is amongst the most energetic AGN known. This figure shows that AGN can affect structures on galaxy and cluster scales.*

ATHENA

*Athena* will make a major contribution to our understanding of cosmic feedback, through the measurement of the volume-filling component of both the energy outflow from supermassive black holes and of the surrounding hot medium in galaxy bulges, groups and clusters.

## Thermal regulation of gas in galaxy bulges and cluster cores

Mechanical feedback dominates in elliptical galaxies, groups, and clusters at late times, as shown by X-ray observations of gas in the bulges of massive galaxies and the cores of galaxy clusters and groups (e.g. Figure 2.23). The energy transfer process is surprisingly subtle. The radiative cooling time of the hot gas in these regions is often much shorter than the age of the system, so that without any additional heating, the gas would cool and flow into the centre. For giant ellipticals the resulting mass cooling rates would be of order 1 solar mass per year, but at the centres of clusters and groups, cooling rates would range from a few to more than a thousand solar masses per year. Spectroscopic evidence from *Chandra* and *XMM-Newton* show that some radiative cooling does occur, but nowhere near the extent predicted by simple cooling (Peterson & Fabian 2006). Limits on cool gas and star formation rates confirm this. Mechanical power from the central AGN acting through jets must be compensating for the energy lost by cooling across scales of tens to hundreds of kpc (McNamara & Nulsen 2007).

The gross energetics of AGN feedback in clusters are reasonably well established. It is directly relevant to feedback in galaxies as well, as the most massive galaxies occur and evolve in these cluster hot halo environments. Remarkably, relatively weak radio sources at the centres of clusters often have mechanical power comparable to the radiative output of a quasar, which is sufficient to prevent hot atmospheres from cooling (McNamara & Nulsen 2007). The coupling between the mechanical power and the surrounding medium is, however, poorly understood. Moreover, it is not clear how such a fine balance can be established and maintained.

The heat source – the accreting black hole – is roughly the size of the Solar System, yet the heating rate must be tuned to conditions operating over scales 9 or more decades larger. The short radiative cooling time of the gas means that the feedback must be more or less continuous. How the jet power, which is highly collimated to begin with, is isotropically spread to the surrounding gas is glimpsed only in the nearest clusters. The obvious signs of heating include bubbles blown in the intracluster gas by the jets (Figure 2.23) and nearly quasi-spherical ripples in the X-ray emission that are interpreted as sound waves and weak shocks. The disturbances found in the hot gas carry enough energy flux to offset cooling, but the microphysics of how such energy is dissipated in the gas is not understood. *Athena* measurements of the velocity field of the hot gas, enabled for the first time via the combination of spectral and spatial resolution offered by the XMS instrument, will enable the viscosity and dissipation to be determined.

The persistence of steep abundance gradients in the cluster gas, imprinted by the ejecta of supernovae in the central galaxy, means that the feedback is gentle, in the sense that it does not rely on violent shock heating or supersonic turbulence. The velocities in long filaments of optical line-emitting gas, as well as *XMM-Newton* RGS spectra from a handful of objects, in some objects also suggest low levels of turbulence. Yet the continuous streams of radio bubbles made by the jets, the movement of member galaxies and occasional infall of subclusters must make for a complex velocity field.

With high resolution imaging and moderate resolution spectroscopy *Chandra* and *XMM-Newton* have established AGN feedback as a fundamental astrophysical process in nature. Evidence has accumulated that gas in cluster centres cools by emitting the X-radiation we see down to temperatures of up to a factor 10 below that of the outer parts (e.g., Sanders et al. 2008). Little gas is seen at X-ray energies at lower temperatures, presumably due to the AGN feedback depositing energy into the cluster gas. *Athena* will, for example, determine the amount of gas at a temperature of 0.25 keV through its OVII emission at 21.6 Å. Most cool cluster cores do however have a central pool of cold atomic and molecular gas where star formation is sometimes seen. Mixing of the hot and cold gases may be an important additional coolant of gas below 5 million K. The net result is that gas does cool from the hot to the cold phase but only 10-20% of what would have cooled without the AGN feedback. *Athena* will map the coolest X-ray emitting gas phases showing where it occurs relative to the cold molecular gas.

At present we only have tantalising glimpses of the dynamics and mechanisms operating in these powerful outflows. *Athena* will open the field up through spatially-resolved, high spectral resolution spectral imaging.

Using spatially resolved, high resolution spectroscopy, *Athena* will measure how mechanical energy from supermassive black holes is distributed in galaxy bulges and clusters, via the process of cosmic feedback which generates flows and turbulence in a wide range of environments across the Universe.

Understanding the dynamics demands a leap in spectral resolution by more than one order of magnitude above that of *Chandra* and *XMM-Newton*. The *Athena* spectral resolution and sensitivity is needed to understand how the bulk kinetic energy is converted to heat. Its capabilities are essential in order to measure and map the gas velocity to an accuracy of tens of km/s, revealing how the mechanical energy is spread and dissipated (Figure 2.24). From accurate measurements of line profiles and from the variations of the line centroid over the image it is possible to deduce the characteristic spatial scales and the velocity amplitude of large (> kpc) turbulent eddies, while the total width of the line provides a measure of the total kinetic energy stored in the stochastic gas motions at all spatial scales. Such data will provide crucial insight into the intracluster medium heating mechanisms and track the mechanical energy flow. Observations of the kinematics of the hot gas phase, which contains the bulk of the gaseous mass, and absorbs the bulk of the mechanical energy in massive elliptical galaxies, are only possible at X-ray wavelengths.

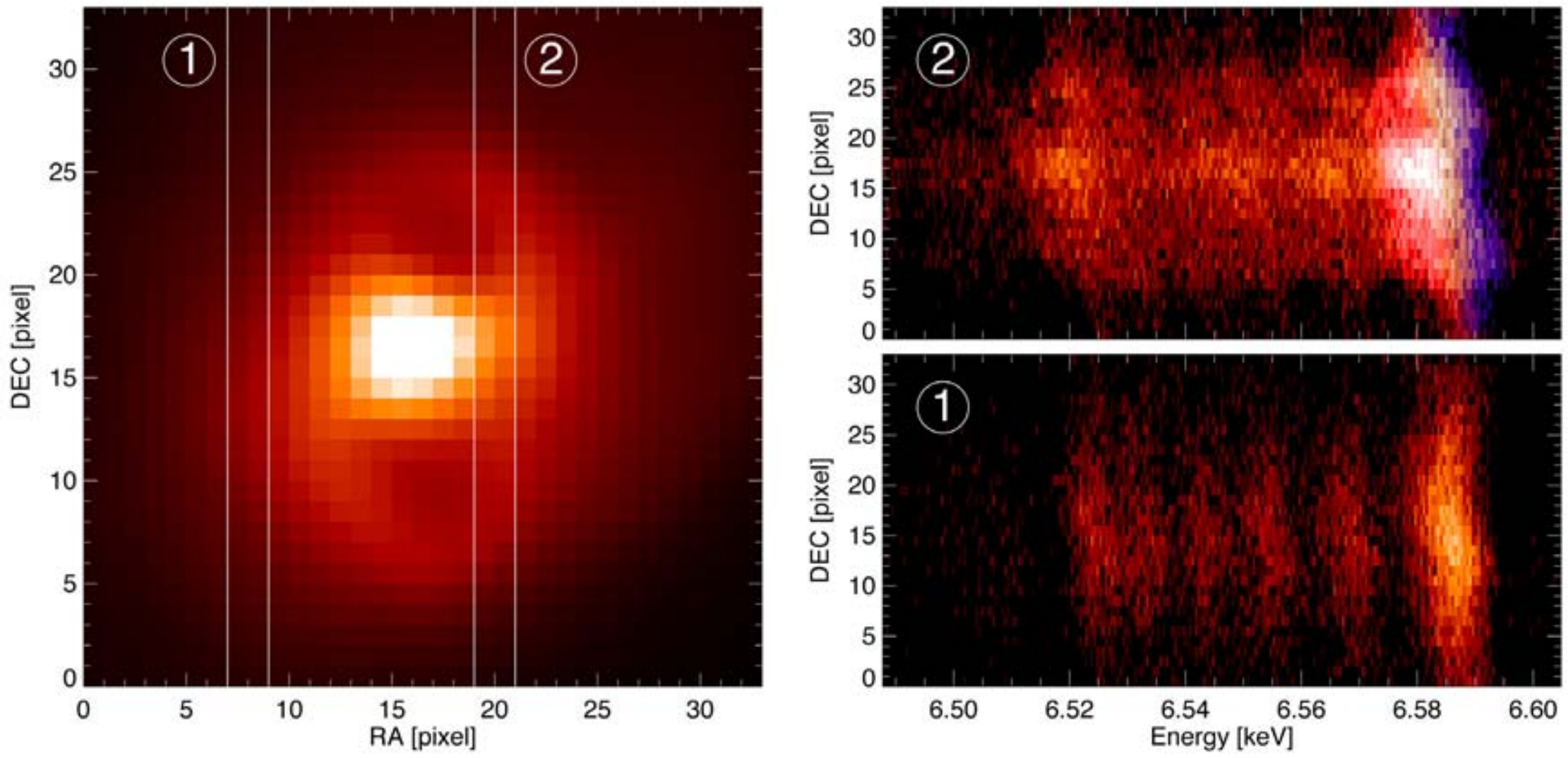

**Figure 2.24.** *Simulated high-resolution X-ray spectra from the shells and X-ray cavities in the Perseus cluster. The left panel shows the X-ray image and the chosen spectral slits, while the right shows the spatially resolved spectrum of the Kα lines from Fe XXV and Fe XXVI (both lines are multiplets), as will be observed by the Athena-XMS in an exposure of 250 ks with 10" resolution. At the location of the cavities, each of the lines splits into three components (approaching, restframe and receding), from which the velocity (and age) and therefore the jet power can be derived. Hydrodynamic simulations of jets in galaxy clusters with parameters appropriate for Perseus (jet power* $10^{45}$ *erg* $s^{-1}$*; Heinz, Brügen & Morsony, 2010) have been used.*

## Feedback from AGN winds

Spectroscopic observations have shown that highly ionised and energetically-significant outflows are present in most, if not all, AGN. These winds are an apparently inevitable consequence of the accretion process and must provide at least part of the link between the processes close to the black hole event horizon and their eventual effects on larger scales. The winds can be particularly dramatic in the outflows from some luminous quasars. UV and X-ray absorption lines indicate that outflows reaching 0.1-0.2c are present in some quasars and may be present in many. X-ray observations are required to determine the total column density and hence the kinetic energy flux. Current work on a small number of objects implies that this can be com-parable to the radiative luminosity (see, e.g. Figure 2.25). To diagnose the energetics of quasars we need large samples of quasars, comparing them with the less energetic (but still substantial) outflows from lower luminosity AGN, which can reach speeds of several thousand km/s. *Athena*'s high spectroscopic throughput will yield detailed evidence of how both radiative and mechanical AGN feedback operates to redshifts $z$=1-3, where the majority of galaxy growth is occurring. *Athena* will be sensitive to all ionisation states from Fe I – Fe XXVI, allowing us to study how feedback affects all phases of interstellar and intergalactic gas, from ten-thousand degree photoionised clouds to million-degree collisionally-ionised plasmas, and to measure the velocities and energetics of galactic outflows.

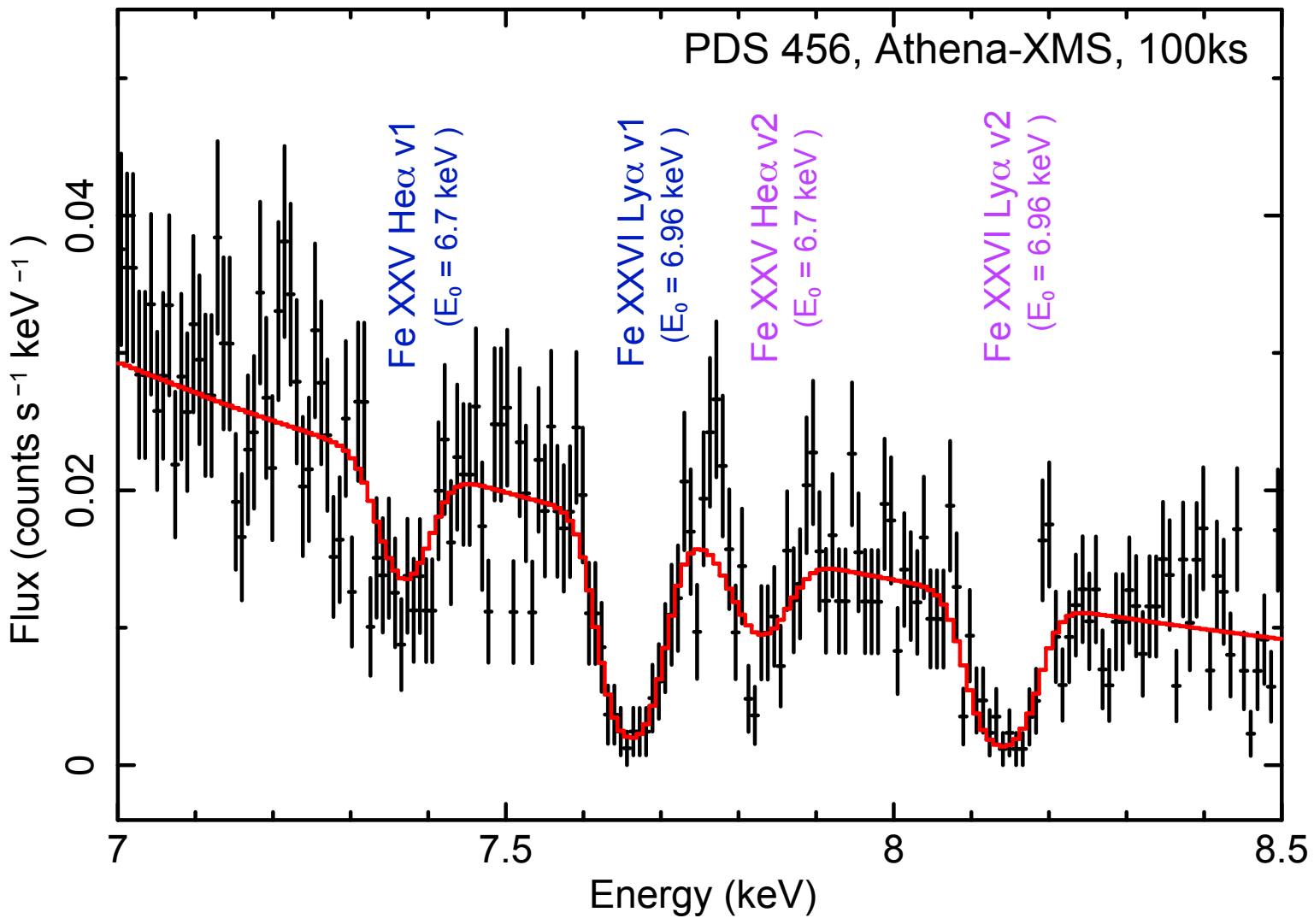

**Figure 2.25.** *Blue-shifted absorption lines are seen in the X-ray spectrum of the nearest powerful quasar, PDS 456, indicating a massive fast outflow. This quasar may be in the "blowout phase", as seen by Athena-XMS in just 100 ks. Such powerful winds might actually pinpoint a key phase in the post-merger galaxy evolution, and their study in more typical moderate luminosity AGN will be possible with Athena in reasonable integration times and out to high redshifts.*

ATHENA

*Athena* will make spectroscopic measurements of the feedback energy released though winds and outflows in quasi-stellar objects out to z~2, revealing to what extent these winds influence the evolution of galaxies at the peak of the cosmic history of star formation.

## 2.2.2 Obscured growth of supermassive black holes

In addition to revealing the physics of AGN feedback, *Athena* will enable us to determine the importance of AGN feedback in shaping the evolution of galaxies, by revealing which galaxies are growing their black holes, and in which mode. Supermassive (>$10^6$ $M_\odot$) black holes are ubiquitous in the centres of all galaxy bulges and their mass is correlated with the host galaxy properties (e.g. luminosity, bulge mass, stellar velocity dispersion; Ferrarese & Merritt 2000). Furthermore, the evolution of the cosmic X-ray emissivity (from accretion onto SMBH) seems to track that of star formation. This provides clear evidence for co-evolution of SMBHs and their host galaxies, with the connection almost certainly being AGN feedback. These observational findings have triggered major theoretical efforts. A possible scenario has emerged suggesting that SMBH growth over cosmic time occurs in two modes. The bulk of the population with lower masses would grow their SMBH and galaxy masses smoothly, possibly due to disk instabilities and cold flows of baryonic matter from the large scale structure filaments (Dekel et al. 2009) on relatively long ~Gyrs timescales. The most luminous and massive systems would assemble most of their mass on short timescales (~$10^8$ years) through an "evolutionary sequence" which is likely to be controlled by mergers and interactions between gas–rich galaxies. In both cases the gas inflow triggers a starburst and feeds the accreting, growing SMBH. As the SMBH grows, accretion-driven energy and momentum pushes the gas and eventually expels it, thus quenching the star-formation episode. For most of the SMBH growth phase, the nucleus is heavily obscured by gas and dust, and therefore hidden in the optical/UV and soft X-ray wavebands. The accreting SMBH shines as an optically luminous QSO for a brief period until it runs out of gas. The final product of the sequence is a "dead quasar" inside a passively evolving galaxy. Within this general framework, the times-cales associated to each phase and whether all major episodes of BH growth are highly obscured remain unknown; moreover, the relative importance of the two modes (fast and secular) for the SMBH growth as a function of cosmic time is far from being understood.

The key to resolving these issues is to perform a complete census of the AGN population at the major epoch of galaxy growth. *Athena*'s sensitivity will allow us to detect the "smoking gun" X-ray spectral signature of the most obscured AGN growth, the intense iron Kalpha emission produced in both transmission and reflection from the obscuring material. Significant samples of these ultra-obscured Compton-thick AGN will be accessible in galaxies up to $z$~1-3, when both mass accretion and star formation peaked, hence providing a complete census of obscured AGN. This will provide further hard evidence for the SMBH/host galaxy co-evolution, and further handles on the physical mechanisms actually at work which link them.

In some models, the main growth phase for SMBHs in massive galaxies should happen in a circumnuclear environment shrouded by large amounts of gas and dust. This fits well with the known observational fact that most accretion in the Universe occurs in obscured environments. In the past decade, X–ray surveys have revealed a fairly robust picture of AGN evolution up to moderate redshifts ($z$~1) and absorption column densities ($N_H \sim 10^{23}$ cm$^{-2}$). The evolution of the X–ray luminosity function shows that lower luminosity AGN (Seyfert–type galaxies) reach a peak in their space density at later times than more luminous QSOs (e.g. Hasinger et al. 2005), similarly to what happens for star–forming galaxies and usually referred to as "cosmic downsizing" (Cowie et al. 1996). From the luminosity dependent behaviour of the accreting SMBH space density, a "golden epoch" of SMBH activity is identified over the relatively broad $z$~1-3 redshift range. However, the fraction of obscured objects as a function of luminosity, and its cosmic evolution in that range are still hotly debated, crucially depending on detecting absorbed sources and recognizing them as such.

A large population of obscured and Compton-thick AGN over a broad range of redshifts and luminosities is also predicted by X–ray background (XRB) synthesis models (e.g. Gilli et al. 2007). Depending on their numbers, they might make a major contribution to the accretion power of the Universe (Fabian & Iwasawa 1999). High-quality X–ray spectroscopy of nearby AGN suggest that heavy absorption, possibly in the Compton-thick ($N_H > 10^{24}$ cm$^{-2}$) regime are reasonably common in the local Universe, forming about 20% of the total AGN population (e.g. Brightman & Nandra 2011). A census at higher redshifts is missing, because the quality of the X-ray spectra from current facilities is poor due to limited photon statistics. Ultra–deep X–ray surveys with *Chandra* and *XMM-Newton* have started to unveil the tip of the iceberg of the heavily obscured AGN population at high redshifts (Comastri et al. 2011; Feruglio et al. 2011). However they are still limited by a relatively poor combination of small field of view and low sensitivity, which prevent them from assembling large samples of heavily obscured and Compton-thick AGN in the redshift range ($z$~1-3) where the bulk of the action is. As a result, key outstanding questions remain, for example:

- How is the X-ray background composed? To this day the proportion of Compton-thick AGN is poorly understood at moderate to high redshift.
- Do obscured AGN evolve differently to unobscured? This is a fundamental open question which is of major significance to SMBH growth modes.
- Are obscured AGN a distinct phase in the co-evolution of galaxies and SMBH? If the obscuration arises from obscuration in the "blow-out" phase of feedback models, the host galaxies of the most obscured populations may be fundamentally different to those of unobscured objects.

With the first statistically significant samples of the most obscured objects, *Athena* will provide definitive information to answer these questions The key breakthrough for *Athena* is to move from mere detection of these (which is possible with current instrumentation), to complete characterisation, particularly of the spectrum. As demonstrated in Figure 2.26, this is crucial, because hardness ratio analysis cannot distinguish between moderately obscured, transmission dominated objects, and the most heavily obscured, reflection-dominated AGN. *Athena* will move us into the regime where these degeneracies can be broken, and significant samples of Compton-thick AGN assembled. In particular, at z~1-3, the telltale intense Kα fluorescent emission is shifted towards the maximum of the effective area and will be easily detected and characterised.

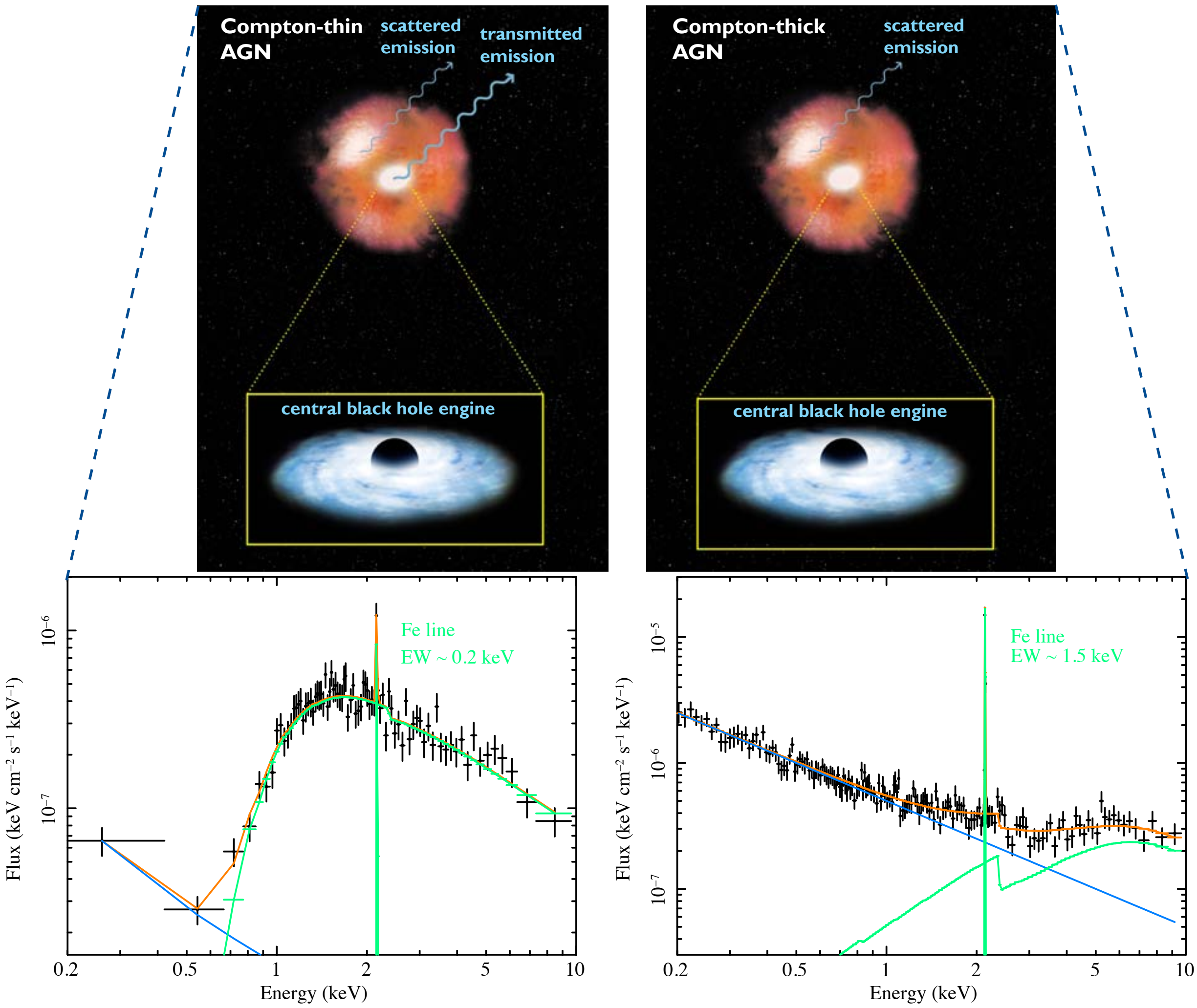

**Figure 2.26.** *Artist's conception of a Compton-thin (**top left**) and Compton-thick (**top right**) AGN. The simulated Athena spectra shown are for two high redshift (z=2) AGN in a deep (1 Ms) WFI exposure. The sources have the same hard (2-10 keV) to soft (0.5-2 keV) flux ratio (~3.2) but a rather different spectral shape that only Athena will be able to reveal, disentangling the amount of obscuration surrounding these growing supermassive black holes.* ***Bottom left:*** *The absorption column density is in the Compton-thin ($N_H$ ~ $10^{23}$) regime and a weak scattered/reflected component is emerging at low energy.* ***Bottom right:*** *The source is obscured by Compton-thick matter and the hard X-ray spectrum is Compton reflection dominated. A prominent soft X-ray component is also present possibly due to a leakage of the nuclear continuum through the absorbing gas.*

Wide area deep surveys with *Athena* will yield a census of the obscured growth of supermassive black holes in the Universe, which provide the power source driving feedback in the evolving galaxy population.

The unique combination of FOV, sensitivity and spectral resolution of the instruments onboard *Athena* will enable the assembly of huge samples of growing SMBH, including >100 spectrally-confirmed Compton-thick objects, over a broad range of luminosities in the redshift range z~1-3. These samples are required to test competing models for SMBH growth, which in turn dictate the importance of feedback in SMBH/

galaxy co-evolution scenarios. For example, merger-driven growth makes different predictions for the X-ray luminosity function evolution than disk instabilities (Bower et al. 2006; Guo et al. 2011). The former predict far more high redshift AGN, specially at low luminosities. Different predictions for the shape of the luminosity function and source counts are obtained for a different combination of model parameters (i.e. the relation between the halo mass and the peak luminosity, the delay between the triggering of AGN activity and unobscured QSO phase, etc.), so all of these can be constrained, provided the X-ray Luminosity Function (XLF) itself has been constructed using a complete census, including the most obscured populations, as will be the case with *Athena*.

## The early growth of supermassive black holes

In addition to understanding the peak of BH and galaxy evolution at $z$=1-3, *Athena* will also be able to shed important light on the epoch where the first galaxies and SMBHs formed at $z$>6. The known AGN population at $z$>6 currently consists of luminous optical quasars (e.g. Fan et al. 2003) hosting SMBHs exceeding a few billion solar masses, with the record-holder at $z$=7.085 (Mortlock et al. 2011). How massive SMBHs formed at such early epochs is a major unknown and the current challenge is to start to probe lower luminosity and obscured AGN at these redshifts to reveal more typical systems contributing more to the total accretion power at this epoch.

Present deep and large area surveys with *Chandra* have yielded a handful of AGN candidates at $z$>5 (e.g. Civano et al. 2011; Fiore et al. 2011), but none is yet confirmed at $z$>6. Even at $z$~4, the present constraints on the shape of the luminosity function are plagued by extremely large uncertainties.

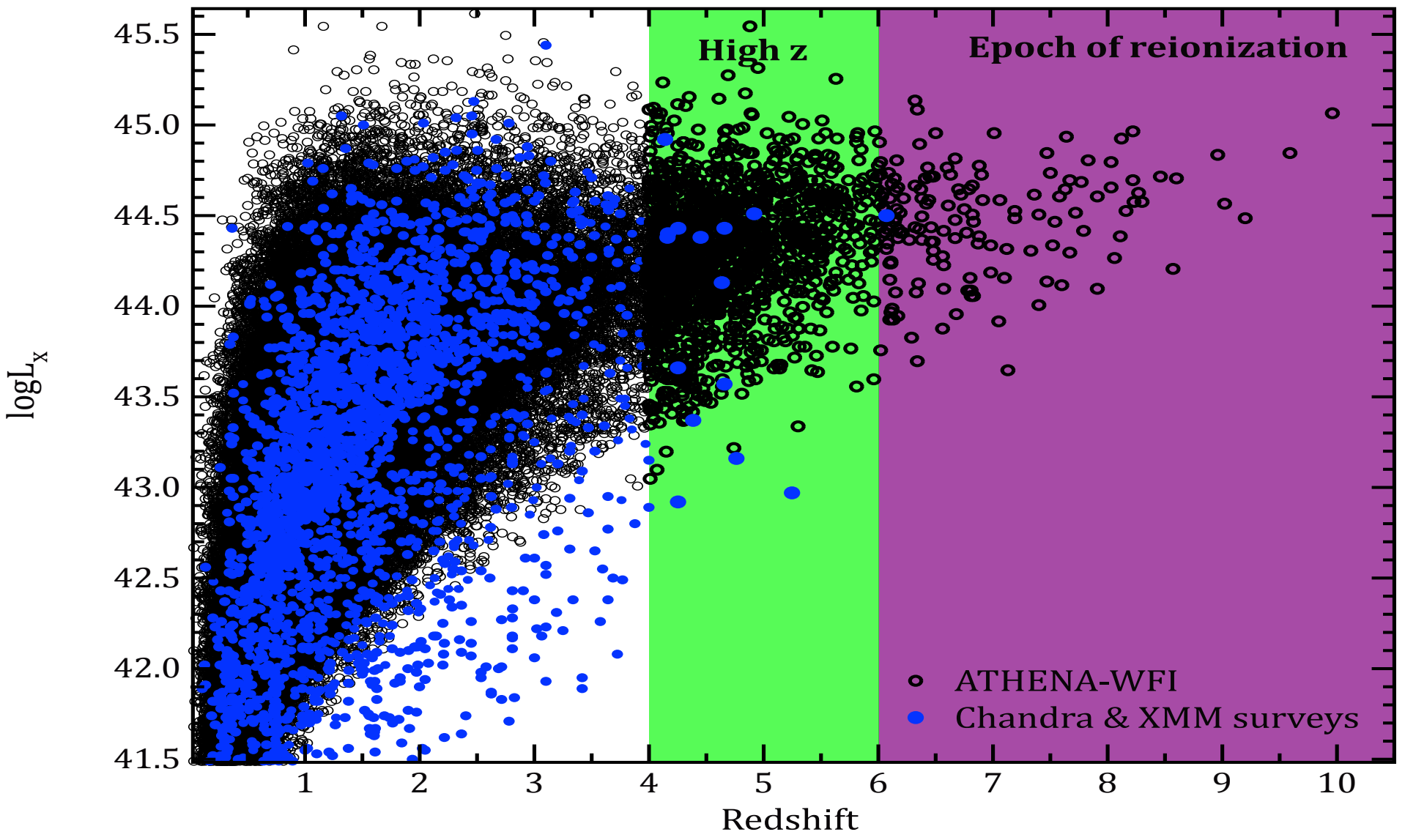

**Figure 2.27.** *Predicted coverage of the luminosity-redshift plane in newly discovered AGN by Athena-WFI in the multi-cone survey planned (black points) as compared to the existing samples constructed with Chandra and XMM-Newton (blue points). The number of 4<z<6 AGN that the Athena will unveil will amplify by a large factor the existing modest numbers of X-ray detected AGN in this redshift regime, enabling population studies of early AGN. In addition, tens of AGN beyond the z>6 epoch and peering into the reionization epoch will be uncovered by Athena.*

The detection of a large population of obscured SMBHs at high (4<$z$<6) and very high ($z$>6) redshifts requires an X-ray observatory able to reach fluxes around the present *Chandra* limit over a much larger area. *Athena* provides such capabilities (see Figure 2.27). A suitably designed "multiple-cone" survey with the *Athena*-WFI, consisting of deep, medium-deep and shallower wide area components will yield 700-1600 X-ray selected AGN, most of them obscured, at $z$~4-6, and from a few tens to a few ten hundreds at $z$>6 depending on the actual luminosity function evolution (Figure 2.28). Numbers will be even larger if the goal WFI FOV can be achieved and with the goal PSF, the highest redshifts ($z$>6) will open up even further. Note that high redshifts cause the least obscured, hard X-ray band, to be shifted into the sensitive part of the *Athena* band.

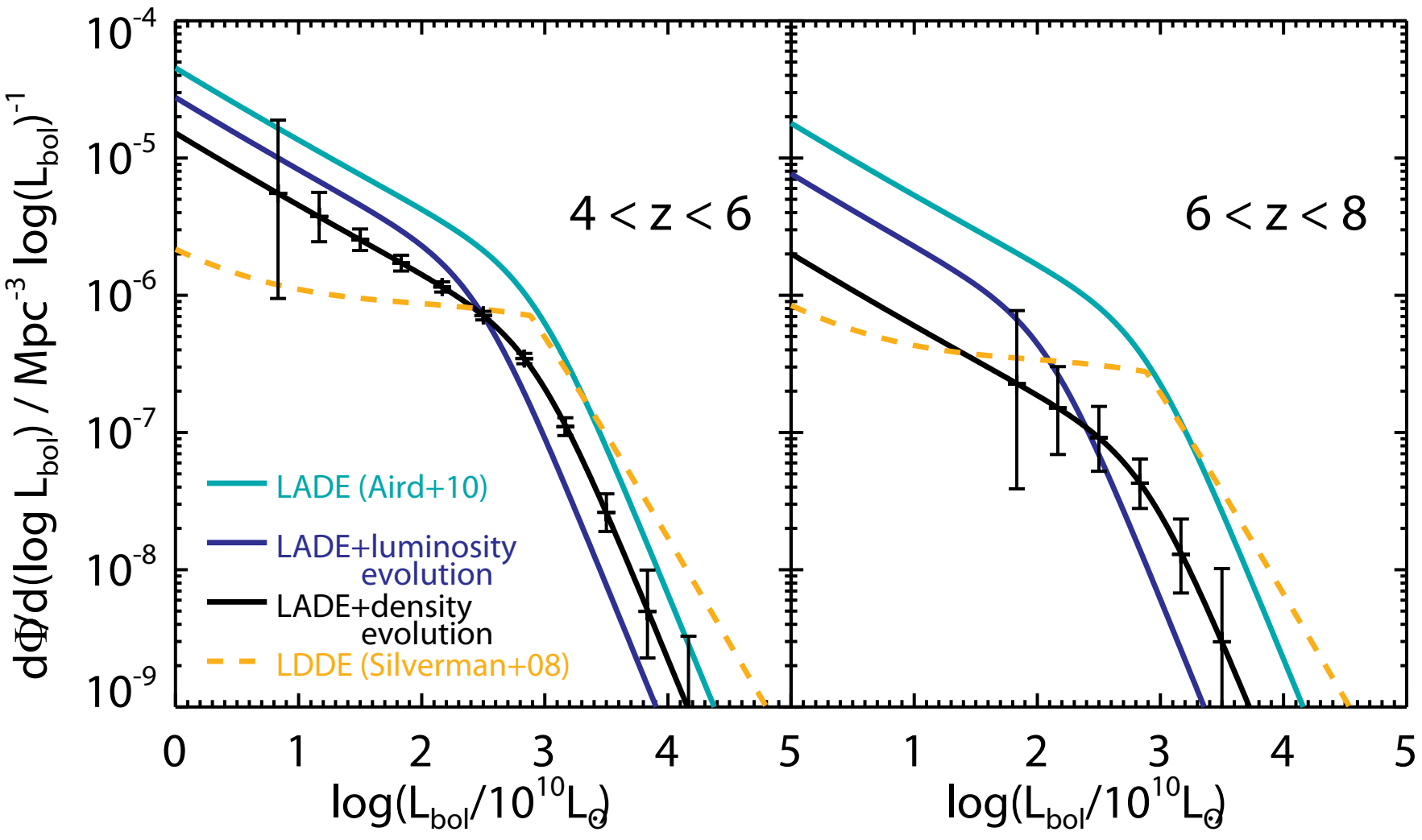

**Figure 2.28.** *The predicted bolometric luminosity function for the 4<z<6 and 6<z<8 redshift intervals for a range of X-ray luminosity function phenomenological extrapolations (LADE - Luminosity And Density Evolution and LDDE - Luminosity Dependent Density Evolution) assuming a 10" PSF. The points show predicted binned estimates for the Aird et al. (2010) model plus negative density evolution to match the observed number counts at lower redshift with their predicted errors (based on the Poisson error in the number of objects). As explained above, the luminosity function at high redshift encodes the information about the relative role of mergers and disk instabilities in supermassive black hole growth, the relationship between halo mass and peak luminosity, and the lifetimes of different phases of black hole growth. Athena has the unique ability to cover large areas of sky and probe down to faint fluxes, which is required to identify moderate luminosity, obscured AGN that dominate black hole growth in the early Universe.*

> Sensitive large area surveys with *Athena* will yield the first samples of X-ray emitting supermassive black holes in the early Universe, stretching back into the recombination era at z>6.

As at lower redshifts, different black hole formation scenarios predict different evolutionary patterns and hence XLFs. For example, rapid growth by mergers predicts a "downsizing" or luminosity-dependent evolution (LDDE) behaviour even at high redshift. Determining the host galaxy properties of high redshift AGN found by *Athena* will add further constraints to formation models. For example, merger-driven growth should be accompanied by strong star-formation, while monolithic collapse models predict none, as the gas cloud from which the black hole forms must collapse without fragmentation. Follow up observations of *Athena* high redshift AGN with the full set of future multi-wavelength facilities (EVLA, ALMA, SPICA, E-ELT, JWST, SKA) will yield their redshift, host galaxy properties, stellar and gas masses, and star formation rates. Conversely, follow up observations of high redshift galaxies identified by these other facilities will be the most efficient (and sometimes the only) way of revealing those with growing SMBHs. The full suite of multi-wavelength observations will yield a rich treasury of information on how AGN activity and feedback relates to galaxy formation at early times. This information will be hopelessly incomplete unless it includes X-ray observations of the sensitivity afforded by *Athena*.

## The supermassive black hole growth mode

A unique probe of the growth and evolution of SMBHs can be derived from *Athena* measurements of black hole spin in nearby AGN (see Section 2.1.1). This is because different growth modes - smooth accretion and major BH merger events representing the most extreme examples - produce very different SMBH spin dis-tributions at late epochs. Steady accretion tends to give high spin, whereas event-driven "chaotic accretion" of gas with random angular momentum vectors gives low spin values, and merger-dominated growth gives values in between (Berti & Volonteri 2008). The spin distribution of AGN in the local Universe is therefore a powerful discriminator between different scenarios which nonetheless may form identical mass-functions (see Figure 2.29 with theoretical distribution at $z$=0, and simulated distribution).

**Figure 2.29.** *Spin as a probe of supermassive black hole growth history. The distribution of black hole spins in the local Universe depends on whether they have accumulated their mass predominantly via mergers, steady accretion or chaotic accretion. The theoretical expectations for each supermassive black hole growth scenario (dotted histograms) is shown (Berti & Volonteri 2008) and compared to simulated Athena measurements (solid histograms), accounting realistically for all observational effects and errors.*

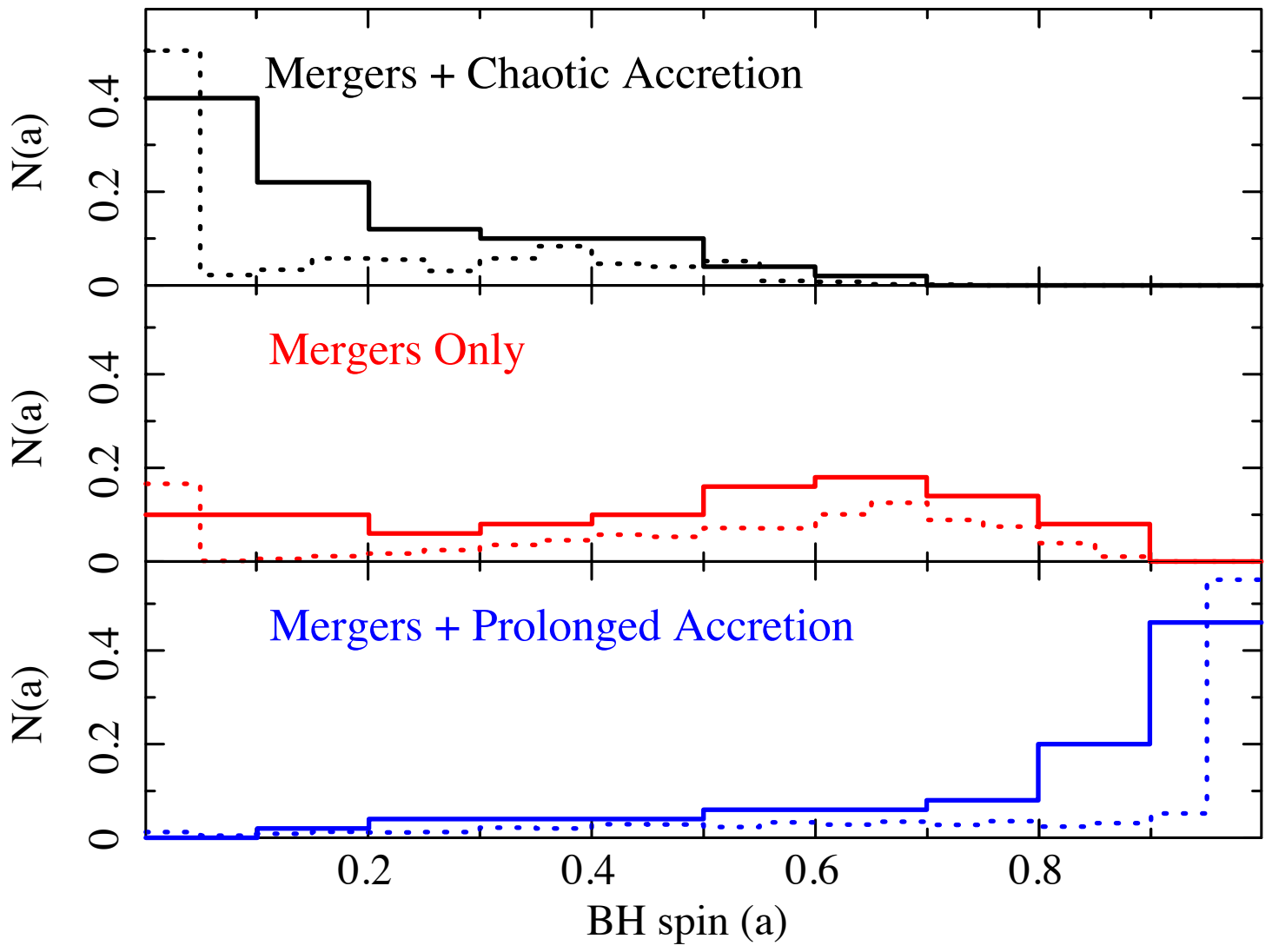

Current X-ray observatories are beginning to measure spins in a handful of BH systems, but large uncertainties and ambiguities still remain. The multiple, complementary ways for *Athena* to measure spin, i.e. through orbital modulations, reverberation, Fe line profile and QPOs (see Section 2.1.1) will provide the necessary self-check and the best means of obtaining the largest possible number of spin measurements.

A survey of the spins of supermassive black holes in nearby AGN with *Athena* will provide a unique probe of the growth mode of supermassive black holes in the Universe, determining the relative importance of the brief cosmic fireworks of major galaxy mergers, versus the quiet, steady buildup of mass over long timescales.

### 2.2.3 Feedback from starburst driven superwinds

In addition to AGN feedback, star formation itself, through supernovae and winds, also influences the ability of galaxies to form further stars. This is specially true in starburst galaxies, which drive galactic scale outflows or superwinds that may be responsible for removing metals from galaxies and enriching the intergalactic medium. Superwinds are powered by the winds of massive stars and by core collapse supernovae which collectively create T<$10^8$ K bubbles of metal-enriched plasma within star forming regions. These over-pressured bubbles expand, sweep up cooler ambient gas, and eventually blow out of the disk into the halo. Tremendous progress has been made in mapping cool entrained gas in outflows through UV/optical imaging and absorption line spectroscopy, demonstrating that superwinds are ubiquitous in galaxies forming stars at high surface densities and that the most powerful starbursts can drive outflows near escape velocity. Twenty percent of massive star formation at low redshifts occurs in starburst galaxies. Starburst/superwind galaxies account for most star formation at redshifts $z$=2-4. Ultra-luminous infrared galaxies show evidence for mass outflows of order 1000 solar masses per year at around 1000 km/s.

Spectroscopic measurements with the *Athena* XMS will enable direct measurements to be made of the rates at which starburst galaxies of all masses eject gas, metals, and energy into the intergalactic medium, through the determination of the composition, energetics and flow rates of the hot gas (Figure 2.30). Through detection of any central obscured AGN, it will also help determine the role of AGN in enhancing superwinds. This will provide crucial observational constraints on theoretical models of galaxy evolution, which have begun to incorporate superwinds, using ad-hoc prescriptions based principally on our knowledge of the cool gas. These efforts are fundamentally impeded by our lack of information about the hot phase of these outflows. The hot X-ray emitting phase of a superwind contains the majority of its energy and newly-synthesised metals, and given its high specific energy and inefficient cooling it is also the component most likely to escape from the galaxy's gravitational potential well. Knowledge of the chemical composition and velocity of the hot gas are crucial to assess the energy and chemical feedback from a starburst. These processes may be responsible for enrichments of the intergalactic medium and the galaxy mass-metallicity relationship.

ATHENA

Spatially resolved high-resolution spectroscopy around starburst galaxies will measure the amount of gas, metals and energy deposited by galactic superwinds into the intergalactic medium.

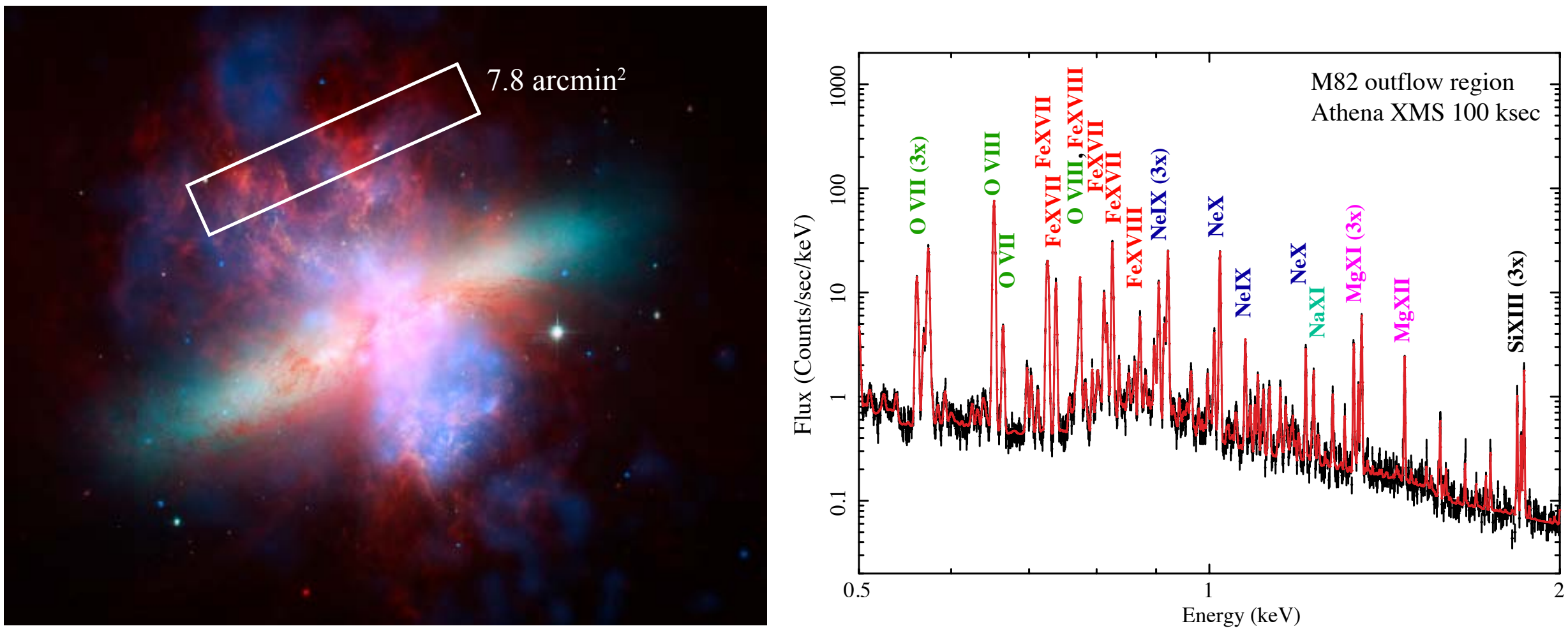

**Figure 2.30.** ***Left:*** *Composite image of Messier 82, which exhibits a starburst-driven superwind. Diffuse thermal X-ray emission as seen by Chandra is shown in blue. Hydrocarbon emission at 8 μm from SPITZER is shown in red. Optical starlight (cyan) and Hα+[NII] (yellow) are from HST-ACS observations.* ***Right:*** *The unique spectral and imaging capabilities of the Athena-XMS will conclusively determine mass outflows from starburst galaxies into the IGM. The mass, temperature, and ionization state of many elements will be determined using diagnostic Fe L-shell, H-like, and He-like lines (which are labeled as 3x).*

## 2.3 Large-scale structure of the Universe

The history of the formation of large-scale structure over cosmic time is encoded in the hot baryons that reside in groups, clusters of galaxies and filamentary structures in the Universe. Clusters of galaxies are the largest gravitationally bound baryon reservoirs in the Universe, enriched by the member galaxies and heated to X-ray emitting temperatures ~$10^7$-$10^8$ K. As a result they can only be accessed via observations in the X-ray band.

Thanks to its capability to perform spatially resolved high-resolution spectroscopy, *Athena*'s XMS will unveil the velocity structure of clusters of galaxies, look for turbulence in the hot intracluster gas and find when in cosmic time the excess energy in clusters was deposited. It will also map the evolution of abundances of chemical elements, through metallicity studies of intracluster gas, which receives the heavy elements deposited by supernovae in the member galaxies. *Athena* will also provide new constraints on cosmology, including dark energy, in a fashion which is highly complementary and largely orthogonal to dedicated dark energy experiments like Pan-Starrs, DES or ESA's M-class mission *Euclid*. Working in concert, observations with the XMS and WFI will beat down systematic errors on mass measurements of clusters – essential if they are to be used as cosmological probes – and give accurate measurements of the baryon fraction in clusters which can be used as a standard candle. At low redshift, the vast majority of the baryons in the Universe (around 40%) are thought to exist in the form of a Warm and Hot Intergalactic Medium (WHIM). This gas, with a density significantly lower than the intracluster medium and at temperatures ranging from $10^5$ to $10^7$ K, is presumed to follow the filamentary structure dictated by the dark matter distribution. Such gas has so far not been convincingly detected due to limitations in current and foreseen X-ray instrumentation. With its combination of high spectral resolution and imaging capability, *Athena* will characterise the WHIM both in absorption towards bright background sources and in emission in high-density filaments. *Athena* will reveal key WHIM properties like spatial distribution, ionization state and chemical abundances.

*Athena*'s goals for large scale structure investigations are to:

- **Determine the dynamical, thermodynamical and chemical evolution of hot baryons**

  The XMS instrument aboard *Athena* provides high spectral resolution over a broad bandpass, with high sensitivity, yielding bulk and turbulent velocities, temperature diagnostics and abundances of clusters. The large field of view of the WFI will measure the energy-resolved surface brightness to far larger radii, out to the virial radius, for a precise characterisation of the entire cluster.

- **Constrain cosmological models, including dark energy, using clusters of galaxies**

  Follow-up observations by *Athena* of known cluster samples (e.g. selected by *eROSITA* especially at large redshift) will provide accurate mass measurements and control of systematics necessary for precision cosmology. Sensitive wide area surveys with the WFI will yield huge samples of new groups and clusters out to the redshifts when they first formed.

- **Complete the census of baryons in the local Universe**

  The high sensitivity and spectral resolution of the *Athena*-XMS should reveal the long searched-for reservoir of missing baryons in the local Universe, expected to be resident in the WHIM. This is accessible via both absorption line studies against bright background objects, and via its emission, which can be mapped using the XMS spatial resolution.

### 2.3.1 The dynamical, thermal & chemical evolution of clusters of galaxies

Recent observational progress, showing that 95% of the total mass-energy of the Universe is contained in cold dark matter and dark energy, has allowed us to robustly define the cosmological framework in which structures form. Much progress has been made in reconstructing the evolution of the dark matter distribution from its initial density fluctuations. In contrast, we still do not understand the evolution of the baryonic visible component of the Universe, trapped in the dark matter potential wells.

Galaxy clusters are key environments to reveal the complex physics of structure formation. Over 80% of their mass (between few times $10^{13}$ $M_\odot$ and up to $10^{15}$ $M_\odot$) is dark matter. The remaining mass is composed of baryons, 85% of which are in the form of a diffuse, hot plasma ($T > 10^7$ K, the Intra-Cluster Medium or ICM) that radiates primarily at X-ray wavelengths. This gas has been enriched with metals and is a key tracer of the chemical evolution of the Universe. Thus, in galaxy clusters, through the radiation from the hot gas and the galaxies, we can observe and study the interplay between the hot and cold components of the baryonic mass budget and the dark matter. Spatially-resolved X-ray spectroscopy with the *Athena*-XMS will provide detailed diagnostics of the ICM in order to trace the gas motions, to study shocks and cold fronts, to evaluate the non-equilibrium ionisation state of merging or supra-thermal plasma, to unveil multi-temperature and entropy structure. With sufficient sensitivity to cover a large enough redshift range, *Athena* will deliver an evolutionary picture of these processes in shaping the thermo-dynamical properties of the hot baryons which are locked within galaxy groups and clusters.

### How do hot baryons dynamically evolve in dark matter potentials?

Clusters form as the end product of the merging of smaller structures of dark and luminous matter accreted along large scale structure filaments. X-ray observations show that many present epoch clusters are indeed not relaxed systems, but are affected by shock fronts and contact discontinuities (Markevitch & Vikhlinin 2007), and that the fraction of unrelaxed clusters likely increases with redshift. Although the intracluster gas evolves in concert with the dark matter potential, this gravitational assembly process is complex, as illustrated by the temporary separations of dark and X-ray luminous matter in massive merging clusters such as the "Bullet Cluster" (Clowe et al. 2006). There are important questions to be answered both to understand the complete story of the cluster formation from first principles and, through a better understanding of cluster physics, to increase the reliability of the constraints on cosmological models derived from cluster observations (see Section 2.3.2). These include:

- How is the gravitational energy that is released during cluster hierarchical formation dissipated in the intra-cluster gas, thus heating the ICM, generating gas turbulence, and producing significant bulk motions?
- What is the total level of non-thermal pressure support, which should be accounted for in the cluster mass measurements, and how does it evolve with the halo mass and time?

To answer these questions, we need to map clusters' structure in velocities and turbulence. At present, only upper limits to these have been obtained (with the RGS aboard *XMM-Newton*, Sanders et al. 2011).

To take the next step requires more than an order of magnitude improvement in spectral resolution, coupled with large effective area and good spatial resolution. These are precisely the capabilities provided by the *Athena*-XMS, which will enable gas flow velocities and gas turbulence to be locally measured in clusters of galaxies using X-ray line width and position measurements with an accuracy of around 50 km/s (see Figure 2.31). This is sufficient to measure the thermal broadening in typical clusters, and hence will reveal any significant non-thermal pressure, as well as bulk velocities. In particular, sub-cluster velocities and directions of motions will be measured by combining redshifts measured from X-ray spectra (which gives relative line-of-sight velocities) and total sub-cluster velocities deduced from temperature and density jumps across merger shocks or cold fronts (Markevitch & Vikhlinin 2007). These measurements will probe how the hot ICM gas reacts in the evolving dark matter potential.

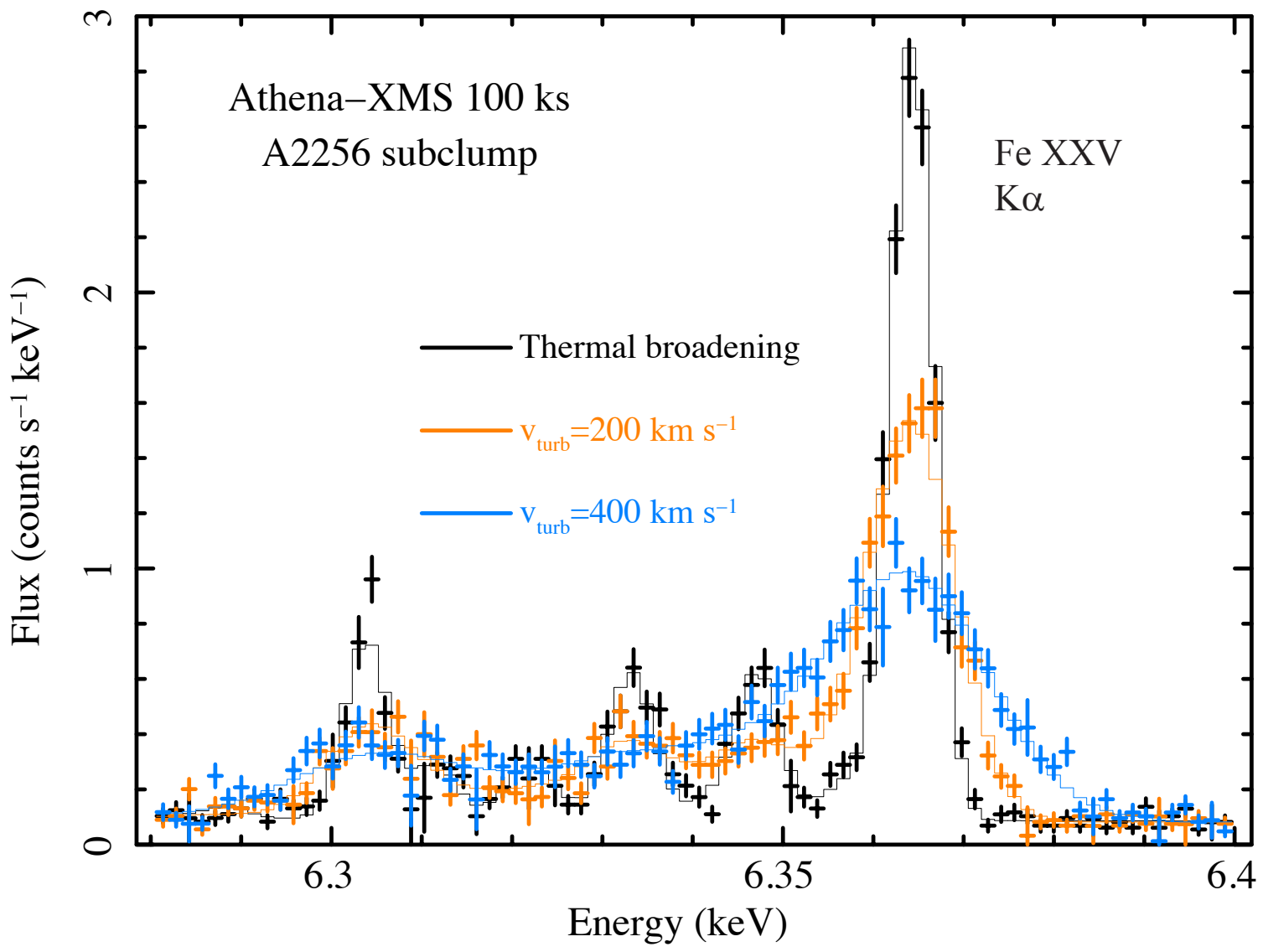

**Figure 2.31.** *Turbulence broadening of the Fe XXV Kα line over a circular region of 1.5 arcmin in radius centred on the subclump accreting onto the main body of A2256 (z=0.0528), from a 100 ks simulation with Athena-XMS. Simulated data with v=200 km/s is shown in orange. Black and blue represent the model with v=0 and v=400 km/s, respectively. The residual bulk motions and gas turbulence above 200 km/s can add significantly to the thermal pressure of the gas and therefore offset significantly cluster mass estimates using the assumption of hydrostatic equilibrium between the intra-cluster medium and the cluster potential. This might be at the root of the discrepant X-ray and lensing cluster mass estimates (e.g. Lau et al. 2009, Zhang et al. 2010, Meneghetti et al. 2010).*

Ions carry most of the kinetic energy involved in accretion shocks or mergers. They will thermalise through Coulomb collisions on a time scale shorter than the one that regulates the energy transfer between ions and electrons. The ion temperature is only estimated from the broadening of the emission lines, while the electron temperature can be obtained from the spectrum's continuum. High-resolution spectroscopy is thus of crucial importance for studying the non-equilibrium ionisation state and the two-temperature structure of the X-ray interacting plasma. Any non-thermal electron distribution can also have an impact on the ionization balance in the ICM. The ratios of the satellite to resonance lines are sensitive to departures from a Maxwellian electron distribution. Such departures are expected, for instance, in stellar flares, supernova remnants, and in shocked regions of the ICM, where an excess of the J He-like Fe XXV satellite line to the resonance line of the H-like Fe XXVI can be significantly detected in case of supra-thermal emission (Kaastra et al. 2009). *Athena*'s high-resolution spectroscopic capability will resolve the emission complex around ~6.7 keV arising from He-like Fe lines and their satellites, allowing these lines to be used as a strong and independent temperature diagnostic of the ICM in the range $10^7$-$10^8$ K (Swartz & Sulkanen 1993, Porquet et al. 2010).

ATHENA

*Athena* will determine intra-cluster gas motions and turbulence via spectroscopic measurements, showing how baryonic gas evolves in the dark matter cluster potential wells.

## How and when was the excess energy in the intracluster medium generated?

As discussed earlier, one of the most important revelations from recent observations is that non-gravitational processes, particularly feedback from outflows created by SMBHs and supernovae (see Section 2.2), must play a fundamental role in the history of all massive galaxies as well as the evolution of groups and clusters as a whole. Cosmic feedback is likely to provide the extra energy required to keep large quantities of gas in cluster cores from cooling all the way down to molecular clouds and to account for the entropy (i.e. energy) excess observed in the gas of groups and clusters.

It is now well established from *XMM-Newton* and *Chandra* observations of local clusters and groups that their hot atmospheres have much more entropy than expected from gravitational heating alone (Pratt et al. 2010; Sun et al. 2009). Determining when and how this non-gravitational excess energy was acquired is an essential goal of *Athena*. Feedback is a suspected source, but understanding whether the energy was introduced early in the formation of the first halos (with further consequences on galaxy formation history), or gradually over time by AGN feedback, SN driven galactic winds, or an as-yet unknown physical process, is crucial to our understanding of how the structures formed and evolved in our Universe. The various feedback processes, as well as cooling, affect the intracluster gas differently, both in terms of the amount of energy modification and of the time-scale over which this occurs. According to current evidence, little star formation (and therefore few supernovae) had occurred since the epoch of cluster formation ($z$~2) to the present time. Measuring the ICM entropy and metallicity (a direct probe of SN feedback) at that epoch and comparing it to that of clusters in the local Universe is key for disentangling and understanding the respective role for each process. Since non-gravitational effects are most noticeable in groups and poor clusters, which in the present bottom-up hierarchical scenario represent the building blocks of today's massive clusters, these systems are of particular interest. A major goal for *Athena* is therefore to study the entropy distribution over the virial region of nearby groups and cool clusters (see Figure 2.32) as well as the properties of the first small clusters emerging at $z$~2 to directly trace the thermodynamic and energetic evolution through the cosmic epoch.

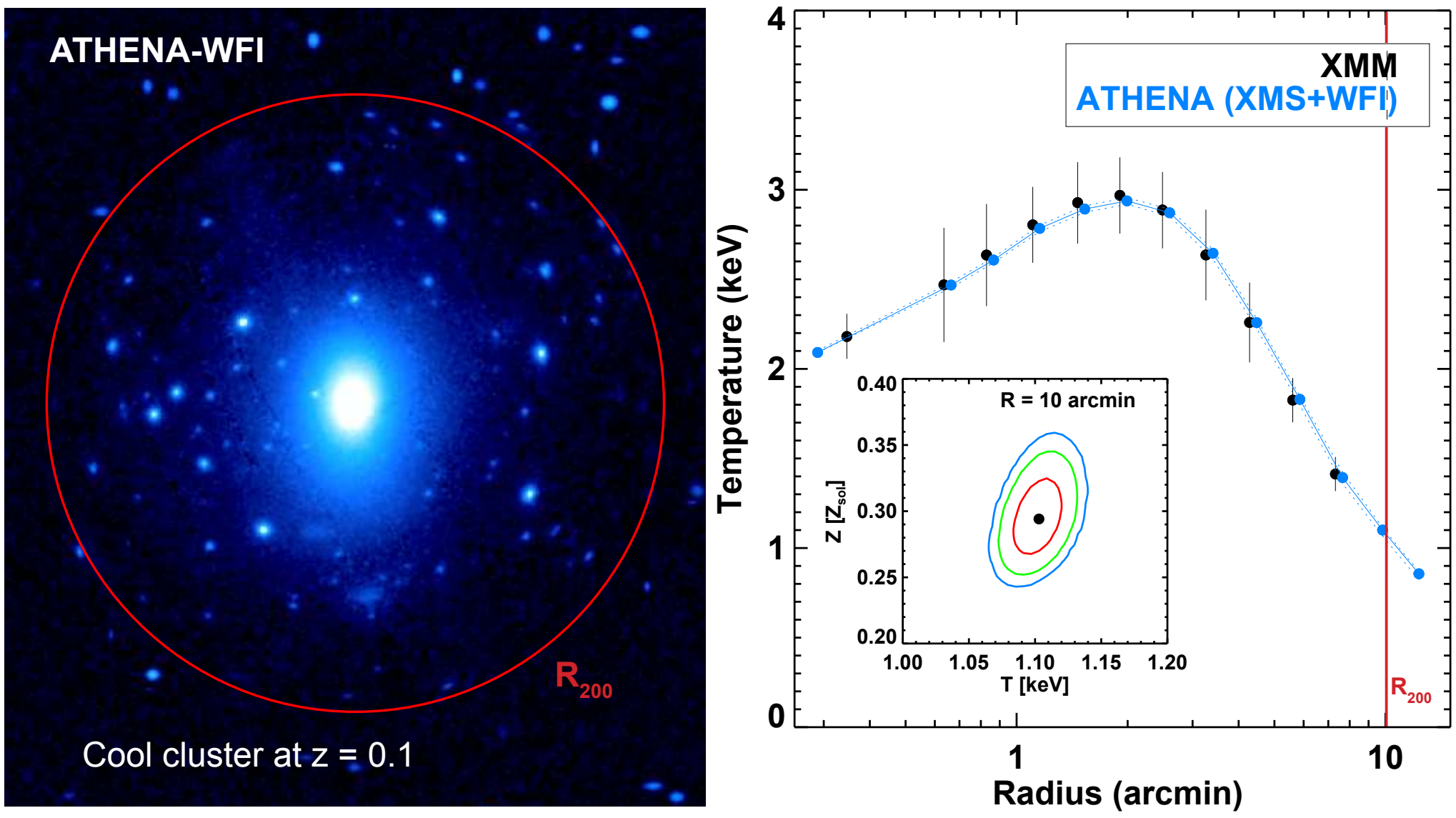

**Figure 2.32.** ***Left:*** *A cool (kT~2.7 keV) and nearby (z=0.1) cluster as it will be seen by Athena-WFI (built from a XMM-Newton observation). The red circle indicates the location of the radius* $R_{200}$*, i.e ~10 arcmin.* ***Right:*** *Associated temperature profile as seen by XMM-Newton (black) and by Athena-XMS & WFI (blue). The red vertical line marks the radius* $R_{200}$*. The inset plot shows the 1,2,3σ confidence level in the temperature-abundance fit parameters plane from an Athena-XMS 25 ks exposure (in cross-scan observation mode). No constraints can be reached at these radii with current X-ray observatories. In these regions, the estimates of the entropy level (derived from spectroscopic temperature and density measurements) and the metal content of the intracluster medium will permit the characterisatio of the physical processes acting on the plasma and their connection to star formation and feedback activities.*

Surface brightnesses and spectra obtained with the XMS and WFI will provide gas density and temperature profiles, and thus, entropy and mass profiles to *z*~1 for low mass clusters (kT~2 keV) with a precision currently achieved only for local systems. Measurements of the global thermal properties of the first poor clusters in the essentially unexplored range *z*=1-2 also will become possible. A proper assessment of the physical processes operating in galaxy clusters during their formation and evolution can be then addressed through the study of the shape and evolution of the scaling relations between global physical properties of the ICM. Present observational measurements are prone to strong selection biases, whereas numerical simulations are limited to a non-exhaustive treatment of feedback effects (see, e.g., Reichert et al. 2011, Short et al. 2010). The physics underpinning these processes are not well understood, and advances are largely driven by observations. *Athena* observations of the hot ICM plasma, combined with radio observations (e.g. SKA), observations of the cold baryons in galaxies (e.g. from JWST, ALMA, and E-ELT) and of the dark matter via lensing data (LSST, *Euclid*) will provide, for the first time, the details for a sufficiently critical comparison. We expect that the major breakthrough of a detailed understanding of structure formation and evolution on cluster scales will come from simulation-assisted interpretation and modelling of this new generation observational data.

By measuring gas density and temperature profiles in groups and clusters out to *z*~2, *Athena* will reveal the evolution of these objects and determine the epoch at which their entropy excess was generated.

## Chemical evolution through cosmic time

The chemical elements from C to Zn play a major role in cooling processes in astrophysical objects through the emission of spectral lines, which allow galaxies, stars, and planets to form. Although the first metals were produced in the interiors of the earliest populations of stars, most elements found in the Universe today are believed to originate from a peak in the star formation around *z*~2-3. Most heavy elements are produced in supernova explosions. Elements from Si to Ni are mainly produced by type Ia supernovae (SNIa), while the lighter elements from O to Si are produced in core-collapse supernovae (SNcc). N, and to a lesser extent C and F, are produced mainly by Asymptotic Giant Branch (AGB) stars (see Werner et al 2008 for an up-to-date review). There is a considerable uncertainty in the metal yields predicted by supernova models, because the progenitor and explosion mechanisms of SNIa's are not well understood (Section 2.4.4). A deep observational study of the chemical abundances in the local Universe and in clusters up to redshifts of *z*=2 will provide insight into the explosion mechanisms and nucleosynthesis processes in supernovae, and knowledge of the stellar populations that produced the bulk of the metals in the Universe. The X-ray band (0.3-10 keV) is uniquely suited to study chemical abundances in hot plasmas, because it contains all the K-shell lines from C to Zn. The superior spectral resolution and effective area of *Athena* allow accurate abundance measurements of all these elements across cosmic time. *Athena* will answer the following key questions regarding chemical evolution:

- Where and when did the elements from carbon to zinc form?
- Where and when were they dispersed in the intra-cluster medium of clusters of galaxies?
- What is the nature of the parent stellar population that enriched the clusters?
- What is the difference between cluster enrichment and enrichment in local field galaxies?

These questions will be addressed primarily by observing clusters of galaxies and AGN using the unprecedented combination of effective area and spectral resolution offered by *Athena*-XMS. Clusters are very suitable objects to study chemical enrichment on large scales, because 85% of the baryons in clusters are in hot X-ray emitting plasma that has been substantially enriched with metals originating from the member galaxies. The gas is in collisional ionisation equilibrium (apart from regions immediately downstream of shocks), which makes abundance determinations from the spectrum very accurate. Due to their deep potential wells, clusters retain their metals and can therefore be considered as closed boxes. The presence of hot gas also prevents the metals to be locked up in dust. In the cluster cores, AGN feedback processes have suppressed star formation in the member galaxies after the star burst which occurred around *z*=2-3 (see Section 2.2.1). Therefore, products from this early stellar population likely dominate the total metallicity. Clusters thus provide a clear fossil record of the era of star formation that created the bulk of metals in the Universe.

The integrated spectrum over the field of view of the XMS in a typical 100 ks local cluster observation will contain lines from trace elements, like Na, Cr, Mn, and Co, which doubles the number of elements detected with respect to *XMM-Newton*. The stacked abundances over a sample of typically 40 nearby clusters will allow the measurement of even the least abundant metals between C and Zn. From the measured abundance ratios, one can infer the contributions from SNIa, SNcc, and AGB stars to the cluster enrichment (Figure 2.33). Stellar population synthesis models and SN nucleosynthesis models can be tested against the measured abundances, which will provide constraints to the Initial Mass Function (IMF) of the parent stellar population and the SNIa and SNcc explosion mechanism.

There are several mechanisms at work that transport the metals from the Inter-Stellar Medium (ISM) of the galaxies, where these elements were produced, into the hot ICM. Feedback from AGN and superwinds from starbursts can drag metals from the galaxies into their environment. In addition, ram-pressure stripping of in-falling galaxies, merger-induced sloshing, and galaxy-galaxy interactions release metals into the surrounding medium. The unique capability for spatially resolved X-ray spectroscopy offered by *Athena*-XMS will capture these enrichment mechanisms in the act. *Athena*-WFI will simultaneously measure the abundances of the most abundant elements over a much larger field of view. This will show how elements are being distributed in the ICM, from the core to the outskirts.

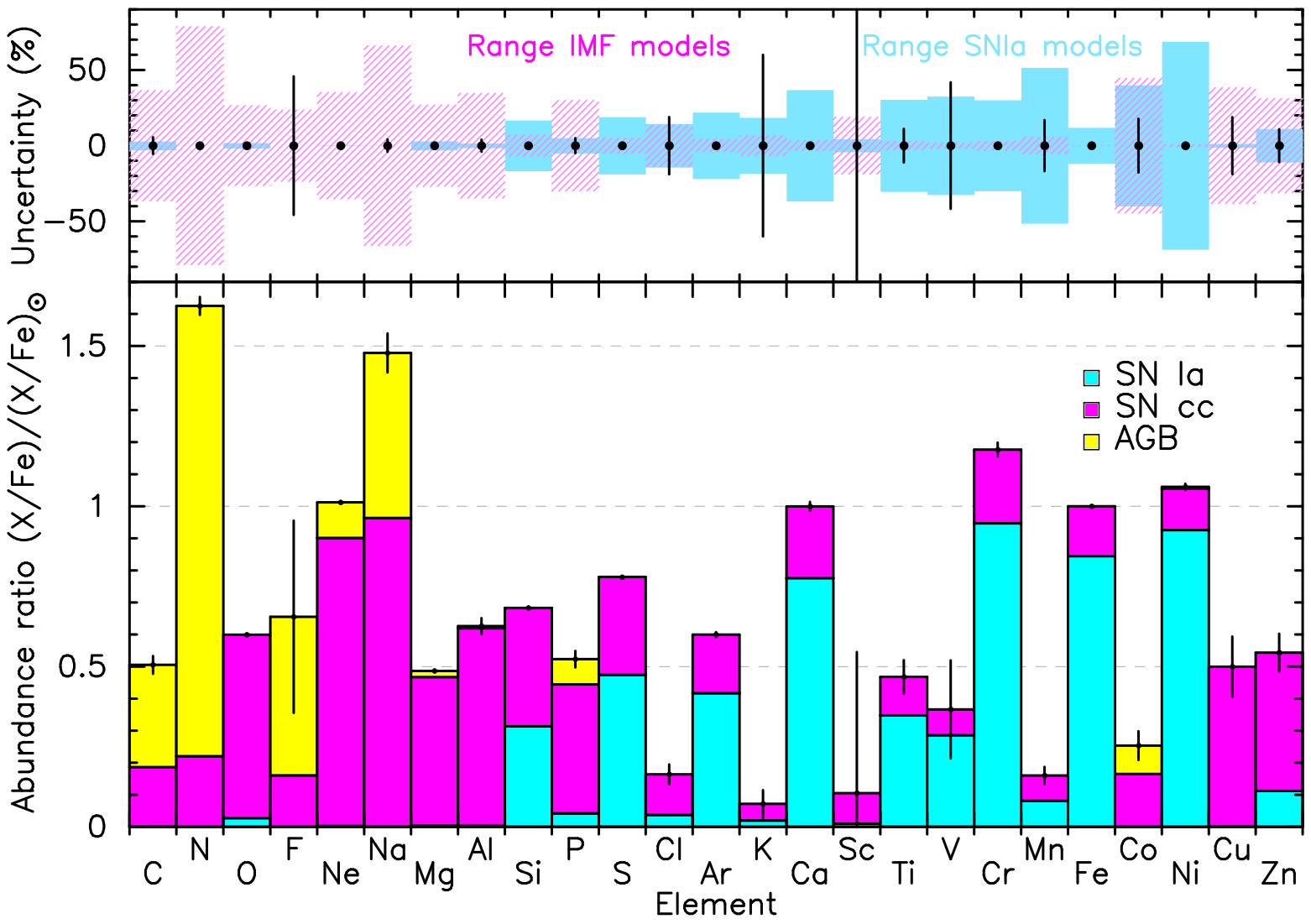

**Figure 2.33.** *Expected abundance measurements for a typical cluster (AS 1101) in a 100 ks Athena-XMS exposure. The lower panel shows the expected abundance ratios relative to solar, where the colours indicate the abundance fraction produced by SNIa, SNcc, and AGB stars. In the upper panel, the blue area shows the range in yields between different SNIa models and the red hatched area the range between a Salpeter and top-heavy IMF. The error bars show the relative precision of the abundance in one 100 ks exposure. Using the twenty elements where the statistical uncertainty is smaller than the model range, supernova models and the IMF will be constrained, revealing details of the star formation history of clusters of galaxies.*

Observations of clusters at redshifts between $z$=0.2-2 will trace cluster chemical evolution through cosmic time. Current instruments can measure the Fe abundance up to z=1.5. *Athena* will extend this by measuring O, Si, S, and Fe up to $z$=1-1.5 within a 100 ks observation per cluster (see Figure 2.34). These observations will show at which rate the ICM was enriched with metals. The current measurements suggest that about half of the metals were released into the ICM in the last 8 Gyr ($z$=1), 2.5-3.7 Gyr after cluster formation at $z$=2-3. By measuring the O/Fe and Si/Fe ratios, *Athena* will show whether this enrichment is due to the release of pre-enriched ISM from the member galaxies or due to ejection of metals by a late-time contribution of SNIa supernovae.

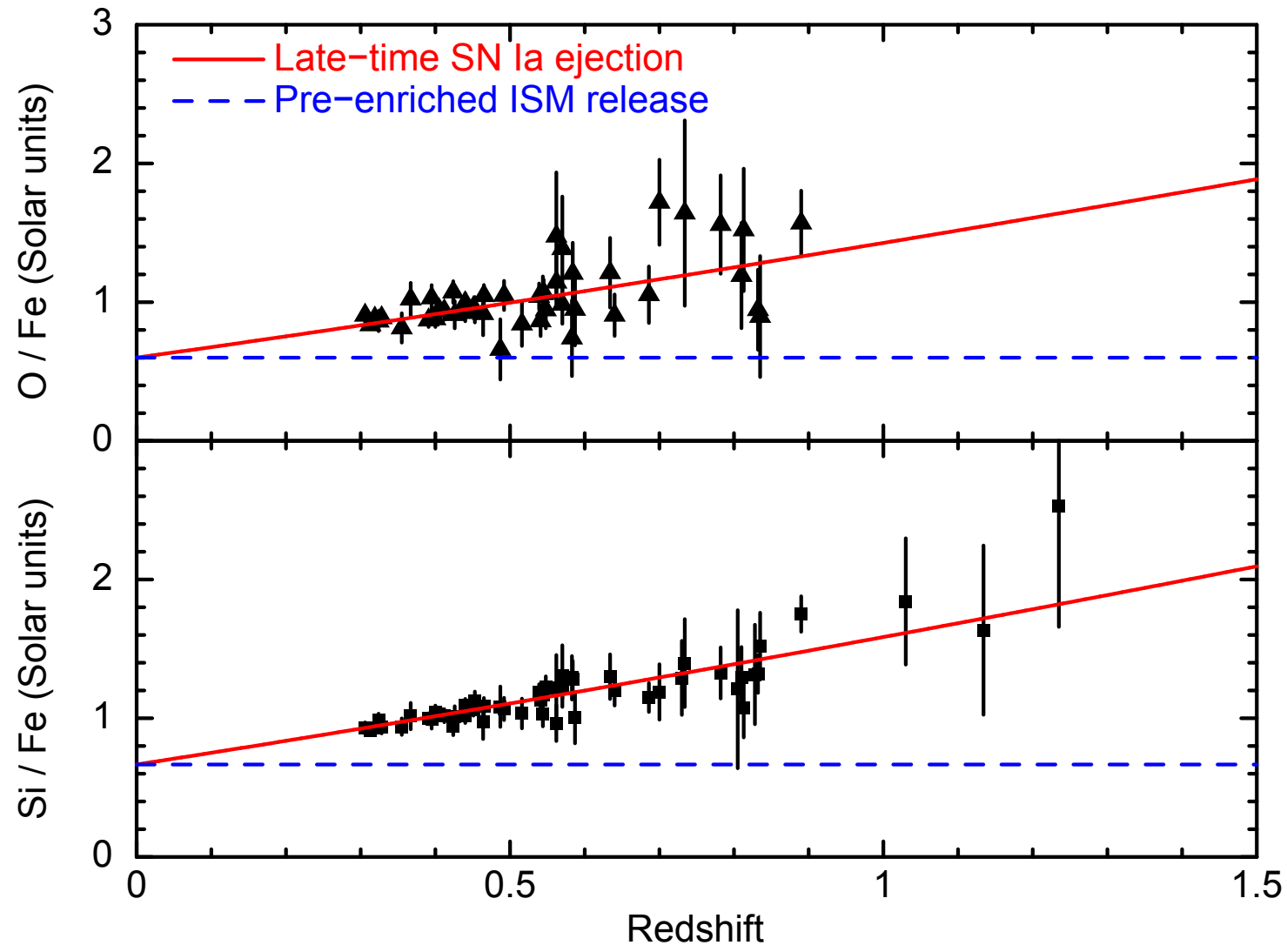

**Figure 2.34.** *O/Fe and Si/Fe ratio for 100 ks Athena exposures of distant clusters in the Balestra et al. (2007) sample as a function of redshift. These abundance ratios will discriminate between a model where recent enrichment is caused by delayed SNIa ejection (red; assumed for these simulations), producing at lower redshift relatively more Fe than O and Si, and a model where the enrichment is due to ISM stripping of already pre-enriched member galaxies (blue), which do not show evolution in redshift. The best-fit model will determine the star formation history of clusters.*

The chemical evolution of clusters is very different from field galaxies, because the feedback processes in cluster cores prevent star formation. In field galaxies, metals are created by several generations of stars with different initial metallicities. Absorption lines in photo-ionised AGN outflows probe the bulge metallicity of field galaxies. Exploiting a deep *XMM-Newton*-RGS observation of Mrk 509, Steenbrugge et al. (2011) derive abundance ratios with respect to iron for C, N, O, Ne, Mg, Si, S, and Ca, which is accurate enough to disentangle the SNcc and SNIa contributions. *Athena* will perform abundance studies for at least the 15 brightest AGN ($z$<0.1) within exposure times of 100 ks, providing the first sample of field galaxies with accurately determined bulge abundances.

*Athena* has the unique capability to measure the abundances of all elements from C to Zn in the hot intracluster medium of clusters, revealing the supernova history and initial mass function of the parent stellar populations.

### 2.3.2 Galaxy cluster cosmology

Galaxy clusters form an integral part of the cosmic large-scale structure and as such lend themselves to accurate tests of cosmological models. X-ray observations provide the most robust and detailed observables for cluster characterisation, and indeed have played an important role in establishing the current cosmological paradigm. For example, measurements of the cluster number density (Henry & Arnaud 1991) and their baryon mass fraction compared to Big Bang nucleosynthesis (White et al. 1993) both provided early evidence for a low matter density relative to the critical density. More recently, X-ray observations of clusters have been used to confirm independently the evidence for dark energy, and constrain the cosmological parameters in a way which is competitive with and entirely complementary to other methods (e.g. Rapetti et al. 2009, Vikhlinin et al. 2009). The launch of *eROSITA* will provide the next major step forward in X-ray cluster cosmology, yielding enormous samples out to $z$~1. Galaxy clusters studies will therefore continue to form an indispensable part of the future precision cosmology efforts.

The accuracy of cosmological constraints from clusters in the future will be limited primarily by systematic errors, predominantly in the measurement of cluster masses. Furthermore, current X-ray cluster surveys e.g. with *XMM-Newton*, and even future ones with *eROSITA*, deliver limited sample sizes in the crucial high redshift regime at $z>1$. *Athena* will address both problems, yielding extremely accurate measurements of the cosmological parameters via two complementary approaches:

- The measurement of the growth of large-scale structure via the mass function of galaxy clusters
- The measurement of the cosmic expansion history via the baryon fraction in clusters.

Both types of cosmological tests are required, for example to break degeneracies between quintessence-type models of dark energy and modified General Relativity.

Both also require robust cluster mass measurements. Key issues in mass determinations include the nature and evolution of the cluster scaling relations, the degree of turbulence in the ICM and any other non-gravitational source of heating. These will be revealed via high resolution spectroscopy with the *Athena*-XMS. With the understanding of the physics in hand, robust mass proxies with minimal systematics can be developed and applied out to high redshift. Here *Athena's* spatial resolution is critical, as it allows the determination of mass proxies, while excising the cluster core, which is more complex due, e.g., to the effects of feedback (Section 2.2.1) and can bias mass measurements. Via a targetted program of observations of well-controlled, known cluster samples (e.g. from *eROSITA* and *XMM-Newton*), *Athena* will yield extremely accurate total and gas mass measurements with full control of systematics, yielding precision cosmological constraints via the cluster mass function and baryon fraction as functions of redshift. Extremely sensitive wide-field surveys with the *Athena*-WFI will yield new samples of clusters and groups out to $z$=2 or even beyond, which are beyond the reach of current or planned X-ray surveys.

*Athena* will provide measurements of cluster total and gas masses out to $z$~2, yielding two independent, precision constraints on the dark energy density and equation of state out to that redshift.

## Growth of Structure: the cluster mass function

The mass function of galaxy clusters is an exponentially sensitive probe of the linear density perturbation amplitude in the dark matter distribution and can be theoretically predicted with high precision for a given cosmological model. Given precise cluster masses, the perturbation growth factor can be determined to 2.5% accuracy from a sample of ~30 clusters per redshift interval. Covering at least three epochs between $z$~1-2 the dark energy equation of state parameter, $w$ (the ratio of dark energy pressure to its density), can be determined to 5-10 % for this cosmic epoch, allowing us to distinguish between major dark energy models involving e.g. currently popular quintessence-type DE or brane world models. These provide a competitive and complementary assessment of cosmic structure growth compared to the two other most promising methods, weak lensing tomography (Amara & Refregier 2008) and redshift-space distortions in galaxy clustering analysis (Guzzo et al. 2008). The critical link, which *Athena* will provide, is the relation between directly observable cluster parameters and the cluster mass.

X-ray observations provide the essential information on the structure and mass of clusters from detailed images and properties of the most massive baryonic cluster component (~85%), the ICM. Recent studies have shown that X-ray observations can provide tight relations between mass and parameters such as X-ray luminosity ($L_X$), ICM temperature ($T_X$), gas mass ($M_{gas}$), and $Y_X$ (= gas mass times temperature) for well-selected cluster samples, with low scatter (10-12%; Kravtsov et al. 2006; Pratt et al. 2009; Vikhlinin et al. 2006; Mantz et al. 2010). The simplest and most applicable proxies (e.g. $L_X$) are improved greatly if, as with *Athena*, one has the angular resolution to excise the complex core region. *Athena* will also be crucial in reducing the intrinsic scatter in these relations, and thus the systematics in mass estimates, by allowing us to include further structure parameters in the scaling relations, like the X-ray spectral line broadening, which measures the dynamical distortion of the clusters (see Section 2.3.1), which affects the mass measurement. Further refinement can be achieved by combining with lensing data (e.g. from *Euclid* or LSST), which yield independent estimates of the total mass. Together these will provide the necessary synergy for precise cluster mass calibration, expected to reach the required uncertainty of 1-2%. A major follow-up programme (~ 8 Msec) to derive masses for ~100 clusters from $z$=1 to 2, combined with the large amount of data at lower redshifts, will provide the necessary accuracy to distinguish clearly between competing cosmological models

## Geometry of the Universe: the baryon fraction

Massive galaxy clusters are large enough to represent fair samples of the Universe. Treating the gas mass fraction ($f_{gas}$) in rich clusters as invariant with redshift and using the $d(z)^{3/2}$ dependence of its measurement one can obtain further constraints on the geometry of the Universe which are competitive to those of SN Ia observations (Allen et al. 2008; Rapetti et al. 2008). *Athena* will address both of the current limitations in these measurements. The first lies in the determination of the total mass, which is obviously a component of the gas mass fraction. The enormous leap forward in mass determinations with *Athena* has been discussed in detail above. The second is the extension of the $f_{gas}$ measurements to higher redshifts, at z>1, extending the lever arm of the standard ruler out to at least $z$~2, where the effects of dark energy are expected to start to emerge. *Athena's* superior photon statistics and high spectral resolution in the central regions, will of course, also greatly improve the statistical accuracy of the $M_{gas}$ measurements for clusters at all redshifts.

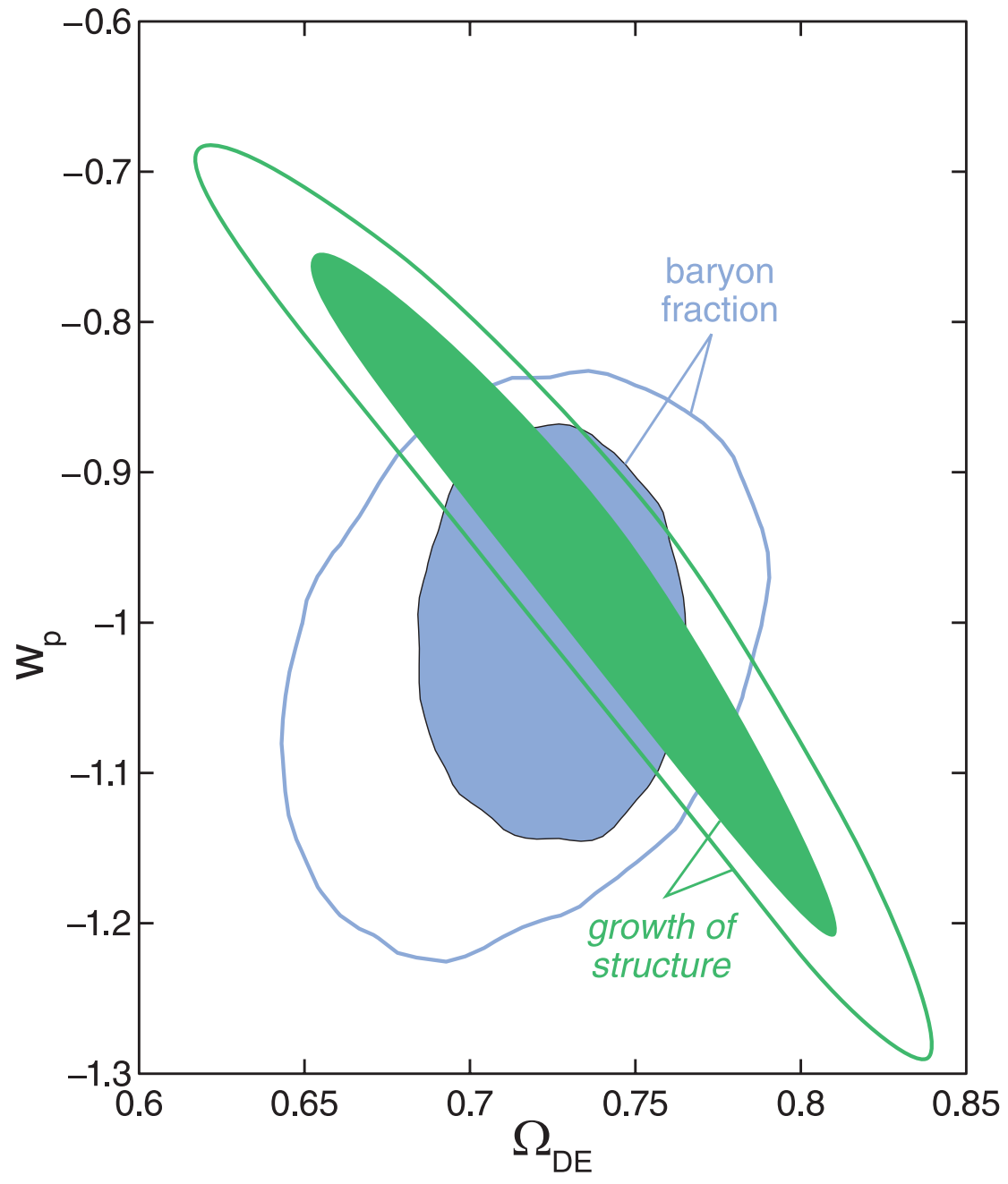

**Figure 2.35.** *Constraints on the density ($\Omega_{DE}$) and equation of state parameter ($w_p$) of dark energy that can be obtained with the proposed Athena galaxy cluster observation programme. Green regions show the 1σ and 2σ constraints resulting from the analysis of the redshift evolution of the cluster mass function while the blue regions show the optimistic and pessimistic constraints from the study of the baryon mass fraction as a function of redshift.*

## Surveys for new, distant groups of clusters

Current X-ray surveys, particularly with *XMM-Newton*, are starting to reveal a handful of clusters at z>1.5, with one apparently at *z*>2 (Gobat et al. 2011). With its superior photon collecting power and large field of view, the *Athena*-WFI will open up the high redshift universe for clusters and groups, revealing the epoch where they first formed. A combination of dedicated blank field surveys (including those which trace the growth of SMBHs; see Section 2.2.2), and serendipitous surveys will yield of order 50,000 galaxy clusters at all redshifts. Among these will be the first significant samples of the distant groups and clusters. While there is considerable uncertainty in the extrapolations, we expect around 200 groups and poor clusters at redshifts ~2.5 and beyond, increasing our horizon for tracing large-scale structure and testing cosmology even further. *eROSITA* is not sufficiently sensitive to detect such objects, but characterising their emission with X-rays will give currently inaccessible information on the evolution of the mass function and scaling relations at high redshift and, as already discussed, further leverage on $f_{gas}$ measurements. In addition to cosmology, the structural and dynamical evolution of clusters, the feedback history and the chemical evolution (see Section 2.3.1) can be extended using these high redshift samples into an unexplored regime. As an example of the feasibility of such studies we show in Figure 2.36 a deep *Athena* observation of a group at *z*=2.

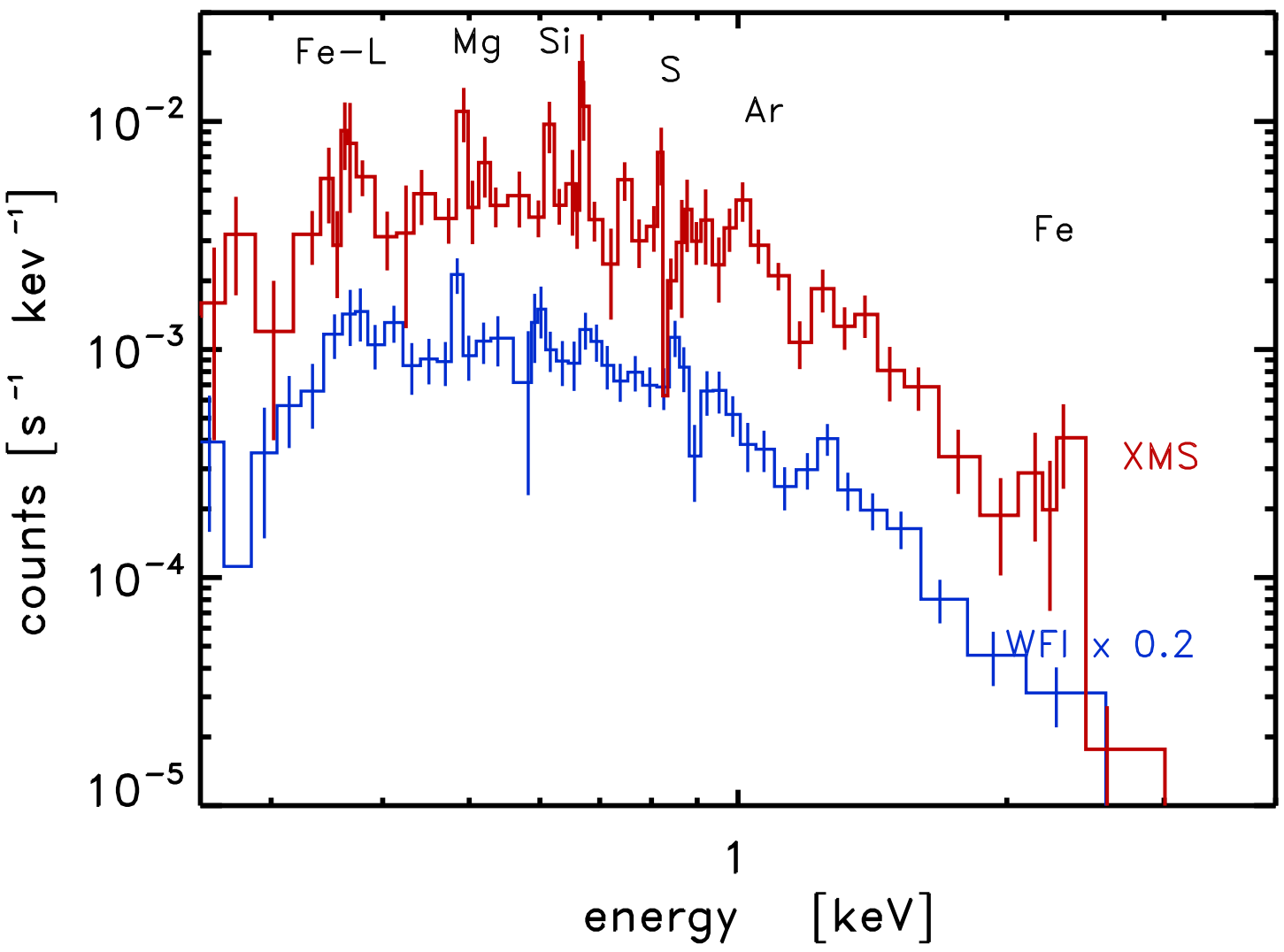

**Figure 2.36.** *Simulated, background subtracted spectrum of a galaxy group at z=2 with flux (0.5-2 keV) =$10^{-15}$ erg $s^{-1}$ $cm^{-2}$ and temperature =2 keV with a deep exposure of 400 ks with the Athena-XMS and WFI detectors (WFI spectrum is scaled down by a factor of 5 to aid visibility). The temperature and abundances can be measured with good precision (ΔT < 0.1 keV).*

### 2.3.3 Missing baryons and the Warm-Hot Intergalactic Medium

Despite enormous progress achieved in recent years in understanding the cosmological framework, the distribution of dark matter and the gross features of dark energy, we still do not know much about the evolution of most of the baryonic 'visible' component of the Universe. The extent of the issue is visualized in Figure 2.37, using the phase diagram of the baryons at $z$=0 as predicted by a cosmological hydrodynamical simulation. Current observations are effectively limited to the cooler parts of this diagram (T<$10^5$ K) associated with the local Ly-α forest and with the hottest and densest parts (T>$10^7$ K, δ>300), where the majority of the hot baryons that we see reside in clusters of galaxies. However a substantial amount of ordinary matter at $z$<2 remains undetected. Cosmological hydrodynamical simulations suggest that these missing baryons are contained in a diffuse, low density, warm-hot intergalactic medium, preferentially distributed in large scale filaments connecting clusters and groups in the nearby Universe. Furthermore they indicate that this is fundamentally a multiscale problem, involving a complex interplay between gravitational and poorly understood non-gravitational processes. Galaxy formation depends on the large scale environment and on the physical and chemical properties of the intergalactic gas from which they form, which in turn is affected by galaxy feedback through energy released from star explosions and AGN (see Section 2.2). These "missing baryons" can only be observed through X-ray studies, but even detecting the very first of these filaments through X-ray absorption spectroscopy of distant sources is likely to be at the edge of (if not beyond) the capabilities of current X-ray observatories like *Chandra* and *XMM-Newton*. The integral field spectroscopy capability of *Athena* will enable absorption and emission observations of the WHIM, yielding precision measurements of hundreds of filaments. This rich data base will allow to measure the mass of the missing baryons, to constrain the formation and evolution of cosmic filaments and their relationship with metal enrichment and heating from galaxy outflows, and ultimately to characterise the environment in which the ordinary matter as we know it, galaxies and stars, formed and evolved.

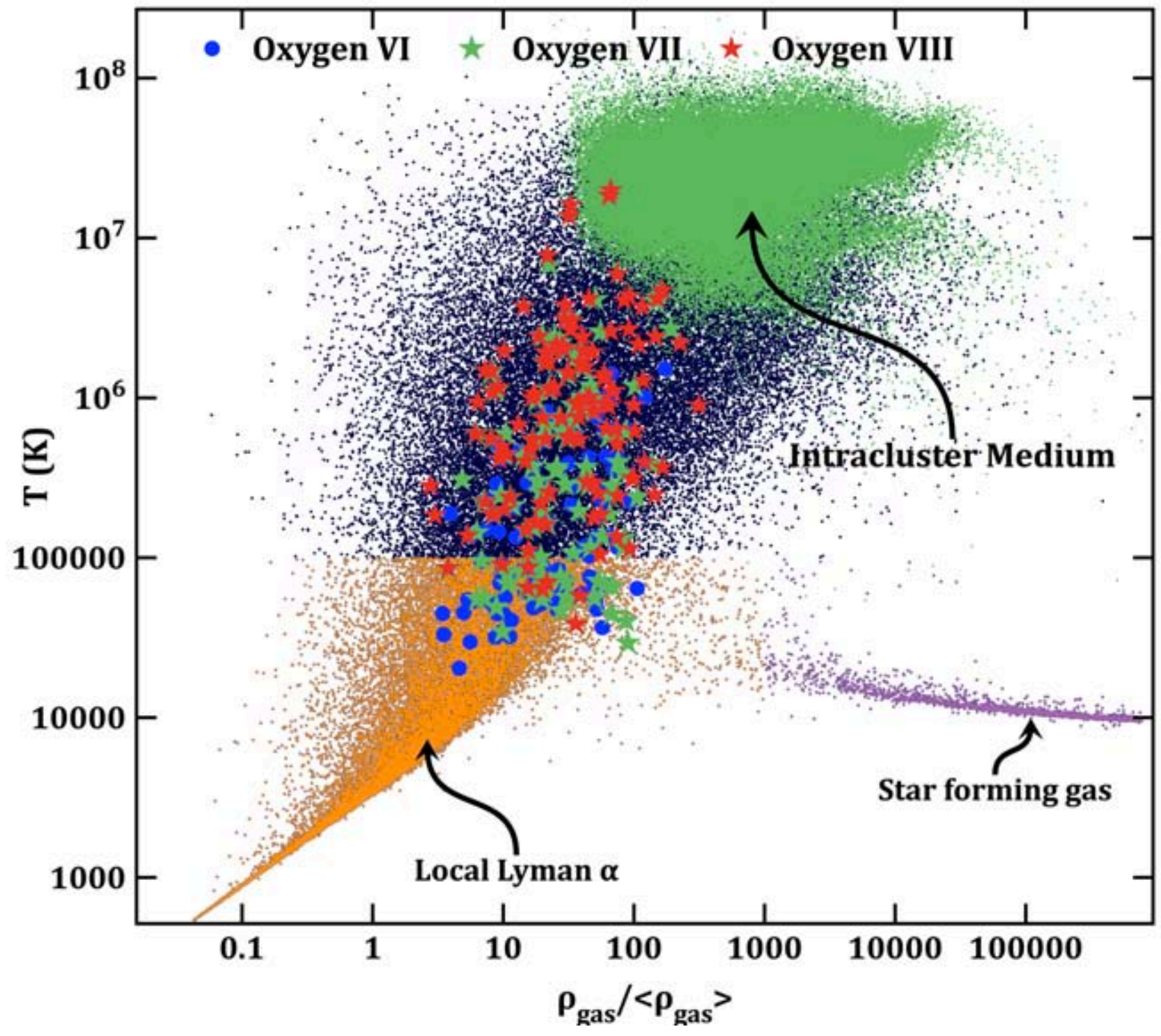

**Figure 2.37.** *Phase diagram of baryons in the nearby Universe (Branchini et al 2009). For gas in systems ranging from the lowest density phase of the intergalactic medium, to the densest gas in star-forming galaxies, we plot normalised density (ρ/<ρ>=δ+1, δ being the overdensity) and gas temperature. Current observations are limited to the central regions of galaxy clusters (green) and have just started to probe their outskirts. Athena will chart the unexplored territory where about half of the total baryons in the local Universe are thought to hide, forming a tenuous and warm-hot phase. Combining emission and absorption spectroscopy with the Athena-XMS spectrometer will provide a detailed characterisation of denser regions of the filaments, while OVIII & OVII absorption line spectroscopy (red & green stars) will extend the measurements down to the regions with lower densities.*

### The Warm-Hot Intergalactic Medium with Athena

Ordinary matter (baryons) represents 4.6% of the total mass/energy density of the local Universe, but less than 10% of this baryonic matter appears in collapsed objects (stars, galaxies, groups; Fukugita & Peebles 2004). Theory predicts that most of the baryons reside in vast unvirialised filamentary structures that connect galaxy groups and clusters (the "Cosmic Web"), and this is confirmed by measurements of the baryon density in the Lyα forest at $z$>2. But at the current epoch about one half of the baryons are missing. Cosmological hydrodynamical simulations suggest that the budget of baryons that seem to have 'disappeared' between $z$=2 and $z$=0, is accounted for by a diffuse, highly ionised, hotter intergalactic medium, the so called Warm-Hot Intergalactic Medium (WHIM). Evidence for a colder and less dense tail, where about 10% of the missing baryons reside, has been obtained via UV-absorption line studies with FUSE and HST-COS (Danforth & Shull 2008), but 40% of the baryons remain unaccounted for.

This gas is extremely hard to detect: H and He are fully ionised, but the thermal continuum emission is much too faint to be detectable against overwhelming backgrounds. The only characteristic radiation from this medium will be in the discrete transitions of highly ionised C, N, O, Ne, and Fe (Figure 2.38).

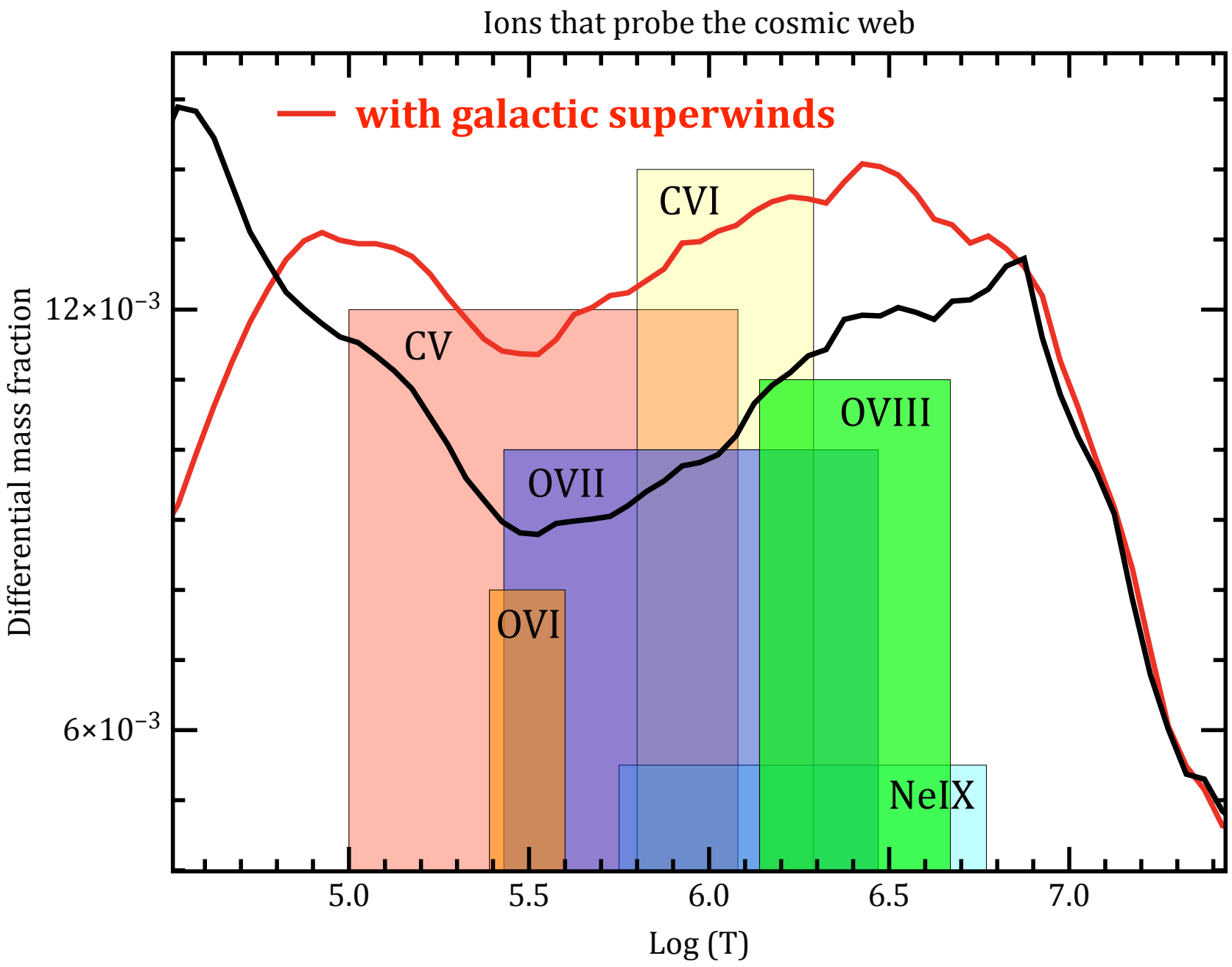

**Figure 2.38.** *The differential gas mass fraction as a function of temperature at low redshift for the ΛCDM cosmological simulation of Cen & Ostriker (2006). This distribution is sensitive to the presence of galactic superwinds (red curve; black curve is without superwinds). The ions with the strongest resonance lines in the $10^5$-$10^7$ K range are shown, and except for OVI (UV line; 1035 Å), the other lines lie in the X-ray band. In all cases, models predict that most of the baryonic mass resides in gas that can only be seen through X-rays.*

> *Athena* will detect the Warm-Hot Intergalactic Medium both in absorption towards bright background sources and in emission in filaments around clusters of galaxies.

*Athena*-XMS has a unique combination of line sensitivity, energy resolution and imaging capabilities for measuring WHIM lines both in emission, and in absorption against a bright background source, as shown in Figure 2.39. Because the number of systems increases linearly with $z$ out to $z$~1, there is a significant advantage in using distant – yet still bright - sources for absorption studies. Thus, in addition to relatively bright AGN (mostly at $z$>0.3), bright afterglows of the more distant GRBs ($z$>1) will extend the measurements closer to the era where the WHIM filaments started forming, yielding a sample of about a hundred of systems with two (OVII and OVIII) or more absorption lines. The sensitivity to weak lines and the number of filaments estimated here are consistent with the assessment by Yao et al. (2011), having taken into account spectral resolution and effective area Taking advantage of the imaging capabilities of the XMS, the angular resolution and the FOV, it will be possible to complement absorption with emission studies. We expect that the property of a filament does not change on a scale of an arcmin. Thus we can simultaneously perform emission line measurements close to the line of sight towards the background source used for absorption studies (Figure 2.39). Extending the same measurement to the full database of XMS observations we expect to detect about 250 emission systems serendipitously. The angular resolution of *Athena* will be crucial not only to excise the central target of the field, but also to get rid of fainter sources that can contaminate the measurement.

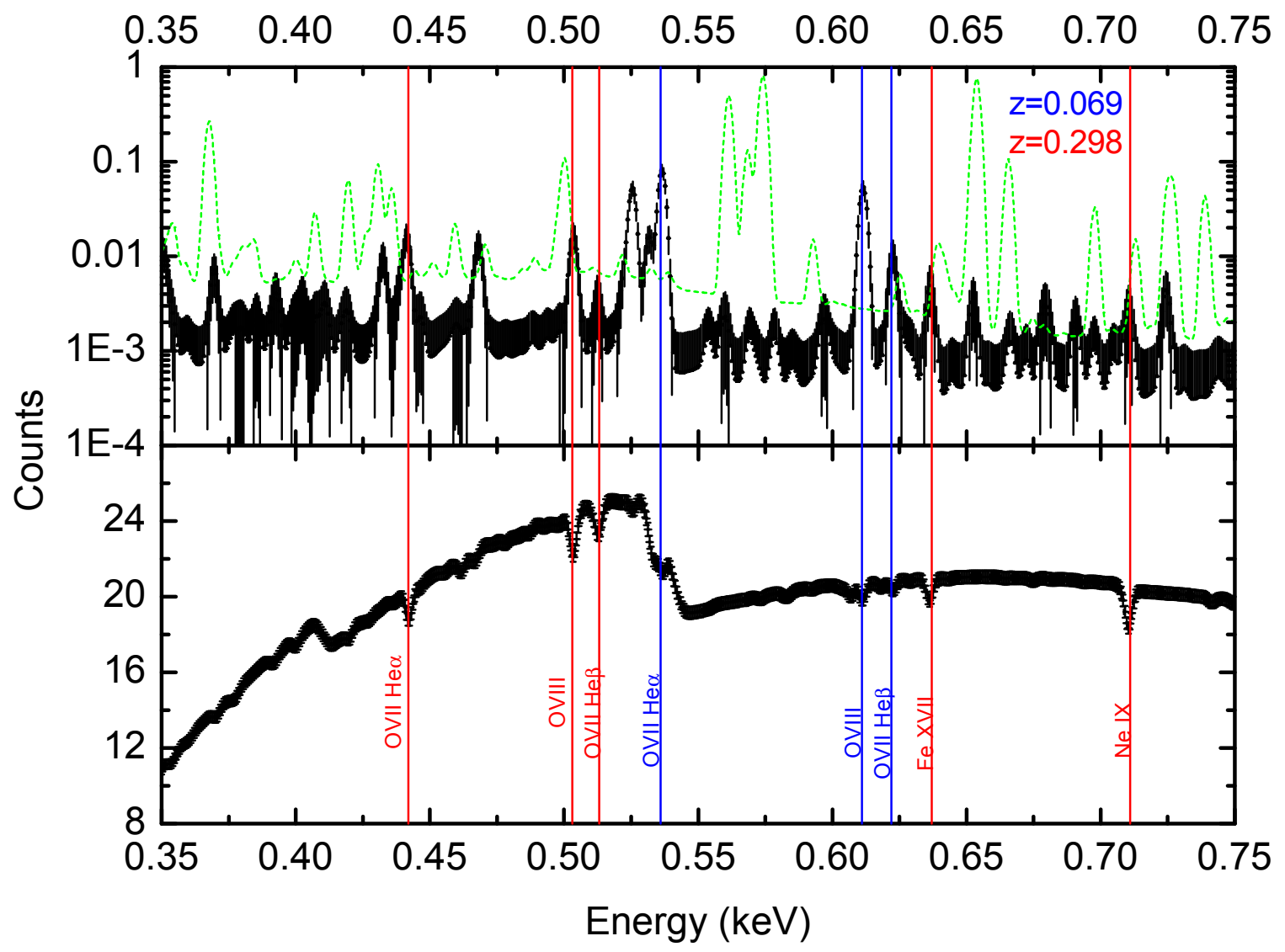

**Figure 2.39.** *The unprecedented combination of spectral and imaging capabilities provided by Athena-XMS allows spectroscopic measurements of WHIM filaments both in absorption (**bottom**) and emission (**top**). The bottom panel shows in black the absorption spectrum from a line of sight where two different filamentary systems are illuminated by a background X-ray source with a soft X-ray fluence of 3x10$^{-6}$ erg cm$^{-2}$. Prominent lines from OVII, OVIII, Ne IX, FeXVII are identified. In the top panel the corresponding emission spectrum of a 4 arcmin$^2$ region from the same filaments is shown in black. Thanks to the fine angular resolution of Athena, the central source can be excised or, in a case of a GRB, the observation can be carried out once the afterglow has faded. The emission of the Galactic foreground, plotted in green, is characterised by several emission lines that can only be disentangled from the WHIM component thanks to the spectral resolution of the Athena-XMS.*

While the mere detection of such lines will reveal the presence of the 'missing baryons", our goal is more ambitious; we plan not just to detect but also to characterise the physical state of the WHIM: its temperature, density, location and metal content which trace the evolution of the medium and its interplay with metal enrichment and non-gravitational heating through galactic super winds driven by star explosions and active nuclei (see Section 2.2). Figure 2.40 illustrates how well *Athena* will discriminate between different WHIM models, involving different metal enrichment and non-gravitational heating processes, independently using absorption and emission spectroscopic data.

The distributions of systems determined from absorption and emission measurements by itself provides a constraint on the models for the WHIM, but with *Athena*-XMS we can combine various independent measurements to constraint the physical and chemical status of the WHIM. We expect that in about 50% of the cases three or more lines will simultaneously be detected, allowing constraints to be placed on the density and temperature of the medium. Targeting UV-bright AGN where broad Lyman-α absorbers can be measured by *HST*-COS will allow to determine the metallicity of the medium. Finally we expect that, for about 30% of the systems detected in absorption, the associated X-ray line emission will be detected. Simultaneous absorption and emission spectroscopy will allow for direct, model-independent measurement of the characteristic density, ρ (emission line strength scales as $\rho^2$, while absorption line equivalent width scales as ρ), length scale, ionisation balance, excitation mechanism (or gas temperature), and element abundance of an intergalactic gas filament.

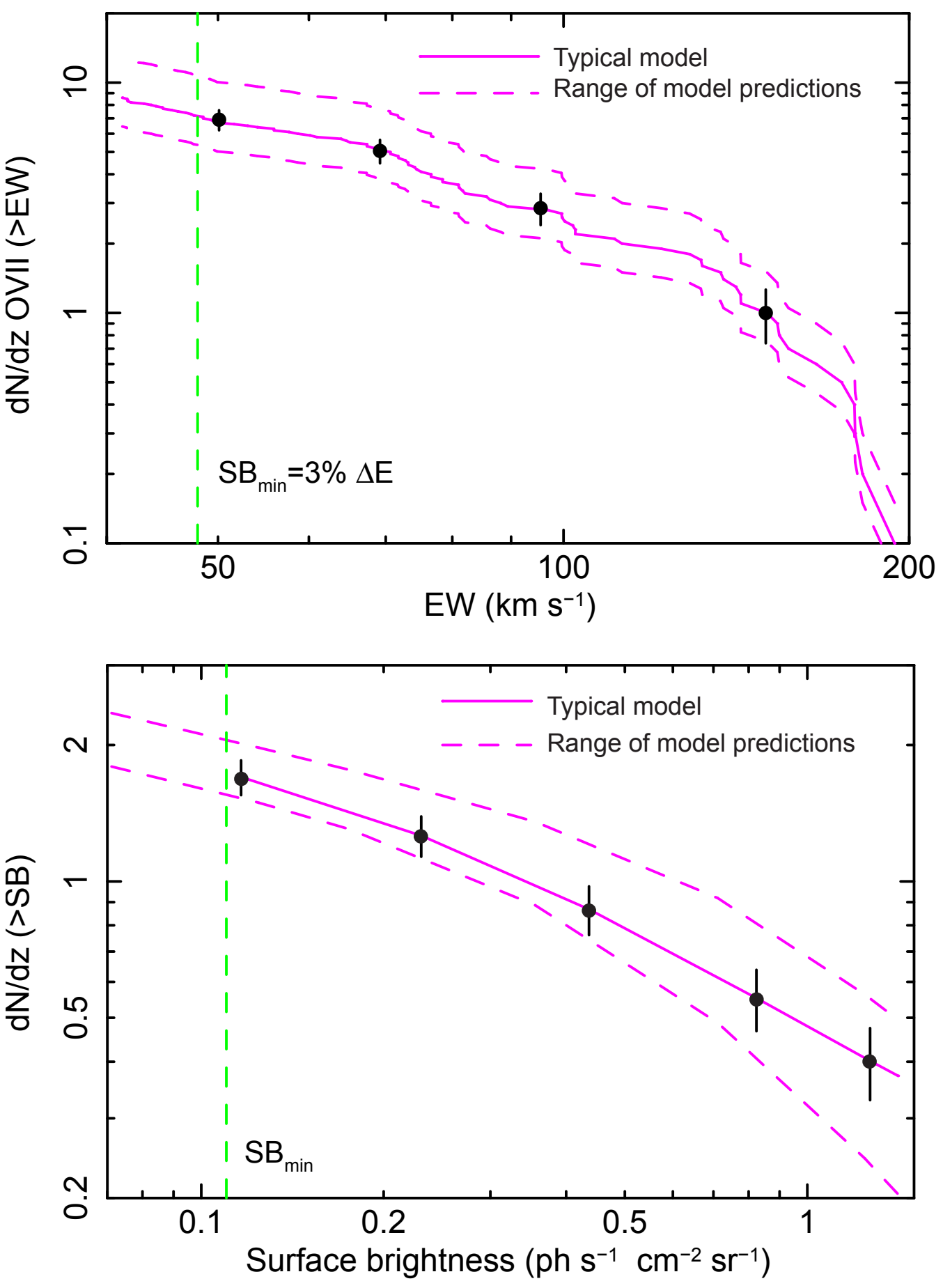

**Figure 2.40.** *Athena will be able to discriminate between models of formation of WHIM filaments in various ways, including measuring the distribution of the number of OVII and OVIII line systems detected in absorption or emission.* ***Top:*** *Solid line shows the predicted mean number of OVII absorption lines per unit redshift as a function of line EW (Cen & Fang 2006, Branchini et al. 2009). The vertical green dashed line represents the minimum EW that can be detected by Athena and is set by systematic effects. We require a simultaneous detection of at least two lines, e.g. OVII and OVIII, at the same redshift. Black data points show the OVII line statistics accumulated in 5 years. Error-bars account for (dominant) shot noise errors and cosmic variance. This program requires observations of a sample composed of AGN and GRB afterglows, totalling about 11 Msec. The dashed line curves are for different WHIM models: highly localised metal diffusion to the region of active star formation (lower curve), and diffusion into the IGM via SN feedback-related processes (upper curve). These two models illustrate the spread in current theoretical predictions (Branchini et al. 2009) and can be clearly distinguished with Athena.* ***Bottom:*** *same line statistics but for filaments detected in emission as a function of line surface brightness (Takei et al. 2011). We expect to detect serendipitously about 1 filament in the field of view of XMS for observations longer than 100 ksec. We have conservatively assumed that for about 30% of the systems detected in absorption, the associated X-ray line emission will be detected.*

Correlating these in-depth studied sites with their environs will permit further constraints on the connection between large scale structure and cosmic feedback. The extent of superwinds and elemental mixing can be determined by studying the spatial relationship between hot gas seen through X-ray absorption or emission and the location of galaxies (Stocke et al. 2006). The emergence of WHIM filaments from clusters, or the hot baryon halos around galaxies that are predicted from formation theories, will be revealed directly in emission because the sensitivity of *Athena* is vastly better than existing instruments. We plan targeted observations with *Athena* towards regions bridging cluster of galaxies or filamentary structures from the 3-D distribution of galaxies. Such distributions are also available in the deep surveys, enabling WHIM studies in deep *Athena* fields.

## 2.4 Astrophysics of hot cosmic plasmas

As already emphasised, *Athena* will open up new observational territory thanks to its unique combination of sensitive wide-field imaging, spatially-resolved high-resolution spectroscopy and timing. *Athena* will explore in detail large portions of the parameter space necessary to understand the very nature of all types of celestial objects, making its most important contributions in areas that have so far remained poorly probed (or are simply not accessible). This section highlights a selection of these science cases, using these to illustrate the prospects for *Athena* to explore hot cosmic plasmas on all astrophysical scales.

These selected science cases are presented in approximate order of distance to us, from the solar wind effects in solar system bodies, through stars from their birth to their death the interstellar medium that permeates our Galaxy. Many of these studies relate to physical mechanisms underpinning phenomena discussed in previous sections, or other astrophysical effects which are important in their own right.

### 2.4.1 Solar System bodies

*Athena* studies of solar system bodies will add enormously to our understanding of the interactions of space plasmas and magnetic fields by giving us deeper insights into the complex workings of planetary magnetospheres and exospheres. This knowledge in turn will be directly applicable to a variety of astrophysical scenarios on larger scales. Short of making X-ray observations in-situ at planets and comets, *Athena* is the only forthcoming opportunity to make significant progress in this field. Including solar system bodies as targets for *Athena* adds a new dimension to the mission's science, a dimension that is in itself one of the themes of ESA's Cosmic Vision 2015-2025 programme: *'How does the Solar System work?'*

X-ray studies of solar system bodies have made great strides in the last decade, thanks to the high sensitivity of *XMM-Newton* and *Chandra*'s high-resolution imaging. On Jupiter, for example, three different processes are known to produce X-rays (Branduardi-Raymont et al. 2007). In soft X-rays (<2 keV), the bright auroral spectrum is dominated by line emission, produced by charge exchange (CX) between highly stripped, energetic ions and $H_2$ molecules of the planet's upper atmosphere (for a review of CX reactions see Dennerl 2010). The origin of the ions – whether from the solar wind, or the inner magnetosphere, i.e. Io's volcanoes – has been matter of debate, and the latter is currently favoured. The X-ray spectrum of Jupiter's disk resembles that of the Sun, with strong iron and magnesium emission lines, implying an origin in scattering and fluorescence of solar X-rays. Figure 2.41 shows the different morphology observed in narrow energy bands centred on the strongest emission lines: the aurorae are very evident in the CX oxygen lines (top panels), while we see a uniform disk if we select iron and magnesium lines (characteristic of the solar coronal spectrum, bottom panels). Above 2–3 keV the Jovian auroral spectrum turns into a smooth continuum, produced by either thermal or non-thermal electron bremsstrahlung, while the disk X-ray emission fades.

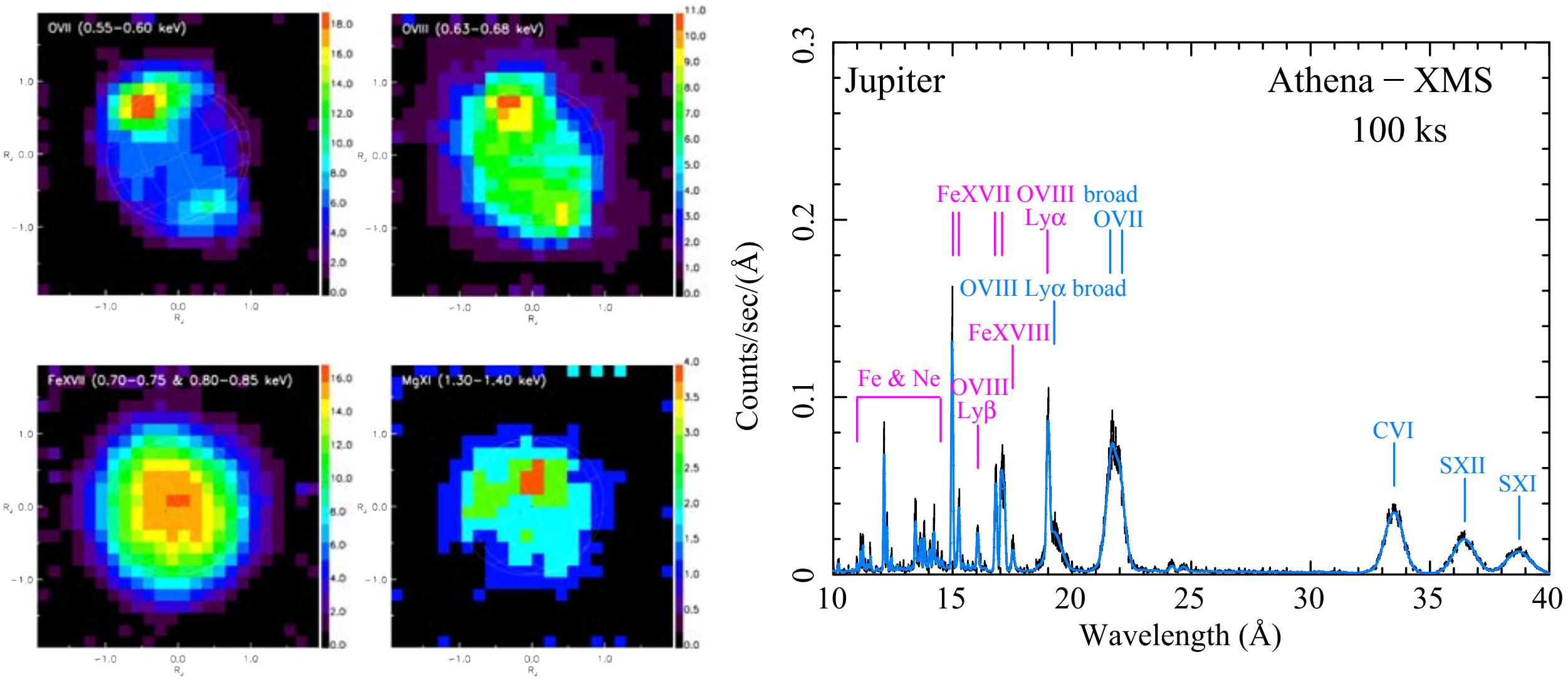

**Figure 2.41.** ***Left***: *XMM-Newton images of Jupiter in narrow energy bands showing the auroral (top) and disk contributions (bottom). Only high sensitivity, high resolution soft X-ray spectroscopy with Athena-XMS can determine the species of ions (C from the solar wind or S from Io) undergoing charge exchange in the Jovian upper atmosphere and producing the X-ray aurorae.* ***Right:*** *Simulated Athena-XMS spectrum for an exposure time of 100 ks: the wavelength coverage extends to the 30-40 Angstrom band allowing the C vs.S ambiguity to be resolved. Results from Athena-XMS spectroscopy will challenge the theories of ion transport and acceleration in Jupiter's magnetosphere.*

Saturn, however, has shown no evidence to date for auroral X-ray emission; a combination of scattering and fluorescence of solar X-rays is thought to be responsible for its disk, polar cap and ring emission (Bhardwaj et al. 2005a,b; Branduardi-Raymont et al. 2010). *Athena*-XMS spectroscopy will push the search for auroral

CX X-ray emission on Saturn to fainter limits, and attempt to establish what differentiates this giant, fast rotating planet, with its complex magnetosphere dominated by the interaction with Titan, from the similar conditions on Jupiter.

In the case of Jupiter, it is still to be conclusively established whether auroral CX ions originate from the solar wind, or from Io's volcanoes. Only high-sensitivity, high-resolution spectroscopy with *Athena*-XMS can determine the ion species (C or S), put to rest this long-standing issue and test the theories of ion transport and acceleration in Jupiter's magnetosphere. Simultaneous measurements of auroral bremsstrahlung at higher energies, where the disk emission has faded away, will characterise the electron energy distribution; observations of variability will show how the ionic and electron populations respond to solar wind changing conditions. With *Athena* spatial resolution, the disk of Jupiter (~40 arcsec across) will be better resolved than in the images obtained with the *XMM-Newton*-pn camera, but the planet's spectrum will still be a superposition of auroral CX and disk coronal emission lines. However, these lines will be well-resolved, with the advantage that the *Athena*-XMS will offer two orders of magnitude improved effective area: Figure 2.41 shows a simulation of the *Athena*-XMS spectrum of Jupiter for an exposure time (100 ks) that includes in the band-pass the critical C and S lines (between 34 and 38 Å) at high spectral resolution.

A similar solar-wind CX (SWCX) origin is attributed to X-rays from comets and the heliosphere. In fact, charge exchange may be an important and ubiquitous soft X-ray line emission mechanism extending from the solar system to interstellar clouds, galactic winds and galaxy clusters (Lallement 2004). Comets, with their extended neutral comae, are ideal targets for SWCX studies: they are spectacular X-ray sources, with extremely line-rich X-ray spectra, and ideal probes of the conditions of the solar wind in proximity to the Sun, especially at high ecliptic latitude where direct observations are challenging (Dennerl et al. 2003, Bodewits et al. 2007).

Comets (one or two of which are passing at ~1 AU on average every year) usually show very extended (several arcmin) X-ray emission. *Athena*-XMS observations over a smaller FOV but with higher spectral resolution will complement the wider FOV, medium resolution spectral images provided by the *Athena*-WFI.

ATHENA

*Athena*'s two instrument complement will measure the solar wind ionic composition and speed at varying distances and ecliptic latitudes via charge exchange X-ray emission in comets.

### 2.4.2 Exoplanet environments

The atmospheric conditions, and eventually the habitability, of planets are likely to be regulated in part by the evolution of the ultraviolet and X-ray emission of their host stars. The intense X-ray emission from the early epochs of solar-type stars, and the associated extreme ultraviolet emission will dissociate and ionize molecules in planetary thermo-spheres and exospheres (cf. Güdel 2007; Penz, Micela & Lammer 2008). Stellar winds and flare particles may also erode the entire atmosphere if no magnetic field is present, while X-rays from the host star will be a likely erosion source as recently concluded by Sanz-Forcada et al. (2010) on the basis of ROSAT, *Chandra* and *XMM-Newton* data. These processes were probably important on Venus, Earth and Mars during the first 100 Myr and are presently leading to hydrodynamic escape of the atmospheres in extra-solar "hot Jupiters".

The effects of X-ray and UV radiation on the atmospheres go beyond their evaporation and chemical processes: the most energetic radiation will penetrate deep into the atmosphere, conditioning the emergence of initial life forms while the radiation is strong, and the early life evolution while the levels are moderate but strong enough to produce genetic alterations.

While EUV radiation of the host star is not directly observable with current or planned missions, detailed X-ray studies are well within the capabilities of *Athena*. Thousands of extra-solar planets are likely to be known by the next decade thanks to ground- and space-based planetary search programmes. *Athena* has the power to measure both the quiescent and flare activity of a significant sample of stars with planets in their habitable zones. *Athena* can thus provide the key measurements which can indicate systems in which the conditions may currently be too hostile for life. *Athena* observations, combined with atmospheric modelling, should conclusively show the effects of the X-ray emission of the host star on planets and provide unique insights into the atmospheric history of the potentially habitable rocky planets.

*Athena* will conclusively show the effects of X-ray stellar emission on planets and provide unique insights into the atmospheric evolution of habitable planets.

### 2.4.3 Stellar X-rays

The Sun was the first cosmic X-ray source ever detected and X-ray studies of the Sun and stars of all types within the solar neighbourhood have contributed vastly to our understanding of stellar processes and their complex interactions. *Athena* observations of stellar systems offer the prospect of making significant advances in a wide range of areas where current observations lack the sensitivity, spectral resolution or indeed time resolution required for all but the brightest systems. Here we highlight a few examples of stellar studies where *Athena* can provide important breakthroughs.

#### Mass-loss from massive stars

Feedback not only operates on the global scales of entire galaxies and galaxy clusters, it also plays an important role on more local scales within our own Galaxy. Owing to their powerful stellar winds, which combine large mass-loss rates and velocities (up to several thousand km/s), massive stars are key players in this process at a local scale as well as the more distant Universe (e.g. Dopita 2008). They interact with the surrounding ISM, thereby impacting the formation of lower-mass stars and they chemically enrich the ISM with the products of previous nucleosynthesis processes.

However, high-quality multi-wavelength observations of massive stars have shown that the classical picture of a smooth, homogeneous stellar wind is far too simplistic. Real stellar winds feature small and large-scale structures (Puls et al. 2008) that have tremendous impact on the conventional diagnostics such as H$\alpha$ emission, UV P-Cygni profiles and free-free radio emission and discrepancies by one or two orders of magnitude in the mass-loss rates are not uncommon (Fullerton et al. 2006; Sundqvist et al. 2010). Resolving these issues is fundamental for a proper understanding of massive star evolution and its impact on the surrounding ISM, both of which are heavily dependent on the mass-loss process. The X-ray emission from these winds provides a sensitive means to probe mass-loss rates, the degree of wind inhomogeneity and the shape of any structures. *Chandra* and *XMM-Newton* have paved the way (Güdel & Nazé 2009 and references therein), but further progress requires an increase in sensitivity (and thus time resolution) in conjunction with high spectral resolution. *Athena* will extend measurements beyond the few brightest objects and enable time-resolved high-resolution spectroscopy, allowing the inner details of the wind to be probed, allowing major advances in the theory of radiation-driven stellar winds.

The head-on collision of winds in massive binary stars produces much harder X-ray emission than in single massive stars. The line profiles of X-ray emission lines formed inside the wind interaction zone of such systems, such as the strong Fe K line seen in many interacting wind systems, provide a unique probe of the immediate post-shock conditions (Henley et al. 2003). The high sensitivity and spectral resolution of *Athena*-XMS will monitor the orbital changes of these lines which reflect the changing orientation of the system and, in the case of eccentric systems, the changes in the wind interaction as the separation between the stars varies. In circular orbit systems, Doppler tomography can be used to map the colliding wind region in velocity space. High-resolution X-ray spectra of these systems, used in conjunction with hydrodynamical models (e.g. Pittard & Parkin 2010), probe the post-shock densities and temperatures in the shock region, hence providing information on the stellar mass-loss rates and wind speeds which are essentially clumping-independent. *Athena*-XMS spectra will also allow us to study processes such as the acceleration of relativistic electrons that are seen through non-thermal radio emission in many massive binary systems (De Becker 2007 and references therein). These studies will provide unique information on the efficiency of electron acceleration in a region of parameter space complementary to that probed by supernova remnants (see Section 2.4.4). *Athena* will further enable us to extend these studies to massive binary systems in the Magellanic Clouds, thereby allowing us to investigate the impact of metallicity on the physics of stellar winds.

## Coronal variability and structures in late-type stars

X-ray studies of late-type stars have proved crucial for our understanding of the complex physics that connects stellar dynamos to their coronae. Magnetic reconnection events occurring in the corona of late type stars are highly dynamic and involve rapid changes in X-ray-emitting plasma composition, temperature, and bulk velocities. Progress in understanding the effects of magnetic reconnection requires constraints on density, elemental abundance, scale length, and velocity of flaring plasma, which are beyond the capability of the high-resolution spectroscopic measurements currently available. The bulk velocities of the impulsive mass motions which take place early in stellar flares are currently unconstrained; based on solar observations, the timescale could be as short as 10-1000 seconds. The velocities associated with bulk flows and turbulence of chromospheric evaporation in the early flare phases are sometimes of the order of 100-400 km/s, accessible with *Athena*-XMS for stellar sources with $F_x \sim 10^{-11}$-$10^{-12}$ erg/sec/cm$^2$.

Studies to date have necessarily concentrated on X-ray bright active stars having high magnetic field filling factors, thus emphasising processes occurring in regions where magnetic fields are highly packed with plasma. However, less active stars have smaller filling factors and so it is not clear how (and if) one can scale X-ray bright active star findings to stars with lower magnetic activity. High resolution spectroscopy of the X-ray brightest stars has determined that:

- stellar coronal plasmas of different temperatures have different densities
- the dominant coronal temperature seen in X-ray spectra increases with X-ray activity
- the coronal abundance anomalies seems to be related to X-ray activity, although recent findings challenge these results (Wood & Linsky 2010).

To progress, a wider range of X-ray luminosities and magnetic field strengths and distributions must be explored. *Athena*-XMS observations will extend these studies to solar-like stars characterised by an average surface flux comparable to that of solar coronal holes (cf. Schmitt & Liefke, 2004). Such observations will enable us to place the Sun in context of other solar-like stars and test the degree and similarity of solar coronal structures to these coronae. The exposure times required to determine the key parameters such as the electron density of oxygen-emitting material to this accuracy range from 2-150 ks.

## Coronae and accretion in low-mass stars and brown dwarfs

Brown dwarfs, in particular those in their early formation phases, have received much attention throughout the last decade. These young substellar objects are low-mass siblings of T Tauri stars, i.e. pre-main sequence stars undergoing accretion from a circumstellar disk. T Tauri accretion disks are sites of planet formation, and X-rays from the central object (be it a star or a brown dwarf) may play a fundamental role in the accretion process via a complex feedback mechanism. X-rays ionize the disk, determining the level of turbulence driven by the magneto-rotational instability, and hence the accretion rate. The shocks that form when the accreted matter impacts onto the star is, in turn, capable of producing X-ray emission that adds to the X-rays from the corona. Moreover, the infalling accretion columns act as absorption screens, determining the shape and flux of the X-ray spectrum that is finally incident on the disk. Finally, photo-evaporation, now seen as a major disk dispersal mechanism during the planet formation stage in low-mass stars, is also strongly affected by both the coronal and accretion X-ray fluxes (Ercolano et al. 2009).

This general picture has been established with the help of a handful of high-resolution X-ray spectra of T Tauri stars yielding density measurements that can distinguish coronal emission from accretion-powered X-rays. *Athena* will provide the first high-resolution spectra for very low mass stars and brown dwarfs, an important test for the continuity of the star formation process across the substellar boundary (e.g. Stelzer et al. 2010). This may not be taken for granted because the magnetic activity of brown dwarfs, responsible for the coronal X-ray emission, may be fundamentally different to that of solar-like stars, as a result of their low atmospheric temperatures and ensuing high electrical resistivity (Mohanty et al. 2002). Sensitive *Athena* -WFI surveys will also yield significant samples of X-ray detected brown dwarfs for the first time, both at young ages in star forming regions and in an evolved stage in the field where currently only a handful of detections are available (Stelzer et al. 2006). Since brown dwarfs cool as they age, this will be essential for understanding the influence of photospheric temperature on the coronal X-ray production.

### 2.4.4 Supernovae and supernova remnants

Supernovae (SNe) are a dominant source of heavy elements in the Universe, and they are the only source of alpha-elements (O, Ne, Mg, Si, S, Ar, Ca) and Fe-group elements (mainly Fe and Ni). These explosions are also the most important source of energy input into the ISM, both in the form of kinetic/thermal energy and in the form of cosmic rays. *Athena* will allow us to study these explosions, their products and their immediate surroundings, either by directly observing X-ray emission from extra-galactic SNe, or by studying supernova remnants (SNRs) in the Galaxy and the Local Group. Understanding SNe/SNRs is important for the rest of the *Athena* science programme, as SNe are of key importance for the chemical evolution of the Universe and often mark the creation of a neutron star or a black hole. Moreover, the physics of collisionless shock waves and cosmic ray acceleration, which can be studied by *Athena* in nearby SNRs, are directly relevant to the large-scale shocks that formed the WHIM and heated clusters of galaxies.

The prime instrument for SNR studies will be *Athena*-XMS, whose combination of excellent spectral and good spatial resolution will enable us to map the abundance patterns of SNRs, but will also add the third dimension by providing accurate Doppler shifts with an accuracy as low as 50 km/s for Si and Fe. *Athena* will be able to measure abundances of rare elements which hold important information on the conditions under which nucleosynthesis occurred. *Athena* observations will provide accurate abundance patterns for a large sample of Galactic and SMC/LMC SNRs with which the abundances in clusters of galaxies can be compared (see Section 2.3.1).

### Type Ia supernova remnants

SNe are divided into thermonuclear or Type Ia supernovae (SN Ia) and core collapse supernovae (SNcc). SNeIa are thought to be exploding C/O white dwarfs, which have accreted material from a companion star or have merged with another white dwarf. A major uncertainty in our understanding of SN Ia is the evolutionary paths which lead to their explosion: population synthesis models of binary stars underproduce SNIa by an order of magnitude. Another uncertainty is the explosion itself. The currently favoured models are delayed detonation and pure deflagration models. There are also models with SN Ia progenitors less massive than the Chandrasekhar limit (1.38 $M_{\odot}$). In these sub-Chandrasekhar models the explosion is induced by a surface layer detonation, triggering a thermonuclear explosion in the core (Fink et al. 2010). These models predict the presence of an outer layer rich in Fe, Cr, and $^{44}$Ti (which decays in 85 yr into $^{44}$Sc/$^{44}$Ca). Recent theoretical studies emphasise the importance of SN ejecta asymmetries. These can be a result of the explosion process itself (Maeda et al. 2010) or of the shielding of ejecta by a nearby donor star (Marietta et al. 2000).

*Athena* will clarify the nature of SN Ia explosions by accurately mapping ejecta abundance patterns, both in the plane of the sky and using Doppler mapping. This will make it possible to investigate the 3-D geometrical properties of SN Ia explosions and link them to specific explosion models. A recent statistical analysis of X-ray data showed that SN Ia SNR shapes are geometrically distinct from those of SNcc SNRs (Lopez et al. 2009), a method that can be expanded to include velocity information. Coupling SN Ia explosion models with hydrodynamical and X-ray emission codes has already enabled the exclusion of several SN Ia explosion models for SN Ia SNRs like Tycho and 0509-69 (Badenes et al., 2006, Badenes et al. 2008a), and the addition of Doppler velocity and other high-resolution spectral information from *Athena* will significantly increase the power of this approach (Figure 2.42).

One of the recent advances in this field was the detection of the rare elements Cr and Mn with *Suzaku* (Tamagawa et al. 2009). Uneven mass number elements like Mn in SN Ia are directly linked to the presence of N in the progenitor star. This can therefore be used to establish whether a SN Ia originates from a Population I or II star (Badenes et al. 2008b). Moreover, the presence of excess Fe, or $^{44}$Ti in the outer ejecta layers can be used to investigate the viability of sub-Chandrasekhar models. *Athena* will be able to routinely measure these elements for young Galactic and LMC/SMC SNRs, thereby providing an important benchmark for abundance studies of clusters of galaxies.

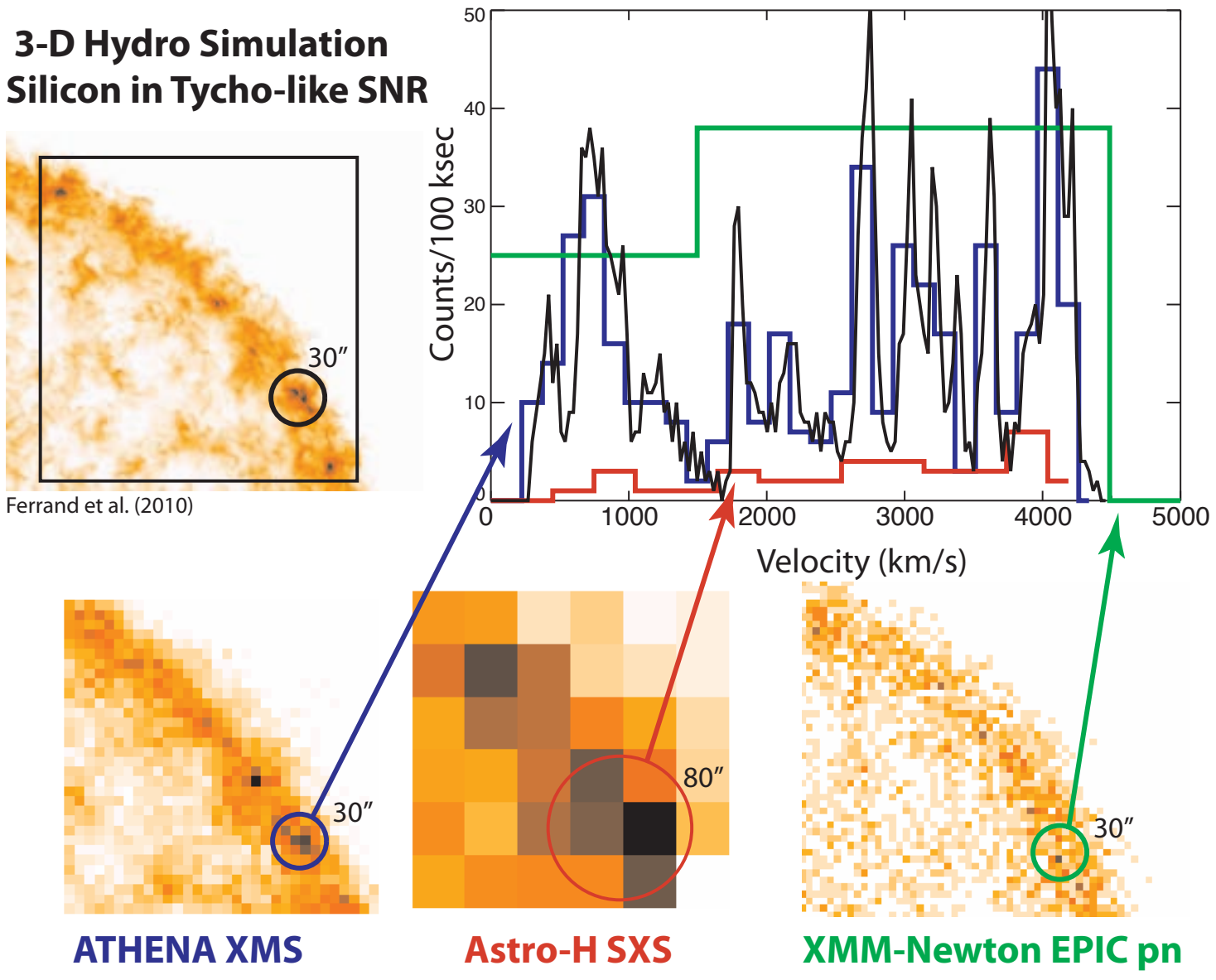

**Figure 2.42.** *Athena XMS, XMM-Newton, and ASTRO-H images and silicon velocity profiles based on a 3D hydrodynamical simulation of Tycho's supernova remnant (Ferrand et al. 2010) are shown with the typical imaging resolution of the instruments (circles). Athena XMS profile is shown in blue, Astro-H SXS is shown in red and XMM-Newton is shown in green Only Athena-XMS has sufficient spectral resolution, sensitivity and imaging resolution to isolate the highlighted knot and retrieve the velocity information that will reveal the 3D dynamics of the supernova remnant, providing constraints on the explosion mechanism through the measurements of asymmetries. Together with accurate temperature measurements, the shock velocity will be used to quantify the supernova remnant's ability to accelerate cosmic rays.*

## Core collapse supernova remnants

SNcc explode due to the collapse of the core of a massive star into a neutron star or black hole. The release of gravitational energy powers the explosion of the rest of the star, but how the collapse energy results in the explosion is not clear and the process has been difficult to recreate with computer models. Neutrino-plasma coupling is thought to be an important ingredient for the explosion. In addition, the infall of plasma onto the core is likely to create various instabilities, which may facilitate the explosion (e.g. Blondin et al. 2003). The association of gamma-ray bursts with core collapse SNe has revived the discussion on whether bipolarity or even jet formation are important for all core collapse SNe (Wheeler et al. 2002). *Athena*-XMS observations will map out the 3-D structure of SNcc SNRs, allowing searches for jet-like emission in any orientation (Figure 2.43). The detection of rare elements in Fe-rich ejecta will constrain various processes affecting the nucleosynthesis processes in the core. Among these is the effect that neutrinos can convert protons into neutrons, providing a nucleosynthesis path to rare elements. These elements therefore hold important clues to the role of neutrinos in SNcc explosions (Wanajo et al. 2011).

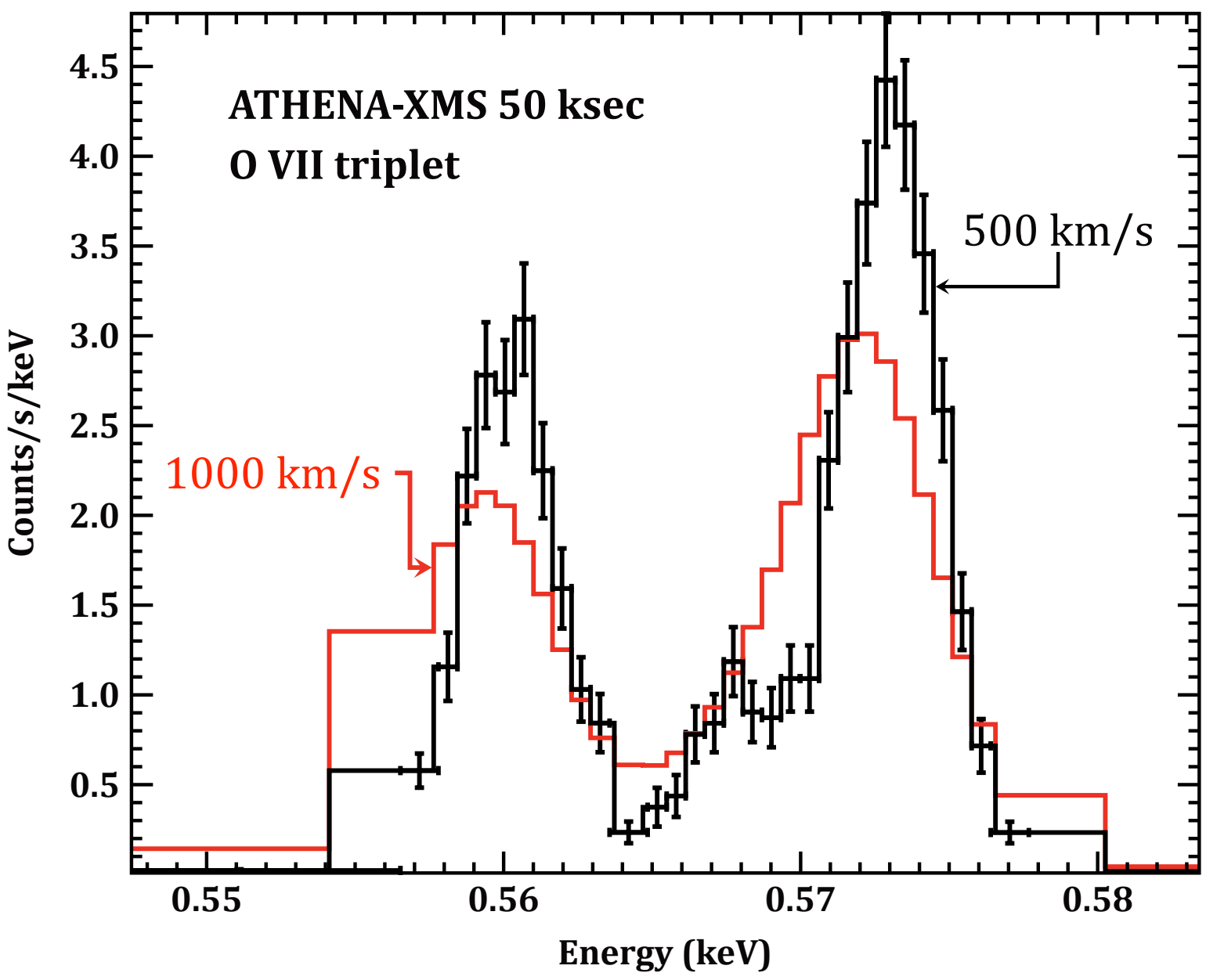

**Figure 2.43.** *Athena-XMS 50 ks simulation of SNR 004327+411830, the brightest SNR in M31 (flux of 6.3x10$^{-14}$ erg cm$^{-2}$ s$^{-1}$). The input model is based on a XMM-Newton spectrum and consists of a two temperature solar abundance plasma at about 0.1 keV and 0.3 keV. The simulated spectrum assumes a line shift and a broadening corresponding to a velocity of 500 km/s (black). In order to illustrate Athena's ability to measure velocities of remnants in local group galaxies the fit model (red) assumes a velocity of 1000 km/s.*

### Shock heating and shock acceleration

Cosmic rays with energies of $10^9$-$10^{15}$ eV originate from SNR shocks. X-ray astronomy has significantly increased our understanding of cosmic-ray acceleration through the detection of X-ray synchrotron emission from SNRs (Koyama et al. 1995) and by showing that the emission is coming from regions very close to the shock front (e.g. Vink & Laming 2003). This established the presence of electrons with energies of $10^{13}$-$10^{14}$ eV. Since these electrons lose their energy on time scales of several years, their presence indicates active and fast particle acceleration. The narrowness of the synchrotron regions also indicates high magnetic fields, probably itself a result of efficient cosmic-ray acceleration (Bell 2004).

*Athena* will continue the successful exploration of cosmic-ray acceleration properties in X-rays by characterising the plasma temperatures close to the shock front. This is key because efficient cosmic-ray acceleration gives rise to plasma temperatures that are much lower than expected from the standard Hugoniot shock-relations (e.g. Vink et al. 2010). Although low spectral resolution X-ray spectroscopy can be used to measure electron temperatures, the thermal properties of the plasma are much better characterised by the ion temperatures, which may differ dramatically from the electron temperatures. These ion temperatures can be measured using thermal Doppler broadening. In order to separate thermal line broadening from bulk motion one needs an X-ray instrument that has a high enough spatial resolution to isolate regions close to the SNR edge, and at the same time provides a spectral resolution to measure line broadening of a few hundred km/s, a unique characteristic of the *Athena*-XMS instrument.

## 2.4.5 The Galactic Centre environment

*Athena* studies of the innermost regions of the Galaxy will have an important role in probing the accretion properties of Sgr A*, the nearest supermassive black hole (Section 2.1.2), but equally the large FOV of the *Athena*-WFI also allow the simultaneous observation of the neighbourhood of Sgr A*, one of the richest regions of the sky with numerous types of extended (e.g. diffuse emission, supernovae remnants) and compact (e.g. X-ray binaries) astrophysical objects. Studies of X-ray reflection of nearby molecular clouds are of particular importance. It is now established that that Sgr A* underwent a giant outburst (by a factor ~$10^6$) a few hundred years ago. Traces of this event are found in the X-ray emission of molecular clouds in the region, as first suggested by Sunyaev et al. (1993), through reprocessed emission producing 6.4 keV Fe K line (Koyama et al. 1996) and non-thermal X-ray continuum radiation. Recent measurements of vari-ability in different molecular clouds of both the neutral iron Kα line at 6.4 keV with *XMM-Newton* (Ponti et al. 2010) and of the high energy continuum above 20 keV with INTEGRAL (Terrier et al. 2010) have given preliminary constraints on the past outburst. The most spectacular variability (with apparent superluminal propagation) of the 6.4 keV line has been measured in a molecular cloud close (15 arcmin) to Sgr A* (Ponti et al. 2010). *Athena*-WFI spectroscopy will allow unambiguous determination of the emission process, distinguishing between hard X-ray photoionisation or by collisional ionisation induced by relativistic particles (Figure 2.44). The correlation between the output of different molecular clouds will allow the 3-D molecular cloud distribution to be determined. This, combined with velocity information from both the fluorescencent X-ray lines and other studies, will provide important constraints on the dynamics in the inner few hundreds parsecs of the Milky Way.

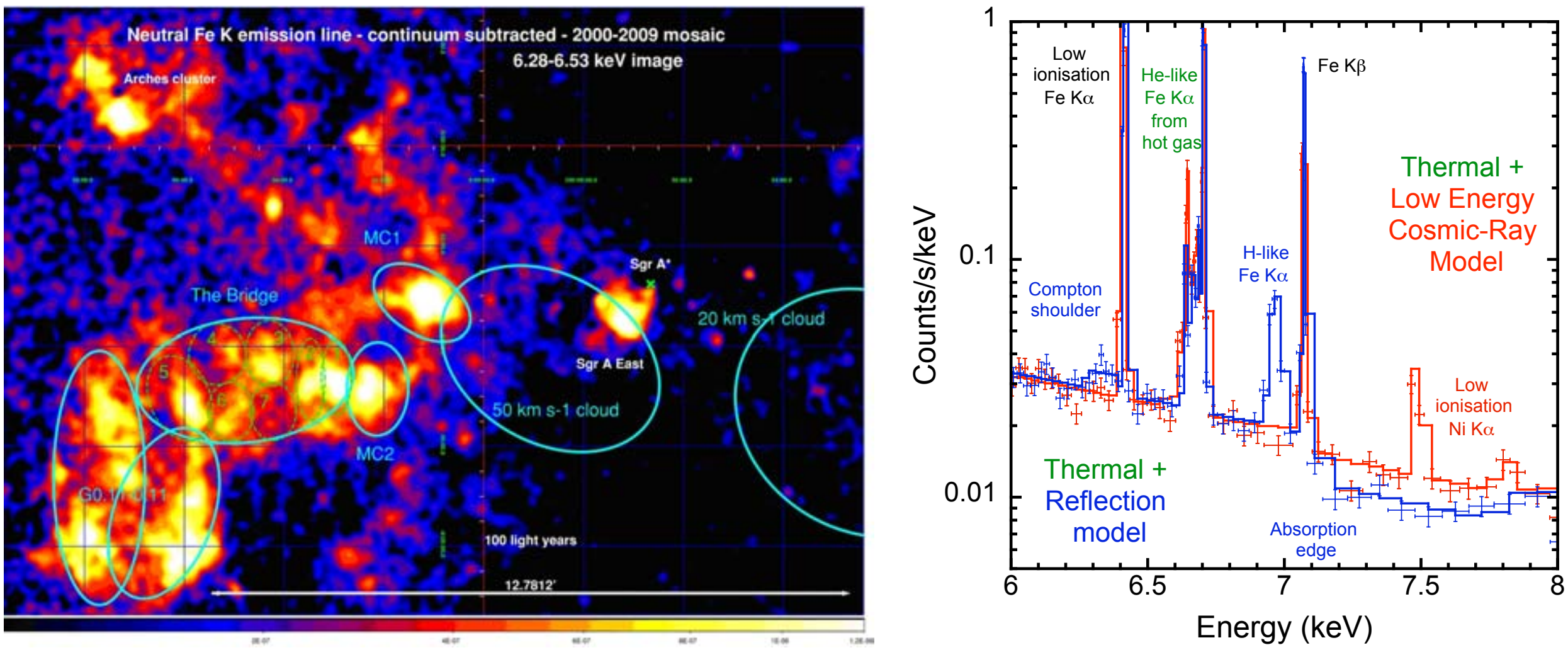

**Figure 2.44.** ***Left:*** *X-ray map of the Galactic Centre region showing the distribution of Fe Kα emission (continuum-subtracted) and the location of various prominent molecular clouds which act as X-ray reflection nebulae (Ponti et al. 2010).* ***Right:*** *Athena-XMS 200 ks simulations of the emission around the Fe K lines expected from molecular clouds in the environment of Sgr A* for two different models: hard X-ray reflection by cold material (in blue) and cosmic ray irradiation (in red). Both models include the thermal emission from the surrounding hot gas. The different components of the models will enable the determination of the origin of the 6.4 keV Fe K emission in the molecular clouds of the Galactic Centre.*

> *Athena* observations of Sgr A* and its environment provide a unique window on the current and past activity of this unique, low luminosity active galaxy in our own back yard.

## 2.4.6 Interstellar gas and dust

The interstellar medium bears the signature of the star formation and evolution in our Galaxy. Dust is efficiently produced by late-type stars and supernova explosions (see Section 2.4.4). In turn dust and gas provide the reservoir of matter for the formation of new stars and planets. The properties of gas and dust in the ISM, especially their abundances, composition and structures impact practically all of astrophysics. Fundamental processes from star formation to stellar winds to galaxy formation all scale with the number of metals. However, significant uncertainties remain in both absolute and relative abundances, as well as how these vary with environment.

X-ray observations can characterise both the gas and dust phase of the diffuse ISM in our Galaxy as the X-ray band covers all ionic transitions of the abundant metals. *Athena*-XMS will access a spectral region up to now relatively unexplored using high signal-to-noise, high spectral-resolution instruments, in particular the Mg K-, Si K- and Fe K-edge region, located between 1.3 and 7.1 keV. These, together with the O K- and Fe L-edges, are crucial to investigate the nature of the silicates in the dust of our Galaxy, and especially the question of Fe inclusion in silicates. The Fe K-edge is also an important metallicity indicator in the densest regions such as the Galactic Centre, where the high column density obliterates all other lower energy edge features. Dust both absorbs and scatters the X-rays. The background light of bright X-ray sources, mainly X-ray binaries, allows us to study interstellar dust using both mechanisms.

The shape of the absorption cross-section close to the edge depends on the environment of the absorbing atom (X-ray atomic fine structure). As shown in Figure 2.45, *Athena*-XMS observations will distinguish between the different types of silicates and iron compounds that have been proposed as dust constituents. Such direct measurements of the dust composition and chemistry will not be possible in any other waveband for the foreseeable future. Since X-ray binaries are mostly located within the Galactic plane, information on the ISM is therefore embedded in most observations carried out on compact objects. This will allow *Athena* to map the diffuse ISM in a variety of environments, with different dust and gas enrichment and mixing history.

X-rays also scatter off dust in the ISM, leading to X-ray halos around bright X-ray sources that can be several arcminutes in size (see, e.g., Xiang et al., 2011). *Athena*'s imaging capabilities will make it possible to combine the information on the dust chemistry derived from the absorption studies with information on the size scale and composition determined from X-ray scattering. Spatially resolved XMS spectroscopy will determine the dust chemistry, while the WFI will detect the full spatial extension of the halo and measure the grain distribution.

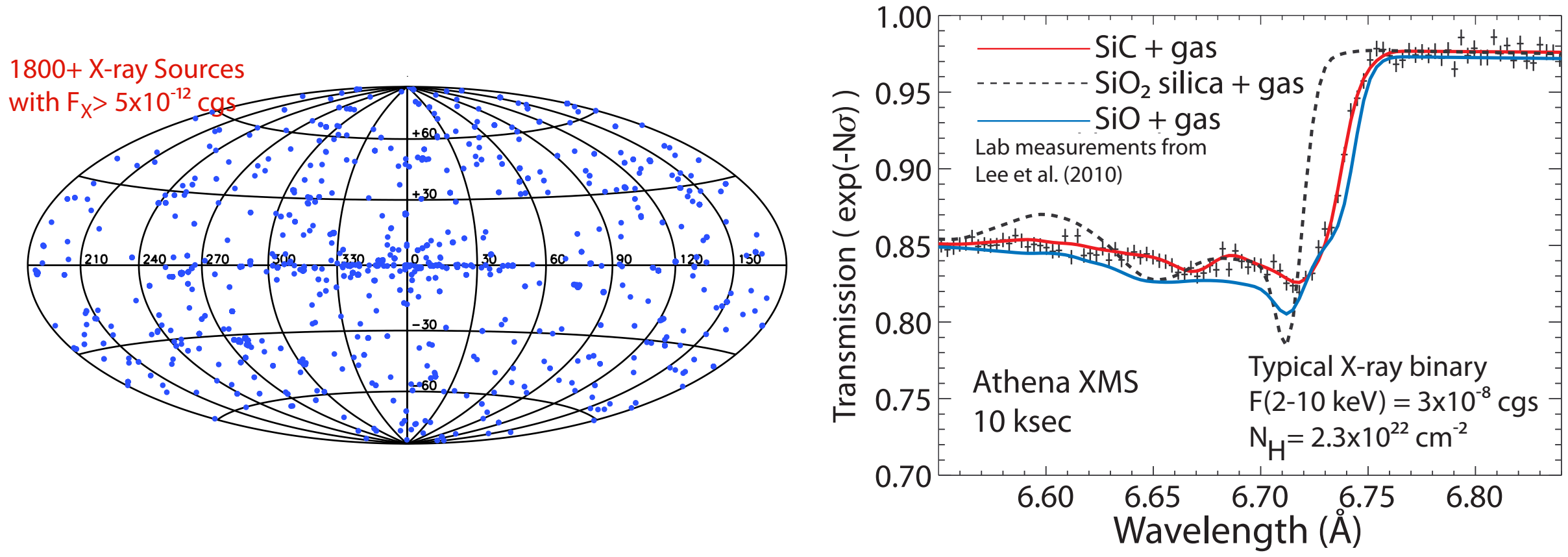

**Figure 2.45.** ***Left:*** *50 ksec Athena-XMS observations can determine Mg, Si, and/or Fe gas and dust-phase interstellar medium abundances to better than 10% along sightlines to any of these bright AGN and Galactic X-ray binaries.* ***Right:*** *Different compounds commonly present in the interstellar medium are easily distinguished via this Athena-XMS absorption spectra of the bandpass around the Si K edge with 3 eV resolution.*

ATHENA

XMS spectroscopy of bright X-ray sources will reveal the physical nature of gas and dust in the diffuse interstellar medium of our Galaxy, via absorption features imprinted on the X-ray spectrum.

# 3 Science Requirements

## 3.1 Summary of Requirements

The performance requirements that flow down from the driving science cases are summarised in Table 3.1. This section describes how these requirements have been derived and how the instrument requirements for the *Athena* payload have been developed. The mission addresses a wide range of science investigations, and no single instrument can provide all the necessary features. The different spectrometer and imager requirements are therefore emphasised. Finally a draft Observing Plan (or Design Reference Mission) has been constructed for the science cases and has been used to estimate the mission observing efficiency, establish the scope of mission planning and Targets of Opportunity, and understand the fraction of time that could be devoted to General Observatory investigations.

| PARAMETER | PERFORMANCE REQUIREMENT | SCIENCE GOALS |
|---|---|---|
| Effective Area (total for 2 telescopes) | 1 $m^2$ @ 1.25 keV (goal of 1.2 $m^2$)<br>0.5 $m^2$ @ 6 keV (goal of 0.7 $m^2$) | -Black hole evolution, large-scale struct.<br>-Strong gravity, equation of state |
| Spectral Resolution (FWHM) | $\Delta E$ = 3 eV (@ 6 keV) in 2 x 2 $arcmin^2$<br>(goal of 2.5 eV and 3 x 3 $arcmin^2$)<br><br>$\Delta E$ = 150 eV (@ 6 keV) in 24 x 24 $arcmin^2$<br>(goal of 125 eV and 28 x 28 $arcmin^2$) | -Cosmic feedback, large-scale structure<br><br>-Black hole evolution |
| Angular resolution | 10" HPD (0.1 - 6 keV)<br>(goal of 5") | -Cosmic feedback, black hole evolution |
| Sensitivity (0.5 - 2 keV) | 4 x $10^{-17}$ erg $cm^{-2}$ $s^{-1}$<br>(goal of 2 x $10^{-17}$ erg $cm^{-2}$ $s^{-1}$) | -Black hole evolution |
| Count Rate (WFI) | 1 Crab with > 90% throughput<br>$\Delta E$ < 200 eV @ 6 keV (0.3 - 10 keV) | -Strong gravity<br>-Equation of state |
| Astrometry | 1.5" at 3$\sigma$ confidence | -Black hole evolution |
| Absolute Timing | 100 $\mu$sec | -Neutron star studies |

**Table 3.1.** *Summary of Athena Science Requirements.*

## 3.2 Performance Requirements

The science case has been developed, coherent with the performance requirements listed. In particular the angular resolution requirement of 10 arcsec has been assumed throughout. However we note below topics where the improvement towards the performance goals provide significantly enhanced science capability.

### 3.2.1 Detection Sensitivity

The limiting sensitivity requirement is determined by the deep field science. The requirement is to detect low luminosity AGN, especially those which are obscured at other wavelengths, in the deepest fields. *Chandra* deep and large area surveys have only yielded a handful of AGN candidates at $z$>5. The specific require-ments are for detecting $L_*$ ($10^{44}$ erg/s) AGN at *z*~7 in shallow survey fields, and to detect 0.1 $L_*$ AGN to *z*~5 in the deepest fields. Therefore *Athena* is required to achieve a sensitivity in the 0.5- 2 keV band of <$2\times10^{-16}$ erg cm$^{-2}$ s$^{-1}$ in a 100 ks observation that allows efficient field coverage. The ultimate sensitivity (in a 1 Ms deep field), will be ~$4\times10^{-17}$ erg cm$^{-2}$ s$^{-1}$ (0.5-2 keV), which improves significantly on *Chandra* survey speed. The goal is to increase the number of lower luminosity AGN at higher redshifts in the deepest fields, i.e. 0.1 $L_*$ to *z*~7, requiring <$2\times10^{-17}$ erg cm$^{-2}$ s$^{-1}$ which motivates the improved sensitivity with goal angular resolution of 5 arcsec.

As discussed in Section 2.2.2 the sampling of the lower luminosity AGN (~$10^{43}$ erg s$^{-1}$ or lower) at high redshifts (*z*>4) requires that a PSF below 10 arcsec and as close to the goal of 5 arcsec as possible is reached. Figure 3.1 illustrates the number of high-*z* AGN that *Athena* would detect in deep surveys at a range of luminosities very significantly larger than those detectable with *Chandra* or *XMM-Newton*. This will open up for the first time the possibility of mapping the population of young growing SMBHs at the earliest epochs.

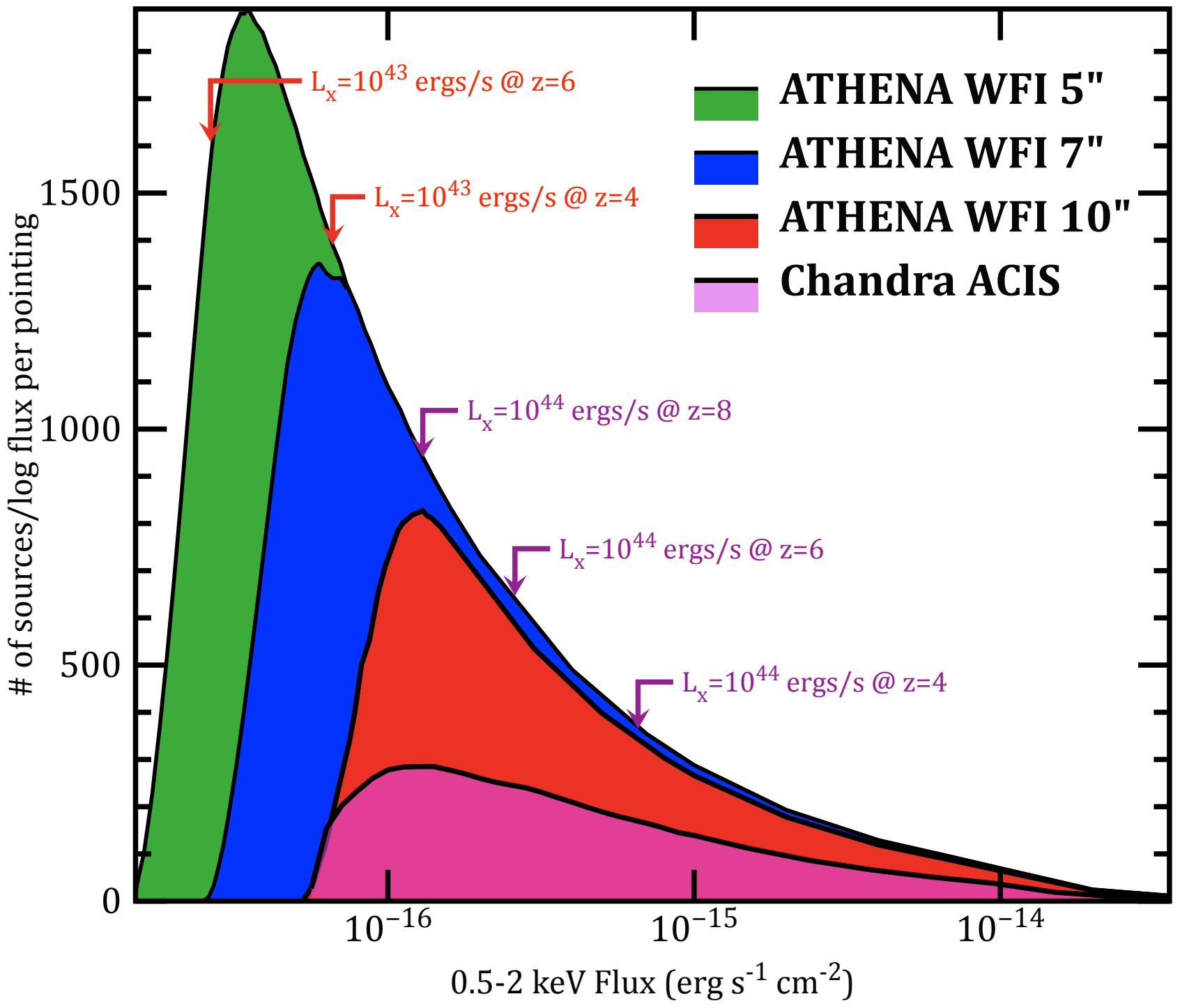

**Figure 3.1.** *Expected number of sources per pointing and per decade of flux in 1 Ms exposures with Athena-WFI for various PSFs, compared to Chandra-ACIS. Improving the PSF opens up new, so far unexplored, territory in low-luminosity AGN at high redshifts.*

### 3.2.2 Angular Resolution

The overall angular resolution requirement derives from the sensitivity requirements discussed above, but is also driven by two other considerations, source confusion limits and source feature extraction. These requirements provide comparable requirements for the angular resolution:

**1)** The increased background that occurs with larger detection cells can be compensated partly by increasing exposure time. This trade-off eventually is limited by source confusion, which in turn is set by source density. The ultimate detectable source density depends upon the relative strengths of luminosity and density evolution with look-back time. This has been investigated with simulations of many scenarios. A simple estimate for the increasing confusion between sources with decreasing flux can take as an example the Moretti et al. (2003) extrapolation of CDF source density. For a confusion limit of one source per 30 beams, a 10 arcsec beam is confused for a source density ~$4\times10^{-17}$ erg cm$^{-2}$ s$^{-1}$ (0.5-2 keV). This is compatible with the ultimate deep fields sensitivity requirement noted above.

**2)** The mean separation of sources in deep IR ($J_{AB}$~26) fields is ~1-1.5 arcsec (Maihara et al. 2001). For a source detection limited with 40 counts, then the error circle on the centroid of the *Athena* PSF is statistically ~0.5 arcsec for a 10 arcsec HEW angular resolution. This is for example also compatible with the catalogue uncertainties in the VISTA VHS survey sources (or with those in other deeper surveys in smaller areas) that might be used for post-facto enhancement of pointing knowledge and source cross-identification. This simple model has been confirmed with detailed simulations.

**3)** Galaxy cluster observations will almost always require spatially resolved spectral data. In nearby clusters, cavities on a scale <30 kpc diameter should be resolvable. For the 10 brightest clusters, for example, the mean (max) red shifts of 0.03 (0.05) require a resolution of 15 (10) arcsec. The identification of cool cores (typically ~50 kpc) will be important to exclude such regions from high redshift clusters to allow more robust analysis of the various scaling relations. With 10 arcsec resolution this excision is feasible up to z~0.4, covering a large sample space where cluster evolution starts to become apparent.

***Motivation for the Goal Angular Resolution***

**1)** The main motivation for reaching the PSF goal of 5 arcsec is to open up the high redshift Universe with *Athena*. If this can be achieved, deep surveys will essentially never be confusion limited in feasible exposure times, enabling *Athena* to reach the deepest limits currently attainable (by *Chandra*) but with a much larger field of view, in shorter exposure times, and with a larger number of photons, essential for source characterisation. An *Athena* survey for high redshift AGN will thus be more efficient by a factor >10. Detecting fainter AGN at very high redshift ($z$ = 6-10) is extremely important as it allows us to probe well below $L_*$, the knee in luminosity function and hence measure the total accretion power of the Universe at the epoch of reionisation and first galaxy formation. These highest redshift AGN are expected to be exceptionally faint at other wavebands as well, with AB magnitudes in the NIR of 25-27. Their identification and characterisation will therefore also be greatly eased by a smaller PSF, which gives improved positional accuracy, thus yielding far less ambiguity in the identification of their counterparts (see Figure 3.2).

Sections 4 and 6 highlight the planned development route towards achieving this resolution goal.

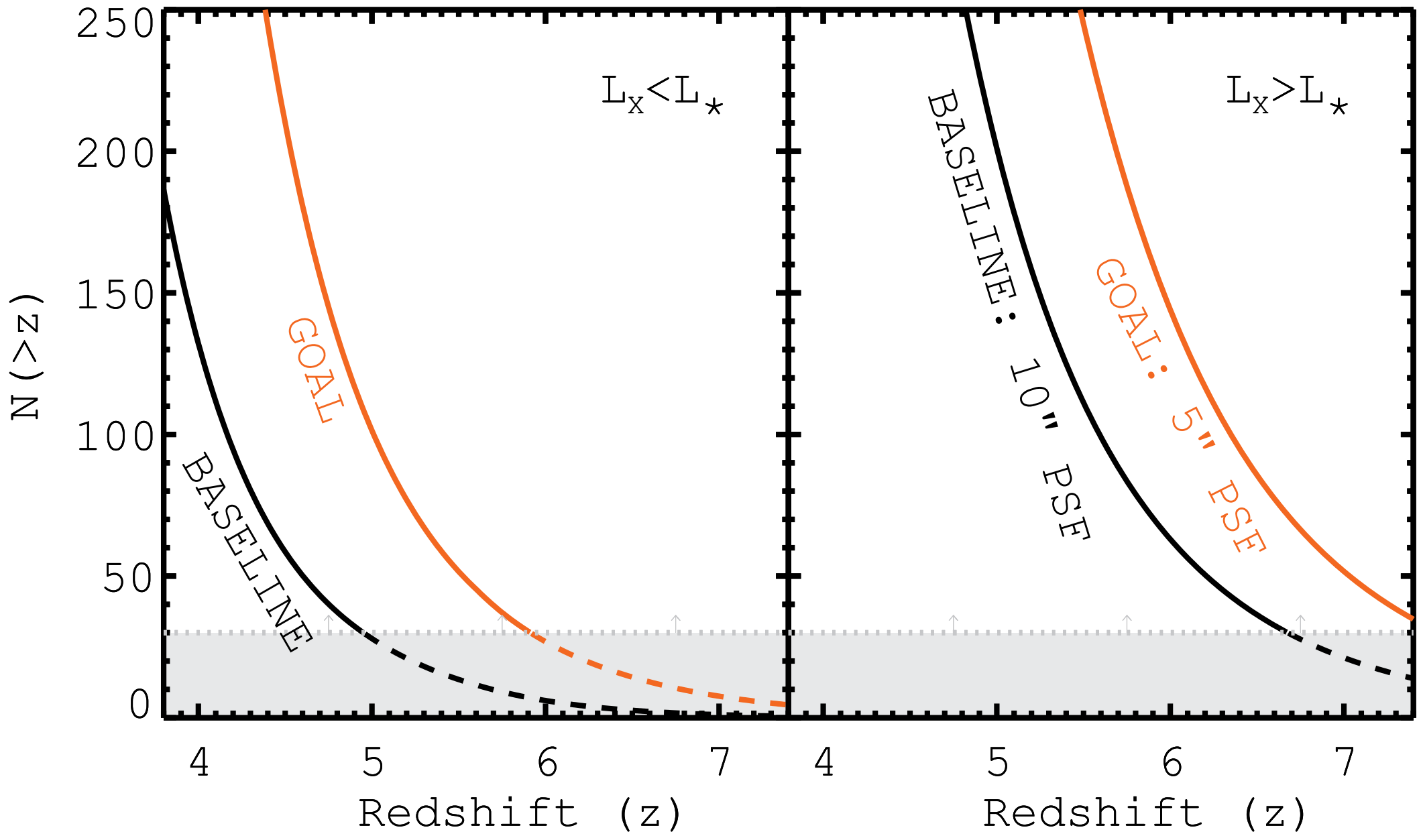

**Figure 3.2.** *Effect of the Athena PSF on the predicted number of AGN above a given redshift, for the Athena baseline survey programme described in the text. The goal PSF of 5 arcsec will open up the study of QSOs ($L>L_*$) at redshifts z~8 and higher, well into the "reionisation era" in significant numbers, and at the same time will enable tracing the growth of supermassive black holes in the lower luminosity AGN out to redshifts z~6, where surveys at longer wavelengths suffer from strong contamination effects from the host galaxies.*

**2)** A number of important areas would benefit significantly from improvement in angular resolution. These include the observations of the Galactic Centre, stellar clusters and of extragalactic ULXs. In the Galactic Centre region there are multiple point sources, and several filamentary arcs, with angular size ~20 arcsec. The goal angular resolution would significantly improve the ability to resolve filamentary structures from point sources, and discriminate more clearly the emission from the central black hole source. ULXs observed in galaxies up to 10 Mpc (e.g. NGC 4490, Holmberg X-1) are barely resolved by *XMM-Newton* from the surrounding point and diffuse emission of those galaxies. Spectral measurements are significantly improved by the reduction in aperture size through reaching the goal angular resolution.

**3)** For studies of the WHIM it is expected that simultaneously performing emission line measurements close to the line of sight towards the background source used for absorption studies will enable the combination of two different measurements of the filament properties to enhance the robustness of measurements. This requires the excision of the emission from the target and any nearby faint point sources. Measurement capability improves massively with increased angular resolution.

**4)** Resolving shock features in SNRs requires the angular resolution to be maximised to discriminate between knots and filamentary structure on one hand, and the general remnant emission on the other, allowing velocity structures to be revealed.

**5)** For robust determination of cluster scaling relations, it is necessary to excise cooling cores (50 kpc) from the emission. To enable this for all practical redshifts, the angular resolution should be ~5 arcsec.

### 3.2.3 Spectral Resolution

The required spectral resolution depends strongly on the science case, and hence with complementary science drivers such as time resolution and field of view, which vary by orders of magnitude between target classes. As a result the different instruments have different spectral resolution requirements.

**1) Narrow Field Imaging:** Spectroscopy of bright AGN and clusters of galaxies is required to study strong gravity and the evolution of feedback in galaxies and clusters. The former requires the centroiding of Fe line emission through different orbital phases of an accretion disc hotspot. For the latter, the measurement of turbulence and velocity-shifted features is needed. Existing *XMM-Newton*-RGS measurements (Sanders et al. 2011) of 1-2 keV gas in cool cores provide an upper limit to turbulence of ~200 km/s, in many clusters. This constrains the energy density to be <20 % in turbulence. Optical filaments in cool cores can be ordered on relatively small scales and while the velocity dispersion between filaments can be higher, the turbulent width in individual structures is ~200 km/s. Simulations to reproduce features such as gas sloshing and cool fronts indicate that velocity structures around 100-300 km/s should be prevalent. This implies a spectral resolution requirement of 3 eV FWHM to ensure velocity features of 100 km/s can be resolved (see Figure 3.3).

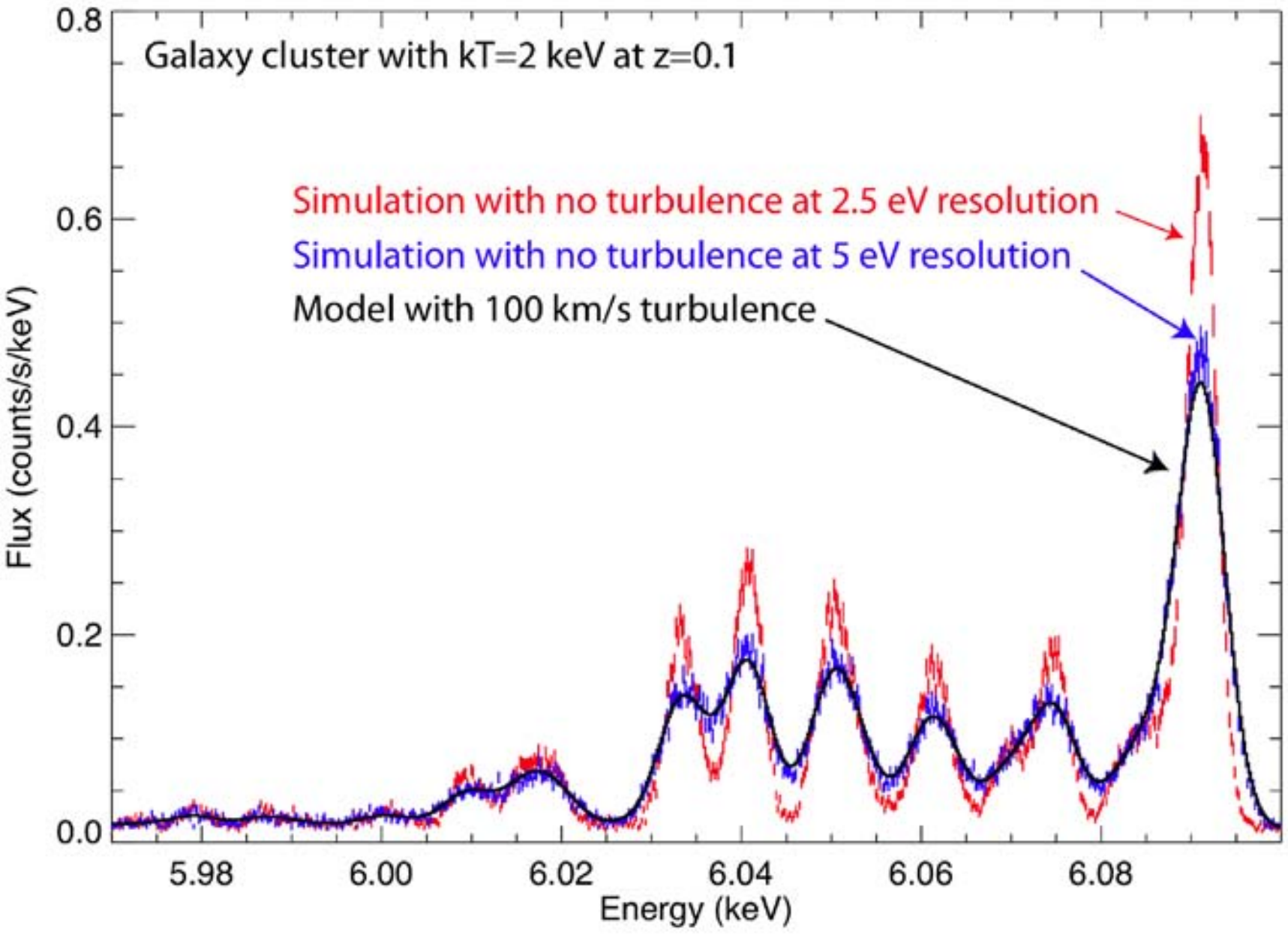

**Figure 3.3.** *Effect of degrading spectral resolution on resolving the iron line complex. The model shown is for a z=0.1 cluster with temperature of 2 keV, and only thermal broadening included. Resolution at the requirement value of 3 eV FWHM will allow turbulent broadening to be unambiguously characterised. Degrading the resolution even to 5 eV FWHM renders 100 km/s turbulent broadening undetectable.*

**2) Wide Field Imaging:** The specific case of the growth of SMBHs requires modest spectral resolution to obtain broad band colours. In wide field surveys to a depth 100 ks, a key goal is to establish a census of the ratio of Compton-thick and thin AGN. Even with spectra with as few as 1000 counts, spectral resolution must be good enough for Fe line diagnostics to break the degeneracy of hardness ratio estimates. The required resolution is 150 eV FWHM at 6 keV to measure iron line redshifts and equivalent widths to 5% accuracy.

### 3.2.4 Effective area

Effective area is related to the sensitivity and resolution parameters, especially energies around 1-2 keV at the peak of the photon count distribution. There are other drivers related to the photon-limited science such as iron line reverberation mapping.

**1) 1-2 keV:** As discussed above, the sensitivity couples the area requirement to angular resolution and survey exposure duration. The need to meet the sensitivity limit with the baseline angular resolution in a time that is commensurate with an observatory programme containing 10 Ms of deep surveys requires an imaging effective area of ~0.5 $m^2$. Over a narrow field of view, the WFI and XMS areas may be combined to provide ~1$m^2$. The soft X-ray high resolution spectroscopy of dynamical effects can be studied if the effective area is ~0.5 $m^2$. For example, in AGN outflows, the variability of the absorber ionisation as a function of the continuum flux constrains the minimum column density that can be meaured on 10 ks timescales on all the nearby bright Seyferts. Likewise the rise time ~100 s in stellar flares of few ks can be studied for non-equilibrium conditions.

**2) 6 keV:** Given typical target fluxes of 1-10 x $10^{-11}$ erg $cm^{-2}$ $s^{-1}$ (2-10 keV), orbital timescales of 1 to 300 ks, and Fe line equivalent widths of 200 eV, a total mirror area of 0.5 $m^2$ at 6 keV will ensure >100 photons in the Fe line per SMBH orbital bin (assuming at least 10 phase bins throughout the orbit). The centroid of narrow and varying iron lines can then be measured for a range of expected SMBH fluxes. This assumes that the combination of WFI and XMS area at 6 keV can be combined for analysis.

**3) Above 10 keV:** Above 10 keV the effective area must be 100 $cm^2$ to perform BH spin measurements (with continuum determination above the Fe line), and for observations of Compton-thick sources above 10 keV (for which the goal should be to 200 $cm^2$ @ 15-20 keV).

### 3.2.5 Field of View

**1) Wide Field:** From the above analysis, the survey science requires a flux limit of $10^{-16}$ erg $cm^{-2}$ $s^{-1}$ (0.5-2 keV) to be reached in 200 ks over an area substantially larger than that provided by the *Chandra*-ACIS, so that deep and wide surveys can be done faster. This is achieved by an optics design which maintains the on-axis PSF of 10 arcsec over >600 $arcmin^2$ (or >25 x 25 $arcmin^2$) with only modest degradation and has moderate vignetting.

Individual objects such as galactic SNRs and low redshift clusters have a typical extent ~10 arcmin, and an additional image region is needed for contemporaneous background extraction with comparable sized field (e.g. SNR diameters Cas A 3´, Crab 4´, Kepler 4´, Tycho 7´, SN1006 25´). For typical $r_{200}$ of ~1 Mpc any cluster more distant than z~0.04 is fully encompassed within a 20 arcmin field.

**2) Narrow Field High Resolution:** The strongest driver for XMS field of view comes from the driving cluster science topics. The nearby bright clusters where the XMS angular resolution is utilised for resolving AGN feedback processes, a field of view ~2 arcminutes is sufficient to cover the full core region greater than 150 kpc for all clusters more distant than the Perseus cluster, for example.

Multi-temperature and velocity mapping in merging clusters at medium redshifts (e.g. Bullet cluster, z~0.3) can be achieved on the scale of merging cluster units with the same field of view. At high redshifts, for abundance evolution studies the spectral resolution of XMS is required for line identification. Then most of the flux in z>0.5 cluster (>$0.5r_{500}$) is encompassed with a 2 arcminute field.

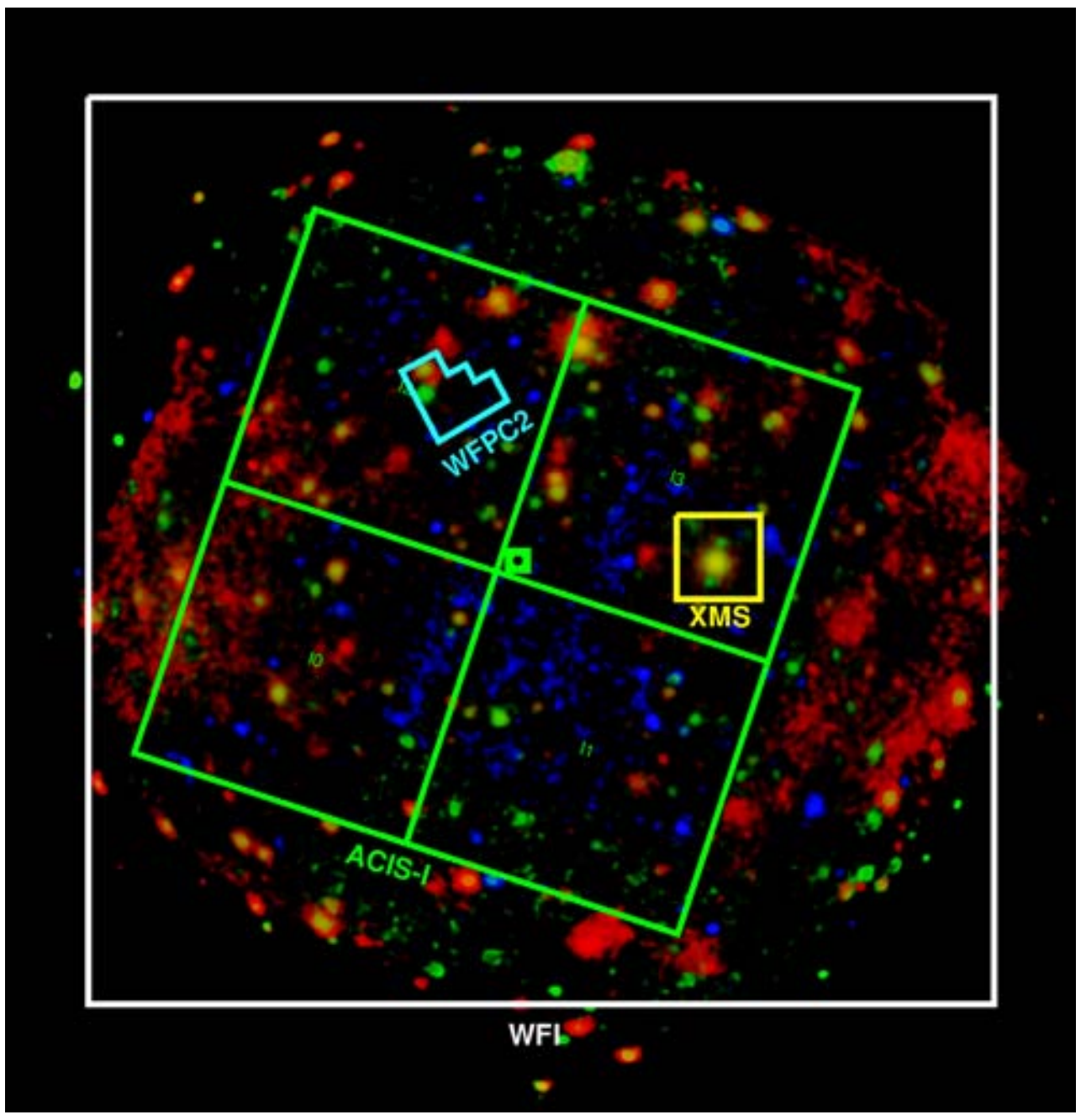

**Figure 3.4.** *XMM-Newton image of the CDF-S field. Superposed are the original WFPC2 (cyan) and Chandra ACIS-I (green) instantaneous fields of view. The baseline fields of view of the Athena-WFI and XMS are shown in white and yellow.*

### 3.2.6 Athena Timing Requirements

Time resolved spectroscopy of persistent or transient X-ray binaries will make it possible to probe accretion physics around the brightest X-ray sources in the sky, offering complementary techniques (relativistic iron line, disk continuum spectral fitting, quasi-periodic oscillations, reverberation mapping) to constrain the spin of black holes and the radius of neutron stars. It will also enable *Athena* to constrain the mass-to-radius ratio (and possibly $M/R^2$) in neutron star systems also using different probes (high-time resolution spec-troscopy of the kHz quasi-periodic oscillations, detection of rotationally broadened absorption line features, iron line variability, etc). The prime instrument for spectral-timing studies will be the *Athena*-WFI, thanks to a count rate capability that is significantly higher than the *Athena*-XMS.

**1) Spectral resolution for timing studies:** The spectral features to be used are relatively broad (e.g. a few keV for the iron line, a few hundreds of eV for the predicted absorption lines). *Athena* will be able to perform spectral and sub-ms timing studies of bright sources simultaneously. When deriving the requirement on the spectral resolution, the critical feature is the iron Kα line. This needs to be resolved to discern its profile, and the variability of that profile, which is shaped by the strong gravity around black holes and neutron stars. To do so we need to identify and separate narrower Kα lines of cold, helium-like and hydrogen-like iron (at 6.4 keV, 6.7 keV and 6.97 keV). The spectro-timing requirements are therefore a spectral resolution of 150 eV at 6 keV (goal 125 eV), simultaneously with the ability to cope with large count rates, up to about 30,000 events per second (or 1 Crab), with low pile-up (<4%) and large throughput (>90%).

**2) Pile-up requirement:** Pile-up, where the charge pattern deposited by a detected photon is confused with that of the next incoming photon, leads to an erroneous energy reconstruction. Pile-up distorts the spectral shape (spectra appear harder) and results in incorrect fluxes. In order to be able to determine the photon index of a power-law continuum to better than 0.1, Monte Carlo simulations appropriate for pixelised X-ray detectors show that a pile-up fraction of <4 % is required.

**3) Maximum count rate capability and throughput:** From the RXTE All Sky Monitor catalog, there are about 50 X-ray binaries with a mean persistent emission of 100 mCrab (1 Crab is a flux unit equivalent to $2.4 \times 10^{-8}$ erg cm$^{-2}$ s$^{-1}$ in the 2-10 keV band). This includes Sco X-1: 20 Crab, some of the Galactic bulge sources, GX5-1, GX17+2 and some interesting black hole systems, such as the microquasar GRS1915+105 or GX339-4. Transient sources, with black hole candidates, can reach a few Crab (XTEJ1550-564: 4.8 Crab, GROJ1655-40: 4.4 Crab, X1543-475: 4 Crab, 4U1630-47: 0.8 Crab). Transient systems with neutron star primaries also reach one Crab or so (Aql X-1, 4U1608-52). There are about 20 persistent and transient sources with a maximum intensity exceeding 1 Crab and all of them, with the exception of Sco X-1, never go above 5 Crab (on daily timescales).

Similarly type I X-ray bursters in neutron star binaries can exceed many times their persistent flux level for a short duration (tens of seconds). Analysis of the RXTE catalogue of type I X-ray bursts (Galloway et al. 2008) indicates that 99% of the bursts have a peak flux below 5 Crab (and ~50% below 1 Crab), over a sample of more than 1170 type I X-ray bursts.

For observations of bright X-ray sources or bright bursts, there is an increase of both the fraction of pile-up events and of the charge patterns in the detector not identifiable as X-ray events. Acceptable pile-up levels can nonetheless be achieved through selection of only good events, at the expense of reducing the throughput (defined as the fraction of source photons detected in valid charge patterns). It is important to limit the losses due to these effects and to provide a sufficient amount of analysable events even at extreme count rates. To yield sufficient signal to noise ratio, *Athena*-WFI is required to provide a throughput of at least 90% for a source brightness of 1 Crab, and a throughput of at least 40% at 5 Crab (or about 150,000 events per second).

**4) Time resolution:** The highest frequency QPO that has been observed is at 1240 Hz. Adequate sampling of the fundamental frequencies therefore requires time resolution less than 500 μs. Our goal is to detect the higher-frequency harmonics, required to constrain the formation and emission scenarios for kHz QPOs, and for modelling of burst oscillation waveforms. Substantial oversampling is required to avoid aliasing, leading to a requirement of 32 μs. This time resolution matches the pile-up and throughput requirements defined above.

**5) Absolute timing accuracy:** The absolute timing accuracy better than 100 μs is required to study the emission processes of rapidly rotating neutron stars (e.g. the ms pulsar PSRJ1937+21, whose X-ray peak has a width of about 30 μs) through comparison of phase folded X-ray and radio light curves.

### 3.2.7 Charged Particle Background

Faint target science drives the requirement on the charged particle background. With an orbit located at L2, the limiting flux of cosmic rays is relatively well known, and physical (GEANT) modelling of typical detector and spacecraft structures is able to predict the quiescent level of background components that can be expected. While this is irreducible, the eventual detector system design must take care to ensure no excess generation of secondaries occurs, as well as implementing rejection techniques to discriminate between the X-ray signal and prompt effects of charged particle background. The simulations performed in compiling the science case have all assumed such performance, where for the wide field imaging, a flat spectrum of unrejected background < $5 \times 10^{-5}$ cts/s/keV/mm$^2$ is required. The equivalent requirement for the narrow field spectrometer has been scaled from low earth orbit measurements of *Suzaku*, and is required to be < $2.4 \times 10^{-4}$ cts/s/keV/mm$^2$.

### 3.2.8 Straylight requirement

Off-axis sources can be falsely imaged by single bounce reflection off the mirrors. In order for the in-field background contribution from this effect to be <10 % of the unresolved AGN background, then the contribution of all sky background in the annulus 0.5-1.5 degrees radius must be rejected by a factor ~$10^{-3}$.

## 3.3 Summary

The most critical requirements for telescope effective area and angular resolution define to a large extent the duration of the core science programme. Achieving the goal angular resolution of 5 arcsec would significantly and qualitatively improve some of the science cases, especially the deep survey redshift limits. All other science requirements have been verified to be achievable with current state of the art telescope and focal plane instrument design and within the available development time.

A draft Observing Plan has been constructed assuming the required instrument performance. Members of the Science Definition Team have constructed a detailed target list and observing strategy that satisfies the scientific investigations described in Section 2, the science case. The required observing durations have been verified by simulations using the predicted instrument performance. Finally an initial attempt was made to account for duplicate target or measurement types. The result of this exercise is presented in Table 3.2 which summarises, for the different science topics described in Section 2, the number of targets or observations per topic, and the total observing time. No further attempt has been made to balance the distribution of observing time between cases, nor to round the target and exposure duration data. For a single year of operations the target distribution around the sky was used to define slew durations and instrument overheads. Estimates were also made of the impact of safe mode contingencies, solar flare losses, etc. The resulting observational efficiency was predicted to be >85 %. For a 5 year science mission carried out with this observational efficiency, the core science cases that drive the mission performance requirements amounted to ~91 Ms. This leaves time for a large programme of general observatory science of ~43 Ms covering other astrophysical topics or wider coverage of the core science. Such a programme would include provision for ~2 Targets of Opportunity per month in addition to identified ToOs of the core science case (which includes some of the WHIM and transient black hole binaries).

| TOPIC | NUMBER OF TARGETS OR OBSERVATIONS | EXPOSURE TIME (Ms) |
|---|---|---|
| Persistent & transient black hole binaries | 340 | 6.8 |
| QPO & burst spectroscopy | 150 | 4.5 |
| NS Atmosphere & waveform modelling | 69 | 3.0 |
| AGN hotspots, reverberation & spin | 71 | 10.5 |
| Quasars & feedback | 57 | 1.9 |
| Compton-thick Fe line | 30 | 1.7 |
| Deep fields | 54 | 12.0 |
| Starburst galaxies | 40 | 4.1 |
| Chemical evolution of clusters | 61 | 6.1 |
| Cluster cosmology | 107 | 12.5 |
| Warm-hot intergalactic medium | 58 | 12.7 |
| Supernovae & supernova remnants | 326 | 4.2 |
| Galactic Centre | 34 | 2.0 |
| Stellar spectroscopy | 230 | 6.4 |
| Stellar clusters & star forming regions | 31 | 2.3 |
| Planets | 34 | 0.7 |
| **TOTAL CORE SCIENCE** | | **91.4** |
| Available in 5 years | | 134.1 |

**Table 3.2.** *Draft of the Observing Plan for Athena.*

As can be seen in Section 2, the science cases presented need different performance features of the instrumentation. Examining the draft observing plan allows the relative usage of the two instruments to be estimated. About 40% of the driving science cases genuinely require the simultaneous use of both XMS and WFI. The XMS is the prime instrument delivering the necessary spectral resolution for a further 40% of observing time. The remaining 20% requires the imaging or high time resolution performance of WFI. This distribution validates the original design decision for implementing two fixed telescope/focal plane combinations. It must be highlighted that when the XMS is requested for its prime scientific investigations, the WFI will always be enabled to perform a serendipitous survey, making use of its excellent resolution over the wide field of view. Finally there will be the synergistic use of the instruments in parallel to improve the cross-calibration (a feature of *XMM-Newton* that has proved to be invaluable).

# 4 Payload

The science aims of *Athena* are highly demanding and require state of the art instrumentation. First of all a X-ray telescope of at least 1 $m^2$ effective area at 1 keV with a spatial resolution <10 arcsec, and a focal length of 12 m is required. The focal plane plate scale for such an optical system is 58 µm/arcsec. Secondly, the requirements on spatial resolution, field of view (FoV), energy resolution, energy range, quantum efficiency and count rate capability, cannot be met by a single focal plane instrument. Therefore a design has been conceived with two fixed coaligned telescopes each feeding a fixed focal plane. The current design of the two instruments comprises:

- **An X-ray imaging Microcalorimeter Spectrometer (XMS)** that covers the 0.3-10 keV energy range with unprecedent energy resolution, a 2.3 x 2.3 arcmin FoV, and relatively modest count rate capability.
- **A Wide Field Imager (WFI) covering the 0.1-12 keV** energy range with a large 24 x 24 arcmin FoV, excellent spatial resolution and efficiency, good energy resolution, and good count rate capability.

In this section the key elements of the X-ray telescope and the focal plane instruments are described. The mirror assembly is addressed in more detail in Section 5.4. Here the features and performance of the mirror technologies are noted. In the following sections the essential features for each science instrument, their driving requirements, operating principles and outline of implementation and resource requirements are described.

## 4.1 Athena X-ray optics

The *Athena* mission requires a large effective area (>1 $m^2$ at 1.25 keV) and high angular resolution (<10 arcsec HEW, at <7 keV). To allow for such a large telescope area within the mission mass constraints an innovative X-ray optics technology is under advanced development by ESA. This is the silicon pore optics, SPO, which as described below, utlises highly polished silicon wafer reflecting plates in a novel self-supporting structure.

SPO is a highly modular concept; it has demonstrated an excellent effective area-to-mass ratio and is based on a set of compact individual mirror modules (MMs). About 500 MMs are required to populate the *Athena* mirrors, and the total mass is less than 350 kg, including the isostatic interfaces and the sub-system maturity margins. The full production process for SPO has been demonstrated employing a consortium of industrial partners.

The SPO solves the *Athena* mass-area-resolution challenge by introducing a new approach to mounting the X-ray mirrors into a matrix-like structure. The resulting MMs are intrinsically very stiff and robust, keeping the figure of the mounted X-ray mirrors stable to the arc second level. Unlike the conventional and established X-ray optics technologies, which use a limited number of interface points to attach the mirror optics elements to the support structure, the SPO technology relies on a much stronger inter-linkage of the X-ray mirror elements via integrated ribs.

The individual mirror plates are assembled into a matrix-like structure, whereby the ribs of a mirror plate bond to the surface of the preceding mirror plate (see Figure 4.1). The individual mirror plates form a monolithic structure, of mono-crystalline silicon. These mirror plate stacks are therefore very rigid. Each mirror plate in the stack is slightly inclined to the previous one, as required for the Wolter geometry of the telescope design.

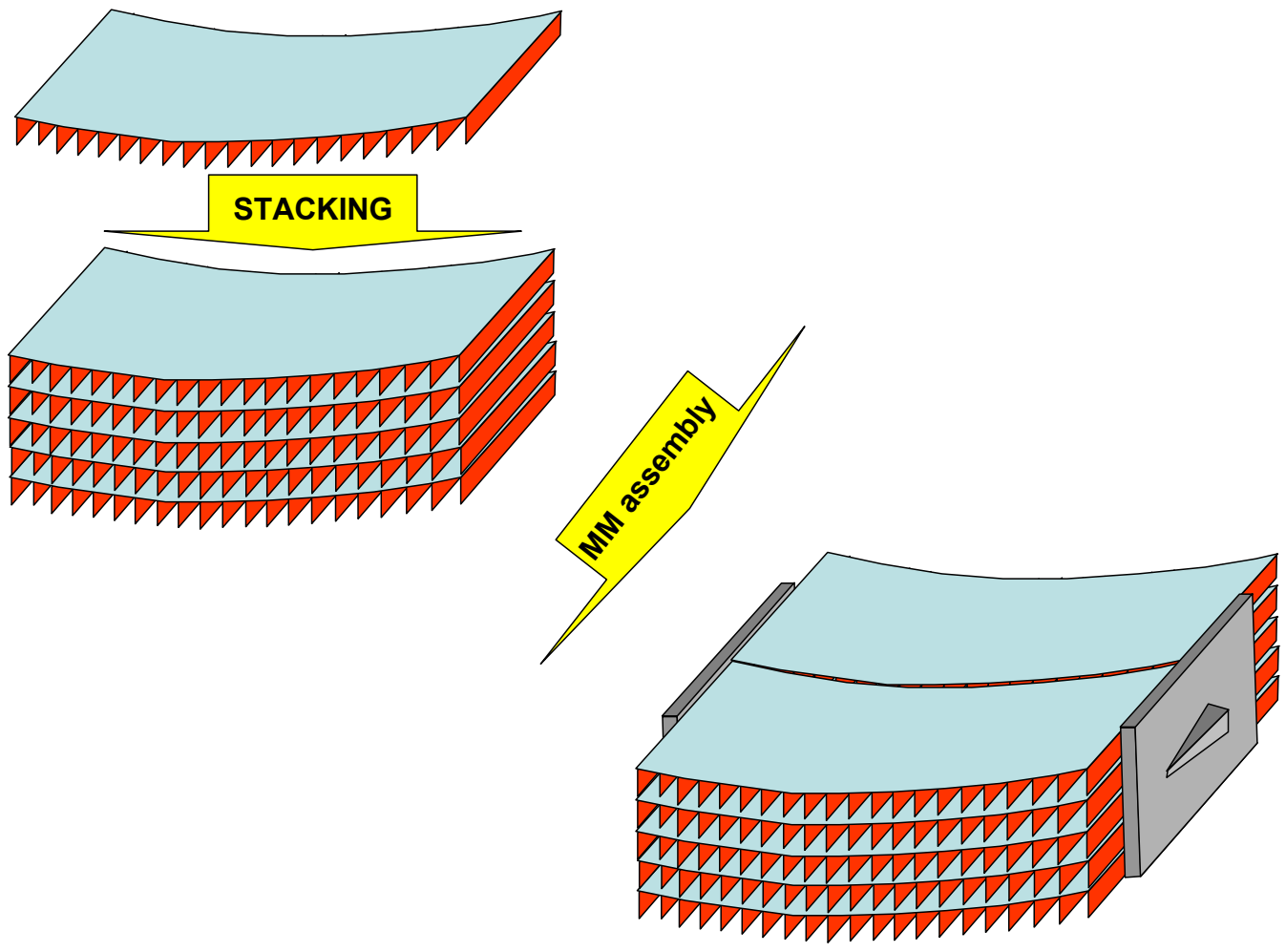

**Figure 4.1.** *Mirror plates are stacked and bonded together to form monolithic blocks of mounted mirror plates. Two such stacks are then assembled into the mirror module, employing two brackets. The mirror modules provide attachment points for the isostatic mounting to the telescope structure.*

Two such mirror plate stacks are then assembled to form a MM, using two brackets. Very lightweight brackets, with matched coefficient of thermal expansion (CTE), are used to provide a stiff and permanent alignment of the mirror plate stacks. These brackets also include the interface elements for the isostatic mounts used to decouple the loads between the MM and the optical bench. Once assembled, a MM effectively forms the equivalent of a "lenslet" and the tolerance requirement for its alignment within the optical bench is much lower and achievable via conventional engineering techniques.

Considering the large number of required optics modules for *Athena*, it was very important to take into account mass production aspects right from the beginning of the technology development. Due to the small size of the SPO modules the production equipment can be kept compact, ensuring the cost effective implementation of a production line, including the associated cleanroom infrastructure.

The mirror plate production begins with 300 mm silicon wafers that are superpolished on both sides. Such wafers are readily available and mass-produced for the semiconductor industry. Furthermore the subsequent processing steps are also tailored from existing industrial processes, including dicing the pores, angular wedging, masking and coating, cleaning and packaging. The silicon plate surfaces are bonded, normally using hydrophilic bonding that occurs between two oxide layers, providing a high bonding strength.

ATHENA

The baseline silicon pore optics concept relies on standard semiconductor industry techniques and equipment. The super-polished silicon plates with X-ray quality reflecting surfaces are commercially available off-the-shelf.

The fully automated assembly robot (Figure 4.2) is specifically developed to stack SPO and is a combination of standard semiconductor systems and newly developed tools. The complete system has a footprint of a few $m^2$ only and is installed in a class 100 clean area. The robot selects a plate for stacking and inspects it for particles. The plate is then handed over to the actual stacking tool, which will elastically bend it into a cylindrical or conical shape, then align and bond it to the underlying mirror plate. The stacking is done from outer radii inwards, thus always exposing the last integrated mirror surface to the metrology tools.

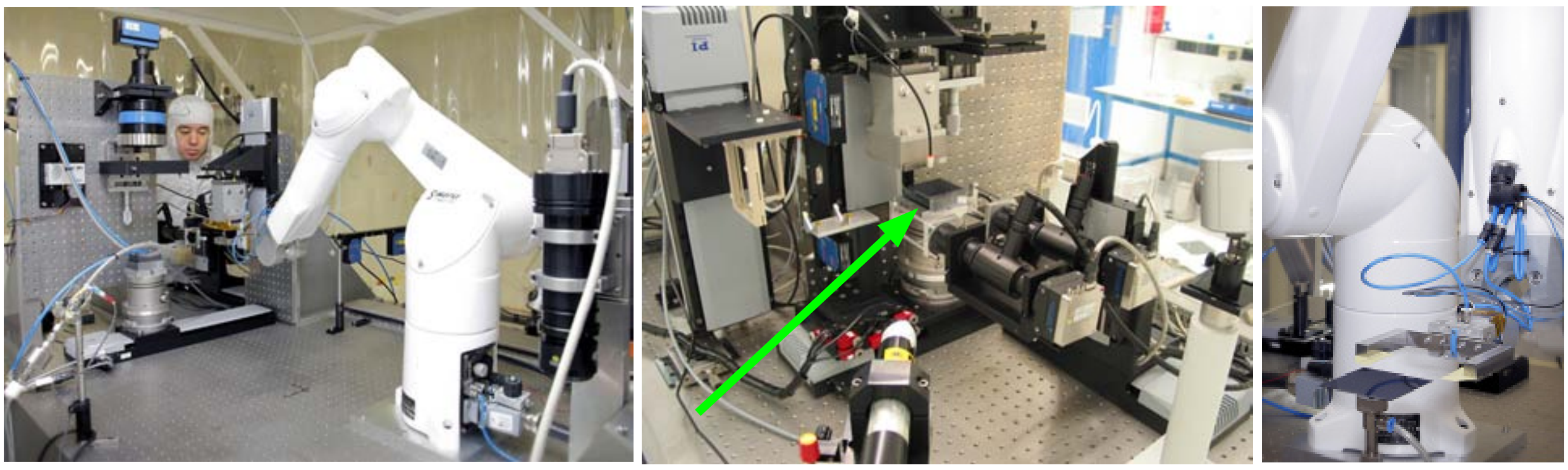

**Figure 4.2.** *Stacking robot inside the class 100 clean area at cosine Research BV. The system is installed on a vibration isolated table, consists of more than 16 axes, is fully automated and is designed to build stacks up to 100 plates high. The plates can be positioned with µm accuracy and automatically be bent into the required shape.*

The integration of two mirror stacks into a MM in flight configuration requires the alignment of the two stacks with an angular accuracy below one arc second. This is achieved using a dedicated integration setup and X-ray beam metrology. Figure 4.3 shows a set of mirror plate stacks and an assembled SPO mirror module. The integration of MMs into a full size petal was already demonstrated within the former XEUS and IXO activities.

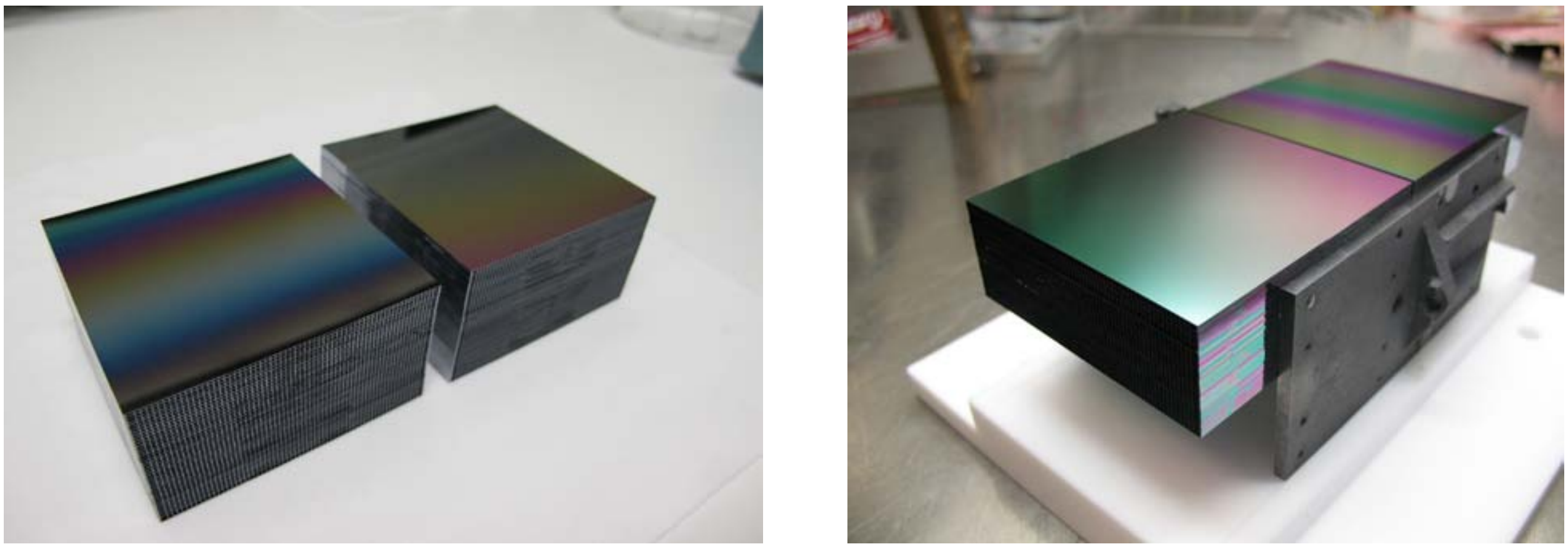

**Figure 4.3.** ***Left****: a pair of SPO stacks, consisting of 45 bonded mirror plates each, exceeding the 35 plate stack required for the Athena telescope baseline design.* ***Right****: A complete SPO mirror module, consisting of a pair of mirror plate stacks permanently joined by two CTE matched silicon carbide brackets (cosine Research BV).*

Since the development of SPO started in 2002, it followed a holistic approach, tackling the important aspects of the entire production chain, including consideration of the eventual mass production in a flight programme. Consequently the processes and equipment relating to production and characterisation of the SPO from plate up to petal level were iterated, with the resulting MMs constantly improving. In parallel, elements like the integrated baffles, electrical grounding issues and the application and performance of reflective coatings (e.g. $B_4C$ on Pt, Ir), have been tackled. In particular the process has been verified with patterned reflecting surfaces (Pt and Ir) between the ribs. Furthermore, coating of multilayers on Si-substrates has been demonstrated and measured for the improved hard X-ray response.

X-ray testing of the MMs, mounted in a flight representative configuration as they will be mounted into the petal, was performed at the X-ray Pencil Beam Facility, a dedicated beamline in the Physikalisch-Technische Bundesanstalt (PTB) laboratory at the synchrotron radiation facility BESSY II and at PANTER. The impressive track record is summarised in Figure 4.4, showing the improvement in measured Half En-ergy Width (in 3 keV X-rays) for 4 representatively mounted mirror plate pairs, as a function of time. The latest measured MM shows a HEW of 7.5 arcseconds for the first 4 mirror pairs, and 9 arcseconds for the first 10 mirror pairs. Note that in Figure 4.4 we arbitrarily display the data for 4 plate pairs (4pp) in each case to track the corresponding improvement with time.

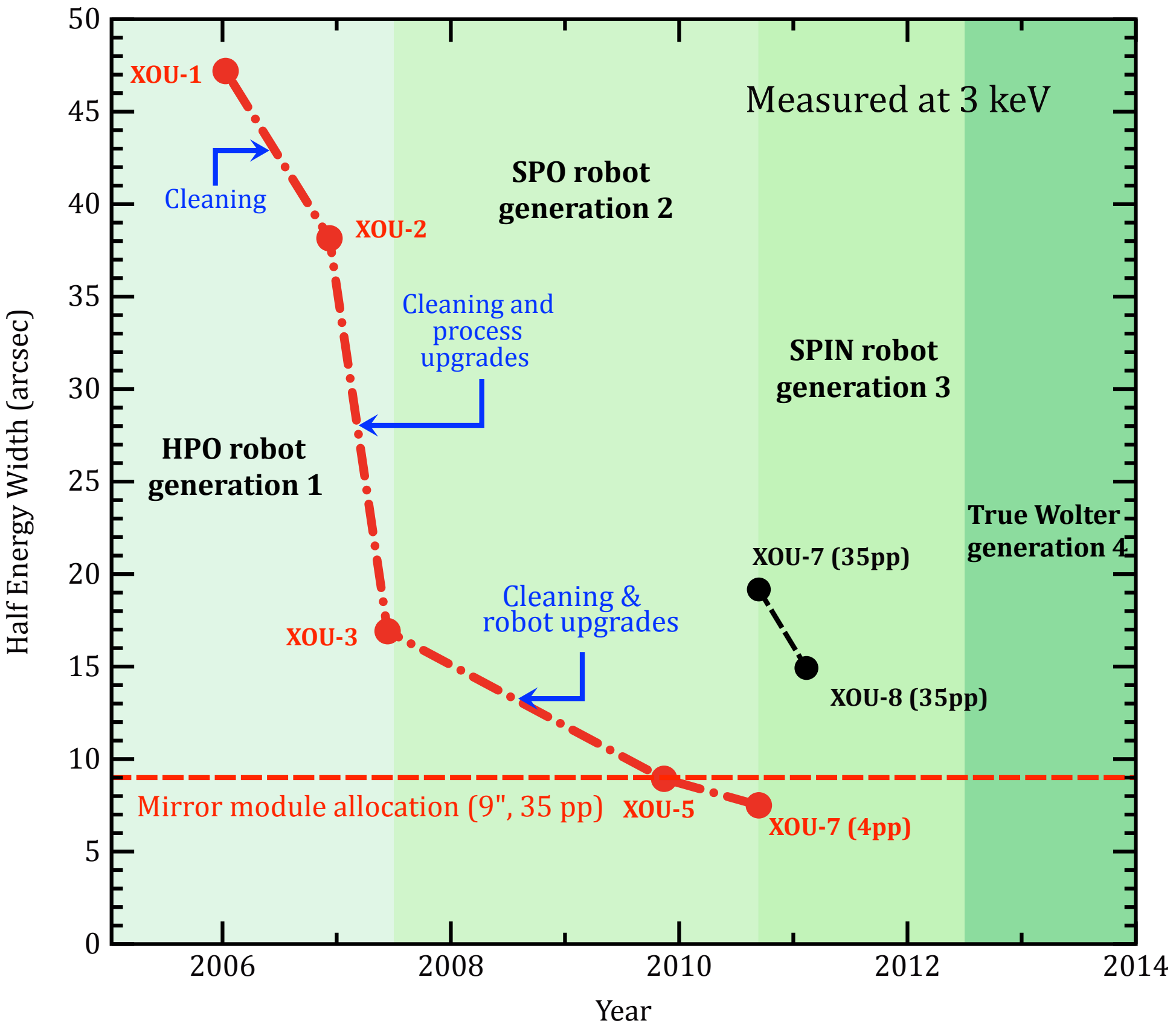

**Figure 4.4.** *The measured performance of the SPO mirror modules as a function of time. The HEW in arcseconds, measured with 3 keV X-rays in double reflection, for the full area of 4 representatively mounted mirror plate pairs (4pp) is shown. With the first generation equipment significant progress was made on the mirror plate production (High Performance Optics contract - HPO), leading to a steep improvement between mirror modules XOU-1/XOU-2 and XOU-3. After that the 2$^{nd}$ generation stacking robot was developed and brought on-line (Silicon Pore Optics contract - SPO), which further improved the performance of the mirror modules produced. For the last module produced, XOU-7, a HEW of 7.5 arcseconds was measured for the first 4 plate pairs.*

The effective area of the optics, measured on uncoated optics from 0.3 to 3 keV, matched the theoretical expectation within 5%, confirming the good surface roughness and alignment achieved in the constructed MMs, satisfying the *Athena* design requirements.

Dedicated optical metrology equipment is employed throughout the plate stacking procedure, and this is verified and complemented with X-ray measurements. The combination of these metrologies identifies the particular parts of the stack process that might fail to meet the 5 arcsecond goal performance. Having a proper understanding of the current limitations, good progress continues to be made towards the *Athena* requirement of 10 arcseconds. A detailed technology development plan was established for this purpose.

As for the previous development contracts, the next phase of SPO performance improvements will simultaneously address a number of modifications, in parallel, to the production process. A non-exhaustive list that has been defined in the Technology Development Plan recently submitted, or already implemented but not yet completed in final MM stacks, is detailed in Table 4.1.

| METHOD | EFFECT |
|---|---|
| Marangoni cleaning station | Remove the effect of incomplete bonding areas at the plate edges. This is now installed and improvements are being quantified. |
| New mandrel coating | Remove effects of Si particles sticking to the mandrel plates. As with cleaning station this is being commissioned. |
| Introduce Wolter I shapes | By introducing a minor secondary curvature on parabola/hyperbola mandrels the conical approximation effect can be removed. |
| Implement reduced inner radii | To partially compensate for the shorter focal length, high energy effective area can be recovered by using the inner radii. Trial stacks have been fabricated to demonstrate that bending does not result in excessive strain. New mandrels and demonstrations of optimised membrane thickness are needed to demonstrate innermost modules' performance. |
| Implement increased outer radii | Outer MM have shorter plate lengths that will require modified handling jigs. The required performance must also be demonstrated on these modules. |
| Modify stack plate numbers | Optimise the number of plates to ensure accumulated stacking errors do not compromise the resolution. Maximise packing efficiency by including more than one stack per bracket. Check for yield performance. |
| Modify reference surfaces | Already started to implement improved QA aspects for traceability. Also improve stacking metrology. |
| Improved wafers | Investigating procurement of lower total-thickness-variation that can result in improved wedge angles and reduce stacking errors. |

**Table 4.1.** *The future activities that have been included in the Technology Development Plan, in order to ensure the SPO reaches the required TRL by 2012, and to ensure sustained improvement in angular resolution performance.*

The future plans for upgrading the production process have already been identified and are in the process of being implemented. Three major technical improvements foreseen for the next two years are: a new cleaning system based on semiconductor industry standard, better wedge profile production and a true Wolter profile that will significantly improve the HEW (Silicon Pore Industrialisation contract - SPIN).

Dedicated activities concentrate on improving the ruggedised mounting system that fullfils the mechanical and thermal requirements for the *Athena* spacecraft. A first design was developed and verified by model calculations. First environmental tests to verify the models were successfully completed in 2010, in a dedicated technology development activity, and it is planned to achieve TRL>5 by the end of the definition phase activities (A/B1). In the 3rd quarter of 2011 a representative MM has successfully passed vibration tests at levels exceeding the qualification requirements (see Figure 4.5). This result allowed the reduction of the optics development risk and gives confidence that the SPO technology will be qualified for flight by the end of the definition phase.

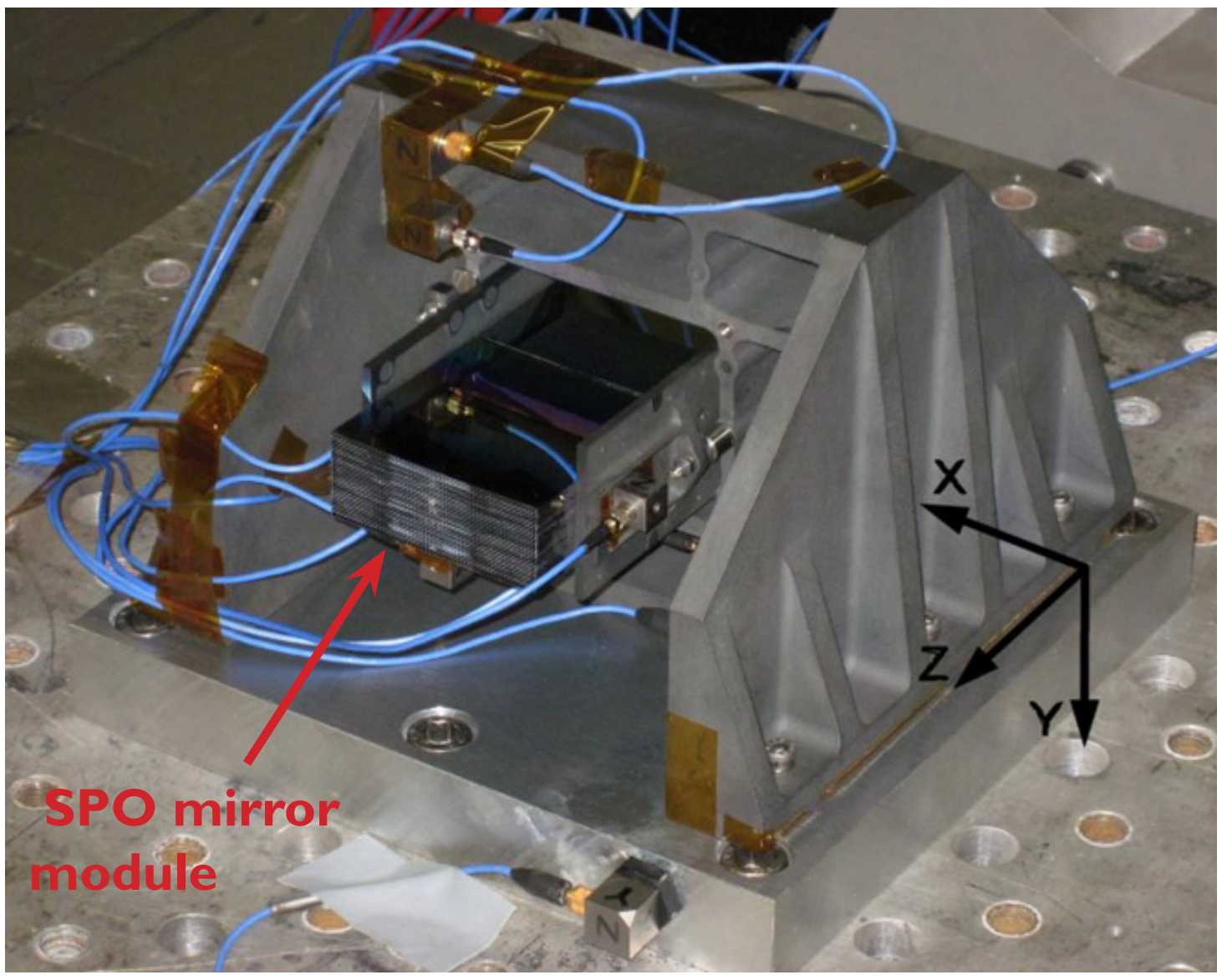

**Figure 4.5.** *A representative silicon pore optics module mounted on a shaker table for the vibration tests. The mirror module passed high level sine sweep tests, at acceleration levels and with durations exceeding the qualification demands. No change in optical performance was subsequently measured.*

The silicon pore optics have always been measured in X-rays with a flight-like configuration. Sustained improvement has been achieved with step-wise developments of cleaning and stacking tools. Further identified development activities will allow the goal 5 arcsecond resolution to be met with the new generation assembly robots under construction.

## 4.2 The X-ray Microcalorimeter Spectrometer (XMS)

The *Athena* X-ray Microcalorimeter Spectrometer (XMS) is being developed by a consortium led by SRON in the Netherlands. The consortium in Europe includes CSL (Belgium), IRAP and CEA/Saclay (France), INAF (Italy), MSSL and Leicester University (UK), ISDC (Switzerland), IFCA (Spain) and Erlangen University (Germany). The work breakdown structure makes full use of existing expertise at these institutes and is based on similar contributions to many earlier instruments. Contributions from NASA (sensor and last stage cooling) and JAXA (cooling chain) are included, based on a continuation of the collaboration between SRON, ISAS and GSFC for IXO (see Section 5.7).

A European-only solution is also feasible, however, and is discussed in Chapter 6.

### 4.2.1 Instrument requirements and specifications

The XMS is a 2-D imaging camera, which allows the identification and characterisation of the different ionisation stages in hot plasmas thanks to its excellent spectral resolution (few eV at 6 keV). This requires that the camera operates at cryogenic temperatures. Complexity and heat load limit the size of the camera to about 1000 pixels. Compared to the planned cryogenic spectrometer on *ASTRO-H*, the first mission which will include a calorimeter, the number of pixels will be increased by a factor 30, the spectral resolution by a factor 2, effective area by a factor 10 and the angular resolution by a factor 8. The instrument requirements are listed in Table 4.2.

| PARAMETER | REQUIREMENT | GOAL |
|---|---|---|
| Energy range | 0.3 - 12 keV | 0.2 - 12 keV |
| Energy resolution E < 7 keV<br>E > 7 keV | ΔE = 3 eV<br>E/ΔE = 2300 | ΔE = 2.5 eV<br>E/ΔE = 2800 |
| Field of View | 2 x 2 arcmin | 3 x 3 arcmin |
| Plate scale (32 x 32 array) | 4.3 arcsec/pixel | 4.3 arcsec/pixel |
| Quantum efficiency @ 1 keV<br>@ 7 keV | > 60%<br>> 80% | > 65 %<br>> 85 % |
| Energy scale stability | 1 eV/h (peak - peak) | 0.5 eV/h (peak-peak) |
| Non X-ray background | 2 x $10^{-2}$ counts/$cm^2$/keV/s | 1.5 x $10^{-2}$ counts/$cm^2$/keV/s |
| Continuous observing time | > 20 hr, regeneration time < 10 % | > 20 hr, regeneration time < 5 % |

**Table 4.2.** *XMS Instrument Requirements.*

ATHENA

The X-ray Microcalorimeter Spectrometer XMS is an integral-field spectrometer operating in the soft X-ray band.

### 4.2.2 Sensor

The operating principle of a calorimeter is shown in Figure 4.6. The detector works by sensing the heat pulses generated by X-ray photons when they are absorbed and thermalised. The temperature increase is directly proportional to the photon energy and is measured by the change in electrical resistance of the sensor. When the sensor is cooled to <100 mK and is biased in its transition between super-conducting and normal resistance resolutions better than 2 eV have been demonstrated.

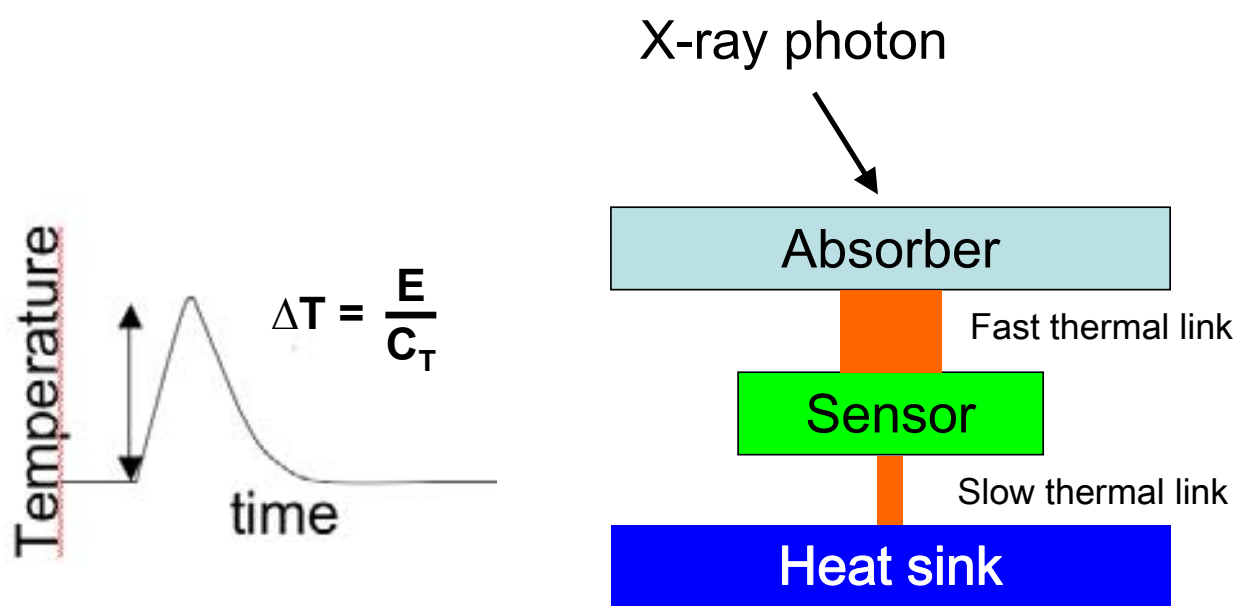

**Figure 4.6.** *Principle of a microcalorimeter. The absorption of an X-ray photon heats both the absorber and the sensor. The resulting signal represents the total energy deposited. The system goes back slowly to its original state through a weak thermal link with a heat sink.*

The detector consists of an array of 32 x 32 absorbers which are 250 µm square composed of 1 µm Au and 4 µm Bi to achieve the correct stopping power at 6 keV and low heat capacitance required for high energy resolution. We use a Mo/Au bilayer Transition Edge Sensor (TES) coupled to a super-conducting quantum interference device (SQUID) for the read-out of the signals. For this configuration an energy resolution of <3 eV has been demonstrated. This sensor is implemented with a 2.3 arcmin FoV with a pixel size of 4.3 arcsecs. For an array with 1024 TESs a multiplexed read-out of the pixels is used, so that the number of connections to the warmer stage through the harness is minimised, and the thermal load reduced. The time domain multiplexing (TDM) scheme, where 16 pixels in the same read-out channel are sequentially sampled, is the baseline. Underneath the sensor an anti-coincidence detector is positioned to identify signals from charged particles. We have selected a segmented Si absorber (300-500 µm thick), each read-out by its own TESs. This maximises the commonality between the electronics for the sensor and the anti-coincidence detector. In the digital electronics the signals of the sensor and anti-coincidence detector are combined and only good events will be transmitted to the ground.

The focal plane assembly (FPA) provides the thermal and mechanical support to the sensor and the anti-coincidence detector. In addition it accommodates the cold electronics and provides the appropriate magnetic shielding. A magnetic field attenuation of 1.6 x $10^5$ has been achieved by two shields: a super-conducting Nb shield and a cryo-perm shield at 4 K. The design, shown in Figure 4.7, is based on flight-qualified components (harness and interconnections and connectors) and technology. In Figure 4.8 the measured spectra of a $Fe^{55}$ source are shown for an 8 x 8 pixel array of which 2 x 8 pixels are read-out by 2 independent channels. The average resolution is <3 eV with a very small spread between the pixels. The time constants of the detectors (300 μs) together with the noise levels of the TDM allow for the observation of a 10 mCrab source with more than 80% of the events detected with the nominal resolution of <3 eV. The instrument also includes a filter wheel, allowing for additional suppression of the optical load on the detectors for optically bright sources. By inserting a diffuser for defocussing of the beam, point sources up to ~few hundred mCrab can be observed with 80% of the events with high spectral resolution but at the expense of a reduced throughput (~30%). Stability of the gain will be monitored by an intense, time-modulated, X-ray source (duty factor 1%).

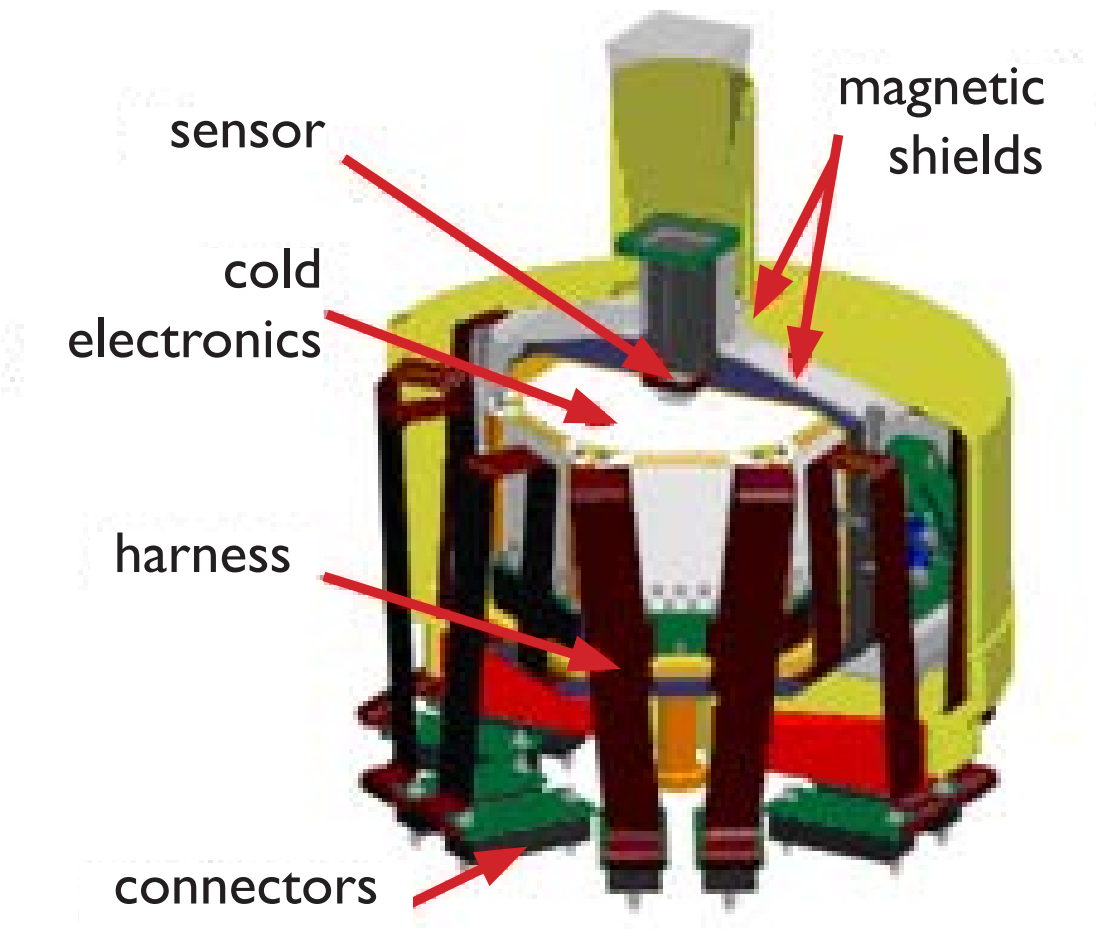

**Figure 4.7.** *Design of the focal plane assembly. The FPA has a hexagonal shape. The TES-sensor and its anti-coincidence detectors are mounted on the upper horizontal plane. Along the sides of the assembly the cold electronics are mounted (white) and connected to the electrical harness. The central unit is surrounded by thermal and magnetic shields at the various temperature levels. Kevlar suspension is used for thermal isolation of these temperature stages.*

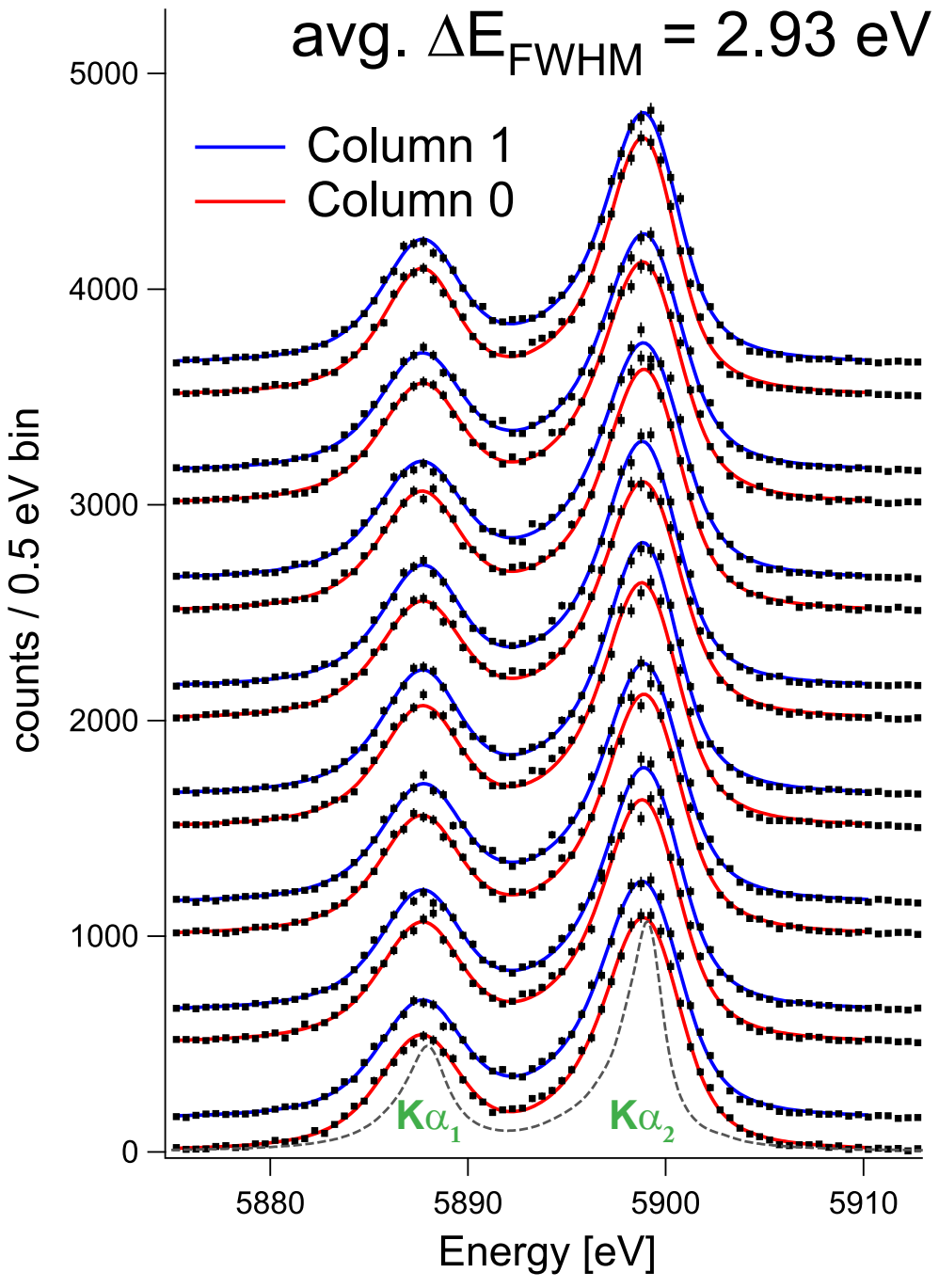

**Figure 4.8.** *Results of a 8 x 8 pixel array using time domain multiplexing to read out 2 x 8 pixels. An average energy resolution of 2.93 eV for these 16 pixels has been obtained. The uniformity of the array is very good with a maximum resolution of 3.1 eV.*

### 4.2.3 Coolers

To achieve the required spectral resolution it is necessary to couple the sensor to a 50 mK thermal bath. With a combination of different cooling techniques it is possible to achieve this temperature with a cryogen-free system, thus not limiting the instrument lifetime by cryogenic consumables. The cooling chain is split into two subsystems: the prime cooling chain for cooling part of the instrument from room temperature to 4 K and the last stage cooler for 4 K to 50 mK. The baseline cooler is based on the *ASTRO-H* design as this is currently the most advanced integrated system. The cooling requirements include a safety factor of about 2 for each T-stage, typically required at the current stage of the programme. We have also chosen a design with full redundancy in any of the mechanical coolers.

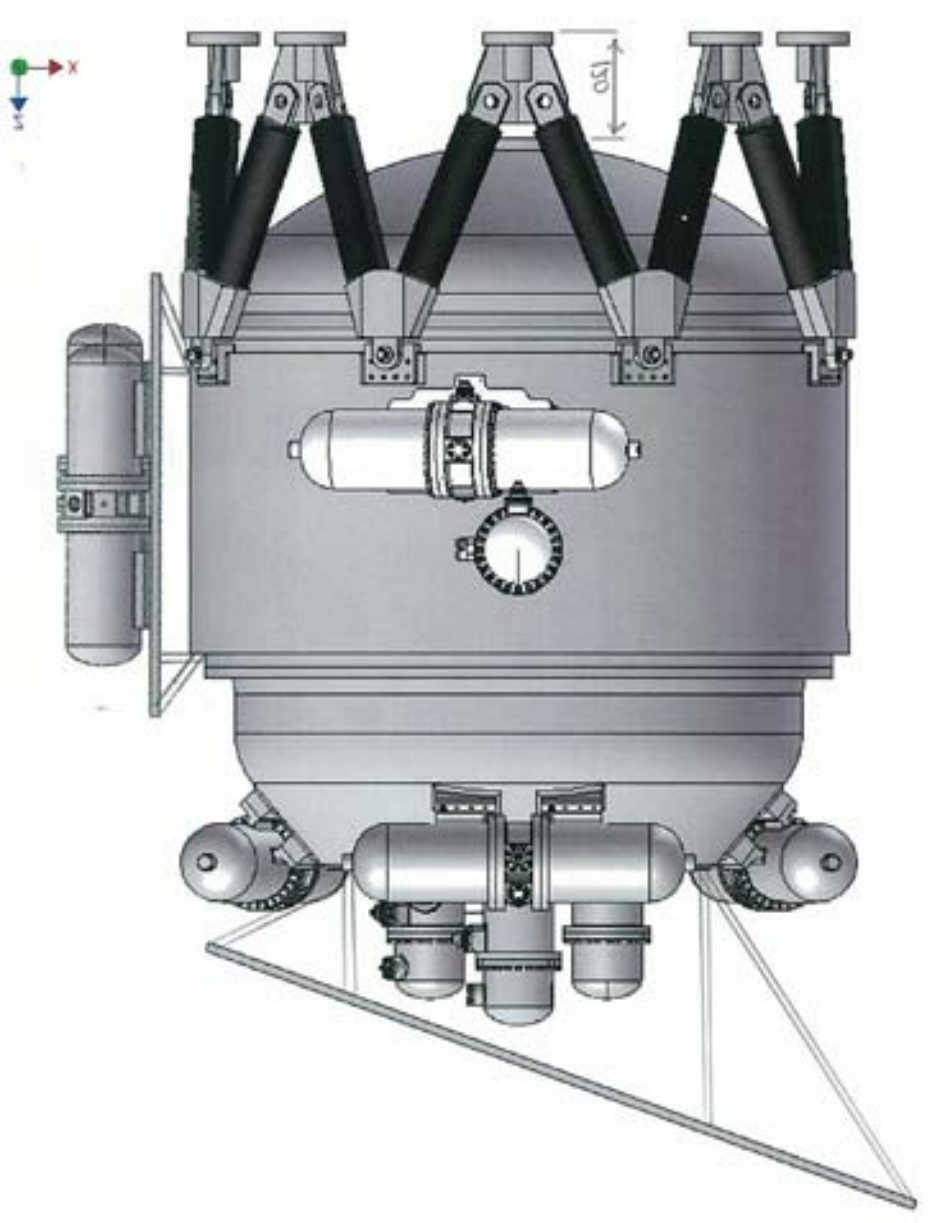

**Figure 4.9.** *Engineering drawing of the Dewar.*

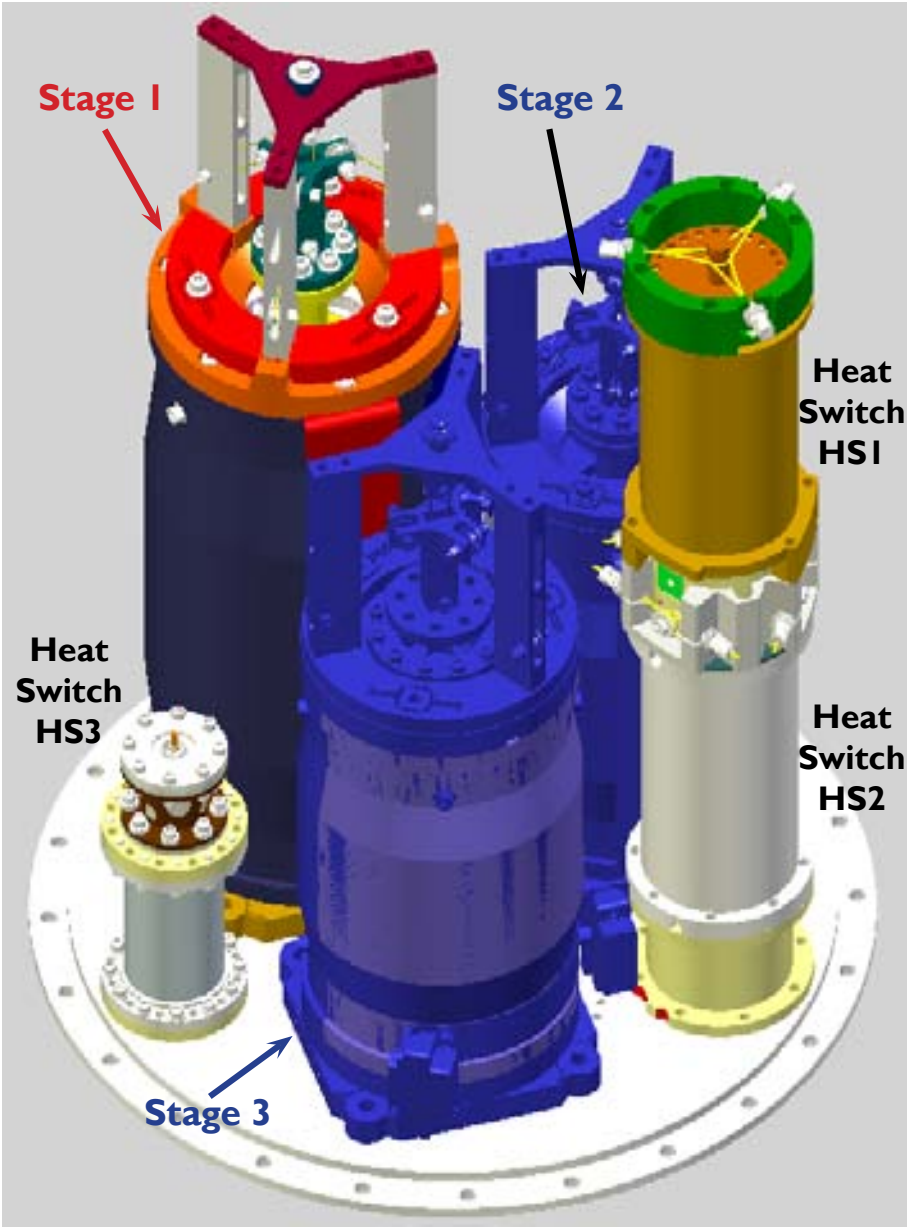

**Figure 4.10.** *Last stage cooler, based on a 3-stage ADR.*

The prime cooling chain provides a continuous cooling power of 20 mW at 4 K (peak). It is a cryogen-free system and uses 2-stage Stirling coolers and $^4$He J-T coolers which all have flight heritage. In addition to full redundancy in the drive electronics each of the mechanical coolers (compressors) are also fully redundant. Initial estimates of the reliability are above 98%.

The last stage cooler has a cooling power of 2 μW at the 50 mK level using a 3-stage adiabatic demagnetization refrigerator (ADR), see Figure 4.10. Following the magnetization of the salt pills, cooling power is provided by a slow relaxation of the spins in the magnetized material. The system has a regeneration time of <2 hours and a hold time of 24 hours. Detailed modelling shows that the FPA (Figure 4.7) and a 3-stage ADR fit within the allocated envelope (200 mm diameter with a height of maximum 37 cm) including the 'snout' of the focal plane assembly which is required to suppress the magnetic fields.

### 4.2.4 Technology development status

The current design of the sensor is based on demonstrated technologies and performance. The present generation employs multiplexing of 8 pixels in a single read-out channel. The increase to 16 pixels is very modest considering that our models indicate that 32 pixels per channel are also feasible and still provide the required performance. It is also expected that the developments for the flexible leads and interconnections will provide additional margin on the FPA design (although not needed for the proposed baseline design). The cooling chain is based on the *ASTRO-H* system, a system which is currently in its critical design review stage. For the definition phase we will demonstrate that the technologies for the FPA (sensor, read-out, anti-coincidence detector, thermal isolation and magnetic shielding) can be tested in a representative environment achieving TRL 5 for this subsystem. In addition a significant level of engineering work is planned to reduce schedule risks at a later stage (e.g. detailed designs of the application-specific integrated circuits (ASICs) and field programmable gate arrays for the demultiplexing, detailed design of the aperture cylinder) is foreseen. Radiation damage is not expected to be an issue but during the definition phase this will be demonstrated.

## 4.3 The Wide Field Imager (WFI)

The *Athena* Wide Field Imager (WFI) is being developed by a collaboration consisting of the MPE (Germany), a consortium of German universities (ECAP, IAAT, TUD), the company PNSensor, partner universities in the United Kingdom (University of Leicester) and IRAP in Toulouse.

### 4.3.1 Instrument requirements and specifications

The *Athena*-WFI is an imaging X-ray spectrometer with a large field of view with solid heritage from IXO-WFI and *BepiColombo*-MIXS. The purpose of the WFI is to provide X-ray images in the energy band of 0.1-15 keV, simultaneously with spectrally and time-resolved photon detection. The WFI is fixed in the focal plane of one of the two telescopes. The instrument requirements are listed in Table 4.3.

| PARAMETER | WFI REQUIREMENT | GOAL |
|---|---|---|
| Quantum efficiency (including optical blocking filter) | 282 eV : 34 %<br>1 keV : 98 %<br>10 keV : 97 % | 282 eV : 40 %<br>(No improvements expected for high-energy quantum efficiency) |
| Pixel size, array format | 130 x 130 μm² (2.2 arcsec)<br>640 x 640 pixel | 120 x 120 μm² (2.1 arcsec)<br>768 x 768 pixel |
| Readout rate, readout multiplicity, frames per second | 4 μs / readout,<br>1 row per hemisphere simultaneous,<br>780 fps full frame | 4 μs / readout,<br>2 rows per hemisphere simultaneous,<br>1300 fps full frame |
| Energy range | 100 eV - 15 keV | |
| Spectral resolution | $\Delta E \leq 150$ eV (FWHM) @ 6 keV | $\Delta E \leq 125$ eV (FWHM) @ 6 keV |
| Read noise | 5 electrons (rms) | 3 electrons (rms) |
| Field of view | monolithic wafer scale detector<br>24 arcmin, 83.2 x 83.2 mm² | monolithic wafer scale detector<br>28 arcmin, 92.2 x 92.2 mm² |
| Angular resolution (HEW) | ≤ 10 arcsec (oversampling by 4.3, SPO resolution requirement) | 5 arcsec (oversampling by 2.2, SPO resolution goal) |
| Fast timing, count rate capability | 32 μs in window mode<br>1 Crab, 4 % pileup, > 90 % throughput<br>5 Crab, 4 % pileup, > 40 % throughput | 16 μs in window mode<br>1 Crab, 2 % pileup, > 90 % throughput<br>5 Crab, 2 % pileup, > 40 % throughput |
| Instrument background at L2 | $1 \times 10^{-3}$ counts per keV/sec/cm² | $5 \times 10^{-4}$ counts per keV/sec/cm² |

**Table 4.3.** *WFI Instrument Requirements.*

*Athena*-WFI is a solid-state imaging X-ray spectrometer with high time-resolution capability, providing essential wide-field survey capabilities for observatory workhorse science. It offers the full imaging resolution allowed by the *Athena* flight mirror assembly, energy resolution at the 100 eV level and time resolution of few tens of microseconds.

### 4.3.2 Instrument description and configuration

The effective area of *Athena* in the WFI energy range is more than a factor of 5 larger than previous X-ray missions, leading to high event rates. This necessitates very high read-out rates and flexible read-out modes. Active pixel sensors (APS) based on depleted P-channel field-effect transistors (DePFETs) are highly suited for this task, as the signal charge does not need to be transferred over macroscopic distances, but is amplified directly in each pixel of the detector. To achieve a FoV of 24 × 24 arcmin, the WFI APS is integrated monolithically onto a single 6-inch wafer, covering almost the whole usable Si wafer surface (see Figure 4.12). The physical sensitive area is 8.32 × 8.32 $cm^2$ without any insensitive gaps. The detector can be operated in a very flexible cascaded-window mode with differently sized simultaneous windows. This mode can be used to further decrease the pile-up from 4% at 1 Crab at over 90% throughput to only 2% pile-up at 1 Crab in cases where the resulting limitations imposed by the Attitude and Orbit Control Subsystem are acceptable. By selecting the single pixel events only, the object brightness can be as high as 5 Crab with a throughput of 40 %. In a WFI option with two simultaneously read-out lines per hemisphere (dual-read-out option), the same performance can be achieved with a single window covering the full PSF, or the performance can be doubled. In this sense, the WFI fully includes the capabilities and thus the science of the High Time Resolution Spectrometer (HTRS) on IXO. Simultaneously to the readout of the window, the rest of the image can be read out at lower frame rates.

The DePFET device is fully depleted using the sideways depletion principle, allowing signal charges generated by radiation entering from the backside of the detector to be detected efficiently. Backside illumination allows the APS to have a geometrical filling factor of 100 %. Furthermore, the use of a very thin entrance window leads to very good quantum efficiency in the energy range below 1 keV. The large thickness of the fully depleted bulk provides for high quantum efficiency, even beyond 10 keV. A beneficial side-effect of backside-illumination is self-shielding. As the radiation absorbed in the bulk does not reach the amplifying structures on the front-side, radiation hardness is intrinsically improved.

As the large effective area X-ray mirror system does not only focus X-rays but also visible and UV light, care must be taken to avoid problematically high optical photon loads. Therefore, an optical blocking filter system consisting of a thin Al-layer on top of an UV blocking system consisting of a silicon oxide/nitride multi-layer coating is foreseen.

In order to achieve ×2 to ×4 oversampling of the expected PSF of *Athena*'s X-ray mirror system, the WFI baseline design foresees a pixel size of 130 × 130 $\mu m^2$ with 640 × 640 pixels covering most of a 6-inch wafer to achieve a 24 arcmin FoV. A FoV extension to 28 arcmin is possible with modest impact by changing the pixel size from 130 μm to 120 μm and increasing the number of pixels to 768 × 768. The resulting increase in mass and power is estimated to be at the 10% level and will be evaluated in detail during the next phase.

To achieve the high speed required, the detector is subdivided into two hemispheres, read out in parallel by 10 read-out ASICs per hemisphere. One option being evaluated halves the effective frame times by independently reading out two rows per hemisphere simultaneously, thereby doubling the number of read-out channels. At the baseline read-out speed of 4 μs per row, the raw datarate produced by WFI is more than 4 Gbit/s (baseline, >8 Gbit/s in the dual-read-out option), necessitating a very efficient on-board data-reduction and compression scheme. WFI incorporates a multi-staged data-reduction pipeline based on several field programmable gate arrays that reduces the raw data to a nominal data-rate of less than 450 kbit/s.

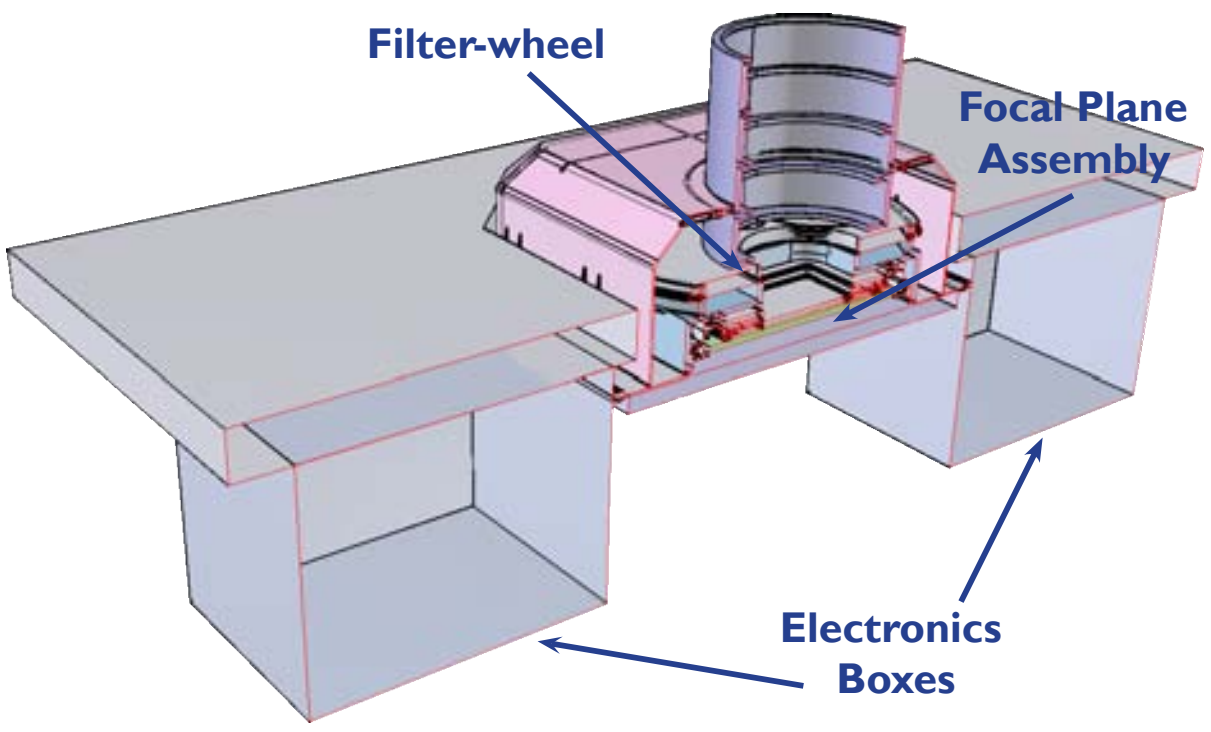

**Figure 4.11.** *Overview over the WFI mechanical layout. The X-ray radiation enters from the top. X-rays in the 0.1-15 keV band are detected by WFI.*

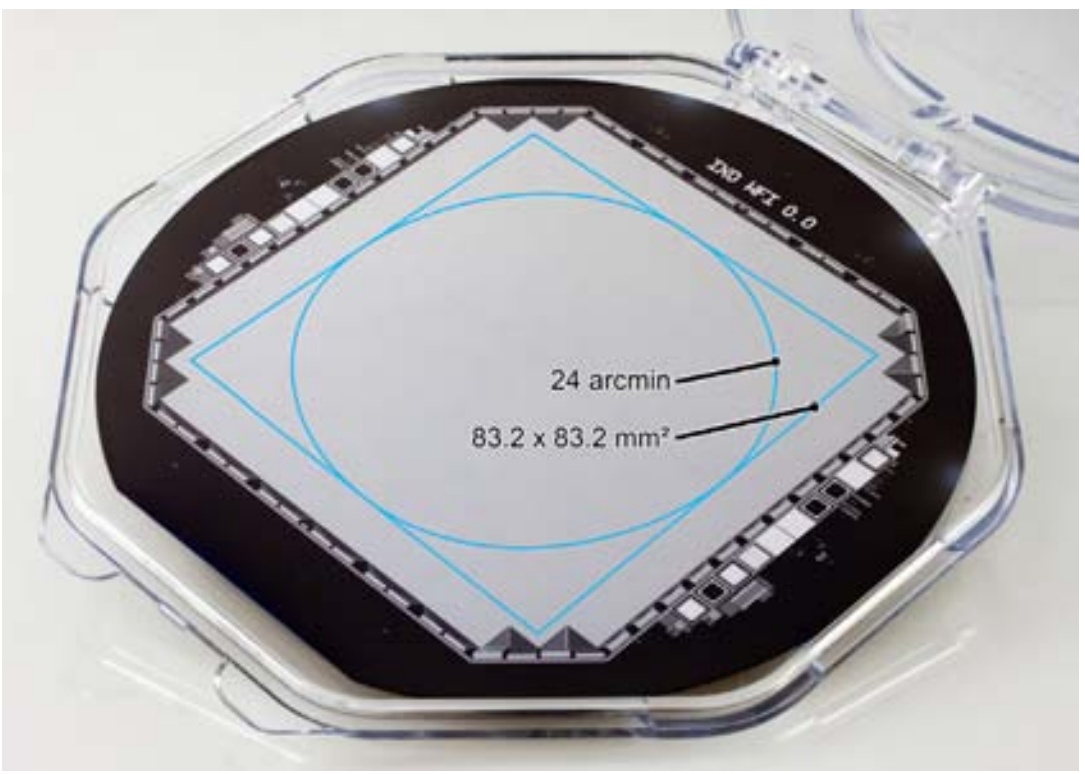

**Figure 4.12.** *Coverage of a 6-inch wafer by the 24 arcmin field of view of the WFI including the electronic periphery of the sensor.*

### 4.3.3 Technology Development Status

The current prototypes of DePFET matrices and the associated bread-boards demonstrate high performance, already meeting all WFI science requirements addressable with this detector size (see Figure 4.13 and Figure 4.14). A demonstrator of representative size is currently under test and will show technological readiness in all aspects before the end of the definition phase.

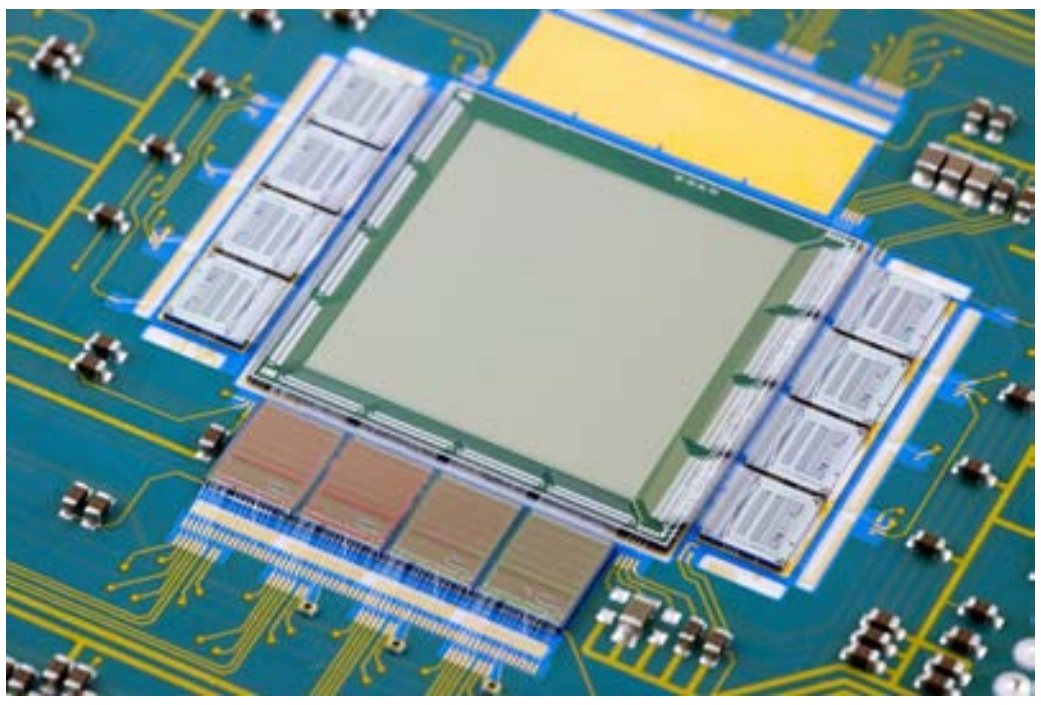

**Figure 4.13.** *DePFET matrix of 256 x 256 pixels of 75 x 75 μm² size. The complete matrix is read out by four ASTEROID ASICs operating in parallel. DePFET arrays with formats from 64 x 64 up to 512 x 512 have already been built.*

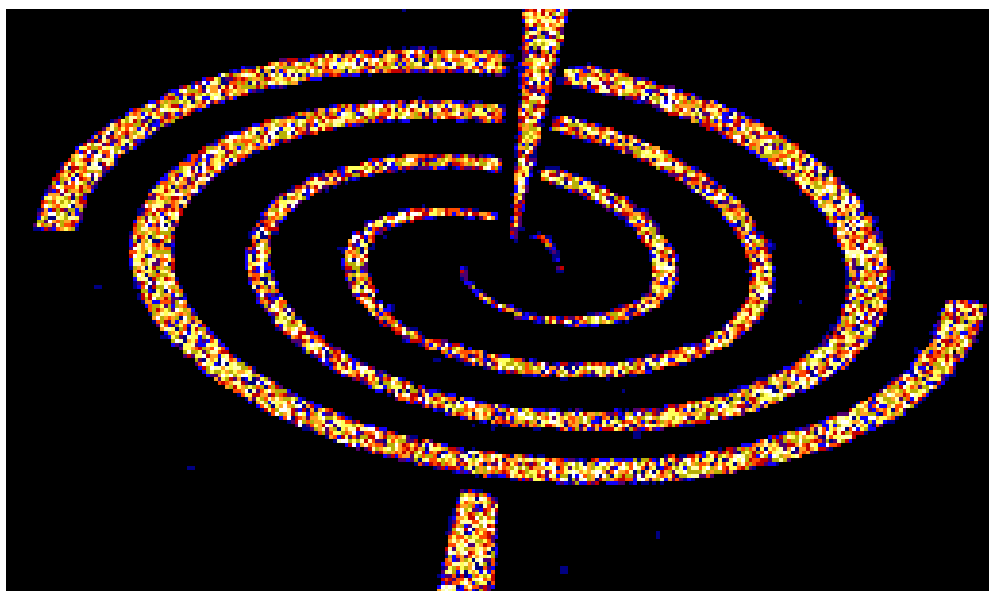

**Figure 4.14.** *Zoom into an X-ray image taken by the DePFET matrix shown in Figure 4.13, demonstrating the high spatial resolution of this prototype.*

The spectral performance was verified by irradiation with X-rays from $C_K$ (282 eV) up to $Mo_K$ (17.48 keV). An example is shown in Figure 4.15 where an $^{55}$Fe source was used with its Mn Kα (5.9 keV) and Mn Kβ (6.4 keV) lines. The detection efficiency at low energies is mainly determined by the quality of the radiation entrance window and the integrated optical/UV light blocking filter. The $C_K$ line can still be resolved with an energy resolution of 50 eV (FWHM), and is determined largely by readout noise performance.

**Figure 4.15.** *Backside-illuminated $^{55}$Fe X-ray spectrum taken with a DePFET matrix. The resolution of 125.4 eV is nearly Fano-limited, and the measurement shows a good Peak:Valley ratio of 2900. The spectral response to the X-ray line source is very clean. The Si escape peak at 4.15 keV can clearly by detected.*

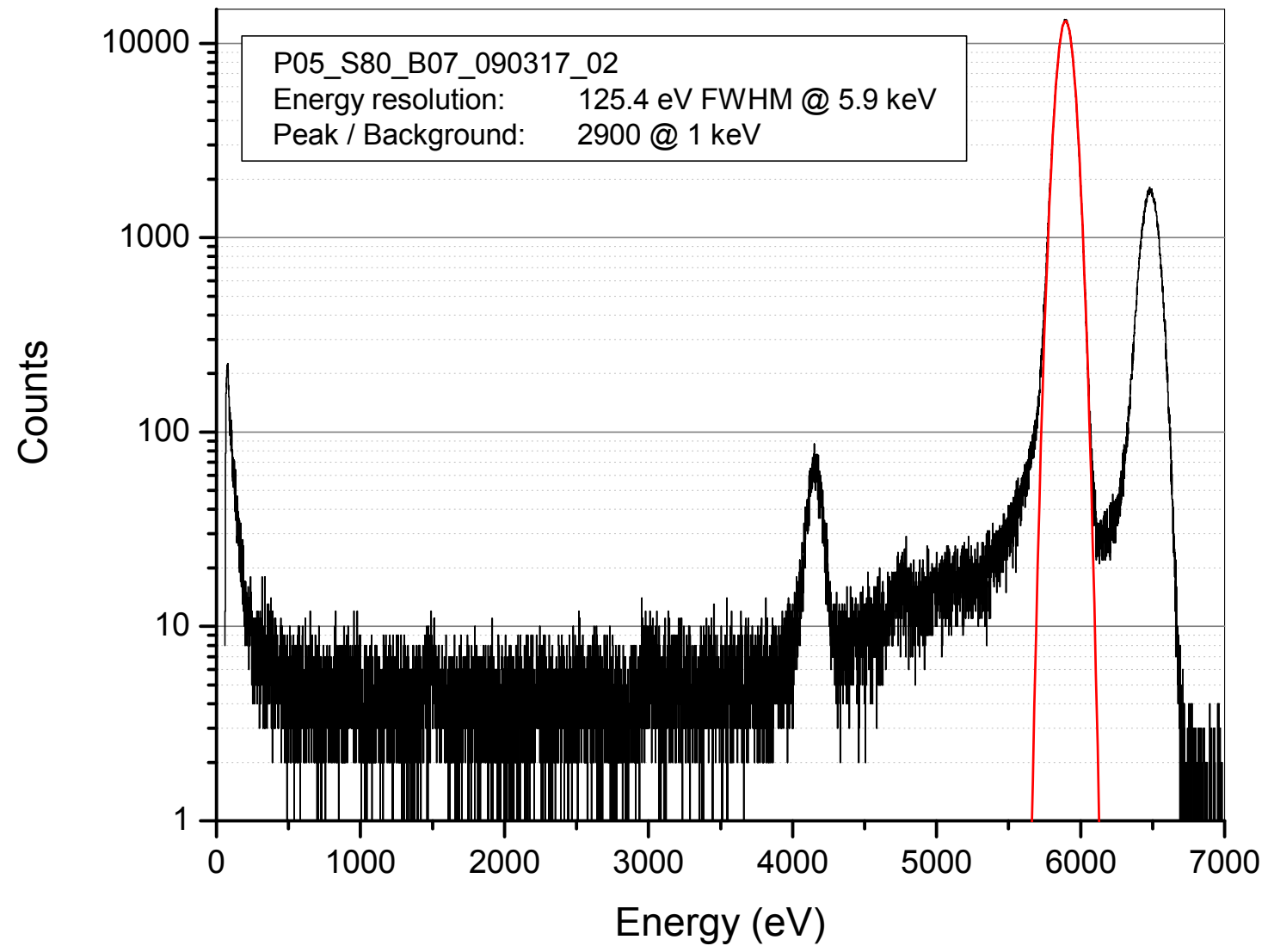

# Appendix A: The history of Athena

In this section we recall the context and previous history of the L-mission reformulation exercise that has led to the current definition of the *Athena* mission. A summary of the changes to the performance and science goals that resulted is also provided.

XEUS, a large X-ray observatory, was selected as one of three candidate missions in response to the first Cosmic Vision 2015-2025 proposal call. This ESA-JAXA proposal was merged with NASA's *Constellation-X* studies under a new mission name: The International X-Ray Observatory (IXO). The merger addressed the recommendation from the SPC to seek international participation to maintain the cost of the mission to ESA within the available funds for an L-class mission. IXO concluded its Assessment Phase successfully, with a high level of technical readiness and impressive scientific capabilities.

An X-ray observatory mission that can be implemented by Europe, with reduced participation from NASA and JAXA, has now been studied. *Athena* (Advanced Telescope for High Energy Astrophysics) embodies many of the key technologies of IXO, with heritage that has allowed a rapid rescoping to an affordable yet high performance mission that addresses most of the goals of the IXO science case (see Table A.1).

The recommendations of the ESA Internal Technical Review of IXO have been absorbed and applied to the new design. The science and performance requirements were reviewed considering the revised boundary conditions. Setting a cost limit to procurement and test of mirror sub-systems was an important feature of the re-design process, and various options were considered before choosing the configuration with broadest science impact and simplest technical implementation. As a consequence major payload simplifications have been introduced: the focal length is reduced to avoid the need of an extendable optical bench, and the number of optics modules reduced to manage the cost and schedule risk. Four instruments have been removed and the two remaining key camera/spectrometer instruments are fed by a fixed telescope each, to avoid a large and complex set of focal plane assembly mechanisms.

*Athena* is a highly capable observatory compared with any other current or planned X-ray astronomy missions. The science case (Section 2) and associated performance requirements have been put together, reviewed and endorsed by a large community of scientists, with a broad range of interests in current astrophysics.

The ESA mission assessment activities started with an internal trade-off and rescoping exercise in April 2011. The outcome of this work has been elaborated and verified first by ESA and then by two parallel industrial studies performed between June and September 2011, with the original competing consortia from the IXO Assessment phase (EADS Astrium and Thales Alenia Space). These industries have suggested a number of payload optimisations and fully endorsed the simplification approach applied to the *Athena* design. Revisions of the IXO payload studies were conducted by European teams that originally responded to the IXO "Declaration of Interest". Meanwhile most of the relevant Technology Development Activities for the optics and instrument components that were directly applicable to *Athena* have also continued. The outcome of all these technical design activities is summarised in Sections 4 and 5. All payload and spacecraft items could be provided with technical solutions from European sources. However the baseline contains modest contributions from NASA and JAXA, which improve instrument heritage (hence technology readiness) and performance. The final instrument complement will be decided through an Announcement of Opportunity.

The instrument studies have benefited greatly from developments for other precursor missions employing related technologies (*BepiColombo*, *Herschel*, *ASTRO-H*, *eROSITA*, etc). The Industrial Assessment Study has concluded that the concepts of *Athena* are rapidly maturing to the necessary TRL and a 2022 launch has been assessed as feasible by the industry studies.

The science case for *Athena* has been freshly developed, leading to the payload performance requirements, and a diverse and detailed observation programme covering a vast range of astrophysics. A reduction x2.5 in low energy area is the most obvious impact of the new design, but can be accommodated in most cases with longer exposure times. The *Athena* programme could be accomplished in 5 years (while preserving a ~25% general programme for topics not contemplated in the science case and not yet identified). Nonetheless, the overlap with the IXO science case is large, and Table A.1 summarises how particular IXO science goals have been impacted by the evolution to *Athena*. The science topics listed are those that were described in the IXO Yellow Book.

Due to the different *Athena* configuration, and specifically due to its larger field of view, new science investigations will be enabled with respect to IXO. In particular astrophysics of clusters of galaxies in the $z$~1-2 range are enabled by *Athena*.

ESA's flagship observatory *XMM-Newton* has placed Europe at the forefront of X-ray astronomy over the past decade. *Athena* is the only facility-class mission that can maintain this success into the 2020s.

| Science Topic | Status for Athena | Athena Prime Instrument | Comment |
|---|---|---|---|
| **Co-evolution of galaxies and their supermassive black holes** | | | |
| First supermassive black holes | Reduced | WFI | Breakthrough to highest redshift precluded with 10" resolution. Will be recovered to a large extent if 5" goal resolution is achieved. |
| Obscured growth | Retained | WFI | Wide field of view survey to moderate depths allows census of obscured AGN. |
| Cosmic feedback | Retained | XMS | Longer exposures required. |
| Supermassive black hole spin survey | Retained | WFI | Longer exposures required. |
| **Large-scale structure and creation of chemical elements** | | | |
| Missing baryons and warm-hot intergalactic medium | Modified | XMS | No grating spectra; achieve a WHIM census without details of velocity structure and evolution. Added capabilities via imaging. |
| Cluster physics and evolution | Retained | XMS | Longer exposures required. |
| Galaxy cluster cosmology | Retained | WFI | Longer exposures required, optimised using other X-ray surveys. |
| Chemical evolution | Retained | XMS | Longer exposures required. |
| **Matter under extreme conditions** | | | |
| Strong gravity | Retained | XMS & WFI | Longer exposures required, benefits from simultaneous XMS and WFI instrument operation. |
| Black hole spin | Reduced redundancy | WFI | Polarisation as additional measurement of spin is lost. |
| Neutron star equation of state | Modified | WFI | WFI time resolution vastly improved to enable sub-millisecond spectral-timing studies of bright X-ray sources. Reduction in sensitivity only due to reduction of effective area. QED probe lost due to lack of polarimeter. |
| Stellar mass black holes | Retained | WFI | |
| **Life cycles of matter and energy** | | | |
| SNR formation of elements | Retained | XMS | Longer exposures required. |
| Shocks and particle acceleration | Modified | WFI | Polarisation diagnostics of shocks lost. Hard X-ray probe of non-thermal processes limited. |
| Interstellar medium | Modified | XMS | Spectroscopic resolution degraded due to loss of grating. |
| Galactic centre | Retained | WFI | Some polarisation measurements lost, but different approach in place. |
| Stars and planets | Retained | XMS | Spectroscopic resolution degraded due to loss of grating, but enough for scientific purposes. |

**Table A.1.** *Summary of the impact of the payload simplifications and effective area reductions with respect to the IXO Yellow Book science case.*

# Acronym List

| | |
|---|---|
| 2MASS | 2 micron All Sky Survey |
| ADR | Adiabatic Demagnetization Refrigerator |
| AGB | Asymptotic Giant Branch |
| AGN | Active Galactic Nuclei |
| AIV/T | Assembly Integration Verification/ Testing |
| ALMA | Atacama Large Millimeter/ Submillimeter Array |
| AO | Announcement of Opportunity |
| AOCS | Attitude and Orbit Control Subsystem |
| AOT | Astronomical Observation Template |
| APS | Active Pixel Sensors |
| ASA | Athena Science Archive |
| ASDPC | Athena Science Data Processing Centre |
| ASGS | Athena Science Ground Segment |
| ASIC | Application-Specific Integrated Circuit |
| BBM | Bread-Board Model |
| BEE | Back End Electronics |
| BFL | Back Focal Length |
| BH | Black Hole |
| BHB | Black Hole Binary |
| Bw | Bandwidth |
| CaC | Cost at Completion |
| CAT | Critical Angle Transmission |
| CCD | Charge Coupled Device |
| CDF | Chandra Deep Field |
| CDR | Critical Design Review |
| CE | Control Electronics |
| CEA | Commissariat à l'énergie atomique et aux énergies alternatives |
| CFRP | Carbon Fibre Reinforced Plastic |
| CSL | Centre Spatial de Liège |
| CTA | Cherenkov Telescope Array |
| CTE | Coefficient of Thermal Expansion |
| CV | Cosmic Vision |
| CX | Charge Exchange |
| DE | Dark Energy |
| DEA | Detector Electronics Assembly |
| DePFET | Depleted P-channel Field-Effect Transistor |
| DES | Dark Energy Survey |
| DLV | Delivery |
| DMM | Design Maturity Margin |
| DP | Data Processing |
| DPA | Digital Processing Assembly |
| DTCP | Daily Telecommunications Period |
| DSN | Deep Space Network |
| D-SRE | Directorate of Science and Robotic Exploration |
| DSSD | Double-sided Si Strip Detector |
| E-ELT | European Extremely Large Telescope |
| EQM | Engineering Qualification Model |
| eROSITA | Extended Röntgen Survey Imaging Telescope Array |
| ESA | European Space Agency |
| ESAC | European Space Astronomy Centre |
| ESOC | European Space Operations Centre |
| ESTEC | European Space & Technology research Centre |
| FEE | Front End Electronics |
| FEM | Finite Element Model |
| FIR | Far Infrared |
| FM | Flight Model |
| FMA | Flight Mirror Assembly |
| FMS | Fixed Metering Structure |
| FoV | Field of View |
| FPA | Focal Plane Assembly |
| FPI | Focal Plane Instrument |
| FW | Filter Wheel |
| GAS | Grating Array Structure |
| GBH | Galactic Black Hole |
| GEM | Gas Electron Multiplier |
| GPD | Gas Pixel Detector |
| GR | General Relativity |
| GRB | Gamma Ray Burst |
| GSE | Ground Support Equipment |
| GTO | Guaranteed Time Observation |
| HDRM | Hold Down Release Mechanism |
| HEW | Half Energy Width |
| HGA | High Gain Antenna |
| HPD | Half Power Diameter |
| HST | Hubble Space Telescope |
| HTRS | High Time Resolution Spectrometer (IXO) |
| HXI | Hard X-ray Imager (IXO) |
| HW | Hardware |
| IA | Interactive analysis |
| ICM | Intracluster Medium |
| I/F | Interface |
| ILT | Instrument Level Test |
| IM | Instrument Module |
| IMBH | Intermediate Mass Black Hole |
| IFCA | Instituto de Física de Cantabria |
| IMF | Initial Mass Function |
| INAF | Istituto Nazionale di AstroFisica |
| IRAP | Institut de Recherche en Astrophysique et Planétologie |
| ISDC | *INTEGRAL* Astrophysics Data Centre |
| ITC | Instrument Team Centre |
| ITT | Invitation To Tender |
| IXO | International X-ray Observatory |
| JAXA | Japan Aerospace Exploration Agency |
| JDEM | Joint Dark Energy Mission |
| JT | Joule-Thomson cooler |
| JWST | James Webb Space Telescope |
| KO | Kick-Off |
| L2 | Sun-Earth Lagrangian Point 2 |
| LEOP | Launch and Early Orbit Phase |
| LMC | Large Magellanic Cloud |
| LISA | Laser Interferometer Space Antenna |
| LSST | Large Synoptic Survey Telescope |
| LV | Launcher Vehicle |
| M1,M2.... | Mirror 1 (primary mirror), Mirror 2 (secondary mirror), etc |
| MA | Mirror Assembly |
| MIP | Movable Instrument Platform |
| MIR | Medium Infrared |
| MIS | Metal-Insulator Sensor |
| MLA | Multi-Lateral Agreement |
| MLI | Multi-Layer Insulation |
| MM | Mirror Module |
| MOC | Mission Operations Centre |
| MOR | Mission Operations Readiness |
| MSSL | Mullard Space Science Laboratory |
| NASA | National Aeronautics and Space Agency |
| NIST | National Institute of Standartds and Technology |
| NLS1 | Narrow Line Seyfert 1 |
| NuSTAR | Nuclear Spectroscopic Telescope Array |
| OD | Operational Day |
| OGSE | Optical Ground Support Equipment |
| ORR | Operational Readiness Review |
| PanSTARRS | Panoramic Survey Telescope Rapid Response System |
| PCA | Proportional Counter Array (RXTE) |
| PCDU | Power Control and Distribution Unit |
| PDR | Preliminary Design Review |
| PFM | Proto-Flight Model |
| PI | Principal Investigator |
| POF | Planned Observation File |
| PS | Project Scientist |
| PSC | Partner Science Centre |
| PSF | Point Spread Function |
| PTB | Physikalisch-Technische Bundesanstalt |
| QCD | Quantum Chromodynamics |
| QM | Qualification Model |
| RAL | Rutherford Appleton Laboratory |
| RF | Radio Frequency |
| RIAF | Radiatively Inefficient Accretion Flow |
| ROSAT | Röntgen Satellite |
| RSS | Root Sum Square |
| RXTE | Rossi X-ray Timing Explorer |
| S/C | Spacecraft |
| SDD | Silicon Drift Detector |
| SEL-2 | Sun-Earth Lagrangian Point 2 (L2) |
| SFR | Star Forming Region |
| SGO | Segmented Glass Optics |
| SGS | Science Ground Segment |
| SKA | Square Kilometer Array |
| SLI | Single layer Insulation |
| SMBH | Super massive Black Hole |
| SMC | Small Magellanic Cloud |
| SNe | Supernovae |
| SN Ia | Supernova Type Ia |
| SNcc | Core Colapse Supernova |
| SNR | Supernova Remnant |
| SOC | Science Operations Centre |
| SPI | Star-Planet Interaction |
| SPICA | Space IR Telescope for Cosmology and Astrophysics |
| SPO | Silicon Pore Optics |
| SQUID | Superconducting Quantum Interference Device |
| SRR | System Requirements Review |
| S(T)M | Structural (Thermal) Model |
| SVM | Service Module |
| SVT | System Verification Test |
| SW | Software |
| SWCX | Solar Wind Charge Exchange |
| TAC | Time Allocation Committee |
| TBC | To Be Confirmed |
| TC | Telecommand |
| TDA | Technology Development Activity |
| TDM | Time Domain Multiplexing |
| ToO | Target of Opportunity |
| TRL | Technology Readiness Level |
| TRP | Technology Research Programme |
| TT&C | Tracking Telemetry and Command |
| WFI | Wide Field Imager (Athena) |
| WFIRST | Wide Field Infrared Survey Telescope |
| WHIM | Warm-Hot Intergalactic Medium |
| XEUS | X-ray Evolving Universe Explorer |
| XGS | X-ray Grating Spectrometer |
| XMS | X-ray Microcalorimeter Spectrometer (Athena) |
| XPOL | X-ray Polarimeter (IXO) |
| YSO | Young Stellar Object |